\documentclass[a4paper,11pt]{article}

\usepackage{amsmath,amsthm,amsbsy,amssymb,amsfonts,graphicx,mathrsfs,fancybox,bm,cite,xcolor,slashbox}
\usepackage{colortbl}
\usepackage{hyperref}
\usepackage[utf8]{inputenc}
\usepackage[all]{xy}
\usepackage{longtable}

\makeatletter
\catcode`\@=11
\@addtoreset{equation}{section}
\renewcommand{\theequation}{\thesection.\arabic{equation}}
\makeatother

\makeatletter
\renewcommand{\thetable}{\thesection.\arabic{table}}
\@addtoreset{table}{section}
\makeatother

\renewcommand{\thefootnote}{\arabic{footnote}}

\renewcommand{\baselinestretch}{1.35}

\newcommand{\Exp}[1]{\operatorname{e}^{#1}}
\newcommand{\diag}{\operatorname{diag}}
\newcommand{\abs}[1]{\lvert {#1} \rvert}
\newcommand{\ket}[1]{\lvert {#1} \rangle}
\newcommand{\rmd}{{\mathrm{d}}}
\newcommand{\nn}{\nonumber}
\newcommand{\Lie}{\pounds}
\newcommand{\lp}{l_p}
\newcommand{\gs}{g_s}
\newcommand{\ls}{l_s}
\newcommand{\ii}{i}

\newcommand{\cA}{\mathcal A}\newcommand{\cB}{\mathcal B}
\newcommand{\cC}{\mathcal C}\newcommand{\cD}{\mathcal D}
\newcommand{\cE}{\mathcal E}\newcommand{\cF}{\mathcal F}
\newcommand{\cG}{\mathcal G}\newcommand{\cH}{\mathcal H}
\newcommand{\cI}{\mathcal I}\newcommand{\cJ}{\mathcal J}
\newcommand{\cK}{\mathcal K}\newcommand{\cL}{\mathcal L}
\newcommand{\cM}{\mathcal M}\newcommand{\cN}{\mathcal N}
\newcommand{\cO}{\mathcal O}\newcommand{\cP}{\mathcal P}
\newcommand{\cQ}{\mathcal Q}\newcommand{\cR}{\mathcal R}
\newcommand{\cS}{\mathcal S}\newcommand{\cT}{\mathcal T}
\newcommand{\cU}{\mathcal U}\newcommand{\cV}{\mathcal V}
\newcommand{\cY}{\mathcal Y}\newcommand{\cZ}{\mathcal Z}

\newcommand{\sfa}{\mathsf{a}}
\newcommand{\sfd}{\mathsf{d}}
\newcommand{\sfg}{\mathsf{g}}
\newcommand{\sfi}{\mathsf{i}}
\newcommand{\sfj}{\mathsf{j}}
\newcommand{\sfm}{\mathsf{m}}
\newcommand{\sfn}{\mathsf{n}}
\newcommand{\sfp}{\mathsf{p}}
\newcommand{\sfq}{\mathsf{q}}
\newcommand{\sfr}{\mathsf{r}}
\newcommand{\sfs}{\mathsf{s}}
\newcommand{\sft}{\mathsf{t}}
\newcommand{\sfx}{\mathsf{x}}
\newcommand{\sfy}{\mathsf{y}}

\newcommand{\sfA}{\mathsf{A}}
\newcommand{\sfB}{\mathsf{B}}
\newcommand{\sfC}{\mathsf{C}}
\newcommand{\sfD}{\mathsf{D}}
\newcommand{\sfE}{\mathsf{E}}
\newcommand{\sfF}{\mathsf{F}}
\newcommand{\sfG}{\mathsf{G}}
\newcommand{\sfH}{\mathsf{H}}
\newcommand{\sfI}{\mathsf{I}}
\newcommand{\sfJ}{\mathsf{J}}
\newcommand{\sfK}{\mathsf{K}}
\newcommand{\sfL}{\mathsf{L}}
\newcommand{\sfM}{\mathsf{M}}
\newcommand{\sfR}{\mathsf{R}}
\newcommand{\sfY}{\mathsf{Y}}
\newcommand{\sfZ}{\mathsf{Z}}
\newcommand{\sfPhi}{\mathsf{\Phi}}

\makeatletter
\newcommand*{\rmT}{{\mathpalette\@transpose{}}}
\newcommand*{\@transpose}[2]{\raisebox{\depth}{$\m@th#1\intercal$}}
\makeatother

\newcommand{\FF}{\text{F}}
\newcommand{\GL}{\text{GL}}
\newcommand{\SL}{\text{SL}}
\newcommand{\SO}{\text{SO}}
\newcommand{\SU}{\text{SU}}
\newcommand{\UU}{\text{U}}
\newcommand{\KK}{\text{KK}}
\newcommand{\NS}{\text{NS}}
\newcommand{\OO}{\text{O}}
\newcommand{\PP}{\text{P}}
\newcommand{\WZ}{\text{WZ}}

\newcommand{\bdelta}{{\boldsymbol\delta}}

\newcommand{\NN}{\mathsf{n}}
\newcommand{\tr}{\text{tr}}
\renewcommand{\Im}{\text{Im}}
\newcommand{\rmM}{\text{\tiny M}}
\newcommand{\rmD}{\text{\tiny D}}
\newcommand{\rmS}{\text{\tiny S}}

\newcommand{\By}{\mathsf{y}}
\newcommand{\Ay}{y}
\newcommand{\Az}{z}

\newcommand{\GT}{\mathbb{T}}

\allowdisplaybreaks[3]

\renewcommand{\topfraction}{.85}
\renewcommand{\bottomfraction}{.60}
\renewcommand{\textfraction}{.15}
\renewcommand{\floatpagefraction}{.6}

\newcommand{\Pbox}[2]{\parbox{#1}{\vspace{.2\baselineskip}\raggedright #2 \vspace{.2\baselineskip}}}
\newcommand{\SUSY}[1]{\textcolor{red}{\underline{\textcolor{black}{#1}}}}

\usepackage{array}
\newcolumntype{L}[1]{>{\raggedright\let\newline\\\arraybackslash\hspace{0pt}}m{#1}}
\newcolumntype{R}[1]{>{\raggedleft\let\newline\\\arraybackslash\hspace{0pt}}m{#1}}
\newcolumntype{C}[1]{>{\centering}m{#1}}

\definecolor{Gray}{gray}{0.9}

\begin{document}

\begin{titlepage}
\renewcommand{\thefootnote}{\fnsymbol{footnote}}

\vspace*{1.0cm}

\begin{center}
\Large\textbf{Exotic branes and mixed-symmetry potentials I: \linebreak predictions from $E_{11}$ symmetry}
\end{center}

\vspace{1.0cm}

\centerline{
{\large Jos\'e J.~Fern\'andez-Melgarejo$^{a}$}%
\footnote{E-mail address: \texttt{jj.fernandezmelgarejo@um.es}},
{\large Yuho Sakatani$^{b}$}%
\footnote{E-mail address: \texttt{yuho@koto.kpu-m.ac.jp}}, 
{\large Shozo Uehara$^{b}$}%
\footnote{E-mail address: \texttt{uehara@koto.kpu-m.ac.jp}}
}

\vspace{0.2cm}

\begin{center}
${}^a${\it Departamento de F\'isica, Universidad de Murcia,}\\
{\it Campus de Espinardo, 30100 Murcia, Spain}

\vspace*{1mm}

${}^b${\it Department of Physics, Kyoto Prefectural University of Medicine,}\\
{\it Kyoto 606-0823, Japan}
\end{center}

\vspace*{1mm}

\begin{abstract}
Type II string theory or M-theory contains a broad spectrum of gauge potentials. 
In addition to the standard $p$-form potentials, various mixed-symmetry potentials have been predicted, which may couple to exotic branes with non-standard tensions.
Together with $p$-forms, mixed-symmetry potentials turn out to be essential to build the multiplets of the $U$-duality symmetry in each dimension.
In this paper, we systematically determine the set of mixed-symmetry potentials and exotic branes on the basis of the $E_{11}$ conjecture. 
We also study the decompositions of $U$-duality multiplets into $T$-duality multiplets and determine which mixed-symmetry tensors are contained in each of the $U$-/$T$-duality multiplets.
\end{abstract}

\thispagestyle{empty}
\end{titlepage}

\tableofcontents

\setcounter{footnote}{0}

\newpage

\section{Introduction and summary}

Maximal supergravities in 10 and 11 dimensions have the $E_{n}$ $U$-duality symmetry when they are toroidally compactified to $d\,(=11-n)$ dimensions. 
The $E_{n}$ symmetry can be clearly seen by packaging all of the $d$-dimensional scalar fields into the generalized metric $\cM_{IJ}$, which parameterizes the coset space $E_{n}/K_n$ ($K_n$\,: the maximal compact subgroup of $E_{n}$). 
A coset representative $g$ can be parameterized by using the positive Borel subalgebra of the $E_{n}$ algebra. 
For example, in $d=4$ (or $n=7$), the Borel subalgebra is spanned by
\begin{align}
 \bigl\{K^i{}_j\ (i<j),\, R^{i_1i_2i_3},\,R^{i_1\cdots i_6} \bigr\} \qquad (i=1,\dotsc,n)\,,
\end{align}
where multiple indices are antisymmetric. 
Indeed, using the scalar fields $\{\hat{G}_{ij},\,\hat{A}_{i_1i_2i_3},\,\hat{A}_{i_1\cdots i_6}\}$ resulting from the compactification, we can parameterize the coset representative as
\begin{align}
 g = \Exp{\sum_{i<j} \hat{h}_i{}^j\,K^i{}_j} \Exp{\frac{1}{3!}\,\hat{A}_{i_1i_2i_3}\,R^{i_1i_2i_3}} \Exp{\frac{1}{6!}\,\hat{A}_{i_1\cdots i_6}\,R^{i_1\cdots i_6}}.
\end{align}
Here, $\hat{h}_i{}^j$ is the logarithm of the vielbein associated with $\hat{G}_{ij}$ and $\hat{A}_{i_1i_2i_3}$ and $\hat{A}_{i_1\cdots i_6}$ are 3-form and 6-form potentials in 11D supergravity. 
Then, we can obtain the generalized metric $\cM_{IJ}$ from a matrix representation of $g$\,. 
However, when we consider the case $d=3$\,, we find that the standard scalar fields are not enough to parameterize $E_{8}/K_8$\,. 
As pointed out in \cite{hep-th:0104081}, an additional generator $R^{i_1\cdots i_8,i}$ (satisfying $R^{[i_1\cdots i_8,i]}=0$) appears in the Borel subalgebra. 
Then, we need to introduce a non-standard field, namely a mixed-symmetry potential $\hat{A}_{i_1\cdots i_8,i}$ (or simply $\hat{A}_{8,1}$), which is called the dual graviton. 

Mixed-symmetry potentials are indispensable for realizing the duality symmetry manifest. 
In addition to the scalars $\cM_{IJ}$, $p$-form fields are also forming some $U$-duality multiplets. 
For example, in 11D supergravity compactified on a $7$-torus, the graviphoton, 3- and 6-form potentials, and the dual graviton give rise to the 1-form fields $\{\hat{A}_\mu^i,\,\hat{A}_{\mu i_1i_2},\,\hat{A}_{\mu i_1\cdots i_5},\,\hat{A}_{\mu i_1\cdots i_7,j}\}$, respectively ($\mu=0,\dotsc,d-1$). 
In fact, they form the 56-dimensional representation of $E_7$. 
In order to construct the $U$-duality multiplets for higher $p$-form fields, we need to introduce more non-standard potentials. 
In general, the new potentials can be decomposed into some irreducible representations of $\SL(n)$ characterized by Young tableaux. 
Each of the potentials is denoted as $\hat{A}_{p,q,r,\cdots}$ when the indices have the following Young-tableau symmetry:
\begin{align}
 \hat{A}_{i_1\cdots i_p,j_1\cdots j_q,k_1\cdots k_r,\cdots} \quad \leftrightarrow\quad
 \vcenter{\hbox{
\begin{picture}(55,50)
\put(0,50){\line(1,0){30}}
\put(20,30){\line(1,0){10}}
\put(10,15){\line(1,0){10}}
\put(0,0){\line(1,0){10}}
\put(0,0){\line(0,1){50}}
\put(10,0){\line(0,1){50}}
\put(20,15){\line(0,1){35}}
\put(30,30){\line(0,1){20}}
\put(2,20){$p$}
\put(12,29){$q$}
\put(22,38){$r$}
\put(33,38){$\cdots$}
 \end{picture}}}.
\label{eq:tableau-symmetry}
\end{align}
These objects are generally called mixed-symmetry potentials. 
Of course, mixed-symmetry potentials are not new independent fields. 
They are related to the standard supergravity fields through the electric--magnetic duality, much like the 3-form potential $\hat{A}_3$ is related to the 6-form potential $\hat{A}_6$\,. 
For example, the dual graviton $\hat{A}_{8,1}$ is dual to the graviphoton. 

One of the main reasons for introducing the mixed-symmetry potentials is the manifest $E_n$ $U$-duality symmetry. 
Existence of non-standard potentials has been expected in the study of $U$-duality multiplets of branes \cite{hep-th:9707217,hep-th:9712047,hep-th:9712075,hep-th:9712084,hep-th:9809039,hep-th:9908094,hep-th:0012051}. 
The standard $p$-form gauge potentials are known to couple to supersymmetric branes (e.g.~momenta P, M2, and M5-branes) and the associated central charges appear in the supersymmetric algebra in 11D,
\begin{align}
 \{Q_{\hat{\tt{a}}\vphantom{\hat{\tt{b}}}} , \,Q_{\hat{\tt{b}}} \} = (C\Gamma^{\hat{M}})_{\hat{\tt{a}}\hat{\tt{b}}}\,P_{\hat{M}} + \frac{1}{2!}\,(C\Gamma_{\hat{M}_1\hat{M}_2})_{\hat{\tt{a}}\hat{\tt{b}}}\,Z^{\hat{M}_1\hat{M}_2} + \frac{1}{5!}\,(C\Gamma_{\hat{M}_1\cdots\hat{M}_5})_{\hat{\tt{a}}\hat{\tt{b}}}\,Z^{\hat{M}_1\cdots\hat{M}_5}\,,
\end{align}
where $\hat{M},\hat{N}=1,\dotsc,11$ and $\hat{\tt{a}}, \hat{\tt{b}}=1,\dotsc,32$ are spinor indices. 
When M-theory or type II theory is compactified on a torus, the so-called exotic branes, which have non-standard tensions, appear in the $U$-duality multiplets. 
Each of the branes was found to correspond to a certain weight of the $E_n$ algebra, and a formula for the brane tension in terms of the weight was obtained in \cite{hep-th:9707217,hep-th:9712047,hep-th:9712075,hep-th:9712084,hep-th:9809039,hep-th:9908094,hep-th:0012051}. 
More recently, various aspects of exotic branes have been studied in \cite{0805.4451,1004.2521,1209.6056,1109.2025,1102.0934,1106.0212,1108.5067,1201.5819,1109.4484,1205.5549,1303.0221,1303.1413,1304.4061,1305.4439,1309.2653,1310.6163,1404.5442,1402.5972,1411.1043,1412.0635,1412.8769,1601.02175,1601.05589,1607.05450,1612.08738,1710.09740,1712.01739,1803.11087,1805.12117,1806.00430,1810.02169,1903.05601,1904.05365} (see \cite{1210.1422,1311.3578,1402.2557,1501.06895,1608.01436,1610.07975,1704.08566,1708.08066,1710.00642,1803.07023,1907.04040} for exotic branes in non-maximal theories), and their higher-dimensional origin has been discussed. 
By following the notation of \cite{hep-th:9809039,1004.2521,1209.6056}, if a $p$-brane in $d$-dimensions has tension
\begin{align}
\begin{split}
 T_p &= \frac{1}{\lp\,(2\pi\lp)^p} \Bigl(\frac{R_{i_1}\cdots R_{i_{b-p}}}{\lp^{b-p}}\Bigr) \Bigl(\frac{R_{j_1}\cdots R_{j_{c_2}}}{\lp^{c_2}}\Bigr)^2\cdots \Bigl(\frac{R_{k_1}\cdots R_{k_{c_s}}}{\lp^{c_s}}\Bigr)^s\,,
\end{split}
\label{eq:tension-M}
\end{align}
(where $l_p$ is the Planck length, and $R_i$ is the radius along the $x^i$-direction) the object in 11D is denoted as an (exotic) $b^{(c_s,\dotsc,c_2)}$-brane. 
Exotic branes generally require the existence of Killing vectors, but by using the Killing vectors, we can write down the worldvolume actions for the exotic branes propagating in 11D (see \cite{hep-th:9812188} and \cite{1601.05589} for actions of the Kaluza--Klein monopole (KKM) and the $5^3$-brane in M-theory). 
According to the worldvolume actions, exotic branes may couple to some mixed-symmetry potentials in 11D: $\hat{A}_{8,1}$ for the $6^1$-brane (or KKM) and $\hat{A}_{9,3}$ for the $5^3$-brane. 
Moreover, their central charges are also expected to have the tensor structures of the mixed-symmetry type: $Z^{7,1}$ for the $6^1$-brane and $Z^{8,3}$ for the $5^3$-brane. 
For a general exotic $b^{(c_s,\dotsc,c_2)}$-brane, the following correspondence has been suggested (see Table \ref{tab:M-branes}):
\begin{table}[b]
 \centerline{
 \begin{tabular}{|c||c|c|c|c|c||c|}\hline
 branes & P & M2 & M5 & KKM ($=6^1$) & $5^3$ & $b^{(c_s,\dotsc,c_2)}$
\\\hline\hline
 central chares & $P_1$ & $Z^2$ & $Z^5$ & $Z^{7,1}$ & $Z^{8,3}$ & $Z^{b+c_2+\cdots+c_s,c_2+\cdots+c_s,\cdots,c_{s-1}+c_s,c_s}$
\\\hline
 potential & $\hat{A}_1^1$ & $\hat{A}_3$ & $\hat{A}_6$ & $\hat{A}_{8,1}$ & $\hat{A}_{9,3}$ & $\hat{A}_{1+b+c_2+\cdots+c_s,c_2+\cdots+c_s,\cdots,c_{s-1}+c_s,c_s}$ \\\hline
\end{tabular}}
\caption{M-theory branes, their central charges, and the potentials that electrically couple to the branes. 
The graviphoton associated with a Killing direction is denoted as $\hat{A}_1^1$.}
\label{tab:M-branes}
\end{table}
\begin{align}
\begin{split}
 b^{(c_s,\dotsc,c_2)} \quad &\leftrightarrow \quad Z^{b+c_2+\cdots+c_s,c_2+\cdots+c_s,\cdots,c_{s-1}+c_s,c_s}
\\
 &\leftrightarrow \quad \hat{A}_{1+b+c_2+\cdots+c_s,c_2+\cdots+c_s,\cdots,c_{s-1}+c_s,c_s}\,. 
\end{split}
\label{eq:brane-pot-M}
\end{align}

In type II theories, the notation slightly differs.
When a $d$-dimensional $p$-brane has tension
\begin{align}
\begin{split}
 T_p &= \frac{\gs^{-\NN}}{\ls\,(2\pi l_s)^p} \Bigl(\frac{R_{m_1}\cdots R_{m_{b-p}}}{\ls^{b-p}}\Bigr) \Bigl(\frac{R_{n_1}\cdots R_{n_{c_2}}}{\ls^{c_2}}\Bigr)^2\cdots \Bigl(\frac{R_{q_1}\cdots R_{q_{c_s}}}{\ls^{c_s}}\Bigr)^s \,,
\end{split}
\label{eq:tension-typeII}
\end{align}
(where $l_s$ is the string length, and $g_s$ is the string coupling constant) the object in 10D is called an (exotic) $b^{(c_s,\dotsc,c_2)}_{\NN}$-brane. 
In particular, worldvolume actions for the exotic $5^2_2$-brane and $5^2_3$-brane, as studied in \cite{1309.2653,1404.5442}, will couple to potentials with the tensor structure $A_{8,2}$\,. 
The correspondence is the same as in the case of M-theory:
\begin{align}
\begin{split}
 b^{(c_s,\dotsc,c_2)}_{\NN} \quad &\leftrightarrow \quad Z^{b+c_2+\cdots+c_s,c_2+\cdots+c_s,\cdots,c_{s-1}+c_s,c_s}
\\
 &\leftrightarrow \quad A_{1+b+c_2+\cdots+c_s,c_2+\cdots+c_s,\cdots,c_{s-1}+c_s,c_s}\,. 
\end{split}
\label{eq:brane-pot}
\end{align}
By following the notation of \cite{1102.0934,1106.0212,1108.5067,1201.5819}, we sometimes denote the potential $A$ as $B,\,C,\,D,\,\cdots$ depending on the integer $\NN$ appended to the name of exotic branes:
\begin{align}
\begin{tabular}{c||c|c|c|c|c|c|c|c|c|c|c|c|c}
 $\NN$&$0$&$1$&$2$&$3$&$4$&$5$&$6$&$7$&$8$&$9$&$10$&$11$&$\cdots$\\\hline
 $A$&$B$&$C$&$D$&$E$&$F$&$G$&$H$&$I$&$J$&$K$& $L$& $M$&$\cdots$
\end{tabular}\,\,.
\label{eq:level-n}
\end{align}
Then, for example, the $5^2_2$-brane and the $5^2_3$-brane couple to $D_{8,2}$ and $E_{8,3}$\,, respectively. 

Given the existence of a variety of mixed-symmetry potentials/exotic branes in M-theory or type II theories, a natural question is what kind of potentials/branes exist in 11D or 10D. 
A partial answer has been given in Refs.~\cite{hep-th:0104081,hep-th:0307098} based on the $E_{11}$ conjecture. 
The $E_{11}$ conjecture claims that M-theory/type II string theory has the $E_{11}$ symmetry even before compactification, and the standard $E_n$ ($n\leq 8$) $U$-duality groups are realized as the subgroup after the compactification. 
As already mentioned, a gauge potential $\hat{A}_{p,q,r,\cdots}$ corresponds to the $E_{n}$ generator $R^{p,q,r,\cdots}$, and since $E_{11}$ is infinite-dimensional, infinitely many of potentials are predicted by the $E_{11}$ conjecture. 
For example, the predicted fields include \cite{0705.0752}
\begin{align}
\Pbox{0.9\textwidth}{
$\hat{e}_1^1$,\quad 
$\hat{A}_{3}$,\quad 
$\hat{A}_{6}$,\quad 
$\hat{A}_{8,1}$,\quad 
$\hat{A}_{9,3}$,\quad 
$\hat{A}_{9,6}$,\quad 
$\hat{A}_{9,8,1}$,\quad 
$\hat{A}_{10,1,1}$,\quad 
$\hat{A}_{10,4,1}$,\quad 
$\hat{A}_{10,6,2}$,\quad 
$\hat{A}_{10,7,1}$,\quad 
$\hat{A}_{10,7,4}$,\quad 
$\hat{A}_{10,7,7}$,\quad 
$\hat{A}_{10,8}$,\quad 
$\hat{A}_{10,8,2,1}$,\quad 
$\hat{A}_{10,8,3}$,\quad 
$\hat{A}_{10,8,5,1}$,\quad 
$\hat{A}_{10,8,6}$,\quad 
$\hat{A}_{10,8,7,2}$,\quad 
$\hat{A}_{10,8,8,1}$,\quad 
$\hat{A}_{10,8,8,4}$,\quad 
$\hat{A}_{10,8,8,7}$,\quad 
$\hat{A}_{11,1}$,\quad 
$\hat{A}_{11,3,1}$,\quad 
$\hat{A}_{11,4}$,\quad 
$\hat{A}_{11,4,3}$,\quad 
$\hat{A}_{11,5,1,1}$,\quad 
$2\,\hat{A}_{11,6,1}$,\quad 
$\hat{A}_{11,6,3,1}$,\quad 
$\hat{A}_{11,6,4}$,\quad 
$\hat{A}_{11,6,6,1}$,\quad 
$\hat{A}_{11,7}$,\quad 
$\hat{A}_{11,7,2,1}$,\quad 
$2\,\hat{A}_{11,7,3}$,\quad 
$\hat{A}_{11,7,4,2}$,\quad 
$\hat{A}_{11,7,5,1}$,\quad 
$2\,\hat{A}_{11,7,6}$,\quad 
$\hat{A}_{11,7,6,3}$,\quad 
$\hat{A}_{11,7,7,2}$,\quad 
$\hat{A}_{11,7,7,5}$,}
\label{eq:M-all-potentials-RW}
\end{align}
where $\hat{e}_1^1$ denotes the vielbein and integers in front of the potentials represent the multiplicities of potentials with the same tensor structure. 
$\hat A_3$ and $\hat A_6$ are the standard ones and the others are exotic mixed-symmetry potentials. 
After a compactification to $d$ dimensions ($d\geq 3$), the number of $p$-form fields is finite. 
As detailed in \cite{0705.0752}, the 11D potentials appearing in \eqref{eq:M-all-potentials-RW} give rise to all the $d$-dimensional $p$-forms ($d\geq 3$), except for the case $d=p=3$. 
The counting of the $d$-dimensional $p$-forms is summarized in Table \ref{tab:p-form}. 
For the case $d=p=3$, the representation was determined in \cite{0705.1304} using the computer program SimpLie \cite{SimpLie}. 

As discussed in \cite{0705.0752}, Table \ref{tab:p-form} is consistent with the results of gauged supergravities in various dimensions.\footnote{See \cite{1609.09745} for a recent review of gauged supergravity.} 
Indeed, in the language of the embedding tensor formalism \cite{hep-th:0010076,hep-th:0212239}, the $(d-2)$-form and the $(d-1)$-form potentials are dual to, respectively, the scalar fields and the embedding tensors (or deformation parameters), while the $d$-forms correspond to the quadratic constraints (which ensure the consistent deformations of the theories), and their numbers are the same as those given in Table \ref{tab:p-form}.\footnote{This agreement occurs upon discarding the gaugings of the trombone symmetry, which would imply additional deformation parameters and quadratic constraints. We also note that the agreement between the $d$-forms and the quadratic constraints is not perfect: ${\bf 248}$ in $d=p=3$ has not been contained in the representation of the quadratic constraint in $d=3$\,.}
\begin{table}
 \centerline{\scalebox{0.85}{
 \begin{tabular}{|c|| C{0.9cm} | C{1.0cm} | C{1.5cm} |C{1.0cm}|C{1.0cm}|C{0.9cm}|C{0.9cm}|C{1.6cm}|C{0.9cm}||c|}\hline
 \backslashbox{$d$}{$p$} & $1$ & $2$ & $3$ & $4$ & $5$ & $6$ & $7$ & $8$ & $9$ & $U$-duality \\\hline\hline
 9 & $\textcolor{red}{{\bf 2}}$\\ $\textcolor{red}{{\bf 1}}$ & $\textcolor{red}{{\bf 2}}$ & $\textcolor{red}{{\bf 1}}$ & $\textcolor{red}{{\bf 1}}$ & $\textcolor{red}{{\bf 2}}$ & $\textcolor{red}{{\bf 2}}$ \\ $\textcolor{red}{{\bf 1}}$ & $\textcolor{red}{{\bf 3}}$ \\ ${\bf 1}$ & $\textcolor{red}{{\bf 3}}$ \\ ${\bf 2}$ & $\textcolor{red}{{\bf 4}}$ \\ $2\times {\bf 2}$ & $\SL(2)$ \\\hline
 8 & $\textcolor{red}{({\bf \overline{3}},{\bf 2})}$ & $\textcolor{red}{({\bf 3},{\bf 1})}$ & $\textcolor{red}{({\bf 1},{\bf 2})}$ & $\textcolor{red}{({\bf \overline{3}},{\bf 1})}$ & $\textcolor{red}{({\bf 3},{\bf 2})}$ & $\textcolor{red}{({\bf 8},{\bf 1})}$ \\ $\textcolor{red}{({\bf 1},{\bf 3})}$ & $\textcolor{red}{({\bf 6},{\bf 2})}$ \\ $({\bf \overline{3}},{\bf 2})$ & $\textcolor{red}{({\bf 15},{\bf 1})}$ \\ $({\bf 3},{\bf 3})$ \\ $2\times({\bf 3},{\bf 1})$ & & $\SL(3)\times\SL(2)$ \\\hline
 7 & $\textcolor{red}{{\bf \overline{10}}}$ & $\textcolor{red}{{\bf 5}}$ & $\textcolor{red}{{\bf \overline{5}}}$ & $\textcolor{red}{{\bf 10}}$ & $\textcolor{red}{{\bf 24}}$ & $\textcolor{red}{{\bf \overline{40}}}$ \\ $\textcolor{red}{{\bf \overline{15}}}$ & $\textcolor{red}{{\bf 70}}$ \\ ${\bf 45}$ \\ ${\bf 5}$ & & & $\SL(5)$ \\\hline
 6 & $\textcolor{red}{{\bf 16}}$ & $\textcolor{red}{{\bf 10}}$ & $\textcolor{red}{{\bf \overline{16}}}$ & $\textcolor{red}{{\bf 45}}$ & $\textcolor{red}{{\bf 144}}$ & $\textcolor{red}{{\bf 320}}$ \\ $\textcolor{red}{{\bf 126}}$ \\ ${\bf 10}$ & & & & $\SO(5,5)$ \\\hline
 5 & $\textcolor{red}{{\bf 27}}$ & $\textcolor{red}{{\bf \overline{27}}}$ & $\textcolor{red}{{\bf 78}}$ & $\textcolor{red}{{\bf 351}}$ & $\textcolor{red}{{\bf \overline{1728}}}$ \\ ${\bf \overline{27}}$ & & & & & $E_6$ \\\hline
 4 & $\textcolor{red}{{\bf 56}}$ & $\textcolor{red}{{\bf 133}}$ & $\textcolor{red}{{\bf 912}}$ & $\textcolor{red}{{\bf 8645}}$ \\ ${\bf 133}$ & & & & & & $E_7$ \\\hline
 3 & $\textcolor{red}{{\bf 248}}$ & $\textcolor{red}{{\bf 3875}}$ \\ ${\bf 1}$ & $\textcolor{red}{{\bf 147250}}$ \\ ${\bf 3875}$ \\ ${\bf 248}$ & & & & & & & $E_8$ \\\hline
\end{tabular}
}}
\caption{Number of $p$-form fields in $d$-dimensional maximal supergravities ($3\leq d\leq 9$) predicted by the $E_{11}$ conjecture \cite{0705.0752,0705.1304}. $U$-duality representations which contain a potential that couples to a supersymmetric brane are marked in red.}
\label{tab:p-form}
\end{table}
In this sense, the $E_{11}$ conjecture is consistent with the requirements of supersymmetry.\footnote{See \cite{1110.4892,1503.00015} for relevant studies based on the Borcherds superalgebras.}

In addition to the mixed-symmetry potentials, the $E_{11}$ conjecture also predicts a variety of brane charges. 
In \cite{hep-th:0307098,hep-th:0406150}, central charges for all of the branes were identified with weight vectors in the vector representation of $E_{11}$\,. 
Despite this representation also being infinite-dimensional, by introducing a parameter called level $\ell$ \cite{hep-th:0212291} the charges can be ordered. 
The low-level charges become
\begin{align}
\begin{split}
 &P_1\ (\ell=0);\quad Z^{2}\ (\ell=1);\quad Z^{5}\ (\ell=2);\quad Z^{7,1},\quad Z^8\ (\ell=3); 
\\
 &Z^{8,3},\quad 
 Z^{9,1,1},\quad 
 Z^{9,2},\quad 
 2\,Z^{10,1},\quad 
 Z^{11}\ (\ell=4);\dotsc \,.
\end{split}
\label{eq:low-level-Z}
\end{align}
The first several charges, $P_1$, $Z^2$, and $Z^5$, are associated with the standard central charges for momenta, M2-brane, and M5-brane, respectively, and the higher-level charges will be associated with central charges for exotic branes. 
In this way, the $E_{11}$ conjecture is consistent with the known results and also exhibits a strong predictive power on the non-standard mixed-symmetry potentials and brane central charges. 

\subsection{Motivation and results}

Recently, duality-manifest formulations of supergravity and brane actions have been intensively studied \cite{hep-th:0104081,hep-th:0307098,hep-th:0412336,0902.1509,1008.1763,1009.4657,1111.0459,1206.7045,1208.5884,1308.1673}. 
In those cases, mixed-symmetry potentials/exotic branes become necessary. 
Formal aspects of the $U$-duality-covariant expressions have been studied well, but in order to make the physical picture more transparent, it is useful to get back to the standard description in which the manifest duality symmetry is broken. 
For example, the $U$-duality-covariant description for the Wess--Zumino term has been studied in \cite{1009.4657}. 
The brane action is described by the $U$-duality-covariant $p$-form fields $\cA_p^{I_p}$\,, where $I_p$ is transforming in the $p$-form multiplets described in Table \ref{tab:p-form}. 
However, in order to reproduce the standard brane actions, we need to decompose the covariant objects $\cA_p^{I_p}$ into the standard potentials and the mixed-symmetry potentials. 
While we will postpone the parameterizations to a subsequent paper \cite{1909.01335}, here we instead concentrate on more algebraic aspects based on the $E_{11}$ conjecture: \emph{what kind of mixed-symmetry potential exist in M-theory/type II theory, and which potential is a member of each $U$- or $T$-duality representation?} 

Using the high predictability of the $E_{11}$ conjecture, we can find new mixed-symmetry potentials and exotic branes as long as we carry out the corresponding computation. 
Because there are so many previous studies, it is convenient to clarify both the current status of this topic and the new results appearing in this paper. 

\paragraph*{M-theory potentials:}

Regarding the mixed-symmetry potentials in 11D, potentials up to level $\ell=10$ (and some of $\ell=11$) are determined in \cite{hep-th:0104081,hep-th:0212291,hep-th:0307098,hep-th:0301017,0705.0752}. 
According to the recent developments in duality-manifest supergravities or brane actions, it is useful to know the constituents for all of the $p$-forms $\cA_p^{I_p}$ given in Table \ref{tab:p-form}. 
In order to know all of the potentials contained in $\cA_p^{I_p}$\,, in fact, we need to determine the $E_{11}$ generators up to level $\ell=17$\,. 
However, through a level decomposition using the program SimpLie, the maximal we can reach with current personal computers is $\ell=10$. 
Here, a slightly indirect approach is taken to determine the constituents of $\cA_p^{I_p}$\,, which is explained in Appendix \ref{app:M-potential}.
In this work, we determine all of the mixed-symmetry potentials that contribute to the counting of the $p$-forms listed in Table \ref{tab:p-form}. 
The result is given in Table \ref{tab:MB-potentials}, and we show that they indeed reproduce the counting of Table \ref{tab:p-form}. 
This list extends the previous results, given in \eqref{eq:M-all-potentials-RW}.
\begin{table}[t]
\centerline{
\begin{tabular}{|c|}\hline
M-theory potentials
\\\hline
\Pbox{\textwidth}{{\scriptsize $\SUSY{\hat{e}_1^1}$,\ \,%
$\SUSY{\hat{A}_{3}}$,\ \,%
$\SUSY{\hat{A}_{6}}$,\ \,%
$\SUSY{\hat{A}_{8,1}}$,\ \,%
$\SUSY{\hat{A}_{9,3}}$,\ \,%
$\SUSY{\hat{A}_{10,1,1}}$,\ \,%
$\hat{A}_{11,1}$,\ \,%
$\SUSY{\hat{A}_{9,6}}$,\ \,%
$\SUSY{\hat{A}_{10,4,1}}$,\ \,%
$\hat{A}_{11,3,1}$,\ \,%
$\hat{A}_{11,4}$,\ \,%
$\SUSY{\hat{A}_{9,8,1}}$,\ \,%
$\SUSY{\hat{A}_{10,6,2}}$,\ \,%
$\hat{A}_{10,7,1}$,\ \,%
$\hat{A}_{10,8}$,\ \,%
$\SUSY{\hat{A}_{11,4,3}}$,\ \,%
$\SUSY{\hat{A}_{11,5,1,1}}$,\ \,%
$2\,\hat{A}_{11,6,1}$,\ \,%
$\hat{A}_{11,7}$,\ \,%
$\SUSY{\hat{A}_{10,7,4}}$,\ \,%
$\SUSY{\hat{A}_{10,8,2,1}}$,\ \,%
$\hat{A}_{10,8,3}$,\ \,%
$\SUSY{\hat{A}_{11,6,3,1}}$,\ \,%
$\hat{A}_{11,6,4}$,\ \,%
$\hat{A}_{11,7,2,1}$,\ \,%
$2\,\hat{A}_{11,7,3}$,\ \,%
$3\,\hat{A}_{11,8,1,1}$,\ \,%
$3\,\hat{A}_{11,8,2}$,\ \,%
$\SUSY{\hat{A}_{10,7,7}}$,\ \,%
$\SUSY{\hat{A}_{10,8,5,1}}$,\ \,%
$\hat{A}_{10,8,6}$,\ \,%
$\SUSY{\hat{A}_{11,6,6,1}}$,\ \,%
$\SUSY{\hat{A}_{11,7,4,2}}$,\ \,%
$\hat{A}_{11,7,5,1}$,\ \,%
$2\,\hat{A}_{11,7,6}$,\ \,%
$\SUSY{\hat{A}_{11,8,3,1,1}}$,\ \,%
$\hat{A}_{11,8,3,2}$,\ \,%
$4\,\hat{A}_{11,8,4,1}$,\ \,%
$3\,\hat{A}_{11,8,5}$,\ \,%
$\SUSY{\hat{A}_{10,8,7,2}}$,\ \,%
$\hat{A}_{10,8,8,1}$,\ \,%
$\SUSY{\hat{A}_{11,7,6,3}}$,\ \,%
$\hat{A}_{11,7,7,2}$,\ \,%
$\SUSY{\hat{A}_{11,8,4,4}}$,\ \,%
$\SUSY{\hat{A}_{11,8,5,2,1}}$,\ \,%
$\hat{A}_{11,8,5,3}$,\ \,%
$2\,\hat{A}_{11,8,6,1,1}$,\ \,%
$4\,\hat{A}_{11,8,6,2}$,\ \,%
$6\,\hat{A}_{11,8,7,1}$,\ \,%
$2\,\hat{A}_{11,8,8}$,\ \,%
$\SUSY{\hat{A}_{10,8,8,4}}$,\ \,%
$\SUSY{\hat{A}_{11,7,7,5}}$,\ \,%
$\SUSY{\hat{A}_{11,8,6,4,1}}$,\ \,%
$\hat{A}_{11,8,6,5}$,\ \,%
$\SUSY{\hat{A}_{11,8,7,2,2}}$,\ \,%
$2\,\hat{A}_{11,8,7,3,1}$,\ \,%
$4\,\hat{A}_{11,8,7,4}$,\ \,%
$\SUSY{\hat{A}_{11,8,8,1,1,1}}$,\ \,%
$4\,\hat{A}_{11,8,8,2,1}$,\ \,%
$4\,\hat{A}_{11,8,8,3}$,\ \,%
$\SUSY{\hat{A}_{10,8,8,7}}$,\ \,%
$\SUSY{\hat{A}_{11,8,7,5,2}}$,\ \,%
$2\,\hat{A}_{11,8,7,6,1}$,\ \,%
$3\,\hat{A}_{11,8,7,7}$,\ \,%
$\SUSY{\hat{A}_{11,8,8,4,1,1}}$,\ \,%
$2\,\hat{A}_{11,8,8,4,2}$,\ \,%
$4\,\hat{A}_{11,8,8,5,1}$,\ \,%
$4\,\hat{A}_{11,8,8,6}$,\ \,%
$\SUSY{\hat{A}_{11,8,7,7,3}}$,\ \,%
$\SUSY{\hat{A}_{11,8,8,5,4}}$,\ \,%
$\SUSY{\hat{A}_{11,8,8,6,2,1}}$,\ \,%
$\hat{A}_{11,8,8,6,3}$,\ \,%
$\hat{A}_{11,8,8,7,1,1}$,\ \,%
$5\,\hat{A}_{11,8,8,7,2}$,\ \,%
$4\,\hat{A}_{11,8,8,8,1}$,\ \,%
$\SUSY{\hat{A}_{11,8,8,7,4,1}}$,\ \,%
$2\,\hat{A}_{11,8,8,7,5}$,\ \,%
$\SUSY{\hat{A}_{11,8,8,8,2,2}}$,\ \,%
$\hat{A}_{11,8,8,8,3,1}$,\ \,%
$3\,\hat{A}_{11,8,8,8,4}$,\ \,%
$\SUSY{\hat{A}_{11,8,8,7,7,1}}$,\ \,%
$\SUSY{\hat{A}_{11,8,8,8,5,2}}$,\ \,%
$\hat{A}_{11,8,8,8,6,1}$,\ \,%
$3\,\hat{A}_{11,8,8,8,7}$,\ \,%
$\SUSY{\hat{A}_{11,8,8,8,7,3}}$,\ \,%
$\hat{A}_{11,8,8,8,8,2}$,\ \,%
$\SUSY{\hat{A}_{11,8,8,8,8,5}}$,\ \,%
$\SUSY{\hat{A}_{11,8,8,8,8,8}}$.}}
\\\hline\hline
Type IIB potentials
\\\hline
\Pbox{\textwidth}{{\scriptsize
$\SUSY{e_{1}^{1\,[0]}}$,\ \,%
$\SUSY{A_{2}^{\alpha\,[0\text{-}1]}}$,\ \,%
$\SUSY{A_{4}^{[1]}}$,\ \,%
$\SUSY{A_{6}^{\alpha\,[1\text{-}2]}}$,\ \,%
$\SUSY{A_{7,1}^{[2]}}$,\ \,%
$\SUSY{A_{8}^{\alpha\beta\,[1\text{-}3]}}$,\ \,%
$\SUSY{A_{8,2}^{\alpha\,[2\text{-}3]}}$,\ \,%
$A_{9,1}^{\alpha\,[2\text{-}3]}$,\ \,%
$\SUSY{A_{10}^{\alpha\beta\gamma\,[1\text{-}4]}}$,\ \,%
$A_{10}^{\alpha\,[2\text{-}3]}$,\ \,%
$\SUSY{A_{8,4}^{[3]}}$,\ \,%
$\SUSY{A_{9,2,1}^{[3]}}$,\ \,%
$\SUSY{A_{9,3}^{\alpha\beta\,[2\text{-}4]}}$,\ \,%
$A_{10,2}^{\alpha\beta\,[2\text{-}4]}$,\ \,%
$2\,A_{10,2}^{[3]}$,\ \,%
$\SUSY{A_{8,6}^{\alpha\,[3\text{-}4]}}$,\ \,%
$\SUSY{A_{9,4,1}^{\alpha\,[3\text{-}4]}}$,\ \,%
$A_{9,5}^{\alpha\,[3\text{-}4]}$,\ \,%
$\SUSY{A_{10,2,2}^{\alpha\,[3\text{-}4]}}$,\ \,%
$A_{10,3,1}^{\alpha\,[3\text{-}4]}$,\ \,%
$\SUSY{A_{10,4}^{\alpha\beta\gamma\,[2\text{-}5]}}$,\ \,%
$2\,A_{10,4}^{\alpha\,[3\text{-}4]}$,\ \,%
$\SUSY{A_{8,7,1}^{[4]}}$,\ \,%
$\SUSY{A_{9,5,2}^{[4]}}$,\ \,%
$\SUSY{A_{9,6,1}^{\alpha\beta\,[3\text{-}5]}}$,\ \,%
$A_{9,6,1}^{[4]}$,\ \,%
$A_{9,7}^{\alpha\beta\,[3\text{-}5]}$,\ \,%
$2\,A_{9,7}^{[4]}$,\ \,%
$\SUSY{A_{10,4,1,1}^{[4]}}$,\ \,%
$\SUSY{A_{10,4,2}^{\alpha\beta\,[3\text{-}5]}}$,\ \,%
$A_{10,4,2}^{[4]}$,\ \,%
$A_{10,5,1}^{\alpha\beta\,[3\text{-}5]}$,\ \,%
$2\,A_{10,5,1}^{[4]}$,\ \,%
$3\,A_{10,6}^{\alpha\beta\,[3\text{-}5]}$,\ \,%
$2\,A_{10,6}^{[4]}$,\ \,%
$\SUSY{A_{9,6,3}^{\alpha\,[4\text{-}5]}}$,\ \,%
$\SUSY{A_{9,7,1,1}^{\alpha\,[4\text{-}5]}}$,\ \,%
$2\,A_{9,7,2}^{\alpha\,[4\text{-}5]}$,\ \,%
$\SUSY{A_{10,4,4}^{\alpha\,[4\text{-}5]}}$,\ \,%
$\SUSY{A_{10,5,2,1}^{\alpha\,[4\text{-}5]}}$,\ \,%
$A_{10,5,3}^{\alpha\,[4\text{-}5]}$,\ \,%
$A_{10,6,1,1}^{\alpha\,[4\text{-}5]}$,\ \,%
$\SUSY{A_{10,6,2}^{\alpha\beta\gamma\,[3\text{-}6]}}$,\ \,%
$4\,A_{10,6,2}^{\alpha\,[4\text{-}5]}$,\ \,%
$A_{10,7,1}^{\alpha\beta\gamma\,[3\text{-}6]}$,\ \,%
$6\,A_{10,7,1}^{\alpha\,[4\text{-}5]}$,\ \,%
$\SUSY{A_{9,6,5}^{[5]}}$,\ \,%
$\SUSY{A_{9,7,3,1}^{[5]}}$,\ \,%
$\SUSY{A_{9,7,4}^{\alpha\beta\,[4\text{-}6]}}$,\ \,%
$A_{9,7,4}^{[5]}$,\ \,%
$\SUSY{A_{10,5,4,1}^{[5]}}$,\ \,%
$\SUSY{A_{10,6,2,2}^{[5]}}$,\ \,%
$\SUSY{A_{10,6,3,1}^{\alpha\beta\,[4\text{-}6]}}$,\ \,%
$A_{10,6,3,1}^{[5]}$,\ \,%
$2\,A_{10,6,4}^{\alpha\beta\,[4\text{-}6]}$,\ \,%
$3\,A_{10,6,4}^{[5]}$,\ \,%
$\SUSY{A_{10,7,1,1,1}^{[5]}}$,\ \,%
$2\,A_{10,7,2,1}^{\alpha\beta\,[4\text{-}6]}$,\ \,%
$4\,A_{10,7,2,1}^{[5]}$,\ \,%
$4\,A_{10,7,3}^{\alpha\beta\,[4\text{-}6]}$,\ \,%
$5\,A_{10,7,3}^{[5]}$,\ \,%
$\SUSY{A_{9,7,5,1}^{\alpha\,[5\text{-}6]}}$,\ \,%
$2\,A_{9,7,6}^{\alpha\,[5\text{-}6]}$,\ \,%
$\SUSY{A_{10,6,4,2}^{\alpha\,[5\text{-}6]}}$,\ \,%
$2\,A_{10,6,5,1}^{\alpha\,[5\text{-}6]}$,\ \,%
$\SUSY{A_{10,6,6}^{\alpha\beta\gamma\,[4\text{-}7]}}$,\ \,%
$3\,A_{10,6,6}^{\alpha\,[5\text{-}6]}$,\ \,%
$\SUSY{A_{10,7,3,1,1}^{\alpha\,[5\text{-}6]}}$,\ \,%
$2\,A_{10,7,3,2}^{\alpha\,[5\text{-}6]}$,\ \,%
$\SUSY{A_{10,7,4,1}^{\alpha\beta\gamma\,[4\text{-}7]}}$,\ \,%
$6\,A_{10,7,4,1}^{\alpha\,[5\text{-}6]}$,\ \,%
$A_{10,7,5}^{\alpha\beta\gamma\,[4\text{-}7]}$,\ \,%
$7\,A_{10,7,5}^{\alpha\,[5\text{-}6]}$,\ \,%
$\SUSY{A_{9,7,6,2}^{[6]}}$,\ \,%
$\SUSY{A_{9,7,7,1}^{\alpha\beta\,[5\text{-}7]}}$,\ \,%
$A_{9,7,7,1}^{[6]}$,\ \,%
$\SUSY{A_{10,6,5,3}^{[6]}}$,\ \,%
$\SUSY{A_{10,6,6,2}^{\alpha\beta\,[5\text{-}7]}}$,\ \,%
$A_{10,6,6,2}^{[6]}$,\ \,%
$\SUSY{A_{10,7,4,2,1}^{[6]}}$,\ \,%
$\SUSY{A_{10,7,4,3}^{\alpha\beta\,[5\text{-}7]}}$,\ \,%
$A_{10,7,4,3}^{[6]}$,\ \,%
$\SUSY{A_{10,7,5,1,1}^{\alpha\beta\,[5\text{-}7]}}$,\ \,%
$2\,A_{10,7,5,1,1}^{[6]}$,\ \,%
$2\,A_{10,7,5,2}^{\alpha\beta\,[5\text{-}7]}$,\ \,%
$5\,A_{10,7,5,2}^{[6]}$,\ \,%
$6\,A_{10,7,6,1}^{\alpha\beta\,[5\text{-}7]}$,\ \,%
$8\,A_{10,7,6,1}^{[6]}$,\ \,%
$4\,A_{10,7,7}^{\alpha\beta\,[5\text{-}7]}$,\ \,%
$5\,A_{10,7,7}^{[6]}$,\ \,%
$\SUSY{A_{9,7,7,3}^{\alpha\,[6\text{-}7]}}$,\ \,%
$\SUSY{A_{10,6,6,4}^{\alpha\,[6\text{-}7]}}$,\ \,%
$\SUSY{A_{10,7,5,3,1}^{\alpha\,[6\text{-}7]}}$,\ \,%
$2\,A_{10,7,5,4}^{\alpha\,[6\text{-}7]}$,\ \,%
$3\,A_{10,7,6,2,1}^{\alpha\,[6\text{-}7]}$,\ \,%
$\SUSY{A_{10,7,6,3}^{\alpha\beta\gamma\,[5\text{-}8]}}$,\ \,%
$6\,A_{10,7,6,3}^{\alpha\,[6\text{-}7]}$,\ \,%
$\SUSY{A_{10,7,7,1,1}^{\alpha\beta\gamma\,[5\text{-}8]}}$,\ \,%
$5\,A_{10,7,7,1,1}^{\alpha\,[6\text{-}7]}$,\ \,%
$A_{10,7,7,2}^{\alpha\beta\gamma\,[5\text{-}8]}$,\ \,%
$9\,A_{10,7,7,2}^{\alpha\,[6\text{-}7]}$,\ \,%
$\SUSY{A_{9,7,7,5}^{[7]}}$,\ \,%
$\SUSY{A_{10,6,6,6}^{[7]}}$,\ \,%
$\SUSY{A_{10,7,5,5,1}^{[7]}}$,\ \,%
$\SUSY{A_{10,7,6,3,2}^{[7]}}$,\ \,%
$\SUSY{A_{10,7,6,4,1}^{\alpha\beta\,[6\text{-}8]}}$,\ \,%
$2\,A_{10,7,6,4,1}^{[7]}$,\ \,%
$2\,A_{10,7,6,5}^{\alpha\beta\,[6\text{-}8]}$,\ \,%
$4\,A_{10,7,6,5}^{[7]}$,\ \,%
$\SUSY{A_{10,7,7,2,1,1}^{[7]}}$,\ \,%
$A_{10,7,7,2,2}^{[7]}$,\ \,%
$3\,A_{10,7,7,3,1}^{\alpha\beta\,[6\text{-}8]}$,\ \,%
$5\,A_{10,7,7,3,1}^{[7]}$,\ \,%
$5\,A_{10,7,7,4}^{\alpha\beta\,[6\text{-}8]}$,\ \,%
$6\,A_{10,7,7,4}^{[7]}$,\ \,%
$\SUSY{A_{9,7,7,7}^{\alpha\,[7\text{-}8]}}$,\ \,%
$\SUSY{A_{10,7,6,5,2}^{\alpha\,[7\text{-}8]}}$,\ \,%
$2\,A_{10,7,6,6,1}^{\alpha\,[7\text{-}8]}$,\ \,%
$\SUSY{A_{10,7,7,3,3}^{\alpha\,[7\text{-}8]}}$,\ \,%
$\SUSY{A_{10,7,7,4,1,1}^{\alpha\,[7\text{-}8]}}$,\ \,%
$2\,A_{10,7,7,4,2}^{\alpha\,[7\text{-}8]}$,\ \,%
$\SUSY{A_{10,7,7,5,1}^{\alpha\beta\gamma\,[6\text{-}9]}}$,\ \,%
$7\,A_{10,7,7,5,1}^{\alpha\,[7\text{-}8]}$,\ \,%
$A_{10,7,7,6}^{\alpha\beta\gamma\,[6\text{-}9]}$,\ \,%
$8\,A_{10,7,7,6}^{\alpha\,[7\text{-}8]}$,\ \,%
$\SUSY{A_{10,7,6,6,3}^{[8]}}$,\ \,%
$\SUSY{A_{10,7,7,5,2,1}^{[8]}}$,\ \,%
$\SUSY{A_{10,7,7,5,3}^{\alpha\beta\,[7\text{-}9]}}$,\ \,%
$2\,A_{10,7,7,5,3}^{[8]}$,\ \,%
$\SUSY{A_{10,7,7,6,1,1}^{\alpha\beta\,[7\text{-}9]}}$,\ \,%
$A_{10,7,7,6,1,1}^{[8]}$,\ \,%
$2\,A_{10,7,7,6,2}^{\alpha\beta\,[7\text{-}9]}$,\ \,%
$5\,A_{10,7,7,6,2}^{[8]}$,\ \,%
$5\,A_{10,7,7,7,1}^{\alpha\beta\,[7\text{-}9]}$,\ \,%
$6\,A_{10,7,7,7,1}^{[8]}$,\ \,%
$\SUSY{A_{10,7,7,5,5}^{\alpha\,[8\text{-}9]}}$,\ \,%
$\SUSY{A_{10,7,7,6,3,1}^{\alpha\,[8\text{-}9]}}$,\ \,%
$2\,A_{10,7,7,6,4}^{\alpha\,[8\text{-}9]}$,\ \,%
$2\,A_{10,7,7,7,2,1}^{\alpha\,[8\text{-}9]}$,\ \,%
$\SUSY{A_{10,7,7,7,3}^{\alpha\beta\gamma\,[7\text{-}10]}}$,\ \,%
$5\,A_{10,7,7,7,3}^{\alpha\,[8\text{-}9]}$,\ \,%
$\SUSY{A_{10,7,7,6,5,1}^{[9]}}$,\ \,%
$A_{10,7,7,6,6}^{[9]}$,\ \,%
$\SUSY{A_{10,7,7,7,3,2}^{[9]}}$,\ \,%
$A_{10,7,7,7,4,1}^{[9]}$,\ \,%
$\SUSY{A_{10,7,7,7,4,1}^{\alpha\beta\,[8\text{-}10]}}$,\ \,%
$2\,A_{10,7,7,7,5}^{\alpha\beta\,[8\text{-}10]}$,\ \,%
$4\,A_{10,7,7,7,5}^{[9]}$,\ \,%
$\SUSY{A_{10,7,7,7,5,2}^{\alpha\,[9\text{-}10]}}$,\ \,%
$2\,A_{10,7,7,7,6,1}^{\alpha\,[9\text{-}10]}$,\ \,%
$\SUSY{A_{10,7,7,7,7}^{\alpha\beta\gamma\,[8\text{-}11]}}$,\ \,%
$3\,A_{10,7,7,7,7}^{\alpha\,[9\text{-}10]}$,\ \,%
$\SUSY{A_{10,7,7,7,6,3}^{[10]}}$,\ \,%
$\SUSY{A_{10,7,7,7,7,2}^{\alpha\beta\,[9\text{-}11]}}$,\ \,%
$A_{10,7,7,7,7,2}^{[10]}$,\ \,%
$\SUSY{A_{10,7,7,7,7,4}^{\alpha\,[10\text{-}11]}}$,\ \,%
$\SUSY{A_{10,7,7,7,7,6}^{[11]}}$.}} 
\\\hline
\end{tabular}}
\caption{All the M-theory/type IIB potentials that contribute to the counting of Table \ref{tab:p-form}. The underlined ones can couple to a supersymmetric brane.}
\label{tab:MB-potentials}
\end{table}
As noted for example in \cite{hep-th:0612001}, the level $\ell$ is related to the number of indices as
\begin{align}
 \ovalbox{ \quad $\displaystyle 3\,\ell = \text{(\# of lower indices)} - \text{(\# of upper indices)}$ \quad }\,,
\end{align}
and we can see that the last one, $\SUSY{\hat{A}_{11,8,8,8,8,8}}$, in Table \ref{tab:MB-potentials} indeed has level $\ell=17$\,. 

\paragraph*{Type IIB potentials:}
When we discuss type IIB theory, it is useful to introduce another level $\ell_9$ \cite{hep-th:0309198}. 
The low-level potentials are determined in \cite{hep-th:0309198} up to level $\ell_9=7$ (and some of level $\ell_9=8$), which is consistent with an earlier work \cite{hep-th:0107181}. 
We have continued the analysis using SimpLie, and potentials up to level $\ell_9=14$ have been determined (see Table \ref{tab:IIB-potentials-14}). 
However, this is still not enough to complete Table \ref{tab:p-form} by means of the type IIB potentials. 
Similar to the case of M-theory, neglecting potentials which do not contribute to Table \ref{tab:p-form}, we have determined all of the type IIB potentials that are necessary to complete Table \ref{tab:p-form}. 
The result is summarized in Table \ref{tab:MB-potentials} (see also Table \ref{tab:IIA-IIB-list}). 
In the case of type IIB potentials, the level $\ell_9$ is related to the number of indices of potentials as \cite{hep-th:0612001}
\begin{align}
 \ovalbox{ \quad $\displaystyle 2\,\ell_9 = \text{(\# of lower indices)} - \text{(\# of upper indices)}$ \quad }\,. 
\label{eq:B-index-level}
\end{align}
Therefore, the highest level $l_9=22$ corresponds to the potential $A_{10,7,7,7,7,6}^{[11]}$. 
We note that some potentials in Table \ref{tab:MB-potentials} have some totally symmetric upper indices ${}^{\alpha_1\cdots\alpha_s}$ ($\alpha=1,2$)\,. 
They denote that the potential transforms as an $\SL(2)$ $S$-duality $(s+1)$-plet. 
Moreover, integers in square brackets indicate the power $\NN$ in the tension formula \eqref{eq:tension-typeII}. 
For example, $A_{10}^{\alpha\beta\gamma\,[1\text{-}4]}$ is an $\SL(2)$ quadruplet consisting of the potentials $\{C_{10},\,D_{10},\,E_{10},\,F_{10}\}$\,. 
An interesting observation which is clear from Table \ref{tab:MB-potentials} is
\begin{align}
 \ovalbox{ \quad $\displaystyle \ell_9/2 = \text{(the average of the integer $\NN$ specified in the square brackets)}$ \quad }\,.
\label{eq:average-n}
\end{align}
Then, for a generator $A_{p,q,\cdots}^{\alpha_1\cdots\alpha_s}$ with $2\,\ell_9$ lower indices, information about the square bracket can be easily determined as $A_{p,q,\cdots}^{\alpha_1\cdots\alpha_s\,[\frac{\ell_9-s}{2}\text{-}\frac{\ell_9+s}{2}]}$\,, so we will omit the square bracket in the following. 
Moreover, this simple rule \eqref{eq:average-n} readily leads to the following one:
\begin{align}
 \ovalbox{\Pbox{0.7\textwidth}{ \quad a type IIB potential with even/odd level $\ell_9$ transforms \quad \newline 
 \text{\ \, in an odd/even-dimensional representation under $S$-duality.}}}
\end{align}
For example, a potential with odd level $\ell_9$ (whose number of indices is 2 mod 4) cannot be an $S$-duality singlet. 

\paragraph*{Type IIA potentials:}
As usual, type IIA fields can be obtained from M-theory potentials by making the $(10+1)$ decomposition, where the M-theory direction is denoted as $x^z$\,. 
In other words, type IIA potentials are characterized by two levels $\ell$ (the same one as in M-theory) and $\NN$ (the one appearing in the tension formula). 
If indices of an M-theory potential is restricted to 10D in type IIA theory, the integer $\NN$ becomes $\NN=\ell$\,. 
On the other hand, $\NN$ decrease as the number of the M-theory direction $z$ is increased, and we have $\NN\leq \ell$\,. 
In the type IIA case, the relation between the levels and the number of indices of potentials is \cite{1102.0934}
\begin{align}
 \ovalbox{ \quad $\displaystyle 2\,\ell + \NN = \text{(\# of lower indices)} - \text{(\# of upper indices)}$ \quad }\,. 
\label{eq:A-index-level}
\end{align}
In \cite{hep-th:0104081,hep-th:0309198}, the set of potentials were determined up to level $\ell\leq 5$\,. 
By decomposing the M-theory potentials, at least, we can determine the potentials with $\ell\leq 10$\,. 
In order to reproduce Table \ref{tab:p-form} in terms of type IIA potentials, we need to know potentials up to level $\ell\leq 17$\,. 
By neglecting potentials that do not contribute to Table \ref{tab:p-form}, the full set of type IIA potentials are summarized in Table \ref{tab:IIA-IIB-list}. 
We note that the potential $M_{10,7,7,7,7,7}$ has the highest level $(\ell,\NN)=(17,11)$\,. 

\begin{table}[htbp]
{\tiny
\centerline{
\begin{tabular}{|c||c|c||c|}\hline
 $\NN$ & IIA & IIB & summary
\\\hline\hline
%%%%%%%%%%%%%%%%%%%%%%%%%%
 0
 &\multicolumn{2}{C{14cm}||}{
 $\SUSY{B^1_{1}}$,
 $\SUSY{B_{2}}$}
 &$\SUSY{B^1_{1}}$,
 $\SUSY{B_{2}}$
\\\hline
 1
 &\Pbox{0.433\textwidth}{
 $\SUSY{C_{1}}$, 
 $\SUSY{C_{3}}$, 
 $\SUSY{C_{5}}$, 
 $\SUSY{C_{7}}$, 
 $\SUSY{C_{9}}$}
 &\Pbox{0.433\textwidth}{
 $\SUSY{C_{2}}$, 
 $\SUSY{C_{4}}$, 
 $\SUSY{C_{6}}$, 
 $\SUSY{C_{8}}$, 
 $\SUSY{C_{10}}$}
 &$\SUSY{C_{q}}$
\\\hline
 2
 &\multicolumn{2}{C{14cm}||}{
 $\SUSY{D_{6}}$, 
 $\SUSY{D_{7,1}}$, 
 $\SUSY{D_{8,2}}$, 
 $\SUSY{D_{9,3}}$, 
 $\SUSY{D_{10,4}}$}
 &$\SUSY{D_{6+n,n}}$
\\\cline{2-4}
 &\multicolumn{2}{c||}{
 $D_{8}$, 
 $D_{9,1}$, 
 $2\,D_{10}$, 
 $D_{10,2}$}
 &N/A
\\\hline
 3
 &\Pbox{0.433\textwidth}{
$\SUSY{E_{8,1}}$, 
$\SUSY{E_{8,3}}$, 
$\SUSY{E_{8,5}}$, 
$\SUSY{E_{8,7}}$, 
$\SUSY{E_{9,1,1}}$, 
$\SUSY{E_{9,3,1}}$, 
$\SUSY{E_{9,5,1}}$, 
$\SUSY{E_{9,7,1}}$, 
$\SUSY{E_{10,3,2}}$, 
$\SUSY{E_{10,5,2}}$, 
$\SUSY{E_{10,7,2}}$}
 &\Pbox{0.433\textwidth}{
$\SUSY{E_{8}}$, 
$\SUSY{E_{8,2}}$, 
$\SUSY{E_{8,4}}$, 
$\SUSY{E_{8,6}}$, 
$\SUSY{E_{9,2,1}}$, 
$\SUSY{E_{9,4,1}}$, 
$\SUSY{E_{9,6,1}}$, 
$\SUSY{E_{10,2,2}}$, 
$\SUSY{E_{10,4,2}}$, 
$\SUSY{E_{10,6,2}}$}
 &$\SUSY{E_{8+n,q,n}}$
\\\cline{2-4}
 &\Pbox{0.433\textwidth}{
 $E_{9,2}$, 
 $E_{9,4}$, 
 $E_{9,6}$, 
 $2\,E_{10,1}$, 
 $3\,E_{10,3}$, 
 $3\,E_{10,5}$, 
 $3\,E_{10,7}$, 
 $E_{10,2,1}$, 
 $E_{10,4,1}$, 
 $E_{10,6,1}$}
 &\Pbox{0.433\textwidth}{
 $E_{9,1}$, 
 $E_{9,3}$, 
 $E_{9,5}$, 
 $E_{9,7}$, 
 $2\,E_{10}$, 
 $3\,E_{10,2}$, 
 $3\,E_{10,4}$, 
 $3\,E_{10,6}$, 
 $E_{10,3,1}$, 
 $E_{10,5,1}$, 
 $E_{10,7,1}$}
 &N/A
\\\hline
 4
 &\multicolumn{2}{C{14cm}||}{
 $\SUSY{F_{9,3}}$, 
 $\SUSY{F_{9,4,1}}$, 
 $\SUSY{F_{9,5,2}}$, 
 $\SUSY{F_{9,6,3}}$, 
 $\SUSY{F_{9,7,4}}$, 
 $\SUSY{F_{10,4,1,1}}$, 
 $\SUSY{F_{10,5,2,1}}$, 
 $\SUSY{F_{10,6,3,1}}$, 
 $\SUSY{F_{10,7,4,1}}$}
 &$\SUSY{F_{9+n,3+m+n,m+n,n}}$
\\\cline{2-4}
 &\multicolumn{2}{c||}{
 $\SUSY{F_{8,6}}$, 
 $\SUSY{F_{8,7,1}}$, 
 $\SUSY{F_{9,7,1,1}}$}
 &$\SUSY{F_{8+n,6+m+n,m+n,n}}$
\\\cline{2-4}
 &\multicolumn{2}{C{14cm}||}{
 $F_{9,5}$, 
 $2\,F_{9,6,1}$, 
 $3\,F_{9,7}$, 
 $2\,F_{9,7,2}$, 
 $F_{10,3,1}$, 
 $2\,F_{10,4}$, 
 $F_{10,4,2}$, 
 $3\,F_{10,5,1}$, 
 $F_{10,5,3}$, 
 $4\,F_{10,6}$, 
 $F_{10,6,1,1}$, 
 $4\,F_{10,6,2}$, 
 $F_{10,6,4}$, 
 $7\,F_{10,7,1}$, 
 $2\,F_{10,7,2,1}$, 
 $4\,F_{10,7,3}$, 
 $F_{10,7,5}$}
 &N/A
\\\cline{2-4}
 &\Pbox{0.433\textwidth}{
 $\SUSY{F_{10,1,1}}$, 
 $\SUSY{F_{10,3,3}}$, 
 $\SUSY{F_{10,5,5}}$, 
 $\SUSY{F_{10,7,7}}$}
 &\Pbox{0.433\textwidth}{
 $\SUSY{F_{10}}$, 
 $\SUSY{F_{10,2,2}}$, 
 $\SUSY{F_{10,4,4}}$, 
 $\SUSY{F_{10,6,6}}$}
 &$\SUSY{F_{10,q,q}}$
\\\cline{2-4}
 &\Pbox{0.433\textwidth}{
 $F_{10,3,1}$, 
 $F_{10,5,1}$, 
 $F_{10,5,3}$, 
 $F_{10,7,1}$, 
 $F_{10,7,3}$, 
 $F_{10,7,5}$}
 &\Pbox{0.433\textwidth}{
 $F_{10,2}$, 
 $F_{10,4}$, 
 $F_{10,4,2}$, 
 $F_{10,6}$, 
 $F_{10,6,2}$, 
 $F_{10,6,4}$}
 &N/A
\\\hline
 5
 &\Pbox{0.433\textwidth}{
 $\SUSY{G_{9,6}}$,
 $\SUSY{G_{9,6,2}}$,
 $\SUSY{G_{9,6,4}}$,
 $\SUSY{G_{9,6,6}}$,
 $\SUSY{G_{9,7,2,1}}$,
 $\SUSY{G_{9,7,4,1}}$,
 $\SUSY{G_{9,7,6,1}}$,
 $\SUSY{G_{10,7,2,1,1}}$,
 $\SUSY{G_{10,7,4,1,1}}$,
 $\SUSY{G_{10,7,6,1,1}}$}
 &\Pbox{0.433\textwidth}{
 $\SUSY{G_{9,6,1}}$,
 $\SUSY{G_{9,6,3}}$,
 $\SUSY{G_{9,6,5}}$,
 $\SUSY{G_{9,7,1,1}}$,
 $\SUSY{G_{9,7,3,1}}$,
 $\SUSY{G_{9,7,5,1}}$,
 $\SUSY{G_{9,7,7,1}}$,
 $\SUSY{G_{10,7,1,1,1}}$,
 $\SUSY{G_{10,7,3,1,1}}$,
 $\SUSY{G_{10,7,5,1,1}}$,
 $\SUSY{G_{10,7,7,1,1}}$}
 &$\SUSY{G_{9+n,6+m,p,m,n}}$
\\
 &\Pbox{0.433\textwidth}{
 $\SUSY{G_{10,4,1}}$,
 $\SUSY{G_{10,4,3}}$,
 $\SUSY{G_{10,5,1,1}}$,
 $\SUSY{G_{10,5,3,1}}$,
 $\SUSY{G_{10,5,5,1}}$,
 $\SUSY{G_{10,6,3,2}}$,
 $\SUSY{G_{10,6,5,2}}$,
 $\SUSY{G_{10,7,3,3}}$,
 $\SUSY{G_{10,7,5,3}}$,
 $\SUSY{G_{10,7,7,3}}$}
 &\Pbox{0.433\textwidth}{
 $\SUSY{G_{10,4}}$,
 $\SUSY{G_{10,4,2}}$,
 $\SUSY{G_{10,4,4}}$,
 $\SUSY{G_{10,5,2,1}}$,
 $\SUSY{G_{10,5,4,1}}$,
 $\SUSY{G_{10,6,2,2}}$,
 $\SUSY{G_{10,6,4,2}}$,
 $\SUSY{G_{10,6,6,2}}$,
 $\SUSY{G_{10,7,4,3}}$,
 $\SUSY{G_{10,7,6,3}}$}
 &$\SUSY{G_{10,4+n,q,n}}$
\\\cline{2-4}
  &\Pbox{0.433\textwidth}{
 $2\,G_{9,7,1}$,
 $2\,G_{9,7,3}$,
 $2\,G_{9,7,5}$,
 $G_{9,7,7}$,
 $G_{10,5,2}$,
 $G_{10,5,4}$,
 $4\,G_{10,6,1}$,
 $5\,G_{10,6,3}$,
 $4\,G_{10,6,5}$,
 $2\,G_{10,6,2,1}$,
 $2\,G_{10,6,4,1}$,
 $G_{10,6,6,1}$,
 $3\,G_{10,7}$,
 $8\,G_{10,7,2}$,
 $9\,G_{10,7,4}$,
 $7\,G_{10,7,6}$,
 $5\,G_{10,7,1,1}$,
 $7\,G_{10,7,3,1}$,
 $7\,G_{10,7,5,1}$,
 $5\,G_{10,7,7,1}$,
 $G_{10,7,2,2}$,
 $2\,G_{10,7,4,2}$,
 $2\,G_{10,7,6,2}$}
 &\Pbox{0.433\textwidth}{
 $G_{9,7}$,
 $2\,G_{9,7,2}$,
 $2\,G_{9,7,4}$,
 $2\,G_{9,7,6}$,
 $G_{10,5,1}$,
 $G_{10,5,3}$,
 $G_{10,6,1,1}$,
 $2\,G_{10,6,3,1}$,
 $2\,G_{10,6,5,1}$,
 $3\,G_{10,6}$,
 $5\,G_{10,6,2}$,
 $5\,G_{10,6,4}$,
 $4\,G_{10,6,6}$,
 $7\,G_{10,7,1}$,
 $9\,G_{10,7,3}$,
 $8\,G_{10,7,5}$,
 $4\,G_{10,7,7}$,
 $6\,G_{10,7,2,1}$,
 $7\,G_{10,7,4,1}$,
 $6\,G_{10,7,6,1}$,
 $2\,G_{10,7,3,2}$,
 $2\,G_{10,7,5,2}$,
 $G_{10,7,7,2}$}
 &N/A
\\\hline
%%%%%%%%%%%%%%%%%%%%%%%%%%
 6
 &\multicolumn{2}{C{14cm}||}{
 $\SUSY{H_{9,7,4}}$,
 $\SUSY{H_{9,7,5,1}}$,
 $\SUSY{H_{9,7,6,2}}$,
 $\SUSY{H_{9,7,7,3}}$}
 &$\SUSY{H_{9,7,4+n,n}}$
\\\cline{2-4}
 &\multicolumn{2}{C{14cm}||}{
 $\SUSY{H_{10,6,2}}$,
 $\SUSY{H_{10,6,3,1}}$,
 $\SUSY{H_{10,6,4,2}}$,
 $\SUSY{H_{10,6,5,3}}$,
 $\SUSY{H_{10,6,6,4}}$,
 $\SUSY{H_{10,7,3,1,1}}$,
 $\SUSY{H_{10,7,4,2,1}}$,
 $\SUSY{H_{10,7,5,3,1}}$,
 $\SUSY{H_{10,7,6,4,1}}$,
 $\SUSY{H_{10,7,7,5,1}}$}
 &$\SUSY{H_{10,6+n,2+m+n,m+n,n}}$
\\\cline{2-4}
 &\multicolumn{2}{C{14cm}||}{
 $H_{10,7,1}$,
 $2\,H_{10,6,4}$,
 $2\,H_{10,7,2,1}$,
 $4\,H_{10,7,3}$,
 $2\,H_{9,7,6}$,
 $2\,H_{10,7,3,2}$,
 $7\,H_{10,7,4,1}$,
 $8\,H_{10,7,5}$,
 $2\,H_{10,6,5,1}$,
 $4\,H_{10,6,6}$,
 $2\,H_{9,7,7,1}$,
 $2\,H_{10,6,6,2}$,
 $2\,H_{10,7,4,3}$,
 $3\,H_{10,7,5,1,1}$,
 $7\,H_{10,7,5,2}$,
 $14\,H_{10,7,6,1}$,
 $9\,H_{10,7,7}$,
 $2\,H_{10,7,5,4}$,
 $3\,H_{10,7,6,2,1}$,
 $7\,H_{10,7,6,3}$,
 $6\,H_{10,7,7,1,1}$,
 $10\,H_{10,7,7,2}$,
 $2\,H_{10,7,6,5}$,
 $3\,H_{10,7,7,3,1}$,
 $5\,H_{10,7,7,4}$,
 $H_{10,7,7,6}$}
 &N/A
\\\hline
%%%%%%%%%%%%%%%%%%%%%%%%%%
 7
 &\Pbox{0.433\textwidth}{
 $\SUSY{I_{9,7,7}}$,
 $\SUSY{I_{9,7,7,2}}$,
 $\SUSY{I_{9,7,7,4}}$,
 $\SUSY{I_{9,7,7,6}}$}
 &\Pbox{0.433\textwidth}{
 $\SUSY{I_{9,7,7,1}}$,
 $\SUSY{I_{9,7,7,3}}$,
 $\SUSY{I_{9,7,7,5}}$,
 $\SUSY{I_{9,7,7,7}}$}
 &$\SUSY{I_{9,7,7,p}}$
\\\cline{2-4}
 &\Pbox{0.433\textwidth}{
 $\SUSY{I_{10,7,4}}$,
 $\SUSY{I_{10,7,4,2}}$,
 $\SUSY{I_{10,7,4,4}}$,
 $\SUSY{I_{10,7,5,2,1}}$,
 $\SUSY{I_{10,7,5,4,1}}$,
 $\SUSY{I_{10,7,6,2,2}}$,
 $\SUSY{I_{10,7,6,4,2}}$,
 $\SUSY{I_{10,7,6,6,2}}$,
 $\SUSY{I_{10,7,7,4,3}}$,
 $\SUSY{I_{10,7,7,6,3}}$}
 &\Pbox{0.433\textwidth}{
 $\SUSY{I_{10,7,4,1}}$,
 $\SUSY{I_{10,7,4,3}}$,
 $\SUSY{I_{10,7,5,1,1}}$,
 $\SUSY{I_{10,7,5,3,1}}$,
 $\SUSY{I_{10,7,5,5,1}}$,
 $\SUSY{I_{10,7,6,3,2}}$,
 $\SUSY{I_{10,7,6,5,2}}$,
 $\SUSY{I_{10,7,7,3,3}}$,
 $\SUSY{I_{10,7,7,5,3}}$,
 $\SUSY{I_{10,7,7,7,3}}$}
 &$\SUSY{I_{10,7,4+n,p,n}}$
\\\cline{2-4}
 &\Pbox{0.433\textwidth}{
 $\SUSY{I_{10,6,6,1}}$,
 $\SUSY{I_{10,6,6,3}}$,
 $\SUSY{I_{10,6,6,5}}$,
 $\SUSY{I_{10,7,7,1,1,1}}$,
 $\SUSY{I_{10,7,7,3,1,1}}$,
 $\SUSY{I_{10,7,7,5,1,1}}$,
 $\SUSY{I_{10,7,7,7,1,1}}$}
 &\Pbox{0.433\textwidth}{
 $\SUSY{I_{10,6,6}}$,
 $\SUSY{I_{10,6,6,2}}$,
 $\SUSY{I_{10,6,6,4}}$,
 $\SUSY{I_{10,6,6,6}}$,
 $\SUSY{I_{10,7,7,2,1,1}}$,
 $\SUSY{I_{10,7,7,4,1,1}}$,
 $\SUSY{I_{10,7,7,6,1,1}}$}
 &$\SUSY{I_{10,6+n,6+n,q,n,n}}$
\\\cline{2-4}
 &\Pbox{0.433\textwidth}{
 $2\,I_{10,7,5,1}$,
 $4\,I_{10,7,6}$,
 $2\,I_{10,7,5,3}$,
 $2\,I_{10,7,6,1,1}$,
 $7\,I_{10,7,6,2}$,
 $9\,I_{10,7,7,1}$,
 $I_{10,7,5,5}$,
 $3\,I_{10,7,6,3,1}$,
 $7\,I_{10,7,6,4}$,
 $7\,I_{10,7,7,2,1}$,
 $11\,I_{10,7,7,3}$,
 $3\,I_{10,7,6,5,1}$,
 $5\,I_{10,7,6,6}$,
 $2\,I_{10,7,7,3,2}$,
 $8\,I_{10,7,7,4,1}$,
 $10\,I_{10,7,7,5}$,
 $2\,I_{10,7,7,5,2}$,
 $7\,I_{10,7,7,6,1}$,
 $5\,I_{10,7,7,7}$,
 $I_{10,7,7,7,2}$}
 &\Pbox{0.433\textwidth}{
 $I_{10,7,5}$, 
 $2\,I_{10,7,5,2}$, 
 $6\,I_{10,7,6,1}$,
 $4\,I_{10,7,7}$,
 $2\,I_{10,7,5,4}$, 
 $3\,I_{10,7,6,2,1}$,
 $7\,I_{10,7,6,3}$,
 $6\,I_{10,7,7,1,1}$,
 $10\,I_{10,7,7,2}$,
 $3\,I_{10,7,6,4,1}$,
 $6\,I_{10,7,6,5}$,
 $I_{10,7,7,2,2}$,
 $8\,I_{10,7,7,3,1}$,
 $11\,I_{10,7,7,4}$,
 $2\,I_{10,7,6,6,1}$,
 $2\,I_{10,7,7,4,2}$,
 $8\,I_{10,7,7,5,1}$,
 $9\,I_{10,7,7,6}$,
 $2\,I_{10,7,7,6,2}$,
 $5\,I_{10,7,7,7,1}$}
 &N/A
\\\hline
%%%%%%%%%%%%%%%%%%%%%%%%%%
 8
 &\multicolumn{2}{C{14cm}||}{
 $\SUSY{J_{9,7,7,7}}$}
 &$\SUSY{J_{9,7,7,7}}$
\\\cline{2-4}
 &\multicolumn{2}{C{14cm}||}{
 $\SUSY{J_{10,7,6,3}}$,
 $\SUSY{J_{10,7,6,4,1}}$,
 $\SUSY{J_{10,7,6,5,2}}$,
 $\SUSY{J_{10,7,6,6,3}}$,
 $\SUSY{J_{10,7,7,4,1,1}}$,
 $\SUSY{J_{10,7,7,5,2,1}}$,
 $\SUSY{J_{10,7,7,6,3,1}}$,
 $\SUSY{J_{10,7,7,7,4,1}}$}
 &$\SUSY{J_{10,7,6+n,3+m+n,m+n,n}}$
\\\cline{2-4}
 &\multicolumn{2}{C{14cm}||}{
 $J_{10,7,7,2}$,
 $2\,J_{10,7,6,5}$,
 $5\,J_{10,7,7,4}$,
 $2\,J_{10,7,7,3,1}$,
 $2\,J_{10,7,7,4,2}$,
 $7\,J_{10,7,7,5,1}$,
 $2\,J_{10,7,6,6,1}$,
 $9\,J_{10,7,7,6}$,
 $7\,J_{10,7,7,6,2}$,
 $5\,J_{10,7,7,7,3}$,
 $10\,J_{10,7,7,7,1}$,
 $2\,J_{10,7,7,5,3}$,
 $2\,J_{10,7,7,6,1,1}$,
 $2\,J_{10,7,7,7,2,1}$,
 $2\,J_{10,7,7,6,4}$,
 $J_{10,7,7,7,5}$}
 &N/A
\\\cline{2-4}
 &\Pbox{0.433\textwidth}{
 $\SUSY{J_{10,7,7}}$,
 $\SUSY{J_{10,7,7,2,2}}$,
 $\SUSY{J_{10,7,7,4,4}}$,
 $\SUSY{J_{10,7,7,6,6}}$}
 &\Pbox{0.433\textwidth}{
 $\SUSY{J_{10,7,7,1,1}}$,
 $\SUSY{J_{10,7,7,3,3}}$,
 $\SUSY{J_{10,7,7,5,5}}$,
 $\SUSY{J_{10,7,7,7,7}}$}
 &$\SUSY{J_{10,7,7,p,p}}$
\\\cline{2-4}
 &\Pbox{0.433\textwidth}{
 $J_{10,7,7,2}$,
 $J_{10,7,7,4}$,
 $J_{10,7,7,4,2}$,
 $J_{10,7,7,6}$,
 $J_{10,7,7,6,2}$,
 $J_{10,7,7,6,4}$}
 &\Pbox{0.433\textwidth}{
 $J_{10,7,7,3,1}$,
 $J_{10,7,7,5,1}$,
 $J_{10,7,7,5,3}$,
 $J_{10,7,7,7,1}$,
 $J_{10,7,7,7,3}$,
 $J_{10,7,7,7,5}$}
 &N/A
\\\hline
%%%%%%%%%%%%%%%%%%%%%%%%%%
 9
 &\Pbox{0.433\textwidth}{
 $\SUSY{K_{10,7,7,5}}$,
 $\SUSY{K_{10,7,7,5,2}}$,
 $\SUSY{K_{10,7,7,5,4}}$,
 $\SUSY{K_{10,7,7,6,2,1}}$,
 $\SUSY{K_{10,7,7,6,4,1}}$,
 $\SUSY{K_{10,7,7,6,6,1}}$,
 $\SUSY{K_{10,7,7,7,2,2}}$,
 $\SUSY{K_{10,7,7,7,4,2}}$,
 $\SUSY{K_{10,7,7,7,6,2}}$}
 &\Pbox{0.433\textwidth}{
 $\SUSY{K_{10,7,7,5,1}}$,
 $\SUSY{K_{10,7,7,5,3}}$,
 $\SUSY{K_{10,7,7,5,5}}$,
 $\SUSY{K_{10,7,7,6,1,1}}$,
 $\SUSY{K_{10,7,7,6,3,1}}$,
 $\SUSY{K_{10,7,7,6,5,1}}$,
 $\SUSY{K_{10,7,7,7,3,2}}$,
 $\SUSY{K_{10,7,7,7,5,2}}$,
 $\SUSY{K_{10,7,7,7,7,2}}$}
 &$\SUSY{K_{10,7,7,5+n,p,n}}$
\\\cline{2-4}
 &\Pbox{0.433\textwidth}{
 $2\,K_{10,7,7,6,1}$,
 $4\,K_{10,7,7,7}$,
 $2\,K_{10,7,7,6,3}$,
 $K_{10,7,7,7,1,1}$,
 $6\,K_{10,7,7,7,2}$,
 $2\,K_{10,7,7,6,5}$,
 $6\,K_{10,7,7,7,4}$,
 $2\,K_{10,7,7,7,3,1}$,
 $5\,K_{10,7,7,7,6}$,
 $2\,K_{10,7,7,7,5,1}$,
 $K_{10,7,7,7,7,1}$}
 &\Pbox{0.433\textwidth}{
 $K_{10,7,7,6}$,
 $2\,K_{10,7,7,6,2}$,
 $5\,K_{10,7,7,7,1}$,
 $2\,K_{10,7,7,6,4}$,
 $2\,K_{10,7,7,7,2,1}$,
 $6\,K_{10,7,7,7,3}$,
 $K_{10,7,7,6,6}$,
 $6\,K_{10,7,7,7,5}$,
 $2\,K_{10,7,7,7,4,1}$,
 $4\,K_{10,7,7,7,7}$,
 $2\,K_{10,7,7,7,6,1}$}
 &N/A
\\\hline
%%%%%%%%%%%%%%%%%%%%%%%%%%
 10
 &\multicolumn{2}{C{14cm}||}{
 $\SUSY{L_{10,7,7,7,3}}$, 
 $\SUSY{L_{10,7,7,7,4,1}}$, 
 $\SUSY{L_{10,7,7,7,5,2}}$, 
 $\SUSY{L_{10,7,7,7,6,3}}$, 
 $\SUSY{L_{10,7,7,7,7,4}}$}
 &$\SUSY{L_{10,7,7,7,3+n,n}}$
\\\cline{2-4}
 &\multicolumn{2}{C{14cm}||}{ 
 $2\,L_{10,7,7,7,5}$,
 $2\,L_{10,7,7,7,6,1}$,
 $4\,L_{10,7,7,7,7}$,
 $2\,L_{10,7,7,7,7,2}$}
 &N/A
\\\hline
%%%%%%%%%%%%%%%%%%%%%%%%%%
 11
 &\Pbox{0.433\textwidth}{
 $\SUSY{M_{10,7,7,7,7,1}}$, 
 $\SUSY{M_{10,7,7,7,7,3}}$, 
 $\SUSY{M_{10,7,7,7,7,5}}$, 
 $\SUSY{M_{10,7,7,7,7,7}}$}
 &\Pbox{0.433\textwidth}{
 $\SUSY{M_{10,7,7,7,7}}$, 
 $\SUSY{M_{10,7,7,7,7,2}}$, 
 $\SUSY{M_{10,7,7,7,7,4}}$, 
 $\SUSY{M_{10,7,7,7,7,6}}$}
 &$\SUSY{M_{10,7,7,7,7,q}}$
\\\hline
\end{tabular}}
\caption{All type II potentials that contribute to Table \ref{tab:p-form}. 
Potentials that may couple to supersymmetric branes are underlined. 
For the underlined potentials, the pattern can be neatly summarized, as given in the fourth column. 
The index $p$/$q$ runs over even/odd (odd/even) numbers in type IIA (IIB) theory while the indices $m,n$ run over non-negative integers. 
For the type IIB potentials, this is a rewriting of Table \ref{tab:MB-potentials}.
\label{tab:IIA-IIB-list}}
}
\end{table}

\paragraph*{$\OO(D,D)$ potentials:}

In compactified $d$-dimensional theories, type IIA and type IIB potentials are related under $T$-duality. 
Hence, it is also useful to study the type II potentials in an $\OO(D,D)$ $T$-duality covariant manner ($D\equiv 10-d$). 
This has been studied in a series of papers \cite{1102.0934,1106.0212,1108.5067,1201.5819}, aimed at the $T$-duality-covariant description of the brane actions. 
For low levels $\NN=0,1,2,3$\,, all of the $\OO(D,D)$-covariant potentials predicted by $E_{11}$ are already determined in each dimension ($d\geq 3$),
\begin{align}
\begin{split}
 &B_{1;A}\,,\qquad B_{2}\,,\quad
  C_{p;a}\quad (p=1,3,5,\cdots)\,,\quad C_{p;\dot{a}}\quad (p:0,2,4,\cdots)\,, 
\\
 &D_{d;A_{1\cdots 4}}\,,\quad D_{d;A_{12}}\,,\quad 2\,D_{d}\,,\quad D_{d-1;A_{123}}\,,\quad D_{d-1;A}\,,
\\
 &D_{d-2}\,,\quad D_{d-2;A_{12}}\,,\quad D_{d-3;A}\,,\quad D_{d-4}\,,
\\
 &E_{d;A_{12}a}\,,\quad E_{d;Aa}\,,\quad 3\,E_{d;\dot{a}}\,,\quad E_{d-1;A\dot{a}}\,,\quad E_{d-1;a}\,,\quad E_{d-2;\dot{a}}\,,
\end{split}
\end{align}
where the first integer $p$ in the subscripts denotes that the potential is a $p$-form in the external $d$ dimensions, and the remaining labels after the semicolon are $\OO(D,D)$ indices\footnote{The index $A=1,\dotsc,2D$ is the vector index and $a,\dot{a}$ are spinor indices. See Table \ref{tab:Odd-irrep} for more details.} (for $\OO(D,D)$ singlets, the semicolon is omitted). 
Obviously, the potentials $B$ contain the graviphoton and the Kalb--Ramond $B$-field, and $C$ fields are the Ramond--Ramond (R--R) potentials. 
On the other hand, $D$ potentials are less trivial. 
The $\OO(D,D)$-covariant $D$ potentials are useful when writing down a manifestly $T$-duality-symmetric brane actions. 
For example, in \cite{1712.01739}, covariant 5-brane actions in $d=6$ and $d=8$ are (partially) written down using the potentials of the form $D_{6;\cdots}$ (see the recent paper \cite{1903.05601} for further details). 
In order to know what kind of 5-brane is described by that action, we need to identify the detailed constituents of the potentials $D$. 
As detailed in \cite{1102.0934}, they are made of the dimensional reductions of the familiar potential $D_6$ that couples to the NS5-brane, the dual graviton $D_{7,1}$ that couples to the KKM, and $D_{8,2}$ that couples to the exotic $5^2_2$-brane, etc.,
\begin{align}
 D_{6}\,,\quad
 D_{7,1}\,,\quad
 D_{8}\,,\quad
 D_{8,2}\,,\quad
 D_{9,1}\,,\quad
 D_{9,3}\,,\quad
 2\,D_{10}\,,\quad
 D_{10,2}\,,\quad
 D_{10,4}\,. 
\label{eq:D-potentials}
\end{align}
Despite the full contents of the potentials $E$ have not been shown in the literature, those which couple to supersymmetric branes have been determined in \cite{1108.5067}:
\begin{align*}
 \begin{tabular}{|c|l|}\hline
 \text{IIA} & \text{\footnotesize
$E_{8,1}$, 
$E_{8,3}$, 
$E_{9,1,1}$, 
$E_{8,5}$, 
$E_{9,3,1}$, 
$E_{9,5,1}$, 
$E_{10,3,2}$, 
$E_{8,7}$, 
$E_{10,5,2}$, 
$E_{9,7,1}$, 
$E_{10,7,2}$, 
$E_{9,9,1}$, 
$E_{10,9,2}$}
\\\hline
 \text{IIB} & \text{\footnotesize
$E_{8}$, 
$E_{8,2}$, 
$E_{8,4}$, 
$E_{9,2,1}$, 
$E_{9,4,1}$, 
$E_{8,6}$, 
$E_{10,2,2}$, 
$E_{9,6,1}$, 
$E_{10,4,2}$, 
$E_{8,8}$, 
$E_{10,6,2}$, 
$E_{9,8,1}$, 
$E_{10,8,2}$, 
$E_{10,10,2}$}
\\\hline
 \end{tabular}\,.
\end{align*}
Brane actions for such exotic branes are discussed in \cite{1903.05601}. 
The $\OO(D,D)$ potentials with higher level ($\NN\geq 4$) also have not been fully identified, but the ones that may couple to supersymmetric branes are fully determined in \cite{1201.5819} up to $\NN=6$\,. 

In this paper, we make a full list of the $\OO(D,D)$ potentials in dimensions $3\leq d\leq 9$\,. 
When we consider a compactification to $d$ dimensions ($3\leq d\leq 9$), all of the $p$-form potentials (which form some $U$-duality multiplets) used are those already given in Table \ref{tab:p-form}. 
Then, we decompose all of the $U$-duality multiplet into the $\OO(D,D)$ multiplets. 
For example, in $d=3$, the 3-form multiplet ${\bf 147250}$ of $E_8$ $U$-duality group contains $\OO(7,7)$ potentials, such as $M_{3;\dot{a}}$ ($\NN=11$). 
The detailed results are given in Appendix \ref{app:En2ODD}.
There, we further decompose the obtained $\OO(D,D)$ potentials into $d$-dimensional mixed-symmetry potentials in type IIA/IIB theory. 

The $E_{11}$ conjecture claims that $E_{11}$ symmetry is a true 11-dimensional symmetry even before the compactification, and it contains the $\OO(10,10)$ ``$T$-duality'' symmetry as a subgroup. 
Indeed, in the $T$-duality manifest supergravity, called double field theory (DFT), the supergravity action has a formal $\OO(10,10)$ symmetry.
Therefore, the mixed-symmetry potentials in type IIA theory and type IIB theory predicted by the $E_{11}$ conjecture can be embedded into $\OO(10,10)$-covariant tensors (see \cite{hep-th:0402140} for the original discussion of $T$-duality in the context of $E_{11}$).\footnote{Here, we consider $d=0$ (rather than $d=10$), and the whole physical space is treated as the internal space.}
For example, all of the potentials $D$ are packaged into two $\OO(10,10)$ tensors, $D_{A_{1\cdots4}}$ and $D$ \cite{1102.0934}. 
For potentials with higher level ($\NN\geq 3$), $\OO(10,10)$ tensors have not been classified, and we determine all of the $\OO(10,10)$ tensors predicted from $E_{11}$ up to level $\NN=6$ (though the multiplicity for the $\OO(10,10)$ singlet $H$ was not determined). 
For higher levels, it is extremely difficult to determine all of the potentials through the standard level decomposition. 
However, by restricting the task to the potentials which couple to supersymmetric branes, we can take another simpler procedure, namely a combination of $T$- and $S$-dualities. 
Through the procedure, we have determined all of the $\OO(10,10)$ potentials that couple to supersymmetric branes up to level $\NN=36$\,. 
The result is perfectly consistent with the results of the standard level decomposition (at least up to $\NN=6$). 
We have also identified which mixed-symmetry potentials in 10D are contained in which $\OO(10,10)$ tensors. 

Here, we summarize some observations associated with $\OO(10,10)$ tensors and mixed-symmetry potentials:
\begin{itemize}
\item When we decompose a level-$\NN$ $\OO(10,10)$ tensor into the mixed-symmetry potentials in type IIA or type IIB theory, a symmetry in the index structure arises. 
For example, an $\OO(10,10)$ spinor $C_{\dot{a}}$ is decomposed into $C_1,\dotsc,C_9$ in type IIA theory, and there is the familiar duality between $C_1\leftrightarrow C_9$, $C_3\leftrightarrow C_7$, $C_4\leftrightarrow C_6$\,. 
The sum of the indices on each pair is $10$ in this case. 
In fact, this symmetry extends to potentials with higher level $\NN$\,. 
In general, we find the following rule:
\begin{align}
 \ovalbox{ \quad The sum of indices on each dual pair is equal to $10\,\NN$\,. \quad }
\label{eq:index-dual1}
\end{align}
For example, the potentials $D$ in \eqref{eq:D-potentials} are members of the $\OO(10,10)$ tensors $D_{A_1\cdots A_4}$ and $D$ (with $\NN=2$). 
One of the $D_{10}$, which corresponds to the singlet $D$, is self-dual, and the summation of indices gives $10\times 2$\,. 
In the multiplet $D_{A_1\cdots A_4}$, the highest weight corresponds to $D_6$ and its dual is $D_{10,4}$\,. 
The total number of indices is again $10\times 2$\,. 
This kind of duality has been noted in \cite{hep-th:9809039,0805.4451} in the context of the duality-invariant mass squared in $U$-duality multiplets.

A stronger rule is as follows (note that $A_{p,\dotsc,q,0,\dotsc,0}\equiv A_{p,\dotsc,q}$):
\begin{align}
 \ovalbox{ \quad \Pbox{0.63\textwidth}{A dual partner of $A_{p_1,\cdots,p_r}$ is $A_{10,\dotsc,10,10-p_r,\dotsc,10-p_1}$\,,\qquad\qquad where the number of 10's can be determined from \eqref{eq:index-dual1}.} \quad }
\label{eq:index-dual2}
\end{align}

The rules \eqref{eq:index-dual1} and \eqref{eq:index-dual2} can be applied also to decompositions of $\OO(D,D)$ tensors by replacing 10 by $D$\,. 
However, when $D$ is odd, if $A_{p_1,\cdots,p_r}$ is a potential in type IIA (IIB) theory, then the partner is in type IIB (IIA) theory. 
Therefore, if we restrict the process to a single theory, we cannot clearly see the duality. 

\item Some potentials, such as $B_2$ or $D_{6}$, are common to both type IIA/IIB theory. 
According to the rule \eqref{eq:B-index-level}, the type IIB potentials always have even number of indices while type IIA potentials satisfy the rule \eqref{eq:A-index-level}. 
Therefore, we obtain the rule,
\begin{align}
 \ovalbox{ A potential can exist in both type II theories only when the level $\NN$ is even. }
\end{align}
Indeed, when $\NN$ is odd, the $\OO(10,10)$ tensors have a spinor index $a$ or $\dot{a}$ and their decompositions into type IIA and IIB yield different $\SL(10)$ tensors. 
This pattern has been noted in \cite{hep-th:0511153} and the odd-$\NN$ sectors are called the generalized R--R sector. 
On the other hand, regarding the even-$\NN$ sectors, some $\OO(10,10)$ tensors give the common tensors in both type IIA and type IIB theory (the generalized NS--NS sector) and some correspond to the generalized R--R sector.\footnote{In a recent paper \cite{1903.10247}, it has been noted that potentials with $\NN = 2 \mod 4$ are only in the NS--NS sector while those with $\NN = 0 \mod 4$ are in the NS--NS sector or the R--R sector. However, this pattern is broken in general. For example, the $S$-dual of $\SUSY{F_{10,10,7,1}}$\,, namely, $\SUSY{L_{10,10,7,1}}$ ($\NN = 10$) is contained in the R--R sector.\label{ref:10247}}
For an $\OO(10,10)$ potential in the (generalized) NS--NS sector, we find the following rule:
\begin{align}
 \ovalbox{ \quad \Pbox{0.75\textwidth}{The highest-weight state in an arbitrary $\OO(10,10)$ multiplet \newline (in the NS--NS sector) corresponds to a mixed-symmetry potential with $\NN/2$ columns, $A_{m_1,\dotsc,m_{n/2}}$\,.} \quad }
\label{eq:highest-rule}
\end{align}
As we have concretely checked, this rule for all possible $\OO(10,10)$ tensors up to level $\NN=26$, we conjecture this is a universal property. 
In general, the potential $A_{m_1,\dotsc,m_{n/2}}$ corresponding to the highest weight has the smallest number of indices, and from rule \eqref{eq:index-dual2}, its dual potential $A_{10,\dotsc,10,10-m_{n/2},\dotsc,10-m_{1}}$ has the largest number of indices in that $\OO(10,10)$ multiplet. 

\item There is an additional pattern for the underlined potentials (i.e.~the ones that may couple to a supersymmetric brane). 
If we want to know whether two potentials belong to the same $\OO(10,10)$ multiplet, the following rule for $\NN\ge2$ applies:
\begin{align}
 \ovalbox{ \Pbox{0.8\textwidth}{For underlined level-$\NN$ potentials $\SUSY{A_{m_1,m_2,\cdots}}$ in the same $\OO(10,10)$ multiplet, their respective integers $\left\{\begin{matrix}
 m_{\frac{n}{2}}-m_{\frac{n+2}{2}} & (n:\text{ even})\\
 m_{\frac{n-1}{2}}-m_{\frac{n+3}{2}} & (n:\text{ odd})
\end{matrix}\right\}$ are the same.} }
\end{align}
For example, let us consider potentials $\SUSY{H_{10,10,9,7,4}}$ and $\SUSY{H_{9,9,7,3,2}}$ (with $\NN=6$). 
Since the integers are $9-7=2$ and $7-3=4$\,, they are not in the same $\OO(10,10)$ multiplet. 
If the integers are the same, in general, we cannot conclude that these potentials are in the same $\OO(10,10)$ multiplet. 
At least up to level $\NN=7$\,, we have checked that all of the level-$\NN$ $\OO(10,10)$ tensors predicted by $E_{11}$ have different values, meaning that if the above integer is the same, the mixed-symmetry potentials are in the same $\OO(10,10)$ multiplet. 
Even for $\NN>7$\,, for all of the mixed-symmetry potentials given in Table \ref{tab:IIA-IIB-list}, the integers are in one-to-one correspondence with the $\OO(10,10)$ multiplets. 

In fact, this rule is equivalent to the following rule \cite{1805.12117} under the correspondence \eqref{eq:brane-pot}:
\begin{align}
 \ovalbox{ \quad \Pbox{0.75\textwidth}{For a supersymmetric exotic brane $b_n^{(c_s,\dotsc,c_2)}$\,, an integer $\left\{\begin{matrix}
 c_{\frac{n}{2}} & (n:\text{ even})\\
 c_{\frac{n-1}{2}}+c_{\frac{n+1}{2}} & (n:\text{ odd})
\end{matrix}\right\}$ is common in an $\OO(10,10)$ multiplet.} }
\end{align}
In particular, the fact that all of the solitonic (i.e.~$\NN=2$) branes are five branes $5_2^m$ (note that $c_1= b$) can be understood from this rule. 

\item Regarding the $S$-duality, the rule is the following:
\begin{align}
 \ovalbox{ \quad \Pbox{0.68\textwidth}{Under an $S$-duality transformation, a level-$\NN$ potential $A^{[\NN]}_{p,q,r,\cdots}$ is mapped to $A^{[\ell_9-\NN]}_{p,q,r,\cdots}$ where $2\,\ell_9=p+q+r+\cdots$.} }
\label{eq:S-rule}
\end{align}
In particular, a potential with $\NN=\ell_9/2$ is a singlet. 
For an underlined potential, we find
\begin{align}
 \ovalbox{ \quad \Pbox{0.6\textwidth}{A level-$\NN$ potential $\SUSY{A^{[\NN]}_{p,q,r,\cdots}}$ is a member of $\SUSY{A^{\alpha_1\cdots\alpha_s}_{p,q,r,\cdots}}$ where $s=\abs{2\,\NN-\ell_9}$ and $2\,\ell_9=p+q+r+\cdots$.} }
\end{align}
\end{itemize}

\paragraph*{Central charges:}

In addition to mixed-symmetry potentials, we have also made a list of the central charges. 
As already explained, low-level charges \eqref{eq:low-level-Z} have been identified in the literature. 
A more detailed analysis was worked out in \cite{0805.4451}, and $U$-duality multiples and tensions of $d$-dimensional $p$-branes were studied for $d\geq 3$\,. 
There, the number of $p$-branes has been counted and a table similar to Table \ref{tab:p-form} has been made. 
However, the multiplicities of the branes have not been taken into account. 
In this paper, by using SimpLie, we have properly determined the multiplicities and the result is summarized in Table \ref{tab:p-brane}. 
\begin{table}[t]
 \centerline{\scalebox{0.82}{
 \begin{tabular}{|c|| C{0.9cm} | C{1.0cm} | C{1.5cm} |C{1.2cm}|C{1.2cm}|C{1.2cm}|C{1.6cm}|C{1.6cm}|C{0.9cm}||c|}\hline
 $d$ & $Z$ & $Z^1$ & $Z^2$ & $Z^3$ & $Z^4$ & $Z^5$ & $Z^6$ & $Z^7$ & $Z^8$ & $U$-duality \\\hline\hline
 9 & $\textcolor{red}{{\bf 2}}$\\ $\textcolor{red}{{\bf 1}}$ & $\textcolor{red}{{\bf 2}}$ & $\textcolor{red}{{\bf 1}}$ & $\textcolor{red}{{\bf 1}}$ & $\textcolor{red}{{\bf 2}}$ & $\textcolor{red}{{\bf 2}}$ \\ $\textcolor{red}{{\bf 1}}$ & $\textcolor{red}{{\bf 3}}$ \\ $2\times {\bf 1}$ & $\textcolor{red}{{\bf 3}}$ \\ $2\times {\bf 2}$ \\ ${\bf 1}$ & $\textcolor{red}{{\bf 4}}$ \\ $4\times {\bf 2}$ \\ ${\bf 1}$ & $\SL(2)$ \\\hline
 8 & $\textcolor{red}{({\bf \overline{3}},{\bf 2})}$ & $\textcolor{red}{({\bf 3},{\bf 1})}$ & $\textcolor{red}{({\bf 1},{\bf 2})}$ & $\textcolor{red}{({\bf \overline{3}},{\bf 1})}$ & $\textcolor{red}{({\bf 3},{\bf 2})}$ & $\textcolor{red}{({\bf 8},{\bf 1})}$ \\ $\textcolor{red}{({\bf 1},{\bf 3})}$ \\ $({\bf 1},{\bf 1})$ & $\textcolor{red}{({\bf 6},{\bf 2})}$ \\ $2\times({\bf \overline{3}},{\bf 2})$ & $\textcolor{red}{({\bf 15},{\bf 1})}$ \\ $({\bf \overline{6}},{\bf 1})$ \\ $2\times({\bf 3},{\bf 3})$ \\ $3\times({\bf 3},{\bf 1})$ & & $\SL(3)\times\SL(2)$ \\\hline
 7 & $\textcolor{red}{{\bf \overline{10}}}$ & $\textcolor{red}{{\bf 5}}$ & $\textcolor{red}{{\bf \overline{5}}}$ & $\textcolor{red}{{\bf 10}}$ & $\textcolor{red}{{\bf 24}}$ \\ ${\bf 1}$ & $\textcolor{red}{{\bf \overline{40}}}$ \\ $\textcolor{red}{{\bf \overline{15}}}$ \\ ${\bf \overline{10}}$ & $\textcolor{red}{{\bf 70}}$ \\ $2\times{\bf 45}$ \\ $2\times{\bf 5}$ & & & $\SL(5)$ \\\hline
 6 & $\textcolor{red}{{\bf 16}}$ & $\textcolor{red}{{\bf 10}}$ & $\textcolor{red}{{\bf \overline{16}}}$ & $\textcolor{red}{{\bf 45}}$ \\ ${\bf 1}$ & $\textcolor{red}{{\bf 144}}$ \\ ${\bf 16}$ & $\textcolor{red}{{\bf 320}}$ \\ $\textcolor{red}{{\bf 126}}$ \\ ${\bf 120}$\\ $2\times{\bf 10}$ & & & & $\SO(5,5)$ \\\hline
 5 & $\textcolor{red}{{\bf 27}}$ & $\textcolor{red}{{\bf \overline{27}}}$ & $\textcolor{red}{{\bf 78}}$ \\ ${\bf 1}$ & $\textcolor{red}{{\bf 351}}$ \\ ${\bf 27}$ & $\textcolor{red}{{\bf \overline{1728}}}$ \\ ${\bf \overline{351}}$ \\ $2\times {\bf \overline{27}}$ & & & & & $E_6$ \\\hline
 4 & $\textcolor{red}{{\bf 56}}$ & $\textcolor{red}{{\bf 133}}$ \\ ${\bf 1}$ & $\textcolor{red}{{\bf 912}}$ \\ ${\bf 56}$ & $\textcolor{red}{{\bf 8645}}$ \\ ${\bf 1539}$ \\ $2\times{\bf 133}$ \\ ${\bf 1}$ & & & & & & $E_7$ \\\hline 
 3 & $\textcolor{red}{{\bf 248}}$ & $\textcolor{red}{{\bf 3875}}$ \\ ${\bf 248}$ \\ ${\bf 1}$ & $\textcolor{red}{{\bf 147250}}$ \\ ${\bf 30380}$ \\ $2\times {\bf 3875}$ \\ $2\times {\bf 248}$ \\ ${\bf 1}$ & & & & & & & $E_8$ \\\hline
\end{tabular}}}
\caption{Number of $p$-branes in $d$ dimensions predicted by the $E_{11}$ conjecture. $U$-duality representations which contain a supersymmetric brane are marked in red. We note that the domain-wall branes are in one-to-one correspondence with the embedding tensors in the presence of the trombone gaugings \cite{0809.5180}. The space-filling branes also correspond to the quadratic constraints in the presence of the trombone gaugings \cite{0809.5180} (an exception is ${\bf 1}$ in $d=4$).}
\label{tab:p-brane}
\end{table}
Detailed contents of each representation are also given in terms of branes in M-theory/type IIB theory. 

\newpage

\subsection{Structure}

The structure of this paper is as follows. 
Section \ref{sec:E11-mixed-potentials} consists of the derivations of the mixed-symmetry potentials predicted from the $E_{11}$ conjecture and their uplifts into $\OO(D,D)$ tensors. 
First, we briefly review the main aspects of $E_{11}$ and the method to obtain the spectra of potentials from M-theory. 
Secondly, we apply the $E_{11}$ conjecture to type IIB theory. 
We demonstrate that the obtained type IIB potentials are more than enough to reproduce Table \ref{tab:p-form}. 
Finally, we find a set of $\OO(10,10)$ potentials as much as possible with current personal computers and SimpLie. 
In lower dimensions, we show that the obtained set of $\OO(D,D)$ potentials are enough to reproduce the $p$-forms in Table \ref{tab:p-form}. 
In these studies, we explicitly distinguish the potentials that may couple to supersymmetric branes by underlining them. 

In Section \ref{sec:E11-branes}, we study a set of central charges in M/IIB theory by searching the vector representation of $E_{11}$\,. 
We also study their reductions to $d$ dimensions, and explain how to determine the central charges of $p$-branes from the $E_{11}$ conjecture. 
The main results of Section \ref{sec:E11-branes} are summarized in Table \ref{tab:p-brane} and Appendices \ref{app:M-branes} and \ref{app:B-branes}.

In Section \ref{sec:conclusions}. we discuss our results and comment on some prospects.

The appendices contain the detailed results of this paper. 
In Appendix \ref{app:convention}, our conventions on Lie algebras are fixed. 
In Appendix \ref{app:M-branes}, we show the embedding of the M-theory branes into every $U$-duality multiplet in any dimension. 
In Appendix \ref{app:B-branes}, we show the embedding of the type IIB branes into every $U$-duality multiplet in any dimension. 
In Appendix \ref{app:M-potential}, we show the explicit decomposition of the M-theory potentials into the $p$-form potentials ($1\leq p \leq d$) for dimensions $3\leq d\leq 9$. 
In Appendix \ref{app:B-potential}, we show the explicit decomposition of the type IIB theory potentials into the $p$-form potentials ($1\leq p \leq d$) for dimensions $3\leq d\leq 9$. 
Appendix \ref{app:O(1010)-G-H} contains a list of the full $\OO(10,10)$ potentials with level $\NN=5,6$ and their decompositions into type II potentials. 
In Appendix \ref{app:En2ODD}, by considering the $E_n\to\OO(D,D)\to\SL(D)$ decomposition chain, we show the contents of each $U$-duality multiplet in terms of type IIA/IIB fields.
Finally, in Appendix \ref{app:BtoZ}, we obtain a full list of type II potentials that couple to supersymmetric branes up to level $\NN=36$\,. 
Since the result is too long, we explicitly show the results up to level $\NN=11$\,, and the results up to level $\NN=24$ (potential $Z$) is given in the ancillary files \cite{anc}. 
The results for higher levels can be generated by using the mathematica notebook, which is also included in the ancillary files.

\newpage

\section{Mixed-symmetry potentials from the $E_{11}$ conjecture}
\label{sec:E11-mixed-potentials}

In this section, we review the basic idea and the results of the $E_{11}$ conjecture. 
We then provide a detailed survey of the mixed-symmetry potentials. 
We also study the $\OO(D,D)$ multiplets of gauge potentials and show how the mixed-symmetry potentials in type IIA/IIB theory are embedded in each $\OO(D,D)$ multiplet. 

\subsection{Mixed-symmetry potentials and the $E_{11}$ generators}
\label{sec:E11-review}

The Kac-Moody algebra $E_{11}$\,, or the very extended $E_8$ algebra $E_8^{+++}$\,, is characterized by the Dynkin diagram,
\begin{align}
\scalebox{0.8}{
 \xygraph{
    *\cir<6pt>{} ([]!{+(0,-.4)} {\alpha_1}) - [r]
    *\cir<6pt>{} ([]!{+(0,-.4)} {\alpha_2}) - [r]
    *\cir<6pt>{} ([]!{+(0,-.4)} {\alpha_3}) - [r]
    *\cir<6pt>{} ([]!{+(0,-.4)} {\alpha_4}) - [r]
    *\cir<6pt>{} ([]!{+(0,-.4)} {\alpha_5}) - [r]
    *\cir<6pt>{} ([]!{+(0,-.4)} {\alpha_6}) - [r]
    *\cir<6pt>{} ([]!{+(0,-.4)} {\alpha_7}) - [r]
    *\cir<6pt>{} ([]!{+(0,-.4)} {\alpha_8}) 
(
        - [u] *\cir<6pt>{} ([]!{+(.5,0)} {\alpha_{11}}),
        - [r] *\cir<6pt>{} ([]!{+(0,-.4)} {\alpha_9})
        - [r] *\cir<6pt>{} ([]!{+(0,-.4)} {\alpha_{10}})
) 
}} .
\label{eq:E11-dynkin}
\end{align}
We denote the Cartan matrix as
\begin{align}
 (A_{ij})= {\footnotesize\begin{pmatrix}
  2 & -1 & & & & \\
  -1 & 2 & \ddots & & & \\
  & \ddots & \ddots & -1 & & -1 \\
  & & -1 & 2 & -1 & 0 \\
  & & & -1 & 2 & 0 \\
  & & -1 & 0 & 0 & 2
\end{pmatrix}},
\end{align}
and the commutation relations for the Chevalley generators $\{H_i,\,E_i,\,F_i\}$ are given by
\begin{align}
 [H_i,\,E_j]=A_{ij}\,E_j\,,\qquad
 [H_i,\,F_j]=-A_{ij}\,F_j\,,\qquad
 [E_i,\,F_j]= \delta_{ij}\,H_j\,, 
\end{align}
which lead to $[H_i,\,H_j]=0$\,. 
By taking the commutators of the simple root generators $E_i$ or $F_i$\,, under the Serre relation,
\begin{align}
 \underbrace{[E_i,[E_i,\cdots [E_i}_{1-A_{ij}},\,E_j]\cdots]]=0\,,\qquad
 \underbrace{[F_i,[F_i,\cdots [F_i}_{1-A_{ij}},\,F_j]\cdots]]=0\,, 
\end{align}
we can uniquely determine the algebra for the infinitely many $E_{11}$ generators. 
An important property of $E_{11}$ is that, if we split the Dynkin diagram by deleting one of the nodes $\alpha_i$ ($3\leq i\leq 11$)\,, the $E_{11}$ decomposes into products of finite dimensional algebras. 
By using this property, it is useful to introduce an ordering, called ``level'' \cite{hep-th:0212291} (see also \cite{hep-th:0207267}). 

Here, we consider the level decomposition by deleting the node $\alpha_{11}$\,, which gives the $\SL(11)$ subalgebra associated with simple roots $\alpha_i$ ($i=1,\dotsc,10$)\,. 
\begin{align}
\scalebox{0.8}{
 \xygraph{
    *\cir<6pt>{} ([]!{+(0,-.4)} {\alpha_1}) - [r]
    *\cir<6pt>{} ([]!{+(0,-.4)} {\alpha_2}) - [r]
    *\cir<6pt>{} ([]!{+(0,-.4)} {\alpha_3}) - [r]
    *\cir<6pt>{} ([]!{+(0,-.4)} {\alpha_4}) - [r]
     \cdots 
    - [r] *\cir<6pt>{} ([]!{+(0,-.4)} {\alpha_7}) 
    - [r] *\cir<6pt>{} ([]!{+(0,-.4)} {\alpha_8}) 
(
        - [u] *{\scalebox{2}{$\times$}}*\cir<6pt>{} ([]!{+(.5,0)} {\alpha_{11}}),
        - [r] *\cir<6pt>{} ([]!{+(0,-.4)} {\alpha_9})
        - [r] *\cir<6pt>{} ([]!{+(0,-.4)} {\alpha_{10}})
) 
}} .
\end{align}
As usual, for the $\SL(11)$ algebra, we introduce the metric
\begin{align}
 &\alpha_i\cdot\alpha_j = a_{ij}\,,\quad 
 (a_{ij})\equiv {\footnotesize\begin{pmatrix}
  2 & -1 & & & \\
  -1 & 2 & \ddots & & \\
  & \ddots & \ddots & -1 & \\
  & & -1 & 2 & -1 \\
  & & & -1 & 2 
\end{pmatrix}},
\\
 &a^{ij} \equiv (a^{-1})^{ij}= \frac{i\,(11-j)}{11} =a^{ji}\quad (i\leq j)\,,
\end{align}
and introduce the fundamental weights as
\begin{align}
 \lambda^i\equiv a^{ij}\,\alpha_j \qquad (\lambda^i\cdot\alpha_j=\delta^i_j)\,.
\end{align}
By using the $\SL(11)$ weights $\lambda^i$ (and recalling $\alpha_8\cdot \alpha_{11}=-1$), the root $\alpha_{11}$ of $E_{11}$ can be expressed as
\begin{align}
 \alpha_{11} = x -\lambda^8 \,,
\end{align}
where $x$ is a certain vector orthogonal to $\lambda^i$ and $\alpha_i$ ($\lambda^i\cdot x=\alpha_i\cdot x=0$). 
From $\alpha_{11}\cdot \alpha_{11}=2$\,, we obtain $x\cdot x = -\frac{2}{11}$\,. 
Then, a general positive/negative root $\alpha$ of $E_{11}$ is expressed as
\begin{align}
 \alpha = \ell\,\alpha_{11} \pm m^i\,\alpha_i 
 = \ell\,x \pm \bigl( m^i\,a_{ij}\,\lambda^j \mp \ell\,\lambda^8\bigr) \,,
\label{eq:general-alpha}
\end{align}
where $m^i$ are non-negative integers and $\ell$ is a non-negative/non-positive integer. 
This integer $\ell$ is called the level associated with the node $\alpha_{11}$\,. 

The generators that are obtained by taking commutators with the simple roots $E_i$, $F_i$ ($i=\mathsf{1},\dotsc,\mathsf{10}$), and the Cartan generators $H_i$ have the level $\ell=0$\,. 
They are nothing but the $\SL(11)$ generators and an $\SL(11)$ singlet $H_{\mathsf{11}}$\,, which are combined to give the $\GL(11)$ generators $K^i{}_j$\,.
By taking a commutator with $E_{\mathsf{11}}$ (or $F_{\mathsf{11}}$), the level is increased (or decreased) by 1, and we obtain a set of generators with $\ell=\pm 1$\,. 
It is important to note that, since the generators with a fixed level $\ell$ form some finite-dimensional representation of $\SL(11)$\,, we can always decompose the $E_{11}$ generators at each level into some irreducible representations of $\SL(11)$\,. 
If we denote the Dynkin label of the irreducible representations as $[p_1,\dotsc,p_{10}]$, from \eqref{eq:general-alpha}, the $E_{11}$ root $\alpha$ associated with the highest-weight state should satisfy
\begin{align}
 p_k\,a^{ki} = \lambda^i\cdot \alpha = \pm \bigl( m^i \mp \ell\, a^{8i} \bigr)\,.
\end{align}
This can be expressed more concretely as
\begin{align}
 \sum_{k\leq i} p_k\, k\,(11-i) + \sum_{k>i} p_k\, i\,(11-k) = \pm 11\,m^i - \ell\times\begin{cases}
 8\, (11-i) & (8\leq i) \\
 3\, i & (i<8) 
\end{cases}\,,
\end{align}
and this should be satisfied by the non-negative integers $p_i$ and $m^i$\,. 
The left-hand side is positive, and by considering the lower sign (where $\ell\leq 0$), it is obvious that there are only a finite number of solutions for non-negative integers $m^i$\,. 
Moreover, since roots of the $E_{11}$ algebra have the norm, $\alpha\cdot\alpha=2,0,-2,-4,\dotsc$\,, the integers $p_i$ should satisfy the condition
\begin{align}
 p_i\,a^{ij}\,p_j - \frac{2}{11}\, \ell^2 =2,\,0\,,-2,\,-4,\dotsc\,. 
\end{align}
Solutions to these conditions at low level were found in \cite{hep-th:0212291},\footnote{At level 3, $[0,1,0,0,0,0,0,0,0,0] \ \leftrightarrow \ R^{i_1\cdots i_9}$ is also a solution to the conditions, but by further considering the Jacobi identities, it was shown that it does not appear in the $E_{11}$ algebra.}
\begin{align}
\begin{split}
 \ell=1:&\quad [0,0,0,0,0,0,0,1,0,0] \ \leftrightarrow \ R^{i_1i_2i_3}\,,
\\
 \ell=2:&\quad [0,0,0,0,1,0,0,0,0,0] \ \leftrightarrow \ R^{i_1\cdots i_6}\,,
\\
 \ell=3:&\quad [0,0,1,0,0,0,0,0,0,1] \ \leftrightarrow \ R^{i_1\cdots i_8,i}\,.
\end{split}
\end{align}
A more detailed analysis was done in \cite{hep-th:0301017} with the help of a computer program, and all $E_{11}$ generators up to level 10 were determined. 
The negative-level generators have the lower indices; for example, the level $-1$ generators are $R_{i_1i_2i_3}$\,. 
The commutation relations between the low-level generators were determined in \cite{hep-th:0307098},
\begin{align}
 \bigl[K^i{}_j,\, K^k{}_l \bigr] 
 = \delta_j^k\,K^i{}_l - \delta_l^i\, K^k{}_j \,, \qquad
 \bigl[K^i{}_j,\, R^{k_1k_2k_3} \bigr] = \frac{1}{2!}\,\bdelta_{jr_1r_2}^{k_1k_2k_3} \,R^{ir_1r_2}\,, \cdots\,,
\end{align}
where $\bdelta_{i_1\cdots i_m}^{j_1\cdots j_m}\equiv m!\,\delta_{[i_1}^{[j_1}\cdots \delta_{i_m]}^{j_m]}$\,.
The Chevalley generators $\{H_i,\,E_i,\,F_i\}$ are chosen as
\begin{align}
\begin{split}
 \{H_i\} &= \bigl\{ K^{\mathsf{0}}{}_{\mathsf{0}} - K^{\mathsf{1}}{}_{\mathsf{1}},\, \dotsc,\, K^{\mathsf{9}}{}_{\mathsf{9}} - K^z{}_z,\,K^{\mathsf{8}}{}_{\mathsf{8}} + K^{\mathsf{9}}{}_{\mathsf{9}} + K^z{}_z - \tfrac{1}{3}\,K^i{}_i \bigr\} \,,
\\
 \{E_i\} &= \{K^{\mathsf{0}}{}_{\mathsf{1}} ,\, \dotsc,\, K^{\mathsf{9}}{}_z,\, R^{\mathsf{8}\mathsf{9}z} \} \,,\qquad
 \{F_i\} = \{K^{\mathsf{1}}{}_{\mathsf{0}} ,\, \dotsc,\, K^z{}_{\mathsf{9}},\,R_{\mathsf{8}\mathsf{9}z} \} \,,
\end{split}
\end{align}
where $z=\mathsf{10}$ corresponds to the M-theory direction. 
We note that the raising/lowering operator associated with the deleted node is $E_{\mathsf{11}}=R^{\mathsf{8}\mathsf{9}z}$/$F_{\mathsf{11}}=R_{\mathsf{8}\mathsf{9}z}$\,, which has three upper/lower indices.
Accordingly, every time we take a commutator with the generator, the number of the upper indices (on $E_{11}$ generators) is increased/decreased by three, and the level $\ell$ satisfies
\begin{align}
 3\,\ell = \text{(\# of the upper indices)} - \text{(\# of the lower indices)}\,. 
\end{align}

As mentioned in the introduction, the $E_{11}$ generators are expected to be in one-to-one correspondence with the gauge potentials. 
By using the results of \cite{hep-th:0307098} (which can be reproduced by using SimpLie), the mixed-symmetry potentials corresponding to generators up to level 10 can be summarized as in Table \ref{table:M-level10}.
\begin{table}[t]
{\scriptsize\centerline{
\begin{tabular}{|c|l|}
 \hline
 {\footnotesize$\ell\vphantom{\Bigl|}$} & \multicolumn{1}{c|}{{\footnotesize Mixed-symmetry potentials in M-theory up to level $10$}} \\\hline\hline
%%%%%%%%%%%%%%%%%%%%%%%%%%
$1$ & \Pbox{\textwidth}{
$\SUSY{\hat{A}_{3}}$}
\\\hline
$2$ & \Pbox{\textwidth}{
$\SUSY{\hat{A}_{6}}$}
\\\hline
$3$ & \Pbox{\textwidth}{
$\SUSY{\hat{A}_{8,1}}$}
\\\hline
$4$ & \Pbox{\textwidth}{
$\SUSY{\hat{A}_{9,3}}$, 
$\SUSY{\hat{A}_{10,1,1}}$, 
$\hat{A}_{11,1}$}
\\\hline
$5$ & \Pbox{\textwidth}{
$\SUSY{\hat{A}_{9,6}}$, 
$\SUSY{\hat{A}_{10,4,1}}$, 
$\hat{A}_{11,3,1}$, 
$\hat{A}_{11,4}$}
\\\hline
$6$ & \Pbox{\textwidth}{
$\SUSY{\hat{A}_{9,8,1}}$, 
$\SUSY{\hat{A}_{10,6,2}}$, 
$\hat{A}_{10,7,1}$, 
$\hat{A}_{10,8}$, 
$\SUSY{\hat{A}_{11,4,3}}$, 
$\SUSY{\hat{A}_{11,5,1,1}}$, 
$2\,\,\hat{A}_{11,6,1}$, 
$\hat{A}_{11,7}$}
\\\hline
$7$ & \Pbox{\textwidth}{
$\SUSY{\hat{A}_{9,9,3}}$, 
$\SUSY{\hat{A}_{10,7,4}}$, 
$\SUSY{\hat{A}_{10,8,2,1}}$, 
$\hat{A}_{10,8,3}$, 
$\hat{A}_{10,9,1,1}$, 
$2\,\,\hat{A}_{10,9,2}$, 
$\hat{A}_{10,10,1}$, 
$\SUSY{\hat{A}_{11,6,3,1}}$, 
$\hat{A}_{11,6,4}$, 
$\hat{A}_{11,7,2,1}$, 
$2\,\,\hat{A}_{11,7,3}$, 
$3\,\,\hat{A}_{11,8,1,1}$, 
$3\,\,\hat{A}_{11,8,2}$, 
$4\,\,\hat{A}_{11,9,1}$, 
$\hat{A}_{11,10}$}
\\\hline
$8$ & \Pbox{\textwidth}{
$\SUSY{\hat{A}_{9,9,6}}$, 
$\SUSY{\hat{A}_{10,7,7}}$, 
$\SUSY{\hat{A}_{10,8,5,1}}$, 
$\hat{A}_{10,8,6}$, 
$\SUSY{\hat{A}_{10,9,3,2}}$, 
$2\,\,\hat{A}_{10,9,4,1}$, 
$2\,\,\hat{A}_{10,9,5}$, 
$\SUSY{\hat{A}_{10,10,2,1,1}}$, 
$2\,\,\hat{A}_{10,10,3,1}$, 
$2\,\,\hat{A}_{10,10,4}$, 
$\SUSY{\hat{A}_{11,6,6,1}}$, 
$\SUSY{\hat{A}_{11,7,4,2}}$, 
$\hat{A}_{11,7,5,1}$, 
$2\,\,\hat{A}_{11,7,6}$, 
$\SUSY{\hat{A}_{11,8,3,1,1}}$, 
$\hat{A}_{11,8,3,2}$, 
$4\,\,\hat{A}_{11,8,4,1}$, 
$3\,\,\hat{A}_{11,8,5}$, 
$\hat{A}_{11,9,2,1,1}$, 
$2\,\,\hat{A}_{11,9,2,2}$, 
$6\,\,\hat{A}_{11,9,3,1}$, 
$7\,\,\hat{A}_{11,9,4}$, 
$2\,\,\hat{A}_{11,10,1,1,1}$, 
$7\,\,\hat{A}_{11,10,2,1}$, 
$6\,\,\hat{A}_{11,10,3}$, 
$3\,\,\hat{A}_{11,11,1,1}$, 
$5\,\,\hat{A}_{11,11,2}$}
\\\hline
$9$ & \Pbox{\textwidth}{
$\SUSY{\hat{A}_{9,9,8,1}}$, 
$\SUSY{\hat{A}_{10,8,7,2}}$, 
$\hat{A}_{10,8,8,1}$, 
$\SUSY{\hat{A}_{10,9,5,3}}$, 
$\SUSY{\hat{A}_{10,9,6,1,1}}$, 
$2\,\,\hat{A}_{10,9,6,2}$, 
$4\,\,\hat{A}_{10,9,7,1}$, 
$3\,\,\hat{A}_{10,9,8}$, 
$\SUSY{\hat{A}_{10,10,4,2,1}}$, 
$\hat{A}_{10,10,4,3}$, 
$\hat{A}_{10,10,5,1,1}$, 
$3\,\,\hat{A}_{10,10,5,2}$, 
$4\,\,\hat{A}_{10,10,6,1}$, 
$4\,\,\hat{A}_{10,10,7}$, 
$\SUSY{\hat{A}_{11,7,6,3}}$, 
$\hat{A}_{11,7,7,2}$, 
$\SUSY{\hat{A}_{11,8,4,4}}$, 
$\SUSY{\hat{A}_{11,8,5,2,1}}$, 
$\hat{A}_{11,8,5,3}$, 
$2\,\,\hat{A}_{11,8,6,1,1}$, 
$4\,\,\hat{A}_{11,8,6,2}$, 
$6\,\,\hat{A}_{11,8,7,1}$, 
$2\,\,\hat{A}_{11,8,8}$, 
$\SUSY{\hat{A}_{11,9,3,3,1}}$, 
$2\,\,\hat{A}_{11,9,4,2,1}$, 
$4\,\,\hat{A}_{11,9,4,3}$, 
$4\,\,\hat{A}_{11,9,5,1,1}$, 
$7\,\,\hat{A}_{11,9,5,2}$, 
$13\,\,\hat{A}_{11,9,6,1}$, 
$9\,\,\hat{A}_{11,9,7}$, 
$\SUSY{\hat{A}_{11,10,3,1,1,1}}$, 
$3\,\,\hat{A}_{11,10,3,2,1}$, 
$3\,\,\hat{A}_{11,10,3,3}$, 
$7\,\,\hat{A}_{11,10,4,1,1}$, 
$11\,\,\hat{A}_{11,10,4,2}$, 
$16\,\,\hat{A}_{11,10,5,1}$, 
$12\,\,\hat{A}_{11,10,6}$, 
$2\,\,\hat{A}_{11,11,2,2,1}$, 
$7\,\,\hat{A}_{11,11,3,1,1}$, 
$7\,\,\hat{A}_{11,11,3,2}$, 
$15\,\,\hat{A}_{11,11,4,1}$, 
$8\,\,\hat{A}_{11,11,5}$}
\\\hline
$10$ & \Pbox{\textwidth}{
$\SUSY{\hat{A}_{9,9,9,3}}$, 
$\SUSY{\hat{A}_{10,8,8,4}}$, 
$\SUSY{\hat{A}_{10,9,6,5}}$, 
$\SUSY{\hat{A}_{10,9,7,3,1}}$, 
$2\,\,\hat{A}_{10,9,7,4}$, 
$3\,\,\hat{A}_{10,9,8,2,1}$, 
$4\,\,\hat{A}_{10,9,8,3}$, 
$3\,\,\hat{A}_{10,9,9,1,1}$, 
$4\,\,\hat{A}_{10,9,9,2}$, 
$\SUSY{\hat{A}_{10,10,5,4,1}}$, 
$\SUSY{\hat{A}_{10,10,6,2,2}}$, 
$2\,\,\hat{A}_{10,10,6,3,1}$, 
$3\,\,\hat{A}_{10,10,6,4}$, 
$\SUSY{\hat{A}_{10,10,7,1,1,1}}$, 
$4\,\,\hat{A}_{10,10,7,2,1}$, 
$6\,\,\hat{A}_{10,10,7,3}$, 
$4\,\,\hat{A}_{10,10,8,1,1}$, 
$10\,\,\hat{A}_{10,10,8,2}$, 
$8\,\,\hat{A}_{10,10,9,1}$, 
$3\,\,\hat{A}_{10,10,10}$, 
$\SUSY{\hat{A}_{11,7,7,5}}$, 
$\SUSY{\hat{A}_{11,8,6,4,1}}$, 
$\hat{A}_{11,8,6,5}$, 
$\SUSY{\hat{A}_{11,8,7,2,2}}$, 
$2\,\,\hat{A}_{11,8,7,3,1}$, 
$4\,\,\hat{A}_{11,8,7,4}$, 
$\SUSY{\hat{A}_{11,8,8,1,1,1}}$, 
$4\,\,\hat{A}_{11,8,8,2,1}$, 
$4\,\,\hat{A}_{11,8,8,3}$, 
$\SUSY{\hat{A}_{11,9,5,3,2}}$, 
$2\,\,\hat{A}_{11,9,5,4,1}$, 
$2\,\,\hat{A}_{11,9,5,5}$, 
$\SUSY{\hat{A}_{11,9,6,2,1,1}}$, 
$\hat{A}_{11,9,6,2,2}$, 
$7\,\,\hat{A}_{11,9,6,3,1}$, 
$8\,\,\hat{A}_{11,9,6,4}$, 
$\hat{A}_{11,9,7,1,1,1}$, 
$11\,\,\hat{A}_{11,9,7,2,1}$, 
$17\,\,\hat{A}_{11,9,7,3}$, 
$14\,\,\hat{A}_{11,9,8,1,1}$, 
$21\,\,\hat{A}_{11,9,8,2}$, 
$17\,\,\hat{A}_{11,9,9,1}$, 
$\SUSY{\hat{A}_{11,10,4,3,1,1}}$, 
$2\,\,\hat{A}_{11,10,4,3,2}$, 
$4\,\,\hat{A}_{11,10,4,4,1}$, 
$2\,\,\hat{A}_{11,10,5,2,1,1}$, 
$3\,\,\hat{A}_{11,10,5,2,2}$, 
$10\,\,\hat{A}_{11,10,5,3,1}$, 
$11\,\,\hat{A}_{11,10,5,4}$, 
$4\,\,\hat{A}_{11,10,6,1,1,1}$, 
$20\,\,\hat{A}_{11,10,6,2,1}$, 
$23\,\,\hat{A}_{11,10,6,3}$, 
$22\,\,\hat{A}_{11,10,7,1,1}$, 
$37\,\,\hat{A}_{11,10,7,2}$, 
$36\,\,\hat{A}_{11,10,8,1}$, 
$16\,\,\hat{A}_{11,10,9}$, 
$\hat{A}_{11,11,3,3,1,1}$, 
$2\,\,\hat{A}_{11,11,3,3,2}$, 
$3\,\,\hat{A}_{11,11,4,2,1,1}$, 
$2\,\,\hat{A}_{11,11,4,2,2}$, 
$11\,\,\hat{A}_{11,11,4,3,1}$, 
$7\,\,\hat{A}_{11,11,4,4}$, 
$4\,\,\hat{A}_{11,11,5,1,1,1}$, 
$19\,\,\hat{A}_{11,11,5,2,1}$, 
$21\,\,\hat{A}_{11,11,5,3}$, 
$24\,\,\hat{A}_{11,11,6,1,1}$, 
$31\,\,\hat{A}_{11,11,6,2}$, 
$38\,\,\hat{A}_{11,11,7,1}$, 
$16\,\,\hat{A}_{11,11,8}$}
\\\hline
\end{tabular}}}
\caption{A list of mixed-symmetry potentials in M-theory up to level 10. The integer in front of the potential represents the multiplicity and the underline represents that the root $\alpha$ associated $E_{11}$ generator has the maximal length squared $\alpha\cdot\alpha=2$\,. 
\label{table:M-level10}}
\end{table}
Given the set of mixed-symmetry potentials in 11D, we can find what kind of $p$-form can appear in lower dimensions after toroidal compactification. 
For example, let us consider the compactification to $d=4$ of the potential $A_{8,1}$, which corresponds to $\overline{\bf 1760}$ of $\SL(11)$. 
We can decompose it into representations of $\SL(4)\times \SL(7)$ as
\begin{align}
 \underset{(\overline{\bf 1760})}{A_{8,1}} &\to 
 \underset{({\bf 4},\,{\bf 7})}{A_{1;7,1}}
 + \underset{({\bf 6},\,{\bf 48})}{A_{2;6,1}}
 + \underset{({\bf 6},\,{\bf 1})}{A_{2;7}}
 + \underset{(\overline{\bf 4},\,\overline{\bf 140})}{A_{3;5,1}}
 + \underset{(\overline{\bf 4},\,\overline{\bf 7})}{A_{3;6}}
 + \underset{({\bf 1},\,\overline{\bf 224})}{A_{4;4,1}}
 + \underset{({\bf 1},\,\overline{\bf 21})}{A_{4;5}}
\nn\\
 &\quad
 + \underset{({\bf 10},\,{\bf 1})}{A_{1,1;7}}
 + \underset{(\overline{\bf 20},\,\overline{\bf 7})}{A_{2,1;6}}
 + \underset{({\bf 15},\,\overline{\bf 21})}{A_{3,1;5}}
 + \underset{({\bf 4},\,\overline{\bf 35})}{A_{4,1;4}}\,,
\end{align}
where the indices before the semicolon represent the external $\SL(4)$ representation while those after the semicolon represent the internal $\SL(7)$ representation. 
Usually, we do not consider the mixed-symmetry potentials in the external space because their interpretation is not clear. 
We thus consider only the $p$-form fields appearing in the first line. 
All of the other mixed-symmetry potentials in 11D can also be similarly decomposed. 
As worked out in \cite{0705.0752}, the number of $p$-form fields in $d$ dimensions coming from the mixed-symmetry potentials can be summarized, as in Table \ref{tab:p-form}. 

In fact, in order to obtain Table \ref{tab:p-form}, it is more efficient to delete a node associated with $\alpha_{d}$ instead of $\alpha_{11}$\,:
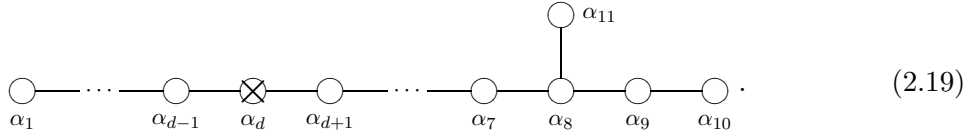
\begin{align}
\scalebox{0.8}{
 \xygraph{
    *\cir<6pt>{} ([]!{+(0,-.4)} {\alpha_1}) - [r]
    \cdots - [r]
    *\cir<6pt>{} ([]!{+(0,-.4)} {\alpha_{d-1}}) - [r]
    *{\scalebox{2}{$\times$}}*\cir<6pt>{} ([]!{+(0,-.4)} {\alpha_d}) - [r]
    *\cir<6pt>{} ([]!{+(0,-.4)} {\alpha_{d+1}}) - [r]
    \cdots 
    - [r] *\cir<6pt>{} ([]!{+(0,-.4)} {\alpha_7}) 
    - [r] *\cir<6pt>{} ([]!{+(0,-.4)} {\alpha_8}) 
(
        - [u] *\cir<6pt>{} ([]!{+(.5,0)} {\alpha_{11}}),
        - [r] *\cir<6pt>{} ([]!{+(0,-.4)} {\alpha_9})
        - [r] *\cir<6pt>{} ([]!{+(0,-.4)} {\alpha_{10}})
) 
}} .
\label{eq:level-d}
\end{align}
This decomposes $E_{11}$ into $\SL(d)\times E_n$\,, where the $\SL(d)$ acting on the external $d$-dimensional spacetime with coordinates $x^\mu$ ($\mu=0,1,\dotsc,d-1$) while $E_n$ is the $U$-duality group. 
Since the raising/lowering operator $E_{d}=K^{d-1}{}_{d}$/$F_{d}=K^d{}_{d-1}$ has one upper/lower external index, the level $\ell_d$ associated with the node $\alpha_{d}$ satisfies
\begin{align}
 \ell_d = \text{(\# of the external upper indices)} - \text{(\# of the external lower indices)}\,. 
\end{align}
Thus, the $p$-form fields in the external spacetime are contained in the $E_{11}$ generators with level $\ell_d=p$\,. 
By taking this into account, all of the $E_{11}$ generators up to level $d$ were determined in \cite{0705.1304}. 
By collecting the $p$-form representations of the $\SL(d)$ subalgebra (with Dynkin index $[\overset{(d-1)\text{th}}{0},\dotsc,0,\overset{p\text{th}}{1},0,\cdots,\overset{1\text{st}}{0}]$), the number of the $p$-form fields predicted by the $E_{11}$ conjecture was determined, even for $d=3$ and $p=3$ as in Table \ref{tab:p-form}. 

By further deleting the node $\alpha_{11}$\,, we can decompose the $E_{n}$ representations into $\SL(n)$ tensors. 
The details are summarized in Appendix \ref{app:En2ODD}. 

\subsection{Treatment of type IIB theory}

In the previous subsection, we obtained mixed-symmetry tensors in M-theory by decomposing the adjoint representation of $E_{11}$ by means of the level $\ell$ associated with the node $\alpha_{11}$\,. 
On the other hand, we can obtain the mixed-symmetry tensors in type IIB theory by deleting the node $\alpha_{9}$\,. 
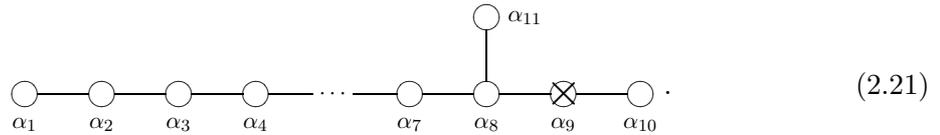
\begin{align}
\scalebox{0.8}{
 \xygraph{
    *\cir<6pt>{} ([]!{+(0,-.4)} {\alpha_1}) - [r]
    *\cir<6pt>{} ([]!{+(0,-.4)} {\alpha_2}) - [r]
    *\cir<6pt>{} ([]!{+(0,-.4)} {\alpha_3}) - [r]
    *\cir<6pt>{} ([]!{+(0,-.4)} {\alpha_4}) - [r]
    \cdots 
    - [r] *\cir<6pt>{} ([]!{+(0,-.4)} {\alpha_7}) 
    - [r] *\cir<6pt>{} ([]!{+(0,-.4)} {\alpha_8}) 
(
        - [u] *\cir<6pt>{} ([]!{+(.5,0)} {\alpha_{11}}),
        - [r] *{\scalebox{2}{$\times$}}*\cir<6pt>{} ([]!{+(0,-.4)} {\alpha_9})
        - [r] *\cir<6pt>{} ([]!{+(0,-.4)} {\alpha_{10}})
) 
}} .
\end{align}
In this case, the adjoint representation of $E_{11}$ is decomposed into representations of $\SL(10)\times \SL(2)$\,, where the $\SL(2)$ is the standard $\SL(2)$ $S$-duality symmetry in type IIB theory. 
We denote the level associated with the deleted node as $\ell_9$\,, and the potentials with level $1\leq \ell_9\leq 14$ are summarized in Table \ref{tab:IIB-potentials-14}.
\begin{table}
{\tiny
\centerline{
\begin{tabular}{|c|l|}
 \hline
 {\footnotesize$\ell_9\vphantom{\Bigl|}$} & \multicolumn{1}{c|}{{\footnotesize mixed-symmetry potentials in type IIB theory}} \\ \hline\hline
%%%%%%%%%%%%%%%%%%%%%%%%%%
$1$ & \Pbox{\textwidth}{
$\SUSY{A_{2}^{\alpha}}$}
\\\hline
$2$ & \Pbox{\textwidth}{
$\SUSY{A_{4}}$}
\\\hline
$3$ & \Pbox{\textwidth}{
$\SUSY{A_{6}^{\alpha}}$}
\\\hline
$4$ & \Pbox{\textwidth}{
$\SUSY{A_{7,1}}$, 
$\SUSY{A_{8}^{\alpha\beta}}$}
\\\hline
$5$ & \Pbox{\textwidth}{
$\SUSY{A_{8,2}^{\alpha}}$, 
$A_{9,1}^{\alpha}$, 
$\SUSY{A_{10}^{\alpha\beta\gamma}}$, 
$A_{10}^{\alpha}$}
\\\hline
$6$ & \Pbox{\textwidth}{
$\SUSY{A_{8,4}}$, 
$\SUSY{A_{9,2,1}}$, 
$\SUSY{A_{9,3}^{\alpha\beta}}$, 
$A_{10,2}^{\alpha\beta}$, 
$2\,\,A_{10,2}$}
\\\hline
$7$ & \Pbox{\textwidth}{
$\SUSY{A_{8,6}^{\alpha}}$, 
$\SUSY{A_{9,4,1}^{\alpha}}$, 
$A_{9,5}^{\alpha}$, 
$\SUSY{A_{10,2,2}^{\alpha}}$, 
$A_{10,3,1}^{\alpha}$, 
$\SUSY{A_{10,4}^{\alpha\beta\gamma}}$, 
$2\,\,A_{10,4}^{\alpha}$}
\\\hline
$8$ & \Pbox{\textwidth}{
$\SUSY{A_{8,7,1}}$, 
$\SUSY{A_{8,8}^{\alpha\beta}}$, 
$\SUSY{A_{9,5,2}}$, 
$\SUSY{A_{9,6,1}^{\alpha\beta}}$, 
$A_{9,6,1}$, 
$A_{9,7}^{\alpha\beta}$, 
$2\,\,A_{9,7}$, 
$\SUSY{A_{10,4,1,1}}$, 
$\SUSY{A_{10,4,2}^{\alpha\beta}}$, 
$A_{10,4,2}$, 
$A_{10,5,1}^{\alpha\beta}$, 
$2\,\,A_{10,5,1}$, 
$3\,\,A_{10,6}^{\alpha\beta}$, 
$2\,\,A_{10,6}$}
\\\hline
$9$ & \Pbox{\textwidth}{
$\SUSY{A_{8,8,2}^{\alpha}}$, 
$\SUSY{A_{9,6,3}^{\alpha}}$, 
$\SUSY{A_{9,7,1,1}^{\alpha}}$, 
$2\,\,A_{9,7,2}^{\alpha}$, 
$\SUSY{A_{9,8,1}^{\alpha\beta\gamma}}$, 
$3\,\,A_{9,8,1}^{\alpha}$, 
$2\,\,A_{9,9}^{\alpha}$, 
$\SUSY{A_{10,4,4}^{\alpha}}$, 
$\SUSY{A_{10,5,2,1}^{\alpha}}$, 
$A_{10,5,3}^{\alpha}$, 
$A_{10,6,1,1}^{\alpha}$, 
$\SUSY{A_{10,6,2}^{\alpha\beta\gamma}}$, 
$4\,\,A_{10,6,2}^{\alpha}$, 
$A_{10,7,1}^{\alpha\beta\gamma}$, 
$6\,\,A_{10,7,1}^{\alpha}$, 
$3\,\,A_{10,8}^{\alpha\beta\gamma}$, 
$6\,\,A_{10,8}^{\alpha}$}
\\\hline
$10$ & \Pbox{\textwidth}{
$\SUSY{A_{8,8,4}}$, 
$\SUSY{A_{9,6,5}}$, 
$\SUSY{A_{9,7,3,1}}$, 
$\SUSY{A_{9,7,4}^{\alpha\beta}}$, 
$A_{9,7,4}$, 
$\SUSY{A_{9,8,2,1}^{\alpha\beta}}$, 
$2\,\,A_{9,8,2,1}$, 
$2\,\,A_{9,8,3}^{\alpha\beta}$, 
$3\,\,A_{9,8,3}$, 
$A_{9,9,1,1}^{\alpha\beta}$, 
$2\,\,A_{9,9,1,1}$, 
$3\,\,A_{9,9,2}^{\alpha\beta}$, 
$2\,\,A_{9,9,2}$, 
$\SUSY{A_{10,5,4,1}}$, 
$\SUSY{A_{10,6,2,2}}$, 
$\SUSY{A_{10,6,3,1}^{\alpha\beta}}$, 
$A_{10,6,3,1}$, 
$2\,\,A_{10,6,4}^{\alpha\beta}$, 
$3\,\,A_{10,6,4}$, 
$\SUSY{A_{10,7,1,1,1}}$, 
$2\,\,A_{10,7,2,1}^{\alpha\beta}$, 
$4\,\,A_{10,7,2,1}$, 
$4\,\,A_{10,7,3}^{\alpha\beta}$, 
$5\,\,A_{10,7,3}$, 
$3\,\,A_{10,8,1,1}^{\alpha\beta}$, 
$3\,\,A_{10,8,1,1}$, 
$\SUSY{A_{10,8,2}^{\alpha_{1\cdots 4}}}$, 
$8\,\,A_{10,8,2}^{\alpha\beta}$, 
$10\,\,A_{10,8,2}$, 
$A_{10,9,1}^{\alpha_{1\cdots 4}}$, 
$9\,\,A_{10,9,1}^{\alpha\beta}$, 
$8\,\,A_{10,9,1}$, 
$2\,\,A_{10,10}^{\alpha_{1\cdots 4}}$, 
$4\,\,A_{10,10}^{\alpha\beta}$, 
$5\,\,A_{10,10}$}
\\\hline
$11$ & \Pbox{\textwidth}{
$\SUSY{A_{8,8,6}^{\alpha}}$, 
$\SUSY{A_{9,7,5,1}^{\alpha}}$, 
$2\,\,A_{9,7,6}^{\alpha}$, 
$\SUSY{A_{9,8,3,2}^{\alpha}}$, 
$3\,\,A_{9,8,4,1}^{\alpha}$, 
$\SUSY{A_{9,8,5}^{\alpha\beta\gamma}}$, 
$4\,\,A_{9,8,5}^{\alpha}$, 
$\SUSY{A_{9,9,2,1,1}^{\alpha}}$, 
$A_{9,9,2,2}^{\alpha}$, 
$\SUSY{A_{9,9,3,1}^{\alpha\beta\gamma}}$, 
$4\,\,A_{9,9,3,1}^{\alpha}$, 
$A_{9,9,4}^{\alpha\beta\gamma}$, 
$5\,\,A_{9,9,4}^{\alpha}$, 
$\SUSY{A_{10,6,4,2}^{\alpha}}$, 
$2\,\,A_{10,6,5,1}^{\alpha}$, 
$\SUSY{A_{10,6,6}^{\alpha\beta\gamma}}$, 
$3\,\,A_{10,6,6}^{\alpha}$, 
$\SUSY{A_{10,7,3,1,1}^{\alpha}}$, 
$2\,\,A_{10,7,3,2}^{\alpha}$, 
$\SUSY{A_{10,7,4,1}^{\alpha\beta\gamma}}$, 
$6\,\,A_{10,7,4,1}^{\alpha}$, 
$A_{10,7,5}^{\alpha\beta\gamma}$, 
$7\,\,A_{10,7,5}^{\alpha}$, 
$2\,\,A_{10,8,2,1,1}^{\alpha}$, 
$\SUSY{A_{10,8,2,2}^{\alpha\beta\gamma}}$, 
$5\,\,A_{10,8,2,2}^{\alpha}$, 
$2\,\,A_{10,8,3,1}^{\alpha\beta\gamma}$, 
$11\,\,A_{10,8,3,1}^{\alpha}$, 
$5\,\,A_{10,8,4}^{\alpha\beta\gamma}$, 
$16\,\,A_{10,8,4}^{\alpha}$, 
$2\,\,A_{10,9,1,1,1}^{\alpha}$, 
$4\,\,A_{10,9,2,1}^{\alpha\beta\gamma}$, 
$16\,\,A_{10,9,2,1}^{\alpha}$, 
$7\,\,A_{10,9,3}^{\alpha\beta\gamma}$, 
$20\,\,A_{10,9,3}^{\alpha}$, 
$2\,\,A_{10,10,1,1}^{\alpha\beta\gamma}$, 
$7\,\,A_{10,10,1,1}^{\alpha}$, 
$\SUSY{A_{10,10,2}^{\alpha_{1\cdots 5}}}$, 
$8\,\,A_{10,10,2}^{\alpha\beta\gamma}$, 
$18\,\,A_{10,10,2}^{\alpha}$}
\\\hline
$12$ & \Pbox{\textwidth}{
$\SUSY{A_{8,8,7,1}}$, 
$\SUSY{A_{8,8,8}^{\alpha\beta}}$, 
$\SUSY{A_{9,7,6,2}}$, 
$\SUSY{A_{9,7,7,1}^{\alpha\beta}}$, 
$A_{9,7,7,1}$, 
$\SUSY{A_{9,8,4,3}}$, 
$\SUSY{A_{9,8,5,1,1}}$, 
$\SUSY{A_{9,8,5,2}^{\alpha\beta}}$, 
$2\,\,A_{9,8,5,2}$, 
$3\,\,A_{9,8,6,1}^{\alpha\beta}$, 
$5\,\,A_{9,8,6,1}$, 
$4\,\,A_{9,8,7}^{\alpha\beta}$, 
$5\,\,A_{9,8,7}$, 
$\SUSY{A_{9,9,3,2,1}}$, 
$\SUSY{A_{9,9,3,3}^{\alpha\beta}}$, 
$\SUSY{A_{9,9,4,1,1}^{\alpha\beta}}$, 
$A_{9,9,4,1,1}$, 
$2\,\,A_{9,9,4,2}^{\alpha\beta}$, 
$4\,\,A_{9,9,4,2}$, 
$5\,\,A_{9,9,5,1}^{\alpha\beta}$, 
$6\,\,A_{9,9,5,1}$, 
$\SUSY{A_{9,9,6}^{\alpha_{1\cdots 4}}}$, 
$5\,\,A_{9,9,6}^{\alpha\beta}$, 
$7\,\,A_{9,9,6}$, 
$\SUSY{A_{10,6,5,3}}$, 
$\SUSY{A_{10,6,6,2}^{\alpha\beta}}$, 
$A_{10,6,6,2}$, 
$\SUSY{A_{10,7,4,2,1}}$, 
$\SUSY{A_{10,7,4,3}^{\alpha\beta}}$, 
$A_{10,7,4,3}$, 
$\SUSY{A_{10,7,5,1,1}^{\alpha\beta}}$, 
$2\,\,A_{10,7,5,1,1}$, 
$2\,\,A_{10,7,5,2}^{\alpha\beta}$, 
$5\,\,A_{10,7,5,2}$, 
$6\,\,A_{10,7,6,1}^{\alpha\beta}$, 
$8\,\,A_{10,7,6,1}$, 
$4\,\,A_{10,7,7}^{\alpha\beta}$, 
$5\,\,A_{10,7,7}$, 
$\SUSY{A_{10,8,3,2,1}^{\alpha\beta}}$, 
$2\,\,A_{10,8,3,2,1}$, 
$A_{10,8,3,3}^{\alpha\beta}$, 
$3\,\,A_{10,8,3,3}$, 
$2\,\,A_{10,8,4,1,1}^{\alpha\beta}$, 
$5\,\,A_{10,8,4,1,1}$, 
$8\,\,A_{10,8,4,2}^{\alpha\beta}$, 
$10\,\,A_{10,8,4,2}$, 
$\SUSY{A_{10,8,5,1}^{\alpha_{1\cdots 4}}}$, 
$13\,\,A_{10,8,5,1}^{\alpha\beta}$, 
$18\,\,A_{10,8,5,1}$, 
$2\,\,A_{10,8,6}^{\alpha_{1\cdots 4}}$, 
$18\,\,A_{10,8,6}^{\alpha\beta}$, 
$17\,\,A_{10,8,6}$, 
$\SUSY{A_{10,9,2,1,1,1}}$, 
$2\,\,A_{10,9,2,2,1}^{\alpha\beta}$, 
$4\,\,A_{10,9,2,2,1}$, 
$5\,\,A_{10,9,3,1,1}^{\alpha\beta}$, 
$7\,\,A_{10,9,3,1,1}$, 
$\SUSY{A_{10,9,3,2}^{\alpha_{1\cdots 4}}}$, 
$11\,\,A_{10,9,3,2}^{\alpha\beta}$, 
$14\,\,A_{10,9,3,2}$, 
$2\,\,A_{10,9,4,1}^{\alpha_{1\cdots 4}}$, 
$24\,\,A_{10,9,4,1}^{\alpha\beta}$, 
$26\,\,A_{10,9,4,1}$, 
$3\,\,A_{10,9,5}^{\alpha_{1\cdots 4}}$, 
$25\,\,A_{10,9,5}^{\alpha\beta}$, 
$25\,\,A_{10,9,5}$, 
$4\,\,A_{10,10,2,1,1}^{\alpha\beta}$, 
$7\,\,A_{10,10,2,1,1}$, 
$A_{10,10,2,2}^{\alpha_{1\cdots 4}}$, 
$11\,\,A_{10,10,2,2}^{\alpha\beta}$, 
$9\,\,A_{10,10,2,2}$, 
$3\,\,A_{10,10,3,1}^{\alpha_{1\cdots 4}}$, 
$21\,\,A_{10,10,3,1}^{\alpha\beta}$, 
$24\,\,A_{10,10,3,1}$, 
$5\,\,A_{10,10,4}^{\alpha_{1\cdots 4}}$, 
$28\,\,A_{10,10,4}^{\alpha\beta}$, 
$20\,\,A_{10,10,4}$}
\\\hline
$13$ & \Pbox{\textwidth}{
$\SUSY{A_{8,8,8,2}^{\alpha}}$, 
$\SUSY{A_{9,7,7,3}^{\alpha}}$, 
$\SUSY{A_{9,8,5,4}^{\alpha}}$, 
$\SUSY{A_{9,8,6,2,1}^{\alpha}}$, 
$3\,\,A_{9,8,6,3}^{\alpha}$, 
$3\,\,A_{9,8,7,1,1}^{\alpha}$, 
$\SUSY{A_{9,8,7,2}^{\alpha\beta\gamma}}$, 
$7\,\,A_{9,8,7,2}^{\alpha}$, 
$2\,\,A_{9,8,8,1}^{\alpha\beta\gamma}$, 
$8\,\,A_{9,8,8,1}^{\alpha}$, 
$\SUSY{A_{9,9,4,3,1}^{\alpha}}$, 
$A_{9,9,4,4}^{\alpha}$, 
$3\,\,A_{9,9,5,2,1}^{\alpha}$, 
$\SUSY{A_{9,9,5,3}^{\alpha\beta\gamma}}$, 
$5\,\,A_{9,9,5,3}^{\alpha}$, 
$\SUSY{A_{9,9,6,1,1}^{\alpha\beta\gamma}}$, 
$5\,\,A_{9,9,6,1,1}^{\alpha}$, 
$2\,\,A_{9,9,6,2}^{\alpha\beta\gamma}$, 
$11\,\,A_{9,9,6,2}^{\alpha}$, 
$5\,\,A_{9,9,7,1}^{\alpha\beta\gamma}$, 
$17\,\,A_{9,9,7,1}^{\alpha}$, 
$4\,\,A_{9,9,8}^{\alpha\beta\gamma}$, 
$11\,\,A_{9,9,8}^{\alpha}$, 
$\SUSY{A_{10,6,6,4}^{\alpha}}$, 
$\SUSY{A_{10,7,5,3,1}^{\alpha}}$, 
$2\,\,A_{10,7,5,4}^{\alpha}$, 
$3\,\,A_{10,7,6,2,1}^{\alpha}$, 
$\SUSY{A_{10,7,6,3}^{\alpha\beta\gamma}}$, 
$6\,\,A_{10,7,6,3}^{\alpha}$, 
$\SUSY{A_{10,7,7,1,1}^{\alpha\beta\gamma}}$, 
$5\,\,A_{10,7,7,1,1}^{\alpha}$, 
$A_{10,7,7,2}^{\alpha\beta\gamma}$, 
$9\,\,A_{10,7,7,2}^{\alpha}$, 
$\SUSY{A_{10,8,4,2,2}^{\alpha}}$, 
$3\,\,A_{10,8,4,3,1}^{\alpha}$, 
$\SUSY{A_{10,8,4,4}^{\alpha\beta\gamma}}$, 
$5\,\,A_{10,8,4,4}^{\alpha}$, 
$\SUSY{A_{10,8,5,1,1,1}^{\alpha}}$, 
$\SUSY{A_{10,8,5,2,1}^{\alpha\beta\gamma}}$, 
$9\,\,A_{10,8,5,2,1}^{\alpha}$, 
$2\,\,A_{10,8,5,3}^{\alpha\beta\gamma}$, 
$14\,\,A_{10,8,5,3}^{\alpha}$, 
$2\,\,A_{10,8,6,1,1}^{\alpha\beta\gamma}$, 
$13\,\,A_{10,8,6,1,1}^{\alpha}$, 
$8\,\,A_{10,8,6,2}^{\alpha\beta\gamma}$, 
$33\,\,A_{10,8,6,2}^{\alpha}$, 
$12\,\,A_{10,8,7,1}^{\alpha\beta\gamma}$, 
$41\,\,A_{10,8,7,1}^{\alpha}$, 
$\SUSY{A_{10,8,8}^{\alpha_{1\cdots 5}}}$, 
$11\,\,A_{10,8,8}^{\alpha\beta\gamma}$, 
$25\,\,A_{10,8,8}^{\alpha}$, 
$\SUSY{A_{10,9,3,2,1,1}^{\alpha}}$, 
$2\,\,A_{10,9,3,2,2}^{\alpha}$, 
$\SUSY{A_{10,9,3,3,1}^{\alpha\beta\gamma}}$, 
$4\,\,A_{10,9,3,3,1}^{\alpha}$, 
$2\,\,A_{10,9,4,1,1,1}^{\alpha}$, 
$3\,\,A_{10,9,4,2,1}^{\alpha\beta\gamma}$, 
$18\,\,A_{10,9,4,2,1}^{\alpha}$, 
$6\,\,A_{10,9,4,3}^{\alpha\beta\gamma}$, 
$23\,\,A_{10,9,4,3}^{\alpha}$, 
$5\,\,A_{10,9,5,1,1}^{\alpha\beta\gamma}$, 
$25\,\,A_{10,9,5,1,1}^{\alpha}$, 
$14\,\,A_{10,9,5,2}^{\alpha\beta\gamma}$, 
$51\,\,A_{10,9,5,2}^{\alpha}$, 
$\SUSY{A_{10,9,6,1}^{\alpha_{1\cdots 5}}}$, 
$25\,\,A_{10,9,6,1}^{\alpha\beta\gamma}$, 
$75\,\,A_{10,9,6,1}^{\alpha}$, 
$A_{10,9,7}^{\alpha_{1\cdots 5}}$, 
$23\,\,A_{10,9,7}^{\alpha\beta\gamma}$, 
$56\,\,A_{10,9,7}^{\alpha}$, 
$A_{10,10,2,2,1,1}^{\alpha}$, 
$\SUSY{A_{10,10,2,2,2}^{\alpha\beta\gamma}}$, 
$3\,\,A_{10,10,2,2,2}^{\alpha}$, 
$3\,\,A_{10,10,3,1,1,1}^{\alpha}$, 
$4\,\,A_{10,10,3,2,1}^{\alpha\beta\gamma}$, 
$18\,\,A_{10,10,3,2,1}^{\alpha}$, 
$4\,\,A_{10,10,3,3}^{\alpha\beta\gamma}$, 
$14\,\,A_{10,10,3,3}^{\alpha}$, 
$6\,\,A_{10,10,4,1,1}^{\alpha\beta\gamma}$, 
$25\,\,A_{10,10,4,1,1}^{\alpha}$, 
$\SUSY{A_{10,10,4,2}^{\alpha_{1\cdots 5}}}$, 
$19\,\,A_{10,10,4,2}^{\alpha\beta\gamma}$, 
$53\,\,A_{10,10,4,2}^{\alpha}$, 
$A_{10,10,5,1}^{\alpha_{1\cdots 5}}$, 
$27\,\,A_{10,10,5,1}^{\alpha\beta\gamma}$, 
$72\,\,A_{10,10,5,1}^{\alpha}$, 
$3\,\,A_{10,10,6}^{\alpha_{1\cdots 5}}$, 
$30\,\,A_{10,10,6}^{\alpha\beta\gamma}$, 
$62\,\,A_{10,10,6}^{\alpha}$}
\\\hline
$14$ & \Pbox{\textwidth}{
$\SUSY{A_{8,8,8,4}}$, 
$\SUSY{A_{9,7,7,5}}$, 
$\SUSY{A_{9,8,6,4,1}}$, 
$\SUSY{A_{9,8,6,5}^{\alpha\beta}}$, 
$2\,\,A_{9,8,6,5}$, 
$\SUSY{A_{9,8,7,2,2}}$, 
$\SUSY{A_{9,8,7,3,1}^{\alpha\beta}}$, 
$3\,\,A_{9,8,7,3,1}$, 
$3\,\,A_{9,8,7,4}^{\alpha\beta}$, 
$5\,\,A_{9,8,7,4}$, 
$\SUSY{A_{9,8,8,1,1,1}}$, 
$3\,\,A_{9,8,8,2,1}^{\alpha\beta}$, 
$5\,\,A_{9,8,8,2,1}$, 
$6\,\,A_{9,8,8,3}^{\alpha\beta}$, 
$7\,\,A_{9,8,8,3}$, 
$\SUSY{A_{9,9,5,3,2}}$, 
$\SUSY{A_{9,9,5,4,1}^{\alpha\beta}}$, 
$2\,\,A_{9,9,5,4,1}$, 
$A_{9,9,5,5}^{\alpha\beta}$, 
$3\,\,A_{9,9,5,5}$, 
$\SUSY{A_{9,9,6,2,1,1}}$, 
$\SUSY{A_{9,9,6,2,2}^{\alpha\beta}}$, 
$A_{9,9,6,2,2}$, 
$3\,\,A_{9,9,6,3,1}^{\alpha\beta}$, 
$6\,\,A_{9,9,6,3,1}$, 
$6\,\,A_{9,9,6,4}^{\alpha\beta}$, 
$7\,\,A_{9,9,6,4}$, 
$\SUSY{A_{9,9,7,1,1,1}^{\alpha\beta}}$, 
$A_{9,9,7,1,1,1}$, 
$8\,\,A_{9,9,7,2,1}^{\alpha\beta}$, 
$11\,\,A_{9,9,7,2,1}$, 
$\SUSY{A_{9,9,7,3}^{\alpha_{1\cdots 4}}}$, 
$12\,\,A_{9,9,7,3}^{\alpha\beta}$, 
$17\,\,A_{9,9,7,3}$, 
$\SUSY{A_{9,9,8,1,1}^{\alpha_{1\cdots 4}}}$, 
$9\,\,A_{9,9,8,1,1}^{\alpha\beta}$, 
$12\,\,A_{9,9,8,1,1}$, 
$2\,\,A_{9,9,8,2}^{\alpha_{1\cdots 4}}$, 
$20\,\,A_{9,9,8,2}^{\alpha\beta}$, 
$20\,\,A_{9,9,8,2}$, 
$2\,\,A_{9,9,9,1}^{\alpha_{1\cdots 4}}$, 
$13\,\,A_{9,9,9,1}^{\alpha\beta}$, 
$14\,\,A_{9,9,9,1}$, 
$\SUSY{A_{10,6,6,6}}$, 
$\SUSY{A_{10,7,5,5,1}}$, 
$\SUSY{A_{10,7,6,3,2}}$, 
$\SUSY{A_{10,7,6,4,1}^{\alpha\beta}}$, 
$2\,\,A_{10,7,6,4,1}$, 
$2\,\,A_{10,7,6,5}^{\alpha\beta}$, 
$4\,\,A_{10,7,6,5}$, 
$\SUSY{A_{10,7,7,2,1,1}}$, 
$A_{10,7,7,2,2}$, 
$3\,\,A_{10,7,7,3,1}^{\alpha\beta}$, 
$5\,\,A_{10,7,7,3,1}$, 
$5\,\,A_{10,7,7,4}^{\alpha\beta}$, 
$6\,\,A_{10,7,7,4}$, 
$\SUSY{A_{10,8,4,4,2}}$, 
$\SUSY{A_{10,8,5,3,1,1}}$, 
$\SUSY{A_{10,8,5,3,2}^{\alpha\beta}}$, 
$2\,\,A_{10,8,5,3,2}$, 
$3\,\,A_{10,8,5,4,1}^{\alpha\beta}$, 
$7\,\,A_{10,8,5,4,1}$, 
$4\,\,A_{10,8,5,5}^{\alpha\beta}$, 
$5\,\,A_{10,8,5,5}$, 
$\SUSY{A_{10,8,6,2,1,1}^{\alpha\beta}}$, 
$2\,\,A_{10,8,6,2,1,1}$, 
$3\,\,A_{10,8,6,2,2}^{\alpha\beta}$, 
$6\,\,A_{10,8,6,2,2}$, 
$10\,\,A_{10,8,6,3,1}^{\alpha\beta}$, 
$16\,\,A_{10,8,6,3,1}$, 
$\SUSY{A_{10,8,6,4}^{\alpha_{1\cdots 4}}}$, 
$17\,\,A_{10,8,6,4}^{\alpha\beta}$, 
$24\,\,A_{10,8,6,4}$, 
$2\,\,A_{10,8,7,1,1,1}^{\alpha\beta}$, 
$5\,\,A_{10,8,7,1,1,1}$, 
$\SUSY{A_{10,8,7,2,1}^{\alpha_{1\cdots 4}}}$, 
$21\,\,A_{10,8,7,2,1}^{\alpha\beta}$, 
$30\,\,A_{10,8,7,2,1}$, 
$2\,\,A_{10,8,7,3}^{\alpha_{1\cdots 4}}$, 
$33\,\,A_{10,8,7,3}^{\alpha\beta}$, 
$38\,\,A_{10,8,7,3}$, 
$A_{10,8,8,1,1}^{\alpha_{1\cdots 4}}$, 
$17\,\,A_{10,8,8,1,1}^{\alpha\beta}$, 
$20\,\,A_{10,8,8,1,1}$, 
$6\,\,A_{10,8,8,2}^{\alpha_{1\cdots 4}}$, 
$41\,\,A_{10,8,8,2}^{\alpha\beta}$, 
$45\,\,A_{10,8,8,2}$, 
$\SUSY{A_{10,9,4,2,2,1}}$, 
$\SUSY{A_{10,9,4,3,1,1}^{\alpha\beta}}$, 
$2\,\,A_{10,9,4,3,1,1}$, 
$3\,\,A_{10,9,4,3,2}^{\alpha\beta}$, 
$5\,\,A_{10,9,4,3,2}$, 
$7\,\,A_{10,9,4,4,1}^{\alpha\beta}$, 
$9\,\,A_{10,9,4,4,1}$, 
$3\,\,A_{10,9,5,2,1,1}^{\alpha\beta}$, 
$7\,\,A_{10,9,5,2,1,1}$, 
$7\,\,A_{10,9,5,2,2}^{\alpha\beta}$, 
$11\,\,A_{10,9,5,2,2}$, 
$\SUSY{A_{10,9,5,3,1}^{\alpha_{1\cdots 4}}}$, 
$21\,\,A_{10,9,5,3,1}^{\alpha\beta}$, 
$30\,\,A_{10,9,5,3,1}$, 
$2\,\,A_{10,9,5,4}^{\alpha_{1\cdots 4}}$, 
$27\,\,A_{10,9,5,4}^{\alpha\beta}$, 
$31\,\,A_{10,9,5,4}$, 
$5\,\,A_{10,9,6,1,1,1}^{\alpha\beta}$, 
$8\,\,A_{10,9,6,1,1,1}$, 
$3\,\,A_{10,9,6,2,1}^{\alpha_{1\cdots 4}}$, 
$45\,\,A_{10,9,6,2,1}^{\alpha\beta}$, 
$56\,\,A_{10,9,6,2,1}$, 
$7\,\,A_{10,9,6,3}^{\alpha_{1\cdots 4}}$, 
$64\,\,A_{10,9,6,3}^{\alpha\beta}$, 
$71\,\,A_{10,9,6,3}$, 
$5\,\,A_{10,9,7,1,1}^{\alpha_{1\cdots 4}}$, 
$52\,\,A_{10,9,7,1,1}^{\alpha\beta}$, 
$57\,\,A_{10,9,7,1,1}$, 
$13\,\,A_{10,9,7,2}^{\alpha_{1\cdots 4}}$, 
$106\,\,A_{10,9,7,2}^{\alpha\beta}$, 
$106\,\,A_{10,9,7,2}$, 
$20\,\,A_{10,9,8,1}^{\alpha_{1\cdots 4}}$, 
$110\,\,A_{10,9,8,1}^{\alpha\beta}$, 
$100\,\,A_{10,9,8,1}$, 
$9\,\,A_{10,9,9}^{\alpha_{1\cdots 4}}$, 
$42\,\,A_{10,9,9}^{\alpha\beta}$, 
$34\,\,A_{10,9,9}$, 
$\SUSY{A_{10,10,3,2,2,1}^{\alpha\beta}}$, 
$A_{10,10,3,2,2,1}$, 
$A_{10,10,3,3,1,1}^{\alpha\beta}$, 
$3\,\,A_{10,10,3,3,1,1}$, 
$3\,\,A_{10,10,3,3,2}^{\alpha\beta}$, 
$3\,\,A_{10,10,3,3,2}$, 
$\SUSY{A_{10,10,4,1,1,1,1}}$, 
$5\,\,A_{10,10,4,2,1,1}^{\alpha\beta}$, 
$8\,\,A_{10,10,4,2,1,1}$, 
$\SUSY{A_{10,10,4,2,2}^{\alpha_{1\cdots 4}}}$, 
$10\,\,A_{10,10,4,2,2}^{\alpha\beta}$, 
$15\,\,A_{10,10,4,2,2}$, 
$2\,\,A_{10,10,4,3,1}^{\alpha_{1\cdots 4}}$, 
$25\,\,A_{10,10,4,3,1}^{\alpha\beta}$, 
$28\,\,A_{10,10,4,3,1}$, 
$4\,\,A_{10,10,4,4}^{\alpha_{1\cdots 4}}$, 
$22\,\,A_{10,10,4,4}^{\alpha\beta}$, 
$26\,\,A_{10,10,4,4}$, 
$6\,\,A_{10,10,5,1,1,1}^{\alpha\beta}$, 
$11\,\,A_{10,10,5,1,1,1}$, 
$5\,\,A_{10,10,5,2,1}^{\alpha_{1\cdots 4}}$, 
$52\,\,A_{10,10,5,2,1}^{\alpha\beta}$, 
$61\,\,A_{10,10,5,2,1}$, 
$8\,\,A_{10,10,5,3}^{\alpha_{1\cdots 4}}$, 
$67\,\,A_{10,10,5,3}^{\alpha\beta}$, 
$63\,\,A_{10,10,5,3}$, 
$6\,\,A_{10,10,6,1,1}^{\alpha_{1\cdots 4}}$, 
$58\,\,A_{10,10,6,1,1}^{\alpha\beta}$, 
$58\,\,A_{10,10,6,1,1}$, 
$21\,\,A_{10,10,6,2}^{\alpha_{1\cdots 4}}$, 
$120\,\,A_{10,10,6,2}^{\alpha\beta}$, 
$119\,\,A_{10,10,6,2}$, 
$\SUSY{A_{10,10,7,1}^{\alpha_{1\cdots 6}}}$, 
$26\,\,A_{10,10,7,1}^{\alpha_{1\cdots 4}}$, 
$142\,\,A_{10,10,7,1}^{\alpha\beta}$, 
$121\,\,A_{10,10,7,1}$, 
$A_{10,10,8}^{\alpha_{1\cdots 6}}$, 
$26\,\,A_{10,10,8}^{\alpha_{1\cdots 4}}$, 
$91\,\,A_{10,10,8}^{\alpha\beta}$, 
$75\,\,A_{10,10,8}$}
\\\hline
\end{tabular}}}
\caption{Mixed-symmetry potentials in type IIB theory up to level $\ell_9=14$.}
\label{tab:IIB-potentials-14}
\end{table}
There, $\alpha,\beta=1,2$ are $\SL(2)$ indices and the multiple $\SL(2)$ indices $\alpha_1\cdots\alpha_s$ are totally symmetrized. 
The underlined potentials corresponds to $E_{11}$ roots $\alpha$ satisfying $\alpha\cdot\alpha=2$\,, which means that the potentials may couple to supersymmetric branes, as discussed in the next subsection. 

In order to construct Table \ref{tab:p-form} in terms of type IIB potentials, higher-level generators are needed, although level $\ell_9=14$ is the maximal we can achieve with current personal computers. 
When we try to construct Table \ref{tab:p-form}, we first perform the level decomposition associated with the node $\alpha_d$ \eqref{eq:level-d}, and then decompose the obtained $p$-form multiplets into type IIB potentials. 
In other words, we consider a level decomposition for the following Dynkin diagram:
\begin{align}
\scalebox{0.8}{
 \xygraph{
    *\cir<6pt>{} ([]!{+(0,-.4)} {\alpha_1}) - [r]
    \cdots - [r]
    *\cir<6pt>{} ([]!{+(0,-.4)} {\alpha_{d-1}}) - [r]
    *{\scalebox{2}{$\times$}}*\cir<6pt>{} ([]!{+(0,-.4)} {\alpha_d}) - [r]
    *\cir<6pt>{} ([]!{+(0,-.4)} {\alpha_{d+1}}) - [r]
    \cdots 
    - [r] *\cir<6pt>{} ([]!{+(0,-.4)} {\alpha_7}) 
    - [r] *\cir<6pt>{} ([]!{+(0,-.4)} {\alpha_8}) 
(
        - [u] *\cir<6pt>{} ([]!{+(.5,0)} {\alpha_{11}}),
        - [r] *{\scalebox{2}{$\times$}}*\cir<6pt>{} ([]!{+(0,-.4)} {\alpha_9})
        - [r] *\cir<6pt>{} ([]!{+(0,-.4)} {\alpha_{10}})
) 
}} .
\end{align}
The results are given in Appendix \ref{app:B-potential}. 
In each dimension $d$, we obtained a list of $d$-dimensional $p$-form potentials which form the $E_n$ $U$-duality multiplets. 
Moreover, by uplifting the potentials to 10D, we have identified that the 10D mixed-symmetry potentials given in Table \ref{tab:MB-potentials} are all of the potentials that contribute to Table \ref{tab:p-form}. 

\subsection{Comments on supersymmetric branes}

Here, for completeness, we make some comments on supersymmetric branes. 
The R--R potentials couple to D-branes, and similarly, standard $p$-form fields couple to certain supersymmetric branes. 
However, as discussed in \cite{hep-th:0611036,0708.2287,1109.2025,1009.4657}, not all of the potentials couple to supersymmetric branes. 
For example, let us consider the $\SL(2)$ S-duality triplet of 8-forms $A_8^{(\alpha\beta)}$ ($\alpha,\beta=1,2$) in type IIB theory.
They are contained in some $U$-duality multiplets in Table \ref{tab:p-form} (e.g.~$\textcolor{red}{\bf 3}$ in $d=9$ and $p=8$). 
Two components, $A_8^{(11)}$ and $A_8^{(22)}$ are known to couple to supersymmetric branes, the D7-brane ($7_1$-brane) and the NS7-brane ($7_3$-brane), respectively. 
However, we cannot write the standard Wess--Zumino term for the remaining component $A_8^{(12)}$ in a gauge-invariant manner \cite{hep-th:0611036}, and it has been considered that there is no supersymmetric brane that corresponds to this potential. 
Another example consists of the 10-forms $A_9^{(\alpha\beta\gamma)}$ and $A_9^\alpha$ in type IIB supergravity, which are predicted in \cite{hep-th:0506013,hep-th:0602280,1004.1348} and are shown to be consistent with the supersymmetry algebra. 
They are also predicted by the $E_{11}$ conjecture \cite{hep-th:0511153}; for example in $d=9$ and $p=9$ of Table \ref{tab:p-form}, the quadruplet $A_9^{(\alpha\beta\gamma)}$ gives $\textcolor{red}{{\bf 4}}$ while the doublet $A_9^\alpha$ gives one of ${\bf 2}$\,. 
Among the quadruplet $A_9^{(\alpha\beta\gamma)}$\,, $A_9^{(111)}$ and $A_9^{(222)}$ couple to supersymmetric branes, D9-brane ($9_1$-brane) and the $9_4$-brane, but the gauge-invariant Wess--Zumino term for the remaining two components cannot be written down. 
Also, from a discussion based on the supersymmetry of brane actions \cite{hep-th:0601128}, it has been concluded that the two 9-forms do not couple to any supersymmetric branes. 

A criterion based on the $E_{11}$ algebra has been discussed in \cite{1109.2025} to elucidate whether a potential couples to a supersymmetric brane:
\emph{A root $\alpha$ of the $E_{11}$ algebra is associated with a supersymmetric brane if it has the length squared $\alpha^2=2$\,. 
Otherwise (i.e.~$\alpha^2=0,-2,-4,\cdots$) it does not couple to any supersymmetric brane.}
For any root $\alpha$ which corresponds to the standard potential (coupling to a supersymmetric brane), $\alpha^2=2$ is indeed satisfied. 
Moreover, the $T$-duality and $S$-duality (which correspond to the Weyl reflections) do not change the norm $\alpha^2$\,, and branes in the Weyl orbits of the standard branes always correspond to roots with $\alpha^2=2$\,. 
On the other hand, $E_{11}$ weights associated with potentials which may not couple to a supersymmetric brane (such as $A_8^{(12)}$) satisfy $\alpha^2<2$\,. 
In Table \ref{tab:p-form}, all representations whose highest weights correspond to $E_{11}$ roots satisfying $\alpha^2=2$ are colored in red. 
The uncolored representations do not contain any root with $\alpha^2=2$\,. 
If we compare Tables \ref{tab:p-form} and \ref{tab:p-brane}, we find that the same colored representations are contained in both tables. 
This means that all of the supersymmetric branes have the corresponding potential as usual. 
On the other hand, the physical meaning of uncolored representation is less clear and is not addressed in this paper. 

Inside the colored representation, the number of $E_{11}$ roots satisfying $\alpha^2=2$ (i.e.~the number of supersymmetric branes) was counted in \cite{1109.2025} (see also \cite{1805.12117}), and the result is summarized in Table \ref{tab:SUSY-brane}. 
\begin{table}[tbp]
 \centerline{\scalebox{0.85}{
 \begin{tabular}{|c||C{0.9cm}|C{1.0cm}|C{1.5cm}|C{1.0cm}|C{1.2cm}|C{1.2cm}|C{1.2cm}|C{1.5cm}|C{0.8cm}||c|}\hline
 \backslashbox{$d$}{$p$} & $0$ & $1$ & $2$ & $3$ & $4$ & $5$ & $6$ & $7$ & $8$ & $U$-duality \\\hline\hline
 9 & $\textcolor{red}{{\bf 2}}$\\ $\textcolor{red}{{\bf 1}}$ & $\textcolor{red}{{\bf 2}}$ & $\textcolor{red}{{\bf 1}}$ & $\textcolor{red}{{\bf 1}}$ & $\textcolor{red}{{\bf 2}}$ & $\textcolor{red}{{\bf 2}}$ \\ $\textcolor{red}{{\bf 1}}$ & $2/\textcolor{red}{{\bf 3}}$ & $2/\textcolor{red}{{\bf 3}}$ & $2/\textcolor{red}{{\bf 4}}$ & $\SL(2)$ \\\hline
 8 & $\textcolor{red}{({\bf \overline{3}},{\bf 2})}$ & $\textcolor{red}{({\bf 3},{\bf 1})}$ & $\textcolor{red}{({\bf 1},{\bf 2})}$ & $\textcolor{red}{({\bf \overline{3}},{\bf 1})}$ & $\textcolor{red}{({\bf 3},{\bf 2})}$ & $6/\textcolor{red}{({\bf 8},{\bf 1})}$ \\ $2/\textcolor{red}{({\bf 1},{\bf 3})}$ & $6/\textcolor{red}{({\bf 6},{\bf 2})}$ & $6/\textcolor{red}{({\bf 15},{\bf 1})}$ & & $\SL(3)\times\SL(2)$ \\\hline
 7 & $\textcolor{red}{{\bf \overline{10}}}$ & $\textcolor{red}{{\bf 5}}$ & $\textcolor{red}{{\bf \overline{5}}}$ & $\textcolor{red}{{\bf 10}}$ & $20/\textcolor{red}{{\bf 24}}$ & $20/\textcolor{red}{{\bf \overline{40}}}$ \\ $5/\textcolor{red}{{\bf \overline{15}}}$ & $20/\textcolor{red}{{\bf 70}}$ & & & $\SL(5)$ \\\hline
 6 & $\textcolor{red}{{\bf 16}}$ & $\textcolor{red}{{\bf 10}}$ & $\textcolor{red}{{\bf \overline{16}}}$ & $40/\textcolor{red}{{\bf 45}}$ & $80/\textcolor{red}{{\bf 144}}$ & $80/\textcolor{red}{{\bf 320}}$ \\ $16/\textcolor{red}{{\bf 126}}$ & & & & $\SO(5,5)$ \\\hline
 5 & $\textcolor{red}{{\bf 27}}$ & $\textcolor{red}{{\bf \overline{27}}}$ & $72/\textcolor{red}{{\bf 78}}$ & $\displaystyle\frac{216}{\textcolor{red}{{\bf 351}}}$ & $\displaystyle\frac{432}{\textcolor{red}{{\bf \overline{1728}}}}$ & & & & & $E_6\vphantom{\Bigg|}$ \\\hline
 4 & $\textcolor{red}{{\bf 56}}$ & $\displaystyle\frac{126}{\textcolor{red}{{\bf 133}}}$ & $\displaystyle\frac{576}{\textcolor{red}{{\bf 912}}}$ & $\displaystyle\frac{2016}{\textcolor{red}{{\bf 8645}}}$ & & & & & & $E_7\vphantom{\Bigg|}$ \\\hline
 3 & $\displaystyle\frac{240}{\textcolor{red}{{\bf 248}}}$ & $\displaystyle\frac{2160}{\textcolor{red}{{\bf 3875}}}$ & $\displaystyle\frac{17280}{\textcolor{red}{{\bf 147250}}}$ & & & & & & & $E_8\vphantom{\Bigg|}$ \\\hline
\end{tabular}
}}
\caption{Number of supersymmetric branes inside $U$-duality multiplets are indicated in black. For $p\leq d-4$, the number is equal to the dimension of the $U$-duality multiplet.}
\label{tab:SUSY-brane}
\end{table}
For $p$-branes with $p\leq d-4$\,, all $E_{11}$ roots contained in each $E_n$ representation satisfy $\alpha^2=2$\,, which means that all weights in that $E_n$ multiplet correspond to supersymmetric branes. 

The situation changes for $p$-branes with $p=d-3$\,, which are known as defect branes. 
In this case, the colored $p$-brane multiplets are the adjoint representation of $E_n$ and $n$ roots may not couple to any supersymmetric brane (see Table \ref{tab:SUSY-brane}). 
In terms of the M-theory/type IIB potentials, the multiplet contains mixed-symmetry potentials in the compactified space or transform in the triplet or higher-dimensional representations of $\SL(2)$\,. 
It is also the case for $p$-branes with $p=d-2$ (domain-wall branes) and $p=d-1$ (space-filling branes). 
For these low-co-dimension branes, only a subset of the colored representations corresponds to supersymmetric branes. 

Now, we explain the correspondence \eqref{eq:brane-pot-M} between exotic branes and mixed-symmetric potentials in more detail. 
The colored representations in Table \ref{tab:p-form} are reproduced from the underlined M-theory/type IIB potentials given in Table \ref{tab:MB-potentials}. 
In M-theory, we can identify which underlined potential couples to which brane as follows \cite{1805.12117}:
\begin{align}
\Pbox{0.9\textwidth}{\footnotesize
$\SUSY{\hat{A}_1^1}$\,\,(P), 
$\SUSY{\hat{A}_{3}}\,\,(2_3)$, 
$\SUSY{\hat{A}_{6}}\,\,(5_6)$, 
$\SUSY{\hat{A}_{8,1}}\,\,(6_{9}^{1})$, 
$\SUSY{\hat{A}_{9,3}}\,\,(5_{12}^{3})$, 
$\SUSY{\hat{A}_{10,1,1}}\,\,(8_{12}^{(1,0)})$, 
$\SUSY{\hat{A}_{9,6}}\,\,(2_{15}^{6})$, 
$\SUSY{\hat{A}_{10,4,1}}\,\,(5_{15}^{(1,3)})$, 
$\SUSY{\hat{A}_{9,8,1}}\,\,(0_{18}^{(1,7)})$, 
$\SUSY{\hat{A}_{10,6,2}}\,\,(3_{18}^{(2,4)})$, 
$\SUSY{\hat{A}_{11,4,3}}\,\,(6_{18}^{(3,1)})$, 
$\SUSY{\hat{A}_{11,5,1,1}}\,\,(5_{18}^{(1,0,4)})$, 
$\SUSY{\hat{A}_{10,7,4}}\,\,(2_{21}^{(4,3)})$, 
$\SUSY{\hat{A}_{10,8,2,1}}\,\,(1_{21}^{(1,1,6)})$, 
$\SUSY{\hat{A}_{11,6,3,1}}\,\,(4_{21}^{(1,2,3)})$, 
$\SUSY{\hat{A}_{10,7,7}}\,\,(2_{24}^{(7,0)})$, 
$\SUSY{\hat{A}_{10,8,5,1}}\,\,(1_{24}^{(1,4,3)})$, 
$\SUSY{\hat{A}_{11,6,6,1}}\,\,(4_{24}^{(1,5,0)})$, 
$\SUSY{\hat{A}_{11,7,4,2}}\,\,(3_{24}^{(2,2,3)})$, 
$\SUSY{\hat{A}_{11,8,3,1,1}}\,\,(2_{24}^{(1,0,2,5)})$, 
$\SUSY{\hat{A}_{10,8,7,2}}\,\,(1_{27}^{(2,5,1)})$, 
$\SUSY{\hat{A}_{11,7,6,3}}\,\,(3_{27}^{(3,3,1)})$, 
$\SUSY{\hat{A}_{11,8,5,2,1}}\,\,(2_{27}^{(1,1,3,3)})$, 
$\SUSY{\hat{A}_{11,8,4,4}}\,\,(2_{27}^{(4,0,4)})$, 
$\SUSY{\hat{A}_{10,8,8,4}}\,\,(1_{30}^{(4,4,0)})$, 
$\SUSY{\hat{A}_{11,7,7,5}}\,\,(3_{30}^{(5,2,0)})$, 
$\SUSY{\hat{A}_{11,8,6,4,1}}\,\,(2_{30}^{(1,3,2,2)})$, 
$\SUSY{\hat{A}_{11,8,7,2,2}}\,\,(2_{30}^{(2,0,5,1)})$, 
$\SUSY{\hat{A}_{11,8,8,1,1,1}}\,\,(2_{30}^{(1,0,0,7,0)})$, 
$\SUSY{\hat{A}_{10,8,8,7}}\,\,(1_{33}^{(7,1,0)})$, 
$\SUSY{\hat{A}_{11,8,7,5,2}}\,\,(2_{33}^{(2,3,2,1)})$, 
$\SUSY{\hat{A}_{11,8,8,4,1,1}}\,\,(2_{33}^{(1,0,3,4,0)})$, 
$\SUSY{\hat{A}_{11,8,7,7,3}}\,\,(2_{36}^{(3,4,0,1)})$, 
$\SUSY{\hat{A}_{11,8,8,5,4}}\,\,(2_{36}^{(4,1,3,0)})$, 
$\SUSY{\hat{A}_{11,8,8,6,2,1}}\,\,(2_{36}^{(1,1,4,2,0)})$, 
$\SUSY{\hat{A}_{11,8,8,7,4,1}}\,\,(2_{39}^{(1,3,3,1,0)})$, 
$\SUSY{\hat{A}_{11,8,8,8,2,2}}\,\,(2_{39}^{(2,0,6,0,0)})$, 
$\SUSY{\hat{A}_{11,8,8,7,7,1}}\,\,(2_{42}^{(1,6,0,1,0)})$, 
$\SUSY{\hat{A}_{11,8,8,8,5,2}}\,\,(2_{42}^{(2,3,3,0,0)})$, 
$\SUSY{\hat{A}_{11,8,8,8,7,3}}\,\,(2_{45}^{(3,4,1,0,0)})$, 
$\SUSY{\hat{A}_{11,8,8,8,8,5}}\,\,(2_{48}^{(5,3,0,0,0)})$, 
$\SUSY{\hat{A}_{11,8,8,8,8,8}}\,\,(2_{51}^{(8,0,0,0,0)})$\,.}
\label{eq:mixed-susy-exotic}
\end{align}
For convenience, we have appended a subscript to the name of each brane; the level $\ell$ multiplied by three. 
Similarly, for type IIB theory, we obtain \cite{1805.12117}:
\begin{align}
\Pbox{0.9\textwidth}{\footnotesize
$\SUSY{A_{1}^{1\,[0]}}\,\,(\text{P})$, 
$\SUSY{A_{2}^{\alpha\,[0\text{-}1]}}\,\,(1_0,\,1_1)$, 
$\SUSY{A_{4}^{[1]}}\,\,(3_1)$, 
$\SUSY{A_{6}^{\alpha\,[1\text{-}2]}}\,\,(5_1,\,5_2)$, 
$\SUSY{A_{7,1}^{[2]}}\,\,(5_2^1)$, 
$\SUSY{A_{8}^{\alpha\beta\,[1\text{-}3]}}\,\,(7_1,\,7_{3})$, 
$\SUSY{A_{8,2}^{\alpha\,[2\text{-}3]}}\,\,(5_{2}^{2},\,5_{3}^{2})$, 
$\SUSY{A_{10}^{\alpha\beta\gamma\,[1\text{-}4]}}\,\,(9_1,\,9_{4})$, 
$\SUSY{A_{8,4}^{[3]}}\,\,(3_{3}^{4})$, 
$\SUSY{A_{9,2,1}^{[3]}}\,\,(6_{3}^{(1,1)})$, 
$\SUSY{A_{9,3}^{\alpha\beta\,[2\text{-}4]}}\,\,(5_{2}^{3},\,5_{4}^{3})$, 
$\SUSY{A_{8,6}^{\alpha\,[3\text{-}4]}}\,\,(1_{3}^{6},\,1_{4}^{6})$, 
$\SUSY{A_{9,4,1}^{\alpha\,[3\text{-}4]}}\,\,(4_{3}^{(1,3)},\,4_{4}^{(1,3)})$, 
$\SUSY{A_{10,2,2}^{\alpha\,[3\text{-}4]}}\,\,(7_{3}^{(2,0)},\,7_{4}^{(2,0)})$, 
$\SUSY{A_{10,4}^{\alpha\beta\gamma\,[2\text{-}5]}}\,\,(5_{2}^{4},\,5_{5}^{4})$, 
$\SUSY{A_{8,7,1}^{[4]}}\,\,(0_{4}^{(1,6)})$, 
$\SUSY{A_{9,5,2}^{[4]}}\,\,(3_{4}^{(2,3)})$, 
$\SUSY{A_{9,6,1}^{\alpha\beta\,[3\text{-}5]}}\,\,(2_{3}^{(1,5)},\,2_{5}^{(1,5)})$, 
$\SUSY{A_{10,4,1,1}^{[4]}}\,\,(5_{4}^{(1,0,3)})$, 
$\SUSY{A_{10,4,2}^{\alpha\beta\,[3\text{-}5]}}\,\,(5_{3}^{(2,2)},\,5_{5}^{(2,2)})$, 
$\SUSY{A_{9,6,3}^{\alpha\,[4\text{-}5]}}\,\,(2_{4}^{(3,3)},\,2_{5}^{(3,3)})$, 
$\SUSY{A_{9,7,1,1}^{\alpha\,[4\text{-}5]}}\,\,(1_{4}^{(1,0,6)},\,1_{5}^{(1,0,6)})$, 
$\SUSY{A_{10,4,4}^{\alpha\,[4\text{-}5]}}\,\,(5_{4}^{(4,0)},\,5_{5}^{(4,0)})$, 
$\SUSY{A_{10,5,2,1}^{\alpha\,[4\text{-}5]}}\,\,(4_{4}^{(1,1,3)},\,4_{5}^{(1,1,3)})$, 
$\SUSY{A_{10,6,2}^{\alpha\beta\gamma\,[3\text{-}6]}}\,\,(3_{3}^{(2,4)},\,3_{6}^{(2,4)})$, 
$\SUSY{A_{9,6,5}^{[5]}}\,\,(2_{5}^{(5,1)})$, 
$\SUSY{A_{9,7,3,1}^{[5]}}\,\,(1_{5}^{(1,2,4)})$, 
$\SUSY{A_{9,7,4}^{\alpha\beta\,[4\text{-}6]}}\,\,(1_{4}^{(4,3)},\,1_{6}^{(4,3)})$, 
$\SUSY{A_{10,5,4,1}^{[5]}}\,\,(4_{5}^{(1,3,1)})$, 
$\SUSY{A_{10,6,2,2}^{[5]}}\,\,(3_{5}^{(2,0,4)})$, 
$\SUSY{A_{10,6,3,1}^{\alpha\beta\,[4\text{-}6]}}\,\,(3_{4}^{(1,2,3)},\,3_{6}^{(1,2,3)})$, 
$\SUSY{A_{10,7,1,1,1}^{[5]}}\,\,(2_{5}^{(1,0,0,6)})$, 
$\SUSY{A_{9,7,5,1}^{\alpha\,[5\text{-}6]}}\,\,(1_{5}^{(1,4,2)},\,1_{6}^{(1,4,2)})$, 
$\SUSY{A_{10,6,4,2}^{\alpha\,[5\text{-}6]}}\,\,(3_{5}^{(2,2,2)},\,3_{6}^{(2,2,2)})$, 
$\SUSY{A_{10,6,6}^{\alpha\beta\gamma\,[4\text{-}7]}}\,\,(3_{4}^{(6,0)},\,3_{7}^{(6,0)})$, 
$\SUSY{A_{10,7,3,1,1}^{\alpha\,[5\text{-}6]}}\,\,(2_{5}^{(1,0,2,4)},\,2_{6}^{(1,0,2,4)})$, 
$\SUSY{A_{10,7,4,1}^{\alpha\beta\gamma\,[4\text{-}7]}}\,\,(2_{4}^{(1,3,3)},\,2_{7}^{(1,3,3)})$, 
$\SUSY{A_{9,7,6,2}^{[6]}}\,\,(1_{6}^{(2,4,1)})$, 
$\SUSY{A_{9,7,7,1}^{\alpha\beta\,[5\text{-}7]}}\,\,(1_{5}^{(1,6,0)},\,1_{7}^{(1,6,0)})$, 
$\SUSY{A_{10,6,5,3}^{[6]}}\,\,(3_{6}^{(3,2,1)})$, 
$\SUSY{A_{10,6,6,2}^{\alpha\beta\,[5\text{-}7]}}\,\,(3_{5}^{(2,4,0)},\,3_{7}^{(2,4,0)})$, 
$\SUSY{A_{10,7,4,2,1}^{[6]}}\,\,(2_{6}^{(1,1,2,3)})$, 
$\SUSY{A_{10,7,4,3}^{\alpha\beta\,[5\text{-}7]}}\,\,(2_{5}^{(3,1,3)},\,2_{7}^{(3,1,3)})$, 
$\SUSY{A_{10,7,5,1,1}^{\alpha\beta\,[5\text{-}7]}}\,\,(2_{5}^{(1,0,4,2)},\,2_{7}^{(1,0,4,2)})$, 
$\SUSY{A_{9,7,7,3}^{\alpha\,[6\text{-}7]}}\,\,(1_{6}^{(3,4,0)},\,1_{7}^{(3,4,0)})$, 
$\SUSY{A_{10,6,6,4}^{\alpha\,[6\text{-}7]}}\,\,(3_{6}^{(4,2,0)},\,3_{7}^{(4,2,0)})$, 
$\SUSY{A_{10,7,5,3,1}^{\alpha\,[6\text{-}7]}}\,\,(2_{6}^{(1,2,2,2)},\,2_{7}^{(1,2,2,2)})$, 
$\SUSY{A_{10,7,6,3}^{\alpha\beta\gamma\,[5\text{-}8]}}\,\,(2_{5}^{(3,3,1)},\,2_{8}^{(3,3,1)})$, 
$\SUSY{A_{10,7,7,1,1}^{\alpha\beta\gamma\,[5\text{-}8]}}\,\,(2_{5}^{(1,0,6,0)},\,2_{8}^{(1,0,6,0)})$, 
$\SUSY{A_{9,7,7,5}^{[7]}}\,\,(1_{7}^{(5,2,0)})$, 
$\SUSY{A_{10,6,6,6}^{[7]}}\,\,(3_{7}^{(6,0,0)})$, 
$\SUSY{A_{10,7,5,5,1}^{[7]}}\,\,(2_{7}^{(1,4,0,2)})$, 
$\SUSY{A_{10,7,6,3,2}^{[7]}}\,\,(2_{7}^{(2,1,3,1)})$, 
$\SUSY{A_{10,7,6,4,1}^{\alpha\beta\,[6\text{-}8]}}\,\,(2_{6}^{(1,3,2,1)},\,2_{8}^{(1,3,2,1)})$, 
$\SUSY{A_{10,7,7,2,1,1}^{[7]}}\,\,(2_{7}^{(1,0,1,5,0)})$, 
$\SUSY{A_{9,7,7,7}^{\alpha\,[7\text{-}8]}}\,\,(1_{7}^{(7,0,0)},\,1_{8}^{(7,0,0)})$, 
$\SUSY{A_{10,7,6,5,2}^{\alpha\,[7\text{-}8]}}\,\,(2_{7}^{(2,3,1,1)},\,2_{8}^{(2,3,1,1)})$, 
$\SUSY{A_{10,7,7,3,3}^{\alpha\,[7\text{-}8]}}\,\,(2_{7}^{(3,0,4,0)},\,2_{8}^{(3,0,4,0)})$, 
$\SUSY{A_{10,7,7,4,1,1}^{\alpha\,[7\text{-}8]}}\,\,(2_{7}^{(1,0,3,3,0)},\,2_{8}^{(1,0,3,3,0)})$, 
$\SUSY{A_{10,7,7,5,1}^{\alpha\beta\gamma\,[6\text{-}9]}}\,\,(2_{6}^{(1,4,2,0)},\,2_{9}^{(1,4,2,0)})$, 
$\SUSY{A_{10,7,6,6,3}^{[8]}}\,\,(2_{8}^{(3,3,0,1)})$, 
$\SUSY{A_{10,7,7,5,2,1}^{[8]}}\,\,(2_{8}^{(1,1,3,2,0)})$, 
$\SUSY{A_{10,7,7,5,3}^{\alpha\beta\,[7\text{-}9]}}\,\,(2_{7}^{(3,2,2,0)},\,2_{9}^{(3,2,2,0)})$, 
$\SUSY{A_{10,7,7,6,1,1}^{\alpha\beta\,[7\text{-}9]}}\,\,(2_{7}^{(1,0,5,1,0)},\,2_{9}^{(1,0,5,1,0)})$, 
$\SUSY{A_{10,7,7,5,5}^{\alpha\,[8\text{-}9]}}\,\,(2_{8}^{(5,0,2,0)},\,2_{9}^{(5,0,2,0)})$, 
$\SUSY{A_{10,7,7,6,3,1}^{\alpha\,[8\text{-}9]}}\,\,(2_{8}^{(1,2,3,1,0)},\,2_{9}^{(1,2,3,1,0)})$, 
$\SUSY{A_{10,7,7,7,3}^{\alpha\beta\gamma\,[7\text{-}10]}}\,\,(2_{7}^{(3,4,0,0)},\,2_{10}^{(3,4,0,0)})$, 
$\SUSY{A_{10,7,7,6,5,1}^{[9]}}\,\,(2_{9}^{(1,4,1,1,0)})$, 
$\SUSY{A_{10,7,7,7,3,2}^{[9]}}\,\,(2_{9}^{(2,1,4,0,0)})$, 
$\SUSY{A_{10,7,7,7,4,1}^{\alpha\beta\,[8\text{-}10]}}\,\,(2_{8}^{(1,3,3,0,0)},\,2_{10}^{(1,3,3,0,0)})$, 
$\SUSY{A_{10,7,7,7,5,2}^{\alpha\,[9\text{-}10]}}\,\,(2_{9}^{(2,3,2,0,0)},\,2_{10}^{(2,3,2,0,0)})$, 
$\SUSY{A_{10,7,7,7,7}^{\alpha\beta\gamma\,[8\text{-}11]}}\,\,(2_{8}^{(7,0,0,0)},\,2_{11}^{(7,0,0,0)})$, 
$\SUSY{A_{10,7,7,7,6,3}^{[10]}}\,\,(2_{10}^{(3,3,1,0,0)})$, 
$\SUSY{A_{10,7,7,7,7,2}^{\alpha\beta\,[9\text{-}11]}}\,\,(2_{9}^{(2,5,0,0,0)},\,2_{11}^{(2,5,0,0,0)})$, 
$\SUSY{A_{10,7,7,7,7,4}^{\alpha\,[10\text{-}11]}}\,\,(2_{10}^{(4,3,0,0,0)},\,2_{11}^{(4,3,0,0,0)})$, 
$\SUSY{A_{10,7,7,7,7,6}^{[11]}}\,\,(2_{11}^{(6,1,0,0,0)})$.} 
\end{align}
We note that, for potentials of $\SL(2)$ $(s+1)$-plets, only the potentials with the highest and the lowest value of $\NN$ in the square bracket have $\alpha^2=2$\,. 
Thus, in general, each $\SL(2)$-covariant type IIB potential can couple to at most two supersymmetric branes. 

Configurations of an M-theory brane can be specified by the following information:
\begin{align}
 b^{(c_s,\dotsc,c_2)}(i_1\cdots i_b,\,j_1\cdots j_{c_2},\cdots,\,k_1\cdots k_{c_s})\,,
\label{eq:exotic-configuration}
\end{align}
where $j,k=1,\dotsc,n$ runs over the internal compactified directions while the first set of indices can run over the whole 11D, and all of them $\{i_1,\dotsc,k_{c_s}\}$ must be different. 
This information specifies the wrapping directions of the brane. 
When the first set of indices $\{i_1,\dotsc, i_b\}$ contains $p$ external uncompactified directions, it behaves as a $p$-brane with tension \eqref{eq:tension-M}. 
The number of configurations (in the internal space) can be easily counted as
\begin{align}
 \frac{n!}{(b-p)!\,c_2!\cdots c_s!\,(n+p-b-c_2-\cdots c_s)!}\,. 
\end{align}
On the other hand, the corresponding central charge $\SUSY{Z^{p;b-p+c_2+\cdots+c_s,c_2+\cdots+c_s,\dotsc,c_{s-1}+c_s,c_s}}$ or the potential $\SUSY{\hat{A}_{p+1;b-p+c_2+\cdots+c_s,c_2+\cdots+c_s,\dotsc,c_{s-1}+c_s,c_s}}$ have a larger number of components. 
As discussed in \cite{1102.0934,1106.0212,1108.5067,1201.5819,1710.00642}, only some restricted components of $\SUSY{\hat{A}_{p;n_1,\dotsc,n_p}}$, or more explicitly $\SUSY{\hat{A}_{p;i_1\cdots i_{n_1},j_1\cdots j_{n_2},\dotsc ,k_1\cdots k_{n_p}}}$\,, will couple to supersymmetric branes:
\begin{align}
 \ovalbox{ \quad $\displaystyle\text{\bf Restriction rule:} \qquad \{i_1\cdots i_{n_1}\} 
 \supset \{j_1\cdots j_{n_2}\}\supset \cdots 
 \supset \{k_1\cdots k_{n_p}\} $\quad } \,.
\label{eq:restriction-rule}
\end{align}
Indeed, the number of the restricted components matches with the number of configurations of the exotic brane. 
The same rule applies also to the type II potentials. 
In the type IIB case, potentials are further restricted as follows in order to couple supersymmetric branes:\footnote{For $A^{\alpha_{1\cdots s}}_{p,q,r,\cdots}$\,, this condition requires that all of the $\SL(2)$ indices $\alpha_1,\dotsc,\alpha_s$ are the same. By using the map between M-theory and type IIB theory \cite{hep-th:0402140,1701.07819,1909.01335}, this rule can be derived from the M-theory rule \eqref{eq:restriction-rule}.}
\begin{align}
 \ovalbox{$\displaystyle\ \,
 \Pbox{0.85\textwidth}{ \text{\bf Restriction rule \ (type IIB):}\newline
 \text{Only the highest/lowest weight states of the $S$-duality multiplets are allowed.}}$}
\label{eq:restriction-rule2}
\end{align}

For the standard form field $\SUSY{\hat{A}_{p;q}}$\,, the restriction is trivial and all components couple to supersymmetric branes. 
Therefore, the restriction occurs only to the mixed-symmetry potentials. 
Indeed, as is clear from Appendix \ref{app:M-potential}, if we decompose the $E_n$ representations of the $p$-brane multiplets ($p\leq d-4$) into representations of $\SL(n)$\,, we find that any $\SL(n)$ representation is a fundamental representation, which behaves as a $p$-form in the compactified space. 
Only in the $p$-brane multiplets with $p\geq d-3$ do mixed-symmetry potentials appear and some components are excluded by the restriction rule \eqref{eq:restriction-rule}. 
The same situation can be observed in type IIB theory from tables in Appendix \ref{app:B-potential}. 

By using the restriction rule \eqref{eq:restriction-rule}, we can easily uplift an underlined type IIA potential to an underlined M-theory potential. 
For example, let us consider an uplift of type IIA potential $\SUSY{E_{8,3}}$ with $\NN=3$\,. 
From \eqref{eq:A-index-level}, this potential has level $\ell=4$\,, and the M-theory uplift has 12 indices. 
Namely, $1\,(=\ell-\NN)$ indices are chosen as the M-theory direction $x^z$\,. 
From the restriction rule \eqref{eq:restriction-rule}, the $z$ should be filled from the left $\SUSY{E_{8z,3}}$\,, which corresponds to $\SUSY{\hat{A}_{9,3}}$\,. 
This potential can be found in Table \ref{table:M-level10}. 
As another example, let us consider $\SUSY{F_{10,9,3,2}}$ with $\NN=4$\,. 
This has level $\ell=10$, and $z$ appears $6\,(=\ell-\NN)$ times. 
By including the $z$-direction six times, we obtain $\SUSY{F_{10z,9z,3z,2z,z,z}}$ which corresponds to $\SUSY{\hat{A}_{11,10,4,3,1,1}}$\,. 

We also comment on the behavior of the restricted components of the underlined potentials under $T$-duality transformations. 
Under a $T$-duality along the $x^y$ direction, the radii $R_i$ and the string coupling constant transform as
\begin{align}
 R_i \to l_s^2/R_i\,,\qquad g_s\to g_s\,l_s/R_i\,,
\end{align}
and the exotic $b_{\NN}^{(c_s,\dotsc,c_2)}$-brane, if $x^y$ is contained in the set of $c_i$ indices, is transformed to the $b'^{(c'_s,\dotsc,c'_2)}_{\NN}$-brane (note that $c_1\equiv b$) with
\begin{align}
 c'_i = c_i -1\,, \qquad c'_{\NN-i} = c_{\NN -i}+1\,,\qquad c'_k = c_k\quad (k:\text{others})\qquad \text{(if $i\neq \NN/2$)} \,, 
\end{align}
and it is invariant if $i=\NN/2$\,. 
If $x^y$ is not contained anywhere, we get a $b'^{(c'_s,\dotsc,c'_2)}_{\NN}$-brane with
\begin{align}
 c'_{\NN} = c_{\NN}+1\,,\qquad c'_k = c_k\quad (k:\text{others}) \,,
\end{align}
which may be understood as the case $i=0$\,. 
In terms of the mixed-symmetry potential, the original brane couples to $A_{\underbrace{\scriptstyle m_1\cdots m_{a_1}y,\dotsc,n_1\cdots n_{a_i}y}_{i\text{ blocks}},p_1\cdots p_{a_{i+1}},\dotsc,q_1\cdots q_{c_s}}$ $(y\not\in\{m_1,\dotsc,q_{c_s}\})$ while the brane after the $T$-duality couples to $A_{\underbrace{\scriptstyle m_1\cdots m_{a_1}y,\dotsc,r_1\cdots r_{a_{\NN-i}}y}_{\NN-i\text{ blocks}},s_1\cdots s_{\NN-i+1},\dotsc,q_1\cdots q_{c_s}}$\,. 
This $T$-duality rule, or more schematically,
\begin{align}
 \ovalbox{\Pbox{0.4\textwidth}{$\displaystyle \qquad\underbrace{y,\dotsc,y}_i \qquad \overset{T_y}{\leftrightarrow} \qquad \underbrace{y,\dotsc,y}_{\NN-i}$}}\,,
\label{eq:T-rule}
\end{align}
has been noted in \cite{1610.07975}. 
For example, let us consider a potential $\SUSY{H_{10,7,7,5,1}}$ that is predicted by $E_{11}$\,. 
Depending on the choice of the integer $i$, we obtain the following potentials:
\begin{align}
\begin{split}
 i=0: & \quad \text{N/A\quad (the first 10 indices contain $y$)}\,,
\\
 i=1: & \quad \SUSY{H_{10,7,7,5,1}} = \SUSY{H_{9y,7,7,5,1}}\quad\overset{T_y}{\leftrightarrow}\quad
  \SUSY{H_{9y,7y,7y,5y,1y}} = \SUSY{H_{10,8,8,6,2}}\,,
\\
 i=2: & \quad \text{N/A\quad (from the restriction rule \eqref{eq:restriction-rule})}\,,
\\
 i=3: & \quad \SUSY{H_{10,7,7,5,1}} = \SUSY{H_{9y,6y,6y,5,1}}\quad\overset{T_y}{\leftrightarrow}\quad
  \SUSY{H_{9y,6y,6y,5,1}} = \SUSY{H_{10,7,7,5,1}}\,,
\\
 i=4: & \quad \SUSY{H_{10,7,7,5,1}} = \SUSY{H_{9y,6y,6y,4y,1}}\quad\overset{T_y}{\leftrightarrow}\quad
  \SUSY{H_{9y,6y,6,4,1}} = \SUSY{H_{10,7,6,4,1}}\,,
\\
 i=5: & \quad \SUSY{H_{10,7,7,5,1}} = \SUSY{H_{9y,6y,6y,4y,y}}\quad\overset{T_y}{\leftrightarrow}\quad
  \SUSY{H_{9y,6,6,4}} = \SUSY{H_{10,6,6,4}}\,.
\end{split}
\end{align}
All of the potentials on the right-hand side are indeed predicted by $E_{11}$ and they are in the same $T$-duality multiplet. 
This rule is simple and useful to know the $T$-duality rule for mixed-symmetry potentials. 

\subsection{$\OO(D,D)$ multiplets of the mixed-symmetry potentials}

We can consider another interesting level decomposition by deleting the node $\alpha_{10}$:
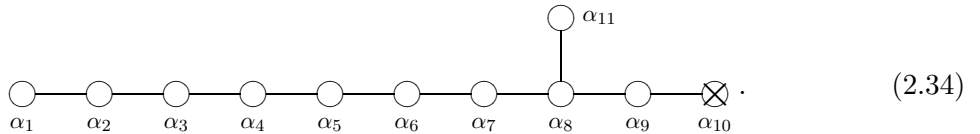
\begin{align}
\scalebox{0.8}{
 \xygraph{
    *\cir<6pt>{} ([]!{+(0,-.4)} {\alpha_1}) - [r]
    *\cir<6pt>{} ([]!{+(0,-.4)} {\alpha_2}) - [r]
    *\cir<6pt>{} ([]!{+(0,-.4)} {\alpha_3}) - [r]
    *\cir<6pt>{} ([]!{+(0,-.4)} {\alpha_4}) - [r]
    *\cir<6pt>{} ([]!{+(0,-.4)} {\alpha_5}) - [r]
    *\cir<6pt>{} ([]!{+(0,-.4)} {\alpha_6}) - [r]
    *\cir<6pt>{} ([]!{+(0,-.4)} {\alpha_7}) - [r]
    *\cir<6pt>{} ([]!{+(0,-.4)} {\alpha_8})
(
        - [u] *\cir<6pt>{} ([]!{+(.5,0)} {\alpha_{11}}),
        - [r] *\cir<6pt>{} ([]!{+(0,-.4)} {\alpha_9})
        - [r] *{\scalebox{2}{$\times$}}*\cir<6pt>{} ([]!{+(0,-.4)} {\alpha_{10}})
) 
}} .
\end{align}
We denote the associated level by $\NN$ and, from the tension formula discussed in \cite{0805.4451}, this level $\NN$ corresponds to the power $T\propto g_s^{-\NN}$ of the associated brane. 
Under this level decomposition, the $E_{11}$ generators are decomposed into representations of the $\OO(10,10)$ $T$-duality group. 
Again by using SimpLie, we can execute the level decomposition. 
The result up to level $\NN=4$ is given in Table \ref{tab:O(10-10)}, and there, further decompositions into the type IIA/IIB tensors are made.
\begin{table}
\centerline{\footnotesize
\begin{tabular}{|c|| C{1.5cm} || L{6.4cm} | L{6.4cm} |}\hline
 $\NN$ & $\OO(10,10)$ tensors & \multicolumn{1}{c|}{Type IIA potentials} & \multicolumn{1}{c|}{Type IIB potentials}
\\\hline\hline
%%%%%%%%%%%%%%%%%%%%%%%%%%%%%%%%%%%%%%%%%%%%%%%%%%%
 0 &$\SUSY{B_{MN}}$\,\,{\tiny$({\bf 190})$}, %0	0 1 0 0 0 0 0 0 0 0		-1 -2 -2 -2 -2 -2 -2 -2 -1 0 -1	2	190	1	1	1	17
 \newline
 $B$ {\tiny$({\bf 1})$}%0	0 0 0 0 0 0 0 0 0 0		0 0 0 0 0 0 0 0 0 0 0	0	1	1	11	1	0
 &\multicolumn{2}{c|}{\tiny 
 $\SUSY{e_1^1}\,\,({\bf 99})$, $e\,\,({\bf 1})$, $\Phi\,\,({\bf 1})$, $\SUSY{B_{2}}\,\,({\bf 45})$, $\SUSY{\beta^{2}}\,\,({\bf \overline{45}})$}
\\\hline
\rowcolor{Gray}
 1 &$\SUSY{C_{\dot{a}}}$ {\tiny$({\bf 512})$}%1	0 0 0 0 0 0 0 0 1 0		0 0 0 0 0 0 0 0 0 1 0	2	512	1	1	1	46
&{\tiny
% 0, 10, True
$\SUSY{C_{1}}\,\,({\bf 10})$, % [0 0 0 0 0 0 0 0 1] 
% 1, 120, True
$\SUSY{C_{3}}\,\,({\bf 120})$, % [0 0 0 0 0 0 1 0 0] 
% 2, 252, True
$\SUSY{C_{5}}\,\,({\bf 252})$, % [0 0 0 0 1 0 0 0 0] 
% 3, 120, True
$\SUSY{C_{7}}\,\,({\bf \overline{120}})$, % [0 0 1 0 0 0 0 0 0] 
% 4, 10, True
$\SUSY{C_{9}}\,\,({\bf \overline{10}})$} % [1 0 0 0 0 0 0 0 0] 
&{\tiny
% 0, 1, True
$\SUSY{C_{0}}\,\,({\bf 1})$, % [0 0 0 0 0 0 0 0 0] 
% 1, 45, True
$\SUSY{C_{2}}\,\,({\bf 45})$, % [0 0 0 0 0 0 0 1 0] 
% 2, 210, True
$\SUSY{C_{4}}\,\,({\bf 210})$, % [0 0 0 0 0 1 0 0 0] 
% 3, 210, True
$\SUSY{C_{6}}\,\,({\bf \overline{210}})$, % [0 0 0 1 0 0 0 0 0] 
% 4, 45, True
$\SUSY{C_{8}}\,\,({\bf \overline{45}})$, % [0 1 0 0 0 0 0 0 0] 
% 5, 1, True
$\SUSY{C_{10}}\,\,({\bf 1})$} % [0 0 0 0 0 0 0 0 0] 
\\\hline
 2 &$\SUSY{D_{A_{1\cdots4}}}$ {\tiny$({\bf 4845})$}%2	0 0 0 1 0 0 0 0 0 0		0 0 0 0 1 2 3 4 3 2 2	2	4845	1	1	1	77
&\multicolumn{2}{c|}{\tiny
% 3, 210, True
$\SUSY{D_{6}}\,\,({\bf \overline{210}})$, % [0 0 0 1 0 0 0 0 0] 
% 4, 1200, True
$\SUSY{D_{7,1}}\,\,({\bf \overline{1155}})$, % [0 0 1 0 0 0 0 0 1] 
$D_{8}\,\,({\bf \overline{45}})$, % [0 1 0 0 0 0 0 0 0] 
% 5, 2025, True
$\SUSY{D_{8,2}}\,\,({\bf 1925})$, % [0 1 0 0 0 0 0 1 0] 
$D_{9,1}\,\,({\bf 99})$, % [1 0 0 0 0 0 0 0 1] 
$D_{10}\,\,({\bf 1})$, % [0 0 0 0 0 0 0 0 0] 
% 6, 1200, True
$\SUSY{D_{9,3}}\,\,({\bf 1155})$, % [1 0 0 0 0 0 1 0 0] 
$D_{10,2}\,\,({\bf 45})$, % [0 0 0 0 0 0 0 1 0] 
% 7, 210, True
$\SUSY{D_{10,4}}\,\,({\bf 210})$} % [0 0 0 0 0 1 0 0 0] 
\\
\rowcolor{Gray}
 &$D$\,\,{\tiny$({\bf 1})$}%2	0 0 0 0 0 0 0 0 0 0		1 2 3 4 5 6 7 8 5 2 4	-2	1	1	46	1	47
&\multicolumn{2}{c|}{\tiny $D_{10}\,\,({\bf 1})$}
\\\hline
 3 &$\SUSY{E_{A_{12}\dot{a}}}$ {\tiny$({\bf 87040})$}%3	0 1 0 0 0 0 0 0 1 0		0 0 1 2 3 4 5 6 4 3 3	2	87040	1	1	1	110
&{\tiny
% 3, 440, True
$\SUSY{E_{8,1}}\,\,({\bf \overline{440}})$, % [0 1 0 0 0 0 0 0 1] 
% 4, 5940, True
$\SUSY{E_{8,3}}\,\,({\bf 4950})$, % [0 1 0 0 0 0 1 0 0] 
$\SUSY{E_{9,1,1}}\,\,({\bf 540})$, % [1 0 0 0 0 0 0 0 2] 
$E_{9,2}\,\,({\bf 440})$, % [1 0 0 0 0 0 0 1 0] 
$E_{10,1}\,\,({\bf 10})$, % [0 0 0 0 0 0 0 0 1] 
% 5, 21240, True
$\SUSY{E_{8,5}}\,\,({\bf 9240})$, % [0 1 0 0 1 0 0 0 0] 
$\SUSY{E_{9,3,1}}\,\,({\bf 9450})$, % [1 0 0 0 0 0 1 0 1] 
$E_{9,4}\,\,({\bf 1980})$, % [1 0 0 0 0 1 0 0 0] 
$E_{10,2,1}\,\,({\bf 330})$, % [0 0 0 0 0 0 0 1 1] 
$2\,E_{10,3}\,\,(2\times {\bf 120})$, % [0 0 0 0 0 0 1 0 0] 
% 6, 31800, True
$\SUSY{E_{9,5,1}}\,\,({\bf 21000})$, % [1 0 0 0 1 0 0 0 1] 
$\SUSY{E_{10,3,2}}\,\,({\bf 3300})$, % [0 0 0 0 0 0 1 1 0] 
$\SUSY{E_{8,7}}\,\,({\bf \overline{3300}})$, % [0 1 1 0 0 0 0 0 0] 
$E_{10,4,1}\,\,({\bf 1848})$, % [0 0 0 0 0 1 0 0 1] 
$E_{9,6}\,\,({\bf \overline{1848}})$, % [1 0 0 1 0 0 0 0 0] 
$2\,E_{10,5}\,\,(2\times {\bf 252})$, % [0 0 0 0 1 0 0 0 0] 
% 7, 21240, True
$\SUSY{E_{10,5,2}}\,\,({\bf \overline{9240}})$, % [0 0 0 0 1 0 0 1 0] 
$\SUSY{E_{9,7,1}}\,\,({\bf \overline{9450}})$, % [1 0 1 0 0 0 0 0 1] 
$E_{10,6,1}\,\,({\bf \overline{1980}})$, % [0 0 0 1 0 0 0 0 1] 
$E_{9,8}\,\,({\bf \overline{330}})$, % [1 1 0 0 0 0 0 0 0] 
$2\,E_{10,7}\,\,(2\times {\bf \overline{120}})$, % [0 0 1 0 0 0 0 0 0] 
% 8, 5940, True
$\SUSY{E_{10,7,2}}\,\,({\bf \overline{4950}})$, % [0 0 1 0 0 0 0 1 0] 
$\SUSY{E_{9,9,1}}\,\,({\bf \overline{540}})$, % [2 0 0 0 0 0 0 0 1] 
$E_{10,8,1}\,\,({\bf \overline{440}})$, % [0 1 0 0 0 0 0 0 1] 
$E_{10,9}\,\,({\bf \overline{10}})$, % [1 0 0 0 0 0 0 0 0] 
% 9, 440, True
$\SUSY{E_{10,9,2}}\,\,({\bf 440})$} % [1 0 0 0 0 0 0 1 0] 
&{\tiny
% 4, 45, True
$\SUSY{E_{8}}\,\,({\bf \overline{45}})$, % [0 1 0 0 0 0 0 0 0] 
% 5, 2025, True
$\SUSY{E_{8,2}}\,\,({\bf 1925})$, % [0 1 0 0 0 0 0 1 0] 
$E_{9,1}\,\,({\bf 99})$, % [1 0 0 0 0 0 0 0 1] 
$E_{10}\,\,({\bf 1})$, % [0 0 0 0 0 0 0 0 0] 
% 6, 12695, True
$\SUSY{E_{8,4}}\,\,({\bf 8250})$, % [0 1 0 0 0 1 0 0 0] 
$\SUSY{E_{9,2,1}}\,\,({\bf 3200})$, % [1 0 0 0 0 0 0 1 1] 
$E_{9,3}\,\,({\bf 1155})$, % [1 0 0 0 0 0 1 0 0] 
$2\,E_{10,2}\,\,(2\times {\bf 45})$, % [0 0 0 0 0 0 0 1 0] 
% 7, 28755, True
$\SUSY{E_{9,4,1}}\,\,({\bf 17280})$, % [1 0 0 0 0 1 0 0 1] 
$\SUSY{E_{8,6}}\,\,({\bf 6930})$, % [0 1 0 1 0 0 0 0 0] 
$E_{9,5}\,\,({\bf 2310})$, % [1 0 0 0 1 0 0 0 0] 
$\SUSY{E_{10,2,2}}\,\,({\bf 825})$, % [0 0 0 0 0 0 0 2 0] 
$E_{10,3,1}\,\,({\bf 990})$, % [0 0 0 0 0 0 1 0 1] 
$2\,E_{10,4}\,\,(2\times {\bf 210})$, % [0 0 0 0 0 1 0 0 0] 
% 8, 28755, True
$\SUSY{E_{9,6,1}}\,\,({\bf \overline{17280}})$, % [1 0 0 1 0 0 0 0 1] 
$\SUSY{E_{10,4,2}}\,\,({\bf \overline{6930}})$, % [0 0 0 0 0 1 0 1 0] 
$E_{10,5,1}\,\,({\bf \overline{2310}})$, % [0 0 0 0 1 0 0 0 1] 
$\SUSY{E_{8,8}}\,\,({\bf \overline{825}})$, % [0 2 0 0 0 0 0 0 0] 
$E_{9,7}\,\,({\bf \overline{990}})$, % [1 0 1 0 0 0 0 0 0] 
$2\,E_{10,6}\,\,(2\times {\bf \overline{210}})$, % [0 0 0 1 0 0 0 0 0] 
% 9, 12695, True
$\SUSY{E_{10,6,2}}\,\,({\bf \overline{8250}})$, % [0 0 0 1 0 0 0 1 0] 
$\SUSY{E_{9,8,1}}\,\,({\bf \overline{3200}})$, % [1 1 0 0 0 0 0 0 1] 
$E_{10,7,1}\,\,({\bf \overline{1155}})$, % [0 0 1 0 0 0 0 0 1] 
$2\,E_{10,8}\,\,(2\times {\bf \overline{45}})$, % [0 1 0 0 0 0 0 0 0] 
% 10, 2025, True
$\SUSY{E_{10,8,2}}\,\,({\bf 1925})$, % [0 1 0 0 0 0 0 1 0] 
$E_{10,9,1}\,\,({\bf 99})$, % [1 0 0 0 0 0 0 0 1] 
$E_{10,10}\,\,({\bf 1})$, % [0 0 0 0 0 0 0 0 0] 
% 11, 45, True
$\SUSY{E_{10,10,2}}\,\,({\bf 45})$} % [0 0 0 0 0 0 0 1 0] 
\\
\rowcolor{Gray}
 &$E_{\dot{a}}$\,\,{\tiny$({\bf 512})$}%3	0 0 0 0 0 0 0 0 1 0		1 2 3 4 5 6 7 8 5 3 4	-2	512	1	46	1	93
&\multicolumn{1}{L{6.41cm}|}{\tiny
% 4, 10, True
$E_{10,1}\,\,({\bf 10})$, % [0 0 0 0 0 0 0 0 1] 
% 5, 120, True
$E_{10,3}\,\,({\bf 120})$, % [0 0 0 0 0 0 1 0 0] 
% 6, 252, True
$E_{10,5}\,\,({\bf 252})$, % [0 0 0 0 1 0 0 0 0] 
% 7, 120, True
$E_{10,7}\,\,({\bf \overline{120}})$, % [0 0 1 0 0 0 0 0 0] 
% 8, 10, True
$E_{10,9}\,\,({\bf \overline{10}})$} % [1 0 0 0 0 0 0 0 0] 
&\multicolumn{1}{L{6.41cm}|}{\tiny
% 5, 1, True
$E_{10}\,\,({\bf 1})$, % [0 0 0 0 0 0 0 0 0] 
% 6, 45, True
$E_{10,2}\,\,({\bf 45})$, % [0 0 0 0 0 0 0 1 0] 
% 7, 210, True
$E_{10,4}\,\,({\bf 210})$, % [0 0 0 0 0 1 0 0 0] 
% 8, 210, True
$E_{10,6}\,\,({\bf \overline{210}})$, % [0 0 0 1 0 0 0 0 0] 
% 9, 45, True
$E_{10,8}\,\,({\bf \overline{45}})$, % [0 1 0 0 0 0 0 0 0] 
% 10, 1, True
$E_{10,10}\,\,({\bf 1})$} % [0 0 0 0 0 0 0 0 0] 
\\\hline
 4 &$\SUSY{F_{A_{1\cdots10}}^+}$ {\tiny$({\bf 92378})$}%4	0 0 0 0 0 0 0 0 2 0		1 2 3 4 5 6 7 8 5 4 4	2	92378	1	1	1	139
&{\tiny
% 4, 55, True
$\SUSY{F_{10,1,1}}\,\,({\bf 55})$, % [0 0 0 0 0 0 0 0 2] 
% 5, 990, True
$F_{10,3,1}\,\,({\bf 990})$, % [0 0 0 0 0 0 1 0 1] 
% 6, 7260, True
$\SUSY{F_{10,3,3}}\,\,({\bf \overline{4950}'})$, % [0 0 0 0 0 0 2 0 0] 
$F_{10,5,1}\,\,({\bf \overline{2310}})$, % [0 0 0 0 1 0 0 0 1] 
% 7, 21945, True
$F_{10,5,3}\,\,({\bf \overline{20790}})$, % [0 0 0 0 1 0 1 0 0] 
$F_{10,7,1}\,\,({\bf \overline{1155}})$, % [0 0 1 0 0 0 0 0 1] 
% 8, 31878, True
$\SUSY{F_{10,5,5}}\,\,({\bf 19404})$, % [0 0 0 0 2 0 0 0 0] 
$F_{10,7,3}\,\,({\bf 12375})$, % [0 0 1 0 0 0 1 0 0] 
$F_{10,9,1}\,\,({\bf 99})$, % [1 0 0 0 0 0 0 0 1] 
% 9, 21945, True
$F_{10,7,5}\,\,({\bf 20790})$, % [0 0 1 0 1 0 0 0 0] 
$F_{10,9,3}\,\,({\bf 1155})$, % [1 0 0 0 0 0 1 0 0] 
% 10, 7260, True
$\SUSY{F_{10,7,7}}\,\,({\bf 4950'})$, % [0 0 2 0 0 0 0 0 0] 
$F_{10,9,5}\,\,({\bf 2310})$, % [1 0 0 0 1 0 0 0 0] 
% 11, 990, True
$F_{10,9,7}\,\,({\bf \overline{990}})$, % [1 0 1 0 0 0 0 0 0] 
% 12, 55, True
$\SUSY{F_{10,9,9}}\,\,({\bf \overline{55}})$} % [2 0 0 0 0 0 0 0 0] 
&{\tiny
% 5, 1, True
$\SUSY{F_{10}}\,\,({\bf 1})$, % [0 0 0 0 0 0 0 0 0] 
% 6, 45, True
$F_{10,2}\,\,({\bf 45})$, % [0 0 0 0 0 0 0 1 0] 
% 7, 1035, True
$\SUSY{F_{10,2,2}}\,\,({\bf 825})$, % [0 0 0 0 0 0 0 2 0] 
$F_{10,4}\,\,({\bf 210})$, % [0 0 0 0 0 1 0 0 0] 
% 8, 7140, True
$F_{10,4,2}\,\,({\bf \overline{6930}})$, % [0 0 0 0 0 1 0 1 0] 
$F_{10,6}\,\,({\bf \overline{210}})$, % [0 0 0 1 0 0 0 0 0] 
% 9, 22155, True
$\SUSY{F_{10,4,4}}\,\,({\bf \overline{13860}''})$, % [0 0 0 0 0 2 0 0 0] 
$F_{10,6,2}\,\,({\bf \overline{8250}})$, % [0 0 0 1 0 0 0 1 0] 
$F_{10,8}\,\,({\bf \overline{45}})$, % [0 1 0 0 0 0 0 0 0] 
% 10, 31626, True
$F_{10,6,4}\,\,({\bf 29700})$, % [0 0 0 1 0 1 0 0 0] 
$F_{10,8,2}\,\,({\bf 1925})$, % [0 1 0 0 0 0 0 1 0] 
$F_{10,10}\,\,({\bf 1})$, % [0 0 0 0 0 0 0 0 0] 
% 11, 22155, True
$\SUSY{F_{10,6,6}}\,\,({\bf 13860''})$, % [0 0 0 2 0 0 0 0 0] 
$F_{10,8,4}\,\,({\bf 8250})$, % [0 1 0 0 0 1 0 0 0] 
$F_{10,10,2}\,\,({\bf 45})$, % [0 0 0 0 0 0 0 1 0] 
% 12, 7140, True
$F_{10,8,6}\,\,({\bf 6930})$, % [0 1 0 1 0 0 0 0 0] 
$F_{10,10,4}\,\,({\bf 210})$, % [0 0 0 0 0 1 0 0 0] 
% 13, 1035, True
$\SUSY{F_{10,8,8}}\,\,({\bf \overline{825}})$, % [0 2 0 0 0 0 0 0 0] 
$F_{10,10,6}\,\,({\bf \overline{210}})$, % [0 0 0 1 0 0 0 0 0] 
% 14, 45, True
$F_{10,10,8}\,\,({\bf \overline{45}})$, % [0 1 0 0 0 0 0 0 0] 
% 15, 1, True
$\SUSY{F_{10,10,10}}\,\,({\bf 1})$} % [0 0 0 0 0 0 0 0 0] 
\\
\rowcolor{Gray}
 &$\SUSY{F_{A_{1\cdots7},B}}$ {\tiny$({\bf 1385670})$}%4	1 0 0 0 0 0 1 0 0 0		0 1 2 3 4 5 6 8 6 4 4	2	1385670	1	1	1	145
&\multicolumn{2}{C{13.23cm}|}{\tiny
% 6, 1155, True
$\SUSY{F_{9,3}}\,\,({\bf 1155})$, % [1 0 0 0 0 0 1 0 0] 
% 7, 20790, True
$\SUSY{F_{9,4,1}}\,\,({\bf 17280})$, % [1 0 0 0 0 1 0 0 1] 
$F_{9,5}\,\,({\bf 2310})$, % [1 0 0 0 1 0 0 0 0] 
$F_{10,3,1}\,\,({\bf 990})$, % [0 0 0 0 0 0 1 0 1] 
$F_{10,4}\,\,({\bf 210})$, % [0 0 0 0 0 1 0 0 0] 
% 8, 122430, True
$\SUSY{F_{9,5,2}}\,\,({\bf \overline{83160}})$, % [1 0 0 0 1 0 0 1 0] 
$\SUSY{F_{10,4,1,1}}\,\,({\bf \overline{9240}'})$, % [0 0 0 0 0 1 0 0 2] 
$F_{9,6,1}\,\,({\bf \overline{17280}})$, % [1 0 0 1 0 0 0 0 1] 
$F_{10,4,2}\,\,({\bf \overline{6930}})$, % [0 0 0 0 0 1 0 1 0] 
$2\,F_{10,5,1}\,\,(2\times {\bf \overline{2310}})$, % [0 0 0 0 1 0 0 0 1] 
$F_{9,7}\,\,({\bf \overline{990}})$, % [1 0 1 0 0 0 0 0 0] 
$F_{10,6}\,\,({\bf \overline{210}})$, % [0 0 0 1 0 0 0 0 0] 
% 9, 325710, True
$\SUSY{F_{9,6,3}}\,\,({\bf \overline{168960}})$, % [1 0 0 1 0 0 1 0 0] 
$\SUSY{F_{10,5,2,1}}\,\,({\bf \overline{63360}})$, % [0 0 0 0 1 0 0 1 1] 
$F_{10,5,3}\,\,({\bf \overline{20790}})$, % [0 0 0 0 1 0 1 0 0] 
$F_{9,7,2}\,\,({\bf \overline{40095}})$, % [1 0 1 0 0 0 0 1 0] 
$F_{10,6,1,1}\,\,({\bf \overline{10395}})$, % [0 0 0 1 0 0 0 0 2] 
$2\,F_{10,6,2}\,\,(2\times {\bf \overline{8250}})$, % [0 0 0 1 0 0 0 1 0] 
$F_{9,8,1}\,\,({\bf \overline{3200}})$, % [1 1 0 0 0 0 0 0 1] 
$2\,F_{10,7,1}\,\,(2\times {\bf \overline{1155}})$, % [0 0 1 0 0 0 0 0 1] 
$F_{9,9}\,\,({\bf \overline{55}})$, % [2 0 0 0 0 0 0 0 0] 
$F_{10,8}\,\,({\bf \overline{45}})$, % [0 1 0 0 0 0 0 0 0] 
% 10, 445500, True
$\SUSY{F_{10,6,3,1}}\,\,({\bf 155925})$, % [0 0 0 1 0 0 1 0 1] 
$\SUSY{F_{9,7,4}}\,\,({\bf \overline{155925}})$, % [1 0 1 0 0 1 0 0 0] 
$F_{10,6,4}\,\,({\bf 29700})$, % [0 0 0 1 0 1 0 0 0] 
$F_{10,7,2,1}\,\,({\bf 35200})$, % [0 0 1 0 0 0 0 1 1] 
$F_{9,8,3}\,\,({\bf \overline{35200}})$, % [1 1 0 0 0 0 1 0 0] 
$2\,F_{10,7,3}\,\,(2\times {\bf 12375})$, % [0 0 1 0 0 0 1 0 0] 
$F_{10,8,1,1}\,\,({\bf 2376})$, % [0 1 0 0 0 0 0 0 2] 
$F_{9,9,2}\,\,({\bf \overline{2376}})$, % [2 0 0 0 0 0 0 1 0] 
$2\,F_{10,8,2}\,\,(2\times {\bf 1925})$, % [0 1 0 0 0 0 0 1 0] 
$2\,F_{10,9,1}\,\,(2\times {\bf 99})$, % [1 0 0 0 0 0 0 0 1] 
% 11, 325710, True
$\SUSY{F_{10,7,4,1}}\,\,({\bf 168960})$, % [0 0 1 0 0 1 0 0 1] 
$\SUSY{F_{9,8,5}}\,\,({\bf 63360})$, % [1 1 0 0 1 0 0 0 0] 
$F_{10,7,5}\,\,({\bf 20790})$, % [0 0 1 0 1 0 0 0 0] 
$F_{10,8,3,1}\,\,({\bf 40095})$, % [0 1 0 0 0 0 1 0 1] 
$F_{9,9,4}\,\,({\bf 10395})$, % [2 0 0 0 0 1 0 0 0] 
$2\,F_{10,8,4}\,\,(2\times {\bf 8250})$, % [0 1 0 0 0 1 0 0 0] 
$F_{10,9,2,1}\,\,({\bf 3200})$, % [1 0 0 0 0 0 0 1 1] 
$2\,F_{10,9,3}\,\,(2\times {\bf 1155})$, % [1 0 0 0 0 0 1 0 0] 
$F_{10,10,1,1}\,\,({\bf 55})$, % [0 0 0 0 0 0 0 0 2] 
$F_{10,10,2}\,\,({\bf 45})$, % [0 0 0 0 0 0 0 1 0] 
% 12, 122430, True
$\SUSY{F_{10,8,5,1}}\,\,({\bf 83160})$, % [0 1 0 0 1 0 0 0 1] 
$F_{10,9,4,1}\,\,({\bf 17280})$, % [1 0 0 0 0 1 0 0 1] 
$\SUSY{F_{9,9,6}}\,\,({\bf 9240'})$, % [2 0 0 1 0 0 0 0 0] 
$F_{10,8,6}\,\,({\bf 6930})$, % [0 1 0 1 0 0 0 0 0] 
$2\,F_{10,9,5}\,\,(2\times {\bf 2310})$, % [1 0 0 0 1 0 0 0 0] 
$F_{10,10,3,1}\,\,({\bf 990})$, % [0 0 0 0 0 0 1 0 1] 
$F_{10,10,4}\,\,({\bf 210})$, % [0 0 0 0 0 1 0 0 0] 
% 13, 20790, True
$\SUSY{F_{10,9,6,1}}\,\,({\bf \overline{17280}})$, % [1 0 0 1 0 0 0 0 1] 
$F_{10,10,5,1}\,\,({\bf \overline{2310}})$, % [0 0 0 0 1 0 0 0 1] 
$F_{10,9,7}\,\,({\bf \overline{990}})$, % [1 0 1 0 0 0 0 0 0] 
$F_{10,10,6}\,\,({\bf \overline{210}})$, % [0 0 0 1 0 0 0 0 0] 
% 14, 1155, True
$\SUSY{F_{10,10,7,1}}\,\,({\bf \overline{1155}})$} % [0 0 1 0 0 0 0 0 1] 
\\
 &$F_{A_{1\cdots6}}$ {\tiny$({\bf 38760})$}%4	0 0 0 0 0 1 0 0 0 0		1 2 3 4 5 6 8 10 7 4 5	-2	38760	1	46	1	133
&\multicolumn{2}{C{13.23cm}|}{\tiny
% 7, 210, True
$F_{10,4}\,\,({\bf 210})$, % [0 0 0 0 0 1 0 0 0] 
% 8, 2520, True
$F_{10,5,1}\,\,({\bf \overline{2310}})$, % [0 0 0 0 1 0 0 0 1] 
$F_{10,6}\,\,({\bf \overline{210}})$, % [0 0 0 1 0 0 0 0 0] 
% 9, 9450, True
$F_{10,6,2}\,\,({\bf \overline{8250}})$, % [0 0 0 1 0 0 0 1 0] 
$F_{10,7,1}\,\,({\bf \overline{1155}})$, % [0 0 1 0 0 0 0 0 1] 
$F_{10,8}\,\,({\bf \overline{45}})$, % [0 1 0 0 0 0 0 0 0] 
% 10, 14400, True
$F_{10,7,3}\,\,({\bf 12375})$, % [0 0 1 0 0 0 1 0 0] 
$F_{10,8,2}\,\,({\bf 1925})$, % [0 1 0 0 0 0 0 1 0] 
$F_{10,9,1}\,\,({\bf 99})$, % [1 0 0 0 0 0 0 0 1] 
$F_{10,10}\,\,({\bf 1})$, % [0 0 0 0 0 0 0 0 0] 
% 11, 9450, True
$F_{10,8,4}\,\,({\bf 8250})$, % [0 1 0 0 0 1 0 0 0] 
$F_{10,9,3}\,\,({\bf 1155})$, % [1 0 0 0 0 0 1 0 0] 
$F_{10,10,2}\,\,({\bf 45})$, % [0 0 0 0 0 0 0 1 0] 
% 12, 2520, True
$F_{10,9,5}\,\,({\bf 2310})$, % [1 0 0 0 1 0 0 0 0] 
$F_{10,10,4}\,\,({\bf 210})$, % [0 0 0 0 0 1 0 0 0] 
% 13, 210, True
$F_{10,10,6}\,\,({\bf \overline{210}})$} % [0 0 0 1 0 0 0 0 0] 
\\
\rowcolor{Gray}
 &$\SUSY{F_{A_{1\cdots4},B_{12}}}$ {\tiny$({\bf 592515})$}%4	0 1 0 1 0 0 0 0 0 0		0 0 1 2 4 6 8 10 7 4 5	2	592515	1	1	1	141
&\multicolumn{2}{C{13.23cm}|}{\tiny
% 7, 6930, True
$\SUSY{F_{8,6}}\,\,({\bf 6930})$, % [0 1 0 1 0 0 0 0 0] 
% 8, 50490, True
$\SUSY{F_{8,7,1}}\,\,({\bf \overline{31185}})$, % [0 1 1 0 0 0 0 0 1] 
$F_{9,6,1}\,\,({\bf \overline{17280}})$, % [1 0 0 1 0 0 0 0 1] 
$F_{8,8}\,\,({\bf \overline{825}})$, % [0 2 0 0 0 0 0 0 0] 
$F_{9,7}\,\,({\bf \overline{990}})$, % [1 0 1 0 0 0 0 0 0] 
$F_{10,6}\,\,({\bf \overline{210}})$, % [0 0 0 1 0 0 0 0 0] 
% 9, 141075, True
$\SUSY{F_{8,8,2}}\,\,({\bf \overline{33880}})$, % [0 2 0 0 0 0 0 1 0] 
$\SUSY{F_{9,7,1,1}}\,\,({\bf \overline{50050}})$, % [1 0 1 0 0 0 0 0 2] 
$F_{9,7,2}\,\,({\bf \overline{40095}})$, % [1 0 1 0 0 0 0 1 0] 
$F_{10,6,2}\,\,({\bf \overline{8250}})$, % [0 0 0 1 0 0 0 1 0] 
$2\,F_{9,8,1}\,\,(2\times {\bf \overline{3200}})$, % [1 1 0 0 0 0 0 0 1] 
$2\,F_{10,7,1}\,\,(2\times {\bf \overline{1155}})$, % [0 0 1 0 0 0 0 0 1] 
$2\,F_{10,8}\,\,(2\times {\bf \overline{45}})$, % [0 1 0 0 0 0 0 0 0] 
% 10, 195525, True
$\SUSY{F_{9,8,2,1}}\,\,({\bf 99099})$, % [1 1 0 0 0 0 0 1 1] 
$F_{10,7,2,1}\,\,({\bf 35200})$, % [0 0 1 0 0 0 0 1 1] 
$F_{9,8,3}\,\,({\bf \overline{35200}})$, % [1 1 0 0 0 0 1 0 0] 
$F_{10,7,3}\,\,({\bf 12375})$, % [0 0 1 0 0 0 1 0 0] 
$F_{9,9,1,1}\,\,({\bf 2925})$, % [2 0 0 0 0 0 0 0 2] 
$F_{10,8,1,1}\,\,({\bf 2376})$, % [0 1 0 0 0 0 0 0 2] 
$F_{9,9,2}\,\,({\bf \overline{2376}})$, % [2 0 0 0 0 0 0 1 0] 
$3\,F_{10,8,2}\,\,(3\times {\bf 1925})$, % [0 1 0 0 0 0 0 1 0] 
$2\,F_{10,9,1}\,\,(2\times {\bf 99})$, % [1 0 0 0 0 0 0 0 1] 
$F_{10,10}\,\,({\bf 1})$, % [0 0 0 0 0 0 0 0 0] 
% 11, 141075, True
$\SUSY{F_{10,8,2,2}}\,\,({\bf 33880})$, % [0 1 0 0 0 0 0 2 0] 
$\SUSY{F_{9,9,3,1}}\,\,({\bf 50050})$, % [2 0 0 0 0 0 1 0 1] 
$F_{10,8,3,1}\,\,({\bf 40095})$, % [0 1 0 0 0 0 1 0 1] 
$F_{10,8,4}\,\,({\bf 8250})$, % [0 1 0 0 0 1 0 0 0] 
$2\,F_{10,9,2,1}\,\,(2\times {\bf 3200})$, % [1 0 0 0 0 0 0 1 1] 
$2\,F_{10,9,3}\,\,(2\times {\bf 1155})$, % [1 0 0 0 0 0 1 0 0] 
$2\,F_{10,10,2}\,\,(2\times {\bf 45})$, % [0 0 0 0 0 0 0 1 0] 
% 12, 50490, True
$\SUSY{F_{10,9,3,2}}\,\,({\bf 31185})$, % [1 0 0 0 0 0 1 1 0] 
$F_{10,9,4,1}\,\,({\bf 17280})$, % [1 0 0 0 0 1 0 0 1] 
$F_{10,10,2,2}\,\,({\bf 825})$, % [0 0 0 0 0 0 0 2 0] 
$F_{10,10,3,1}\,\,({\bf 990})$, % [0 0 0 0 0 0 1 0 1] 
$F_{10,10,4}\,\,({\bf 210})$, % [0 0 0 0 0 1 0 0 0] 
% 13, 6930, True
$\SUSY{F_{10,10,4,2}}\,\,({\bf \overline{6930}})$} % [0 0 0 0 0 1 0 1 0] 
\\
 &$F_{A_{1\cdots4}}$ {\tiny$({\bf 4845})$}%4	0 0 0 1 0 0 0 0 0 0		1 2 3 4 6 8 10 12 8 4 6	-4	4845	1	206	1	124
&\multicolumn{2}{C{13.23cm}|}{\tiny
% 8, 210, True
$F_{10,6}\,\,({\bf \overline{210}})$, % [0 0 0 1 0 0 0 0 0] 
% 9, 1200, True
$F_{10,7,1}\,\,({\bf \overline{1155}})$, % [0 0 1 0 0 0 0 0 1] 
$F_{10,8}\,\,({\bf \overline{45}})$, % [0 1 0 0 0 0 0 0 0] 
% 10, 2025, True
$F_{10,8,2}\,\,({\bf 1925})$, % [0 1 0 0 0 0 0 1 0] 
$F_{10,9,1}\,\,({\bf 99})$, % [1 0 0 0 0 0 0 0 1] 
$F_{10,10}\,\,({\bf 1})$, % [0 0 0 0 0 0 0 0 0] 
% 11, 1200, True
$F_{10,9,3}\,\,({\bf 1155})$, % [1 0 0 0 0 0 1 0 0] 
$F_{10,10,2}\,\,({\bf 45})$, % [0 0 0 0 0 0 0 1 0] 
% 12, 210, True
$F_{10,10,4}\,\,({\bf 210})$} % [0 0 0 0 0 1 0 0 0] 
\\
\rowcolor{Gray}
 &$F_{A_{123},B}$ {\tiny$({\bf 17765})$}%4	1 0 1 0 0 0 0 0 0 0		0 1 2 4 6 8 10 12 8 4 6	-2	17765	1	44	1	127
&\multicolumn{2}{C{13.23cm}|}{\tiny
% 8, 990, True
$F_{9,7}\,\,({\bf \overline{990}})$, % [1 0 1 0 0 0 0 0 0] 
% 9, 4455, True
$F_{9,8,1}\,\,({\bf \overline{3200}})$, % [1 1 0 0 0 0 0 0 1] 
$F_{10,7,1}\,\,({\bf \overline{1155}})$, % [0 0 1 0 0 0 0 0 1] 
$F_{9,9}\,\,({\bf \overline{55}})$, % [2 0 0 0 0 0 0 0 0] 
$F_{10,8}\,\,({\bf \overline{45}})$, % [0 1 0 0 0 0 0 0 0] 
% 10, 6875, True
$F_{10,8,1,1}\,\,({\bf 2376})$, % [0 1 0 0 0 0 0 0 2] 
$F_{9,9,2}\,\,({\bf \overline{2376}})$, % [2 0 0 0 0 0 0 1 0] 
$F_{10,8,2}\,\,({\bf 1925})$, % [0 1 0 0 0 0 0 1 0] 
$2\,F_{10,9,1}\,\,(2\times {\bf 99})$, % [1 0 0 0 0 0 0 0 1] 
% 11, 4455, True
$F_{10,9,2,1}\,\,({\bf 3200})$, % [1 0 0 0 0 0 0 1 1] 
$F_{10,9,3}\,\,({\bf 1155})$, % [1 0 0 0 0 0 1 0 0] 
$F_{10,10,1,1}\,\,({\bf 55})$, % [0 0 0 0 0 0 0 0 2] 
$F_{10,10,2}\,\,({\bf 45})$, % [0 0 0 0 0 0 0 1 0] 
% 12, 990, True
$F_{10,10,3,1}\,\,({\bf 990})$} % [0 0 0 0 0 0 1 0 1] 
\\
 &$2\,F_{A_{12}}$ {\tiny$(2\times{\bf 190})$}%4	0 1 0 0 0 0 0 0 0 0		1 2 4 6 8 10 12 14 9 4 7	-6	190	1	801	2	111
&\multicolumn{2}{c|}{\tiny 
% 9, 45, True
$2\times\bigl[F_{10,8}\,\,({\bf \overline{45}})$, % [0 1 0 0 0 0 0 0 0] 
% 10, 100, True
$F_{10,9,1}\,\,({\bf 99})$, % [1 0 0 0 0 0 0 0 1] 
$F_{10,10}\,\,({\bf 1})$, % [0 0 0 0 0 0 0 0 0] 
% 11, 45, True
$F_{10,10,2}\,\,({\bf 45})\bigr]$} % [0 0 0 0 0 0 0 1 0] 
\\\hline
\end{tabular}
}
\caption{$\OO(10,10)$ tensors in the adjoint representation of $E_{11}$ up to level $\NN=4$\,. }
\label{tab:O(10-10)}
\end{table}
We can proceed further to the potentials with level $\NN=5$\,, and obtained $\OO(10,10)$ tensors,
\begin{align}
\Pbox{0.8\textwidth}{$\SUSY{G_{A_{1\cdots6}\dot{a}}}$,\quad 
$\SUSY{G_{A_{12},B_{12}\dot{a}}}$,\quad 
$\SUSY{G_{A_{1\cdots4},Ba}}$,\quad 
$G_{A_{123},B\dot{a}}$,\quad 
$G_{A_{12},Ba}$,\quad 
$2\,G_{A_{1\cdots4}\dot{a}}$,\quad 
$G_{A,B\dot{a}}$,\quad 
$2\,G_{A_{123}a}$,\quad 
$4\,G_{A_{12}\dot{a}}$,\quad 
$3\,G_{Aa}$,\quad 
$3\,G_{\dot{a}}$\,.}
\end{align}
The table similar to Table \ref{tab:O(10-10)} is relegated to Appendix \ref{app:O(1010)-G-H} because it is extremely long. 
For the next level, $\NN=6$\,, it is difficult to execute the level decomposition because the dimension of the $\OO(10,10)$ representation is extremely high, of the order of $10^9$\,. 
As we explain in Appendix \ref{app:O(1010)-G-H}, we have determined most of the $\OO(10,10)$ tensors with $\NN=6$\,. 
The $\OO(10,10)$ tensors which contribute to Table \ref{tab:p-form} are completely determined as follows:
\begin{align}
\Pbox{0.8\textwidth}{$\SUSY{H_{A_{1\cdots8},B_{1\cdots4}}}$,\quad 
$H_{A_{1\cdots6},B_{1\cdots4}}$,\quad 
$H_{A_{1\cdots9},B_{123}}$,\quad 
$2\,H_{A_{1\cdots7},B_{123}}$,\quad 
$\SUSY{H_{A_{1\cdots6},B_{123},C}}$,\quad 
$H_{A_{1\cdots4},B_{123},C}$,\quad 
$2\,H_{A_{1\cdots4},B_{1\cdots4}}$,\quad 
$2\,H_{A_{1\cdots5},B_{123}}$,\quad 
$2\,H_{A_{123},B_{123}}$\,.}
\end{align}
On the other hand, the $\OO(10,10)$ tensors which do not contribute to Table \ref{tab:p-form} are as follows:
\begin{align}
\begin{split}
&\text{\underline{\textbf{Generalized NS--NS sector:}}}
\\
&\Pbox{0.9\textwidth}{
$\SUSY{H_{A_{1\cdots9},A_{12},B}}$,\quad
$H_{A_{1\cdots 7},B_{12},C}$,\quad
$\SUSY{H_{A_{1\cdots 4},B_{12},C_{12}}}$,\quad
$2\,H_{A_{1\cdots 8},B,C}$,\quad
$3\,H_{A_{1\cdots 8},B_{12}}$,\quad
$2\,H_{A_{1\cdots 5},B_{12},C}$,\quad
$H_{A_{1\cdots 6},B,C}$,\quad
$6\,H_{A_{1\cdots 6},B_{12}}$,\quad
$5\,H_{A_{1\cdots9},A}$,\quad
$6\,H_{A_{1\cdots 7},B}$,\quad
$2\,H_{A_{123},B_{12},C}$,\quad
$3\,H_{A_{1\cdots4},B,C}$,\quad
$6\,H_{A_{1\cdots4},B_{12}}$,\quad
$7\,H_{A_{1\cdots 8}}$,\quad
$8\,H_{A_{1\cdots5},B}$,\quad
$5\,H_{A_{1\cdots 6}}$,\quad
$H_{A_{12},B,C}$,\quad
$5\,H_{A_{12},B_{12}}$,\quad
$7\,H_{A_{123},B}$,\quad
$10\,H_{A_{1\cdots 4}}$,\quad
$3\,H_{A,B}$,\quad
$5\,H_{A_{12}}$,\quad
$\textcolor{blue}{?}\,H$,}
\end{split}
\\
\begin{split}
&\text{\underline{\textbf{Generalized R--R sector:}}}
\\
&\Pbox{0.9\textwidth}{
$H_{A_{1\cdots10},B_{12}}^-$,\quad
$3\,H_{A_{1\cdots10},B_{12}}^+$,\quad
$H_{A_{1\cdots10}}^-$,\quad
$3\,H_{A_{1\cdots10}}^+$.}\end{split}
\end{align}
Only the multiplicity of the singlet $H$ is not determined. 
Regarding the $\OO(10,10)$ tensors with higher level $\NN\geq 7$\,, only the possible $\OO(10,10)$ tensors that may contribute to Table \ref{tab:p-form} are given in Table \ref{tab:O(10-10)-2}, but the multiplicities and the details of the constituting mixed-symmetry potentials are not determined. 
\begin{table}
\centering{\tiny
\begin{tabular}{|c||c||c|c|}\hline
 $\NN$ & $\OO(10,10)$ tensor & IIA & IIB 
\\\hline\hline
%%%%%%%%%%%%%%%%%%%%%%%%%%
 7 & $\SUSY{I_{A_{1\cdots6},B_{123}a}}$ %7	0 0 1 0 0 1 0 0 0 1		1 2 3 5 7 9 12 15 11 7 7	2	7236523008	1	0	0	250
 &$\SUSY{I_{10,7,4}}$, $\cdots$
 &$\SUSY{I_{10,7,4,1}}$, $\cdots$
\\\cline{2-4}
\rowcolor{Gray}
 & $\SUSY{I_{A_{123},B_{123},Ca}} $ %7	1 0 2 0 0 0 0 0 0 1		0 1 2 5 8 11 14 17 12 7 8	2	1673605120	1	0	0	244
 &$\SUSY{I_{9,7,7}}$, $\cdots$
 &$\SUSY{I_{9,7,7,1}}$, $\cdots$
\\\cline{2-4}
 & $\textcolor{blue}{?}\,I_{A_{1\cdots4},B_{123}a}$ %7	0 0 1 1 0 0 0 0 0 1		1 2 3 5 8 11 14 17 12 7 8	-2	827706880	1	0	0	241
 &$\textcolor{blue}{?}\,I_{10,7,6}$, $\cdots$
 &$\textcolor{blue}{?}\,I_{10,7,6,1}$, $\cdots$
\\\cline{2-4}
\rowcolor{Gray}
 & $\SUSY{I_{A_{1\cdots4},B_{1\cdots4}\dot{a}}}$ %7	0 0 0 2 0 0 0 0 1 0		1 2 3 4 7 10 13 16 11 7 8	2	1704102400	1	0	0	247
 &$\SUSY{I_{10,6,6,1}}$, $\cdots$
 &$\SUSY{I_{10,6,6}}$, $\cdots$
\\\cline{2-4}
 & $\textcolor{blue}{?}\,I_{A_{1\cdots5},B_{123}\dot{a}}$ %7	0 0 1 0 1 0 0 0 1 0		1 2 3 5 7 10 13 16 11 7 8	0	2941542912	1	0	0	246
 &$\textcolor{blue}{?}\,I_{10,7,5,1}$, $\cdots$
 &$\textcolor{blue}{?}\,I_{10,7,5}$, $\cdots$
\\\cline{2-4}
\rowcolor{Gray}
 & $\textcolor{blue}{?}\,I_{A_{123},B_{123}\dot{a}}$ %7	0 0 2 0 0 0 0 0 1 0		1 2 3 6 9 12 15 18 12 7 9	-4	134041600	1	0	0	235
 &$\textcolor{blue}{?}\,I_{10,7,7,1}$, $\cdots$
 &$\textcolor{blue}{?}\,I_{10,7,7}$, $\cdots$
\\\hline
%%%%%%%%%%%%%%%%%%%%%%%%%%
 8 & $\SUSY{J_{A_{1\cdots10},B_{123},C_{123}}^-}$ %8	0 0 2 0 0 0 0 0 0 2		1 2 3 6 9 12 15 18 13 8 8	2	18204378192	1	0	0	281
 &$\SUSY{J_{10,7,7}}$, $\cdots$
 &$\SUSY{J_{10,7,7,1,1}}$, $\cdots$
\\\cline{2-4}
\rowcolor{Gray}
 &$\textcolor{blue}{?}\,J_{A_{1\cdots8},B_{123},C_{123}} $ %8	0 0 2 0 0 0 0 1 0 0		1 2 3 6 9 12 15 18 13 8 9	0	23533437620	1	0	0	280
 &\multicolumn{2}{c|}{
 $\textcolor{blue}{?}\,J_{10,7,7,2}$, $\cdots$}
\\\cline{2-4}
 & $\SUSY{J_{A_{1\cdots7},B_{1\cdots4},C_{123}}}$ %8	0 0 1 1 0 0 1 0 0 0		1 2 3 5 8 11 14 18 13 8 9	2	73638625280	1	0	0	284
 &\multicolumn{2}{c|}{
 $\SUSY{J_{10,7,6,3}}$, $\cdots$}
\\\cline{2-4}
\rowcolor{Gray}
 & $\textcolor{blue}{?}\,J_{A_{1\cdots5},B_{1\cdots4},C_{123}}$ %8	0 0 1 1 1 0 0 0 0 0		1 2 3 5 8 12 16 20 14 8 10	0	9096536064	1	0	0	277
 &\multicolumn{2}{c|}{
 $\textcolor{blue}{?}\,J_{10,7,6,5}$, $\cdots$}
\\\cline{2-4}
 & $\textcolor{blue}{?}\,J_{A_{1\cdots6},B_{123},C_{123}}$ %8	0 0 2 0 0 1 0 0 0 0		1 2 3 6 9 12 16 20 14 8 10	-2	5959766670	1	0	0	275
 &\multicolumn{2}{c|}{
 $\textcolor{blue}{?}\,J_{10,7,7,4}$, $\cdots$}
\\\cline{2-4}
\rowcolor{Gray}
 & $\textcolor{blue}{?}\,J_{A_{1\cdots4},B_{123},C_{123}}$ %8	0 0 2 1 0 0 0 0 0 0		1 2 3 6 10 14 18 22 15 8 11	-4	440465410	1	0	0	266
 &\multicolumn{2}{c|}{
 $\textcolor{blue}{?}\,J_{10,7,7,6}$, $\cdots$}
\\\cline{2-4}
 & $\SUSY{J_{A_{123},B_{123},C_{123},D}}$ %8	1 0 3 0 0 0 0 0 0 0		0 1 2 6 10 14 18 22 15 8 11	2	594914320	1	0	0	269
 &\multicolumn{2}{c|}{
 $\SUSY{J_{9,7,7,7}}$, $\cdots$}
\\\hline
%%%%%%%%%%%%%%%%%%%%%%%%%%
\rowcolor{Gray}
 9 & $\SUSY{K_{A_{1\cdots5},B_{123},C_{123}a}}$ %9	0 0 2 0 1 0 0 0 0 1		1 2 3 6 9 13 17 21 15 9 10	2	524675205120	1	0	0	317
 &$\SUSY{K_{10,7,7,5}}$, $\cdots$
 &$\SUSY{K_{10,7,7,5,1}}$, $\cdots$
\\\cline{2-4}
 & $\textcolor{blue}{?}\,K_{A_{1\cdots4},B_{123},C_{123}\dot{a}}$ %9	0 0 2 1 0 0 0 0 1 0		1 2 3 6 10 14 18 22 15 9 11	0	130541199360	1	0	0	312
 &$\textcolor{blue}{?}\,K_{10,7,7,6,1}$, $\cdots$
 &$\textcolor{blue}{?}\,K_{10,7,7,6}$, $\cdots$
\\\cline{2-4}
\rowcolor{Gray}
 & $\textcolor{blue}{?}\,K_{A_{123},B_{123},C_{123}a}$ %9	0 0 3 0 0 0 0 0 0 1		1 2 3 7 11 15 19 23 16 9 11	-2	17070772224	1	0	0	306
 &$\textcolor{blue}{?}\,K_{10,7,7,7}$, $\cdots$
 &$\textcolor{blue}{?}\,K_{10,7,7,7,1}$, $\cdots$
\\\hline
%%%%%%%%%%%%%%%%%%%%%%%%%%
 10 & $\SUSY{L_{A_{1\cdots7},B_{123},C_{123},D_{123}}}$ %10	0 0 3 0 0 0 1 0 0 0		1 2 3 7 11 15 19 24 17 10 12	2	1472062367490	1	0	0	349
 &\multicolumn{2}{c|}{
$\SUSY{L_{10,7,7,7,3}}$, 
$\cdots$}
\\\cline{2-4}
\rowcolor{Gray}
 & $\textcolor{blue}{?}\,L_{A_{1\cdots5},B_{123},C_{123},D_{123}}$ %10	0 0 3 0 1 0 0 0 0 0		1 2 3 7 11 16 21 26 18 10 13	0	196799781750	1	0	0	342
 &\multicolumn{2}{c|}{ 
$\textcolor{blue}{?}\,L_{10,7,7,7,5}$, $\cdots$
}
\\\cline{2-4}
 & $\textcolor{blue}{?}\,L_{A_{123},B_{123},C_{123},D_{123}}$ %10	0 0 4 0 0 0 0 0 0 0		1 2 3 8 13 18 23 28 19 10 14	-2	4334375760	1	0	0	331
 &\multicolumn{2}{c|}{ 
$\textcolor{blue}{?}\,L_{10,7,7,7,7}$, $\cdots$
}
\\\hline
%%%%%%%%%%%%%%%%%%%%%%%%%%
\rowcolor{Gray}
 11 & $\SUSY{M_{A_{123},B_{123},C_{123},D_{123}\dot{a}}}$ %11	0 0 4 0 0 0 0 0 1 0		1 2 3 8 13 18 23 28 19 11 14	2	1249944883200	1	0	0	377
 &$\SUSY{M_{10,7,7,7,7,1}}$, $\cdots$
 &$\SUSY{M_{10,7,7,7,7}}$, $\cdots$
\\\hline
\end{tabular}
\caption{List of the $\OO(10,10)$ tensors with level $7\leq \NN\leq 11$ that contribute to Table \ref{tab:p-form}. Only the mixed-symmetry potentials corresponding to the highest weight are shown.}
\label{tab:O(10-10)-2}
}
\end{table}
It is possible that some of the multiplicities are zero. 

In $d$ dimensions, these $\OO(10,10)$ tensors are reduced to $\OO(D,D)$ tensors. 
In order to determine the low-level $\OO(D,D)$ tensors, rather than directly decomposing $\OO(10,10)$ tensors, it is easier to perform the level decomposition first by deleting the node $\alpha_d$ \eqref{eq:level-d} and then decompose further by deleting the node $\alpha_{10}$\,:
\begin{align}
\scalebox{0.8}{
 \xygraph{
    *\cir<6pt>{} ([]!{+(0,-.4)} {\alpha_1}) - [r]
    \cdots - [r]
    *\cir<6pt>{} ([]!{+(0,-.4)} {\alpha_{d-1}}) - [r]
    *{\scalebox{2}{$\times$}}*\cir<6pt>{} ([]!{+(0,-.4)} {\alpha_d}) - [r]
    *\cir<6pt>{} ([]!{+(0,-.4)} {\alpha_{d+1}}) - [r]
    \cdots 
    - [r] *\cir<6pt>{} ([]!{+(0,-.4)} {\alpha_7}) 
    - [r] *\cir<6pt>{} ([]!{+(0,-.4)} {\alpha_8}) 
(
        - [u] *\cir<6pt>{} ([]!{+(.5,0)} {\alpha_{11}}),
        - [r] *\cir<6pt>{} ([]!{+(0,-.4)} {\alpha_9})
        - [r] *{\scalebox{2}{$\times$}}*\cir<6pt>{} ([]!{+(0,-.4)} {\alpha_{10}})
) 
}} .
\end{align}
The resulting potentials with level $\NN=1,2$ in any dimensions can be summarized as \cite{1102.0934,1108.5067,1201.5819}
\begin{align}
 B_{1;A}\,,\qquad B_{2}\,,\qquad
 C_{p;a}\quad (p=1,3,5,\cdots)\,,\qquad C_{p;\dot{a}}\quad (p:0,2,4,\cdots)\,.
\end{align}
They are obviously coming from the $\OO(10,10)$ tensors $B_{MN}$, $C_a$\,. 
The results for $\NN=2,3,4$ are summarized in Table \ref{tab:DEFG-list}. 
\begin{table}[htbp]
{\scriptsize
\begin{tabular}{|c|| p{1.5cm} | p{1.4cm} | p{1.4cm} | p{1.2cm} | p{1.1cm} | p{1.1cm} | p{1.cm} | p{1.cm} |}\hline
 $d$ & \multicolumn{1}{c|}{$1$} & \multicolumn{1}{c|}{$2$} & \multicolumn{1}{c|}{$3$} & \multicolumn{1}{c|}{$4$} & \multicolumn{1}{c|}{$5$} & \multicolumn{1}{c|}{$6$} & \multicolumn{1}{c|}{$7$} & \multicolumn{1}{c|}{$8$} \\\hline\hline
 $d$-form & $\SUSY{D_{1;A_{1\cdots4}}}$, $D_{1;A_{12}}$,\newline $D_{1}$ & $\SUSY{D_{2;A_{1\cdots4}}}$, $D_{2;A_{12}}$, $2\,D_{2}$ & $\SUSY{D_{3;A_{1\cdots 4}}}$, $D_{3;A_{12}}$, $2\,D_{3}$ & $\SUSY{D_{4;A_{1\cdots4}}}$, $D_{4;A_{12}}$, $2\,D_{4}$ & $\SUSY{D_{5;A_{1\cdots4}}}$, $D_{5;A_{12}}$, $2\,D_{5}$ & $\SUSY{D_{6;A_{1\cdots4}}^\pm}$, $D_{6;A_{12}}$, $2\,D_{6}$ & $D_{7;A_{12}}$, $D_{7;A_{12}}$, $2\,D_{7}$ & $D_{8}$, $D_{8;A_{12}}^\pm$, $2\,D_{8}$
\\\hline
 $(d-1)$-form & & $\SUSY{D_{1;A_{123}}}$, $D_{1;A}$ & $\SUSY{D_{2;A_{123}}}$, $D_{2;A}$ & $\SUSY{D_{3;A_{123}}}$, $D_{3;A}$ & $\SUSY{D_{4;A_{123}}}$, $D_{4;A}$ & $\SUSY{D_{5;A_{123}}}$, $D_{5;A}$ & $\SUSY{D_{6;A_{123}}^\pm}$, $D_{6;A}$ & $D_{7;A}$, $D_{7;A}$
\\\hline
 $(d-2)$-form & & & $\SUSY{D_{1;A_{12}}}$, $D_{1}$ & $\SUSY{D_{2;A_{12}}}$, $D_{2}$ & $\SUSY{D_{3;A_{12}}}$, $D_{3}$ & $\SUSY{D_{4;A_{12}}}$, $D_{4}$ & $\SUSY{D_{5;A_{12}}}$, $D_{5}$ & $\SUSY{D_{6;A_{12}}^\pm}$, $D_{6}$
\\\hline
 $(d-3)$-form & & & & $\SUSY{D_{1;A}}$ & $\SUSY{D_{2;A}}$ & $\SUSY{D_{3;A}}$ & $\SUSY{D_{4;A}}$ & $\SUSY{D_{5;A}}$
\\\hline
 $(d-4)$-form & & & & & $\SUSY{D_{1}}$ & $\SUSY{D_{2}}$ & $\SUSY{D_{3}}$ & $\SUSY{D_{4}}$ 
\\\hline\hline
 $d$-form & $\SUSY{E_{1;A_{12}\dot{a}}}$, $E_{1;Aa}$, $2\,E_{1;\dot{a}}$ & $\SUSY{E_{2;A_{12}\dot{a}}}$, $E_{2;Aa}$, $3\,E_{2;\dot{a}}$ & $\SUSY{E_{3;A_{12}\dot{a}}}$, $E_{3;Aa}$, $3\,E_{3;\dot{a}}$ & $\SUSY{E_{4;A_{12}\dot{a}}}$, $E_{4;Aa}$, $3\,E_{4;\dot{a}}$ & $\SUSY{E_{5;A_{12}\dot{a}}}$, $E_{5;Aa}$, $3\,E_{5;\dot{a}}$ & $\SUSY{E_{6;A_{12}\dot{a}}}$, $E_{6;Aa}$, $3\,E_{6;\dot{a}}$ & $\SUSY{E_{7;a\dot{b}\dot{c}}}$, $E_{7;Aa}$, $3\,E_{7;\dot{a}}$ & $\SUSY{E_{8;\dot{a}\dot{b}\dot{c}}}$, $E_{8;ab\dot{c}}$, $3\,E_{8;\dot{a}}$
\\\hline
 $(d-1)$-form & & $\SUSY{E_{1;A\dot{a}}}$, $E_{1;a}$ & $\SUSY{E_{2;A\dot{a}}}$, $E_{2;a}$ & $\SUSY{E_{3;A\dot{a}}}$, $E_{3;a}$ & $\SUSY{E_{4;A\dot{a}}}$, $E_{4;a}$ & $\SUSY{E_{5;A\dot{a}}}$, $E_{5;a}$ & $\SUSY{E_{6;A\dot{a}}}$, $E_{6;a}$ & $\SUSY{E_{7;a\dot{b}\dot{c}}}$, $E_{7;a}$
\\\hline
 $(d-2)$-form & & & $\SUSY{E_{1;\dot{a}}}$ & $\SUSY{E_{2;\dot{a}}}$ & $\SUSY{E_{3;\dot{a}}}$ & $\SUSY{E_{4;\dot{a}}}$ & $\SUSY{E_{5;\dot{a}}}$ & $\SUSY{E_{6;\dot{a}}}$
\\\hline\hline
 $d$-form & $\SUSY{F_{1;A_{1\cdots9}}^+}$, $F_{1;A_{1\cdots7}}$, $\SUSY{F_{1;A_{1\cdots6},B}}$, $2\,F_{1;A_{1\cdots5}}$, $3\,F_{1;A_{123}}$, $F_{1;A_{1\cdots4},B}$, $\SUSY{F_{1;A_{123},B_{12}}}$, $2\,F_{1;A_{12},B}$, $3\,F_{1;A}$ & $\SUSY{F_{2;A_{1\cdots8}}^+}$, $F_{2;A_{1\cdots6}}$, $\SUSY{F_{2;A_{1\cdots5},B}}$, $3\,F_{2;A_{1\cdots4}}$, $4\,F_{2;A_{12}}$, $F_{2;A_{123},B}$, $\SUSY{F_{2;A_{12},B_{12}}}$, $2\,F_{2;A,B}$, $4\,F_{2}$ & $\SUSY{F_{3;A_{1\cdots 7}}^+}$, $F_{3;A_{1\cdots 5}}$, $\SUSY{F_{3;A_{1\cdots4},B}}$, $3\,F_{3;A_{123}}$, $4\,F_{3;A}$, $F_{3;A_{12},B}$ & $\SUSY{F_{4;A_{1\cdots6}}^+}$, $F_{4;A_{1\cdots4}}$, $\SUSY{F_{4;A_{123},B}}$, $3\,F_{4;A_{12}}$, $2\,F_{4}$ & $\SUSY{F_{5;A_{1\cdots5}}^+}$, $F_{5;A_{123}}$, $\SUSY{F_{5;A_{12},B}}$, $2\,F_{5;A}$ & $\SUSY{F_{6;A_{1\cdots4}}^+}$, $F_{6;A_{12}}$, $\SUSY{F_{6;A,B}}$, $2\,F_{6}$ & $\SUSY{F_{7;A_{123}}^+}$, $F_{7;A}$ & $\SUSY{F_{8;A_{12}}^+}$
\\\hline
 $(d-1)$-form & & $\SUSY{F_{1;A_{1\cdots5}}}$, $F_{1;A_{123}}$, $\SUSY{F_{1;A_{12},B}}$, $F_{1;A}$ & $\SUSY{F_{2;A_{1\cdots4}}}$, $F_{2;A_{12}}$, $\SUSY{F_{2;A,B}}$, $2\,F_{2}$ & $\SUSY{F_{3;A_{123}}}$, $F_{3;A}$ & $\SUSY{F_{4;A_{12}}}$ & $\SUSY{F_{5;A}}$ & $\SUSY{F_{6}}$ & 
\\\hline
 $(d-2)$-form & & & $\SUSY{F_{1;A}}$ & $\SUSY{F_{2}}$ & & & & 
\\\hline\hline
 $d$-form & $\SUSY{G_{1;A_{1\cdots5}\dot{a}}}$, $G_{1;A_{1\cdots4}a}$, $4\,G_{1;A_{123}\dot{a}}$, $\SUSY{G_{1;A_{123},Ba}}$, $5\,G_{1;A_{12}a}$, $2\,G_{1;A_{12},B\dot{a}}$, $8\,G_{1;A\dot{a}}$, $2\,G_{1;A,Ba}$, $5\,G_{1;a}$ & $\SUSY{G_{2;A_{1\cdots4}\dot{a}}}$, $G_{2;A_{123}a}$, $5\,G_{2;A_{12}\dot{a}}$, $\SUSY{G_{2;A_{12},Ba}}$, $6\,G_{2;Aa}$, $G_{2;A,B\dot{a}}$, $8\,G_{2;\dot{a}}$ & $\SUSY{G_{3;A_{123}\dot{a}}}$, $G_{3;A_{12}a}$, $4\,G_{3;A\dot{a}}$, $\SUSY{G_{3;A,Ba}}$, $4\,G_{3;a}$ & $\SUSY{G_{4;A_{12}\dot{a}}}$, $G_{4;Aa}$, $3\,G_{4;\dot{a}}$ & $\SUSY{G_{5;A\dot{a}}}$ & $\SUSY{G_{6;\dot{a}}}$ & & 
\\\hline
 $(d-1)$-form & & $\SUSY{G_{1;A_{12}a}}$, $2\,G_{1;A\dot{a}}$, $2\,G_{1;a}$ & $\SUSY{G_{2;Aa}}$, $G_{2;\dot{a}}$ & $\SUSY{G_{3;a}}$ & & & & \\\hline
\end{tabular}}
\caption{$\OO(D,D)$ multiplets of $p$-forms with level $\NN=2,\dotsc,5$ in each dimension.}
\label{tab:DEFG-list}
\end{table}
As found in \cite{1102.0934}, potentials with level $\NN=3$ (in $2\leq d$) can be nicely summarized as\footnote{In $d=7$ or $d=8$, by using the totally antisymmetric tensor $\epsilon_{A_1\cdots A_{2D}}$\,, $\SUSY{D_{d;A_{1\cdots 4}}}$ becomes $D_{d;A_{12}}$ or $D_{d}$ and the underline disappears. Similarly, in $d=8$, $\SUSY{D_{d-1;A_{123}}}$ becomes $D_{d-1;A}$ and the underline disappears.}
\begin{align}
\begin{split}
 &\SUSY{D_{d;A_{1\cdots 4}}}\,,\quad D_{d;A_{12}}\,,\quad 2\,D_{d}\,,\quad \SUSY{D_{d-1;A_{123}}}\,,\quad D_{d-1;A}\,,
\\
 &D_{d-2}\,,\quad D_{d-2;A_{12}}\,,\quad D_{d-3;A}\,,\quad D_{d-4}\,.
\end{split}
\end{align}
All of them arise from the reduction of the 10D potential $\SUSY{D_{A_{1\cdots4}}}$ and $D$\,. 
For level $\NN=3$, as found in \cite{1108.5067}, the $\OO(D,D)$ tensors (in $2\leq d$) can be summarized as
\begin{align}
 \SUSY{E_{d;A_{12}\dot{a}}}\,,\quad E_{d;Aa}\,,\quad 3\,E_{d;\dot{a}}\,,\SUSY{\quad E_{d-1;A\dot{a}}}\,,\quad E_{d-1;a}\,,\quad E_{d-2;\dot{a}}\,.
\end{align}
Again, they are arising from the 10D potentials $\SUSY{E_{A_{12}\dot{a}}}$ and $E_{\dot{a}}$\,. 

For the higher level potentials, the pattern is much more non-trivial. 
In 10D, potentials with level $\NN=4,5$ are
\begin{align}
&\Pbox{0.8\textwidth}{$\SUSY{F_{A_{1\cdots10}}^+}$,\quad %4	0 0 0 0 0 0 0 0 2 0		1 2 3 4 5 6 7 8 5 4 4	2	92378	1	1	1	139
$\SUSY{F_{A_{1\cdots4},B_{12}}}$,\quad %4	0 1 0 1 0 0 0 0 0 0		0 0 1 2 4 6 8 10 7 4 5	2	592515	1	1	1	141
$\SUSY{F_{A_{1\cdots7},B}}$,\quad %4	1 0 0 0 0 0 1 0 0 0		0 1 2 3 4 5 6 8 6 4 4	2	1385670	1	1	1	145
$F_{A_{123},B}$,\quad %4	1 0 1 0 0 0 0 0 0 0		0 1 2 4 6 8 10 12 8 4 6	-2	17765	1	44	1	127
$F_{A_{1\cdots6}}$,\quad %4	0 0 0 0 0 1 0 0 0 0		1 2 3 4 5 6 8 10 7 4 5	-2	38760	1	46	1	133
$F_{A_{1\cdots4}}$,\quad %4	0 0 0 1 0 0 0 0 0 0		1 2 3 4 6 8 10 12 8 4 6	-4	4845	1	206	1	124
$2\,F_{A_{12}}$, %4	0 1 0 0 0 0 0 0 0 0		1 2 4 6 8 10 12 14 9 4 7	-6	190	1	801	2	111
$\SUSY{G_{A_{1\cdots6}\dot{a}}}$,\quad 
$\SUSY{G_{A_{12},B_{12}\dot{a}}}$,\quad 
$\SUSY{G_{A_{1\cdots4},Ba}}$,\quad 
$G_{A_{123},B\dot{a}}$,\quad 
$G_{A_{12},Ba}$,\quad 
$2\,G_{A_{1\cdots4}\dot{a}}$,\quad 
$G_{A,B\dot{a}}$,\quad 
$2\,G_{A_{123}a}$,\quad 
$4\,G_{A_{12}\dot{a}}$,\quad 
$3\,G_{Aa}$,\quad 
$3\,G_{\dot{a}}$\,.}
\end{align}
The general pattern for their reduction is non-trivial, but if restricted to the underlined tensors, we can observe the following pattern \cite{1201.5819}:
\begin{align}
\begin{split}
&\SUSY{F_{d;A_{1\cdots (10-d)}}^+}\,,\quad
 \SUSY{F_{d;A_{1\cdots (7-d)},B}}\,,\quad
 \SUSY{F_{d;A_{1\cdots (4-d)},B_{12}}}\quad (4-d\geq 2)\,,
\\
&\SUSY{F_{d-1;A_{1\cdots (7-d)}}}\,,\quad
 \SUSY{F_{d-1;A_{1\cdots (4-d)},B}}\quad (4-d\geq 1)\,,\quad
 \SUSY{F_{d-2;A_{1\cdots (4-d)}}}\,,
\\
&\SUSY{G_{d;A_{1\cdots(6-d)}\dot{a}}}\,,\quad 
 \SUSY{G_{d;A_{1\cdots(4-d)},Ba}}\quad (4-d\geq 1)\,,\quad 
 \SUSY{G_{d-1;A_{1\cdots(4-d)}a}} \,.
\end{split}
\end{align}
The $p$-form multiplets for higher levels $\NN=6,7,8$ are summarized in Table \ref{tab:HIJ-list}. 
\begin{table}[tbhp]
\centerline{\scriptsize
\begin{tabular}{|c||p{1.9cm}| p{1.5cm} | p{1.2cm} || p{1.6cm} | p{1.1cm} | p{0.5cm} || p{1.4cm} |}\hline
 $d$ & \multicolumn{1}{c|}{$2$} & \multicolumn{1}{c|}{$3$} & \multicolumn{1}{c||}{$4$} & \multicolumn{1}{c|}{$2$} & \multicolumn{1}{c|}{$3$} & \multicolumn{1}{c||}{$4$} & \multicolumn{1}{c|}{$3$} \\\hline\hline
 $d$-form & $\SUSY{H_{2;A_{1\cdots6},B_{12}}}$, %2 6	0 0 1 0 0 0 1 0 0		1 2 3 4 6 8 10 12 9 6 6	2	716040	1	1	1	147
$2\,H_{2;A_{1\cdots4},B_{12}}$, %2 6	0 0 1 0 1 0 0 0 0		1 2 3 4 6 8 11 14 10 6 7	0	141372	1	8	2	142
$4\,H_{2;A_{12},B_{12}}$, %2 6	0 0 2 0 0 0 0 0 0		1 2 3 4 7 10 13 16 11 6 8	-2	5304	1	46	4	133
$2\,H_{2;A_{1\cdots7},B}$, %2 6	0 1 0 0 0 0 0 1 1		1 2 3 5 7 9 11 13 9 6 6	0	162162	1	8	2	142
$4\,H_{2;A_{1\cdots5},B}$, %2 6	0 1 0 0 0 1 0 0 0		1 2 3 5 7 9 11 14 10 6 7	-2	60060	1	46	4	139
$7\,H_{2;A_{123},B}$, %2 6	0 1 0 1 0 0 0 0 0		1 2 3 5 7 10 13 16 11 6 8	-4	7020	1	206	7	132
$6\,H_{2;A,B}$, %2 6	0 2 0 0 0 0 0 0 0		1 2 3 6 9 12 15 18 12 6 9	-6	135	1	789	6	121
$\SUSY{H_{2;A_{1\cdots4},B,C}}$, %2 6	0 2 0 0 1 0 0 0 0		1 2 2 4 6 8 11 14 10 6 7	2	176800	1	1	1	143
$H_{2;A_{12},B,C}$, %2 6	0 2 1 0 0 0 0 0 0		1 2 2 4 7 10 13 16 11 6 8	0	8925	1	8	1	134
$4\,H_{2;A_{1\cdots8}}^+$, %2 6	0 0 0 0 0 0 0 2 0		1 2 4 6 8 10 12 14 9 6 7	-2	6435	1	46	4	135
$2\,H_{2;A_{1\cdots8}}^-$, %2 6	0 0 0 0 0 0 0 0 2		1 2 4 6 8 10 12 14 10 6 6	-2	6435	1	44	2	135
$5\,H_{2;A_{1\cdots6}}$, %2 6	0 0 0 0 0 0 1 0 0		1 2 4 6 8 10 12 14 10 6 7	-4	8008	1	206	5	134
$11\,H_{2;A_{1\cdots4}}$, %2 6	0 0 0 0 1 0 0 0 0		1 2 4 6 8 10 13 16 11 6 8	-6	1820	1	801	11	129
$13\,H_{2;A_{12}}$, %2 6	0 0 1 0 0 0 0 0 0		1 2 4 6 9 12 15 18 12 6 9	-8	120	1	2781	13	120
$8\,H_{2}$ %2 6	0 0 0 0 0 0 0 0 0		1 2 5 8 11 14 17 20 13 6 10	-10	1	1	8885	8	107
 & $\SUSY{H_{3;A_{1\cdots 5},B}}$, 
 $2\,H_{3;A_{123},B}$, 
 $3\,H_{3;A,B}$, 
 $H_{3;A_{1\cdots 6}}$, 
 $3\,H_{3;A_{1\cdots4}}$, 
 $4\,H_{3;A_{12}}$, 
 $3\,H_{3}$ & $\SUSY{H_{4;A_{1\cdots4}}}$, $H_{4;A_{12}}$, $2\,H_{4}$ 
 & $\SUSY{I_{2;A_{12},B_{12}\dot{a}}}$, %2 7	0 0 2 0 0 0 0 1 0		1 2 3 4 7 10 13 16 11 7 8	2	524160	1	1	1	162
 $\SUSY{I_{2;A_{1\cdots4},Ba}}$, %2 7	0 1 0 0 1 0 0 0 1		1 2 3 5 7 9 12 15 11 7 7	2	2036736	1	1	1	165
 $2\,I_{2;A_{123},B\dot{a}}$, %2 7	0 1 0 1 0 0 0 1 0		1 2 3 5 7 10 13 16 11 7 8	0	670208	1	8	2	161
 $6\,I_{2;A_{12},Ba}$, %2 7	0 1 1 0 0 0 0 0 1		1 2 3 5 8 11 14 17 12 7 8	-2	141440	1	46	6	156
 $7\,I_{2;A,B\dot{a}}$, %2 7	0 2 0 0 0 0 0 1 0		1 2 3 6 9 12 15 18 12 7 9	-4	15360	1	206	7	150
 $\SUSY{I_{2;A,B,Ca}}$, %2 7	0 3 0 0 0 0 0 0 1		1 2 2 5 8 11 14 17 12 7 8	2	87040	1	1	1	157
 $\SUSY{I_{2;A_{1\cdots6}\dot{a}}}$, %2 7	0 0 0 0 0 0 1 1 0		1 2 4 6 8 10 12 14 10 7 7	2	465920	1	1	1	163
 $I_{2;A_{1\cdots5}a}$, %2 7	0 0 0 0 0 1 0 0 1		1 2 4 6 8 10 12 15 11 7 7	0	326144	1	8	1	161
 $5\,I_{2;A_{1\cdots4}\dot{a}}$, %2 7	0 0 0 0 1 0 0 1 0		1 2 4 6 8 10 13 16 11 7 8	-2	161280	1	46	5	158
 $8\,I_{2;A_{123}a}$, %2 7	0 0 0 1 0 0 0 0 1		1 2 4 6 8 11 14 17 12 7 8	-4	56320	1	206	8	154
 $19\,I_{2;A_{12}\dot{a}}$, %2 7	0 0 1 0 0 0 0 1 0		1 2 4 6 9 12 15 18 12 7 9	-6	13312	1	801	19	149
 $22\,I_{2;Aa}$, %2 7	0 1 0 0 0 0 0 0 1		1 2 4 7 10 13 16 19 13 7 9	-8	1920	1	2781	22	143
 $22\,I_{2;\dot{a}}$ %2 7	0 0 0 0 0 0 0 1 0		1 2 5 8 11 14 17 20 13 7 10	-10	128	1	8885	22	136
 & $\SUSY{I_{3;A,B\dot{a}}}$, $\SUSY{I_{3;A_{123}a}}$, $I_{3;A_{12}\dot{a}}$, $4\,I_{3;Aa}$, $4\,I_{3;\dot{a}}$
 & $\SUSY{I_{4;\dot{a}}}$
 & $\SUSY{J_{3;A_{1\cdots 7}}^-}$, %3 8	0 0 0 0 0 0 0 0 2		1 2 3 6 9 12 15 18 13 8 8	2	1716	1	1	1	137
$\SUSY{J_{3;A_{1\cdots 4},B}}$, %3 8	0 0 1 0 0 1 0 0 0		1 2 3 5 8 11 14 18 13 8 9	2	11648	1	1	1	140
$J_{3;A_{12},B}$, %3 8	0 0 1 1 0 0 0 0 0		1 2 3 5 8 12 16 20 14 8 10	0	896	1	8	1	133
$J_{3;A_{1\cdots 5}}$, %3 8	0 0 0 0 0 0 1 0 0		1 2 3 6 9 12 15 18 13 8 9	0	2002	1	8	1	136
$3\,J_{3;A_{123}}$, %3 8	0 0 0 0 1 0 0 0 0		1 2 3 6 9 12 16 20 14 8 10	-2	364	1	46	3	131
$4\,J_{3;A}$ %3 8	0 0 1 0 0 0 0 0 0		1 2 3 6 10 14 18 22 15 8 11	-4	14	1	246	4	122
\\\hline
$(d-1)$-form & $\SUSY{H_{1;A_{1\cdots4},B}}$, %1 6	1 1 0 0 1 0 0 0 0		0 1 2 4 6 8 11 14 10 6 7	2	48384	1	1	1	128
$2\,H_{1;A_{12},B}$, %1 6	1 1 1 0 0 0 0 0 0		0 1 2 4 7 10 13 16 11 6 8	0	2688	1	8	2	119
$\SUSY{H_{1;A_{1\cdots7}}}$, %1 6	1 0 0 0 0 0 0 1 1		0 1 3 5 7 9 11 13 9 6 6	2	22880	1	1	1	127
$H_{1;A_{1\cdots5}}$, %1 6	1 0 0 0 0 1 0 0 0		0 1 3 5 7 9 11 14 10 6 7	0	8736	1	8	1	124
$3\,H_{1;A_{123}}$, %1 6	1 0 0 1 0 0 0 0 0		0 1 3 5 7 10 13 16 11 6 8	-2	1120	1	44	3	117
$3\,H_{1;A}$ %1 6	1 1 0 0 0 0 0 0 0		0 1 3 6 9 12 15 18 12 6 9	-4	32	1	192	3	106
 & $\SUSY{H_{2;A_{123}}}$, $H_{2;A}$ & 
 & $\SUSY{I_{1;A,Ba}}$, %1 7	1 2 0 0 0 0 0 0 1		0 1 2 5 8 11 14 17 12 7 8	2	30720	1	1	1	142
 $\SUSY{I_{1;A_{123}\dot{a}}}$, %1 7	1 0 0 1 0 0 0 1 0		0 1 3 5 7 10 13 16 11 7 8	2	112640	1	1	1	146
 $2\,I_{1;A_{12}a}$, %1 7	1 0 1 0 0 0 0 0 1		0 1 3 5 8 11 14 17 12 7 8	0	26624	1	8	2	141
 $4\,I_{1;A\dot{a}}$, %1 7	1 1 0 0 0 0 0 1 0		0 1 3 6 9 12 15 18 12 7 9	-2	3840	1	44	4	135
 $4\,I_{1;a}$ %1 7	1 0 0 0 0 0 0 0 1		0 1 4 7 10 13 16 19 13 7 9	-4	256	1	192	4	128
 & $\SUSY{I_{2;a}}$
 &
 & $\SUSY{J_{2}}$
\\\hline
\end{tabular}}
\caption{List of $p$-form potentials with level $\NN=6,7,8$ in each dimension. \label{tab:HIJ-list}}
\end{table}
In the region of our concern, $d\geq 3$\,, higher-level potentials with $\NN=9,10,11$ appear only in the 3-form multiplet in $d=3$\,:
\begin{align}
 \SUSY{K_{3;A_{12}a}}\,,\quad %3 9	0 0 0 1 0 0 0 0 1		1 2 3 6 9 13 17 21 15 9 10	2	4928	1	1	1	149
 K_{3;A\dot{a}}\,,\quad %3 9	0 0 1 0 0 0 0 1 0		1 2 3 6 10 14 18 22 15 9 11	0	832	1	8	1	144
 3\,K_{3;a}\,,\quad %3 9	0 0 0 0 0 0 0 0 1		1 2 3 7 11 15 19 23 16 9 11	-2	64	1	46	3	138
 \SUSY{L_{3;A_{1\cdots 4}}}\,,\quad %3 10	0 0 0 0 0 1 0 0 0		1 2 3 7 11 15 19 24 17 10 12	2	1001	1	1	1	157
 L_{3;A_{12}}\,,\quad %3 10	0 0 0 1 0 0 0 0 0		1 2 3 7 11 16 21 26 18 10 13	0	91	1	8	1	150
 2\,L_{3}\,,\quad %3 10	0 0 0 0 0 0 0 0 0		1 2 3 8 13 18 23 28 19 10 14	-2	1	1	46	2	139
 \SUSY{M_{3;\dot{a}}}\,. %3 11	0 0 0 0 0 0 0 1 0		1 2 3 8 13 18 23 28 19 11 14	2	64	1	0	0	161
\end{align}

Although the multiplicities are non-trivial, we observe some patterns in the index structure of the potentials with $\NN\geq 2$\,. 
If we find a certain $(d-i)$-form $(i=0,1,2,\cdots)$ in $d$ dimensions, there is a $(d-i)$-form in $(d+1)$ dimensions where the $[\NN/2]-1$ indices $\underbrace{A,\dotsc,B}_{[\NN/2]-1}$ are removed, where the bracket $[\cdots]$ denotes the integer part. 
For example, we find a series of $d$-form potentials $H$ (where $[\NN/2]-1=2$),
\begin{align}
 \underset{(d=0)}{\SUSY{H_{A_{1\cdots 8,B_{1\cdots 4}}}}}\to
 \underset{(d=1)}{\SUSY{H_{1;A_{1\cdots 7,B_{123}}}}}\to
 \underset{(d=2)}{\SUSY{H_{2;A_{1\cdots 6,B_{12}}}}}\to
 \underset{(d=3)}{\SUSY{H_{3;A_{1\cdots 5,B}}}}\to
 \underset{(d=4)}{\SUSY{H_{4;A_{1\cdots 4}}}}\to
 \underset{(d=5)}{0}\,.
\end{align}
As another example, the reduction rule for $\SUSY{L_{A_{1\cdots7},B_{123},C_{123},D_{123}}}$ (where $[\NN/2]-1=4$) is
\begin{align}
 \underset{(d=0)}{\SUSY{L_{A_{1\cdots7},B_{123},C_{123},D_{123}}}} \to
 \underset{(d=1)}{\SUSY{L_{1;A_{1\cdots6},B_{12},C_{12},D_{12}}}} \to
 \underset{(d=2)}{\SUSY{L_{2;A_{1\cdots5},B,C,D}}} \to
 \underset{(d=3)}{\SUSY{L_{3;A_{1\cdots4}}}} \to
 \underset{(d=4)}{0}\,.
\end{align}
Moreover, when we consider a reduction of an $\OO(D,D)$-tensor-valued $p$-form to an $\OO(D-1,D-1)$-tensor-valued $p$-form, we remove $[\NN/2]$ indices $\underbrace{A,\dotsc,B}_{[\NN/2]}$\,. 
For example, for a potential $F$ ($[\NN/2]=2$), we find a series
\begin{align}
 \underset{(d=1)}{\SUSY{F_{1;A_{123},B_{12}}}} \to \underset{(d=2)}{\SUSY{F_{1;A_{12},B}}} \to \underset{(d=3)}{\SUSY{F_{1;A}}} \to \underset{(d=4)}{0}\,. 
\end{align}
The non-underlined potentials also follow these rules, but we do not find any patterns for their multiplicities. 
The underlined potentials always have multiplicity 1. 

We also studied the decompositions of the $E_n$ multiplets into $\OO(D,D)$ representations. 
The results are detailed in Appendix \ref{app:En2ODD}. 

\section{BPS branes and the vector representation}
\label{sec:E11-branes}

We also consider the vector representation of $E_{11}$\,, which contains the momenta $P_i$ and the brane central charges. 
As discussed in \cite{hep-th:0307098}, the multiplet is called the $l^1$ representation because the highest-weight state corresponds to the fundamental weight $l^1$\,. 
Here, the fundamental weights $l^i$ of the $E_{11}$ algebra are expressed as
\begin{align}
 l^i \equiv \lambda^i + \frac{\lambda^8\cdot \lambda^i}{x\cdot x}\, x \quad (i=1,\dotsc,10)\,,\qquad 
 l^{11} \equiv \frac{x}{x\cdot x}\,,
\end{align}
by using the fundamental weights $\lambda^i$ of the subalgebra $\SL(11)$\,. 
The $l^1$ is the highest weight of the $\SL(11)$ multiplet $[1,0,\dotsc,0]$\,, which corresponds to a tensor $P^{i_1\cdots i_{10}}$\,. 
Through the dualization, this tensor corresponds to the momenta $P_i$\,. 
By taking commutators with the positive generators we obtain infinitely many central charges. 
The commutation relations for low-level generators have been determined in \cite{hep-th:0307098,hep-th:0312247,hep-th:0406150} as
\begin{align}
 [K^k{}_l,\,P_i] = - \delta^k_i\,P_l + \frac{1}{2}\,\delta^k_l\,P_i\,,\qquad 
 [R^{k_1k_2k_3},\,P_i] = 3\,\delta^{[k_1}_i\,Z^{k_2k_3]}\,,\cdots\,.
\end{align}
When we take the commutator with $E_{\mathsf{11}}=R^{\mathsf{8}\mathsf{9}z}$\,, the level $\ell$ is increased, and at the same time, the number of upper indices is increased by three. 
Thus, the level $\ell$ satisfies
\begin{align}
 3\,\ell -1 = \text{(\# of the upper indices)} - \text{(\# of the lower indices)} \,. 
\end{align}

In order to obtain the central charges in the vector representation, a more systematic way was found in \cite{hep-th:0312247,hep-th:0406150}. 
By adding a new node $\alpha_*$ to the Dynkin diagram, we consider a level decomposition of the adjoint representation of $E_8^{++++}$,
\begin{align}
\scalebox{0.8}{
 \xygraph{
    *{\scalebox{2}{$\times$}}*\cir<6pt>{} ([]!{+(0,-.4)} {\alpha_*}) - [r]
    *\cir<6pt>{} ([]!{+(0,-.4)} {\alpha_1}) - [r]
    *\cir<6pt>{} ([]!{+(0,-.4)} {\alpha_2}) - [r]
    *\cir<6pt>{} ([]!{+(0,-.4)} {\alpha_3}) - [r]
    *\cir<6pt>{} ([]!{+(0,-.4)} {\alpha_4}) - [r]
    *\cir<6pt>{} ([]!{+(0,-.4)} {\alpha_5}) - [r]
    *\cir<6pt>{} ([]!{+(0,-.4)} {\alpha_6}) - [r]
    *\cir<6pt>{} ([]!{+(0,-.4)} {\alpha_7}) - [r]
    *\cir<6pt>{} ([]!{+(0,-.4)} {\alpha_8})
(
        - [u] *{\scalebox{2}{$\times$}}*\cir<6pt>{} ([]!{+(.5,0)} {\alpha_{11}}),
        - [r] *\cir<6pt>{} ([]!{+(0,-.4)} {\alpha_9})
        - [r] *\cir<6pt>{} ([]!{+(0,-.4)} {\alpha_{10}})
) 
}} .
\end{align}
We then consider the $E_8^{++++}$ generators with level $m_*=1$ (associated with the node $\alpha_*$), and we find that these $E_8^{++++}$ generators precisely correspond to the weights in the vector representation of $E_{11}$ \cite{hep-th:0312247,hep-th:0406150}. 
The low-level central charges up to level $\ell=7$ are given in Table \ref{tab:central-charges}. 
Again, central charges associated with the $E_8^{++++}$ roots $\alpha$ satisfying $\alpha\cdot\alpha=2$ are associated with the supersymmetric branes, and they are underlined. 
In order to find the central charges in type IIB theory, we delete the node $\alpha_9$ instead of $\alpha_{11}$\,, and the result up to level $\ell_9=10$ is given in Table \ref{tab:central-charges}. 
\begin{table}[tbh]
\centerline{\tiny
\begin{tabular}{|c|l|}
 \hline
 $\ell$ & \multicolumn{1}{c|}{Central charges in M-theory up to level $\ell=7$} \\ \hline\hline
%%%%%%%%%%%%%%%%%%%%%%%%%%
 $0$ & $\SUSY{P_1}$
\\\hline
$1$ & \Pbox{\textwidth}{
$\SUSY{Z^{2}}$}
\\\hline
$2$ & \Pbox{\textwidth}{
$\SUSY{Z^{5}}$}
\\\hline
$3$ & \Pbox{\textwidth}{
$\SUSY{Z^{7,1}}$, 
$Z^{8}$}
\\\hline
$4$ & \Pbox{\textwidth}{
$\SUSY{Z^{8,3}}$, 
$\SUSY{Z^{9,1,1}}$, 
$Z^{9,2}$, 
$2\,\,Z^{10,1}$, 
$Z^{11}$}
\\\hline
$5$ & \Pbox{\textwidth}{
$\SUSY{Z^{8,6}}$, 
$\SUSY{Z^{9,4,1}}$, 
$Z^{9,5}$, 
$2\,\,Z^{10,3,1}$, 
$2\,\,Z^{10,4}$, 
$\SUSY{Z^{11,1,1,1}}$, 
$2\,\,Z^{11,2,1}$, 
$3\,\,Z^{11,3}$}
\\\hline
$6$ & \Pbox{\textwidth}{
$\SUSY{Z^{8,8,1}}$, 
$\SUSY{Z^{9,6,2}}$, 
$2\,\,Z^{9,7,1}$, 
$2\,\,Z^{9,8}$, 
$\SUSY{Z^{10,4,3}}$, 
$\SUSY{Z^{10,5,1,1}}$, 
$Z^{10,5,2}$, 
$4\,\,Z^{10,6,1}$, 
$3\,\,Z^{10,7}$, 
$Z^{11,3,3}$, 
$2\,\,Z^{11,4,1,1}$, 
$3\,\,Z^{11,4,2}$, 
$5\,\,Z^{11,5,1}$, 
$5\,\,Z^{11,6}$}
\\\hline
$7$ & \Pbox{\textwidth}{
$\SUSY{Z^{9,7,4}}$, 
$\SUSY{Z^{9,8,2,1}}$, 
$2\,\,Z^{9,8,3}$, 
$Z^{9,9,1,1}$, 
$3\,\,Z^{9,9,2}$, 
$\SUSY{Z^{10,6,3,1}}$, 
$2\,\,Z^{10,6,4}$, 
$2\,\,Z^{10,7,2,1}$, 
$4\,\,Z^{10,7,3}$, 
$5\,\,Z^{10,8,1,1}$, 
$7\,\,Z^{10,8,2}$, 
$8\,\,Z^{10,9,1}$, 
$2\,\,Z^{10,10}$, 
$\SUSY{Z^{11,4,4,1}}$, 
$2\,\,Z^{11,5,3,1}$, 
$2\,\,Z^{11,5,4}$, 
$\SUSY{Z^{11,6,1,1,1}}$, 
$4\,\,Z^{11,6,2,1}$, 
$7\,\,Z^{11,6,3}$, 
$7\,\,Z^{11,7,1,1}$, 
$11\,\,Z^{11,7,2}$, 
$15\,\,Z^{11,8,1}$, 
$8\,\,Z^{11,9}$}
\\\hline
$8$ & \Pbox{\textwidth}{
$\SUSY{Z^{9,7,7}}$, 
$\SUSY{Z^{9,8,5,1}}$, 
$2\,\,Z^{9,8,6}$, 
$\SUSY{Z^{9,9,3,2}}$, 
$2\,\,Z^{9,9,4,1}$, 
$3\,\,Z^{9,9,5}$, 
$\SUSY{Z^{10,6,6,1}}$, 
$\SUSY{Z^{10,7,4,2}}$, 
$2\,\,Z^{10,7,5,1}$, 
$4\,\,Z^{10,7,6}$, 
$\SUSY{Z^{10,8,3,1,1}}$, 
$2\,\,Z^{10,8,3,2}$, 
$7\,\,Z^{10,8,4,1}$, 
$7\,\,Z^{10,8,5}$, 
$2\,\,Z^{10,9,2,1,1}$, 
$3\,\,Z^{10,9,2,2}$, 
$11\,\,Z^{10,9,3,1}$, 
$13\,\,Z^{10,9,4}$, 
$3\,\,Z^{10,10,1,1,1}$, 
$10\,\,Z^{10,10,2,1}$, 
$10\,\,Z^{10,10,3}$, 
$\SUSY{Z^{11,6,3,3}}$, 
$\SUSY{Z^{11,6,4,1,1}}$, 
$2\,\,Z^{11,6,4,2}$, 
$4\,\,Z^{11,6,5,1}$, 
$4\,\,Z^{11,6,6}$, 
$2\,\,Z^{11,7,3,1,1}$, 
$4\,\,Z^{11,7,3,2}$, 
$12\,\,Z^{11,7,4,1}$, 
$11\,\,Z^{11,7,5}$, 
$5\,\,Z^{11,8,2,1,1}$, 
$5\,\,Z^{11,8,2,2}$, 
$22\,\,Z^{11,8,3,1}$, 
$22\,\,Z^{11,8,4}$, 
$7\,\,Z^{11,9,1,1,1}$, 
$28\,\,Z^{11,9,2,1}$, 
$32\,\,Z^{11,9,3}$, 
$22\,\,Z^{11,10,1,1}$, 
$29\,\,Z^{11,10,2}$, 
$16\,\,Z^{11,11,1}$}
\\\hline\hline
 $\ell_9$ & \multicolumn{1}{c|}{Central charges in type IIB theory up to level $\ell_9=11$} \\ \hline\hline
%%%%%%%%%%%%%%%%%%%%%%%%%%
 $0$ & $\SUSY{P_1}$
\\\hline
$1$ & \Pbox{\textwidth}{
$\SUSY{Z^{1}_{\alpha}}$}
\\\hline
$2$ & \Pbox{\textwidth}{
$\SUSY{Z^{3}}$}
\\\hline
$3$ & \Pbox{\textwidth}{
$\SUSY{Z^{5}_{\alpha}}$}
\\\hline
$4$ & \Pbox{\textwidth}{
$\SUSY{Z^{6,1}}$, 
$\SUSY{Z^{7}_{\alpha\beta}}$, 
$Z^{7}$}
\\\hline
$5$ & \Pbox{\textwidth}{
$\SUSY{Z^{7,2}_{\alpha}}$, 
$2\,\,Z^{8,1}_{\alpha}$, 
$\SUSY{Z^{9}_{\alpha\beta\gamma}}$, 
$2\,\,Z^{9}_{\alpha}$}
\\\hline
$6$ & \Pbox{\textwidth}{
$\SUSY{Z^{7,4}}$, 
$\SUSY{Z^{8,2,1}}$, 
$\SUSY{Z^{8,3}_{\alpha\beta}}$, 
$Z^{8,3}$, 
$Z^{9,1,1}$, 
$2\,\,Z^{9,2}_{\alpha\beta}$, 
$3\,\,Z^{9,2}$, 
$3\,\,Z^{10,1}_{\alpha\beta}$, 
$3\,\,Z^{10,1}$}
\\\hline
$7$ & \Pbox{\textwidth}{
$\SUSY{Z^{7,6}_{\alpha}}$, 
$\SUSY{Z^{8,4,1}_{\alpha}}$, 
$2\,\,Z^{8,5}_{\alpha}$, 
$\SUSY{Z^{9,2,2}_{\alpha}}$, 
$2\,\,Z^{9,3,1}_{\alpha}$, 
$\SUSY{Z^{9,4}_{\alpha\beta\gamma}}$, 
$4\,\,Z^{9,4}_{\alpha}$, 
$4\,\,Z^{10,2,1}_{\alpha}$, 
$2\,\,Z^{10,3}_{\alpha\beta\gamma}$, 
$6\,\,Z^{10,3}_{\alpha}$}
\\\hline
$8$ & \Pbox{\textwidth}{
$\SUSY{Z^{7,7,1}}$, 
$\SUSY{Z^{8,5,2}}$, 
$\SUSY{Z^{8,6,1}_{\alpha\beta}}$, 
$2\,\,Z^{8,6,1}$, 
$2\,\,Z^{8,7}_{\alpha\beta}$, 
$3\,\,Z^{8,7}$, 
$\SUSY{Z^{9,4,1,1}}$, 
$\SUSY{Z^{9,4,2}_{\alpha\beta}}$, 
$2\,\,Z^{9,4,2}$, 
$2\,\,Z^{9,5,1}_{\alpha\beta}$, 
$4\,\,Z^{9,5,1}$, 
$5\,\,Z^{9,6}_{\alpha\beta}$, 
$5\,\,Z^{9,6}$, 
$\SUSY{Z^{10,2,2,1}}$, 
$Z^{10,3,1,1}$, 
$2\,\,Z^{10,3,2}_{\alpha\beta}$, 
$3\,\,Z^{10,3,2}$, 
$5\,\,Z^{10,4,1}_{\alpha\beta}$, 
$7\,\,Z^{10,4,1}$, 
$\SUSY{Z^{10,5}_{\alpha_{1\cdots 4}}}$, 
$7\,\,Z^{10,5}_{\alpha\beta}$, 
$8\,\,Z^{10,5}$}
\\\hline
$9$ & \Pbox{\textwidth}{
$\SUSY{Z^{8,6,3}_{\alpha}}$, 
$\SUSY{Z^{8,7,1,1}_{\alpha}}$, 
$3\,\,Z^{8,7,2}_{\alpha}$, 
$\SUSY{Z^{8,8,1}_{\alpha\beta\gamma}}$, 
$4\,\,Z^{8,8,1}_{\alpha}$, 
$\SUSY{Z^{9,4,4}_{\alpha}}$, 
$\SUSY{Z^{9,5,2,1}_{\alpha}}$, 
$2\,\,Z^{9,5,3}_{\alpha}$, 
$2\,\,Z^{9,6,1,1}_{\alpha}$, 
$\SUSY{Z^{9,6,2}_{\alpha\beta\gamma}}$, 
$7\,\,Z^{9,6,2}_{\alpha}$, 
$2\,\,Z^{9,7,1}_{\alpha\beta\gamma}$, 
$12\,\,Z^{9,7,1}_{\alpha}$, 
$4\,\,Z^{9,8}_{\alpha\beta\gamma}$, 
$11\,\,Z^{9,8}_{\alpha}$, 
$3\,\,Z^{10,4,2,1}_{\alpha}$, 
$\SUSY{Z^{10,4,3}_{\alpha\beta\gamma}}$, 
$4\,\,Z^{10,4,3}_{\alpha}$, 
$4\,\,Z^{10,5,1,1}_{\alpha}$, 
$2\,\,Z^{10,5,2}_{\alpha\beta\gamma}$, 
$12\,\,Z^{10,5,2}_{\alpha}$, 
$5\,\,Z^{10,6,1}_{\alpha\beta\gamma}$, 
$20\,\,Z^{10,6,1}_{\alpha}$, 
$8\,\,Z^{10,7}_{\alpha\beta\gamma}$, 
$20\,\,Z^{10,7}_{\alpha}$}
\\\hline
$10$ & \Pbox{\textwidth}{
$\SUSY{Z^{8,6,5}}$, 
$\SUSY{Z^{8,7,3,1}}$, 
$\SUSY{Z^{8,7,4}_{\alpha\beta}}$, 
$2\,\,Z^{8,7,4}$, 
$\SUSY{Z^{8,8,2,1}_{\alpha\beta}}$, 
$2\,\,Z^{8,8,2,1}$, 
$2\,\,Z^{8,8,3}_{\alpha\beta}$, 
$4\,\,Z^{8,8,3}$, 
$\SUSY{Z^{9,5,4,1}}$, 
$Z^{9,5,5}$, 
$\SUSY{Z^{9,6,2,2}}$, 
$\SUSY{Z^{9,6,3,1}_{\alpha\beta}}$, 
$2\,\,Z^{9,6,3,1}$, 
$3\,\,Z^{9,6,4}_{\alpha\beta}$, 
$5\,\,Z^{9,6,4}$, 
$\SUSY{Z^{9,7,1,1,1}}$, 
$3\,\,Z^{9,7,2,1}_{\alpha\beta}$, 
$7\,\,Z^{9,7,2,1}$, 
$7\,\,Z^{9,7,3}_{\alpha\beta}$, 
$10\,\,Z^{9,7,3}$, 
$5\,\,Z^{9,8,1,1}_{\alpha\beta}$, 
$7\,\,Z^{9,8,1,1}$, 
$\SUSY{Z^{9,8,2}_{\alpha_{1\cdots 4}}}$, 
$14\,\,Z^{9,8,2}_{\alpha\beta}$, 
$17\,\,Z^{9,8,2}$, 
$Z^{9,9,1}_{\alpha_{1\cdots 4}}$, 
$13\,\,Z^{9,9,1}_{\alpha\beta}$, 
$12\,\,Z^{9,9,1}$, 
$\SUSY{Z^{10,4,4,1}_{\alpha\beta}}$, 
$2\,\,Z^{10,4,4,1}$, 
$\SUSY{Z^{10,5,2,1,1}}$, 
$\SUSY{Z^{10,5,2,2}_{\alpha\beta}}$, 
$2\,\,Z^{10,5,2,2}$, 
$2\,\,Z^{10,5,3,1}_{\alpha\beta}$, 
$5\,\,Z^{10,5,3,1}$, 
$5\,\,Z^{10,5,4}_{\alpha\beta}$, 
$7\,\,Z^{10,5,4}$, 
$Z^{10,6,1,1,1}$, 
$7\,\,Z^{10,6,2,1}_{\alpha\beta}$, 
$12\,\,Z^{10,6,2,1}$, 
$\SUSY{Z^{10,6,3}_{\alpha_{1\cdots 4}}}$, 
$14\,\,Z^{10,6,3}_{\alpha\beta}$, 
$17\,\,Z^{10,6,3}$, 
$9\,\,Z^{10,7,1,1}_{\alpha\beta}$, 
$15\,\,Z^{10,7,1,1}$, 
$2\,\,Z^{10,7,2}_{\alpha_{1\cdots 4}}$, 
$28\,\,Z^{10,7,2}_{\alpha\beta}$, 
$32\,\,Z^{10,7,2}$, 
$5\,\,Z^{10,8,1}_{\alpha_{1\cdots 4}}$, 
$37\,\,Z^{10,8,1}_{\alpha\beta}$, 
$37\,\,Z^{10,8,1}$, 
$6\,\,Z^{10,9}_{\alpha_{1\cdots 4}}$, 
$27\,\,Z^{10,9}_{\alpha\beta}$, 
$21\,\,Z^{10,9}$}
\\\hline
$11$ & \Pbox{\textwidth}{
$\SUSY{Z^{8,7,5,1}_{\alpha}}$, 
$3\,\,Z^{8,7,6}_{\alpha}$, 
$\SUSY{Z^{8,8,3,2}_{\alpha}}$, 
$3\,\,Z^{8,8,4,1}_{\alpha}$, 
$\SUSY{Z^{8,8,5}_{\alpha\beta\gamma}}$, 
$5\,\,Z^{8,8,5}_{\alpha}$, 
$\SUSY{Z^{9,6,4,2}_{\alpha}}$, 
$3\,\,Z^{9,6,5,1}_{\alpha}$, 
$\SUSY{Z^{9,6,6}_{\alpha\beta\gamma}}$, 
$5\,\,Z^{9,6,6}_{\alpha}$, 
$\SUSY{Z^{9,7,3,1,1}_{\alpha}}$, 
$3\,\,Z^{9,7,3,2}_{\alpha}$, 
$\SUSY{Z^{9,7,4,1}_{\alpha\beta\gamma}}$, 
$10\,\,Z^{9,7,4,1}_{\alpha}$, 
$2\,\,Z^{9,7,5}_{\alpha\beta\gamma}$, 
$14\,\,Z^{9,7,5}_{\alpha}$, 
$3\,\,Z^{9,8,2,1,1}_{\alpha}$, 
$\SUSY{Z^{9,8,2,2}_{\alpha\beta\gamma}}$, 
$7\,\,Z^{9,8,2,2}_{\alpha}$, 
$3\,\,Z^{9,8,3,1}_{\alpha\beta\gamma}$, 
$19\,\,Z^{9,8,3,1}_{\alpha}$, 
$7\,\,Z^{9,8,4}_{\alpha\beta\gamma}$, 
$28\,\,Z^{9,8,4}_{\alpha}$, 
$3\,\,Z^{9,9,1,1,1}_{\alpha}$, 
$5\,\,Z^{9,9,2,1}_{\alpha\beta\gamma}$, 
$22\,\,Z^{9,9,2,1}_{\alpha}$, 
$9\,\,Z^{9,9,3}_{\alpha\beta\gamma}$, 
$29\,\,Z^{9,9,3}_{\alpha}$, 
$\SUSY{Z^{10,5,4,1,1}_{\alpha}}$, 
$3\,\,Z^{10,5,4,2}_{\alpha}$, 
$4\,\,Z^{10,5,5,1}_{\alpha}$, 
$\SUSY{Z^{10,6,2,2,1}_{\alpha}}$, 
$2\,\,Z^{10,6,3,1,1}_{\alpha}$, 
$\SUSY{Z^{10,6,3,2}_{\alpha\beta\gamma}}$, 
$7\,\,Z^{10,6,3,2}_{\alpha}$, 
$3\,\,Z^{10,6,4,1}_{\alpha\beta\gamma}$, 
$19\,\,Z^{10,6,4,1}_{\alpha}$, 
$5\,\,Z^{10,6,5}_{\alpha\beta\gamma}$, 
$22\,\,Z^{10,6,5}_{\alpha}$, 
$7\,\,Z^{10,7,2,1,1}_{\alpha}$, 
$2\,\,Z^{10,7,2,2}_{\alpha\beta\gamma}$, 
$15\,\,Z^{10,7,2,2}_{\alpha}$, 
$7\,\,Z^{10,7,3,1}_{\alpha\beta\gamma}$, 
$39\,\,Z^{10,7,3,1}_{\alpha}$, 
$15\,\,Z^{10,7,4}_{\alpha\beta\gamma}$, 
$52\,\,Z^{10,7,4}_{\alpha}$, 
$7\,\,Z^{10,8,1,1,1}_{\alpha}$, 
$15\,\,Z^{10,8,2,1}_{\alpha\beta\gamma}$, 
$64\,\,Z^{10,8,2,1}_{\alpha}$, 
$\SUSY{Z^{10,8,3}_{\alpha_{1\cdots 5}}}$, 
$29\,\,Z^{10,8,3}_{\alpha\beta\gamma}$, 
$86\,\,Z^{10,8,3}_{\alpha}$, 
$14\,\,Z^{10,9,1,1}_{\alpha\beta\gamma}$, 
$48\,\,Z^{10,9,1,1}_{\alpha}$, 
$2\,\,Z^{10,9,2}_{\alpha_{1\cdots 5}}$, 
$40\,\,Z^{10,9,2}_{\alpha\beta\gamma}$, 
$102\,\,Z^{10,9,2}_{\alpha}$, 
$3\,\,Z^{10,10,1}_{\alpha_{1\cdots 5}}$, 
$30\,\,Z^{10,10,1}_{\alpha\beta\gamma}$, 
$62\,\,Z^{10,10,1}_{\alpha}$}
\\\hline
\end{tabular}
}
\caption{Central charges in M-theory up to level $\ell=8$ and in type IIB potentials up to level $\ell_9=11$. The underlined ones are central charges of supersymmetric branes.
\label{tab:central-charges}}
\end{table}

As it is well known, the momenta and brane charges are mixed under $U$-duality transformations, and this vector representation is the multiplet for such brane charges. 
The standard coordinates $x^i$ are canonical conjugate to the momenta $P_i$, and for the manifest $U$-duality covariance, it is natural to introduce additional coordinates $y_{i_1i_2}$, $y_{i_1\cdots i_5}$, $\cdots$ which are conjugate to the brane central charges $Z^{i_1i_2}$, $Z^{i_1\cdots i_5}$, $\cdots$\,. 
This extension of geometry has been discussed in \cite{hep-th:0307098,hep-th:0312247,hep-th:0406150} (see for example \cite{0712.1795,1008.1763,1110.3930,1111.0459,1308.1673} for further discussion), leading to the recent developments in the $U$-duality manifest formulation of supergravity, known as the exceptional field theory (EFT). 

\subsection{Central charges in $d$ dimensions}

When we discuss the standard $E_n$ $U$-duality $n\leq 8$\,, the spacetime is decomposed into the external $d$-dimensional spacetime and the internal $n$-dimensional space. 
Accordingly, central charges in the vector representations are decomposed into $\SL(d)\times \SL(n)$ tensors. 
For example, in $d=9$\,, since the internal space is two-dimensional, only the following charges can appear
{\footnotesize
\begin{align}
\begin{split}
 &\SUSY{P_1}\to\SUSY{P_1}\,\,({\bf 2}), 
\qquad
 \SUSY{Z^{2}} \to\SUSY{Z^{0;2}}\,\,({\bf 1}),\ \SUSY{Z^{1;1}}\,\,({\bf 3}),\ \SUSY{Z^{2}}\,\,({\bf 1}),
\qquad
 \SUSY{Z^{5}} \to\SUSY{Z^{3;2}}\,\,({\bf 1}),\ \SUSY{Z^{4;1}}\,\,({\bf 2}),\ \SUSY{Z^{5}}\,\,({\bf 1}), 
\\
 &\SUSY{Z^{7,1}} \to\SUSY{Z^{5;2,1}}\,\,({\bf 2}),\ \SUSY{Z^{6;1,1}}\,\,({\bf 3}),\ Z^{6;2}\,\,({\bf 1}),\ Z^{7;1}\,\,({\bf 2}),
\qquad
 \SUSY{Z^{9,1,1}} \to \SUSY{Z^{7;2,1,1}}\,\,({\bf 3}),\ \SUSY{Z^{8;1,1,1}}\,\,({\bf 4}),\ Z^{8;2,1}\,\,({\bf 2}), 
\\
 &Z^{8}\to Z^{6;2}\,\,({\bf 1}),\ Z^{7;1}\,\,({\bf 2}),\ Z^{8}\,\,({\bf 1}),
\qquad
Z^{9,2} \to Z^{7;2,2}\,\,({\bf 1}),\ Z^{8;2,1}\,\,({\bf 2}), 
\qquad
 2\,Z^{10,1} \to 2\,Z^{8;2,1}\,\,({\bf 2}). 
\end{split}
\end{align}}\noindent
By collecting the central charges for $p$-branes, we obtain $d=9$ of Table \eqref{tab:p-brane}. 

For lower-dimensional cases, it is again useful to consider the following level decomposition and consider the central charges with level $m_*=1$:
\begin{align}
\scalebox{0.8}{
 \xygraph{
    *{\scalebox{2}{$\times$}}*\cir<6pt>{} ([]!{+(0,-.4)} {\alpha_*}) - [r]
    *\cir<6pt>{} ([]!{+(0,-.4)} {\alpha_1}) - [r]
    \cdots - [r]
    *\cir<6pt>{} ([]!{+(0,-.4)} {\alpha_{d-1}}) - [r]
    *{\scalebox{2}{$\times$}}*\cir<6pt>{} ([]!{+(0,-.4)} {\alpha_d}) - [r]
    *\cir<6pt>{} ([]!{+(0,-.4)} {\alpha_{d+1}}) - [r]
    \cdots 
    - [r] *\cir<6pt>{} ([]!{+(0,-.4)} {\alpha_7}) 
    - [r] *\cir<6pt>{} ([]!{+(0,-.4)} {\alpha_8}) 
(
        - [u] *\cir<6pt>{} ([]!{+(.5,0)} {\alpha_{11}}),
        - [r] *\cir<6pt>{} ([]!{+(0,-.4)} {\alpha_9})
        - [r] *\cir<6pt>{} ([]!{+(0,-.4)} {\alpha_{10}})
) 
}} .
\end{align}
The same analysis was already done in \cite{0805.4451}, but the multiplicities are not completely determined. 
We here determine the multiplicities as well, and the resulting $p$-brane charges in each dimension $d$ are given in Table \ref{tab:p-brane}. 

In Table \ref{tab:p-brane}, the $E_n$ multiples of the central charges are described, but in terms of the M-theory or type II theories, the $E_n$ multiples can be decomposed into mixed-symmetry tensors in 11D/10D. 
The decompositions into brane charges in M-theory and type IIB theory are studied in detail in Appendix \ref{app:M-branes} and \ref{app:B-branes}, respectively. 
We can see that the momenta $P_i$ appear only in the particle (or 0-brane) multiplet, and in the $E_n$ EFT we introduce generalized coordinates $x^I$ that are canonical conjugate to the central charges in the particle multiplet. 
For example, in the $E_8$ EFT, we introduce 248 generalized coordinates $x^I$ in addition to the three external coordinates $x^\mu$\,. 

\section{Conclusions}
\label{sec:conclusions}

In this work, we have conducted a detailed survey of mixed-symmetry potentials and brane charges in M-theory and type II theories that are predicted from the $E_{11}$ conjecture. 
We also considered their dimensional reductions to $d$ dimensions ($3\le d\le 9$) and checked that all the $p$-form potentials or $p$-brane charges form $U$-duality multiplets. 
We have also given an explicit construction of both $U$-duality and $\OO(D,D)$ $T$-duality multiplets in terms of the mixed-symmetry potentials and charges associated with M-theory and type II theories.
This tour de force calculation recovered all the results in the existing literature, unifying previous studies on mixed-symmetry potentials and central charges, and going far beyond. 
This work also reveals the high predictability of the $E_{11}$ conjecture. 
In principle, we can determine all of the mixed-symmetry potentials/central charges up to an arbitrary level. 

In this paper, we studied only the spectra of the gauge potentials, but in order to clarify the role of such objects in string/M-theory, it is important to study their dynamics. 
As these potentials do not appear explicitly in the standard formulations, it is useful to employ the duality-symmetric theories, such as DFT and EFT. 
Generically, formulations of EFTs in an arbitrary dimension contain $p$-form potentials $\cA_p^{I_p}$\,. 
Here, under $U$-duality transformations, the index $I_p$ transforms in the $p$-form multiplet given in Table \ref{tab:p-form}. 
Then, regardless of the duality frame in which the section condition is solved, we should be able to parameterize such $p$-form fields $\cA_p^{I_p}$ in terms of certain potentials. 
As detailed in this paper, we already know what potentials can enter in the $p$-form multiplet $\cA_p^{I_p}$, but the explicit parameterization (which depends on the convention) still needs to be specified. 
In the companion paper \cite{1909.01335}, explicit examples of parameterizations will be given. 
Once the parameterization is fixed, we can evaluate the supergravity action for the mixed-symmetry potentials as well as the standard supergravity fields. 
Besides supergravities, we can also study brane actions in a $U$-duality covariant manner. 
The $p$-form fields $\cA_p^{I_p}$ enter also in such theories, and it is important to study the role of mixed-symmetry potentials in brane actions. 

If the parameterization of the $\cA_p^{I_p}$ fields is given in terms of M-theory potentials or type II potentials, we can determine the duality transformation rules under $T$- and $S$-duality. 
As discussed in \cite{1805.12117}, duality rules for mixed-symmetry potentials are very useful in generating new supergravity solutions, and they also will be studied in the forthcoming work \cite{1909.01335}. 

\subsection*{Acknowledgments}

We would like to thank Paul Cook and Peter West for helpful comments. 
The work of JJFM is supported by Plan Propio de Investigaci\'on of the University of Murcia R-957/2017 and Fundaci\'on S\'eneca (21257/PI/19 and 20949/PI/18).
The work of YS is supported by JSPS KAKENHI Grant Numbers 18K13540 and 18H01214. 

\newpage

\appendix

\section{Convention}
\label{app:convention}

In this paper, we employ the convention for the Dynkin diagram of $E_{11}$ given in \eqref{eq:E11-dynkin}. 
Then, after level decompositions, we find a certain $\SL(n)$ representation with Dynkin label $[a_{n-1},\dotsc,a_2,a_1]$\,. 
This corresponds to a mixed-symmetry tensor $A_{\smash{\overbrace{\scriptstyle n,\dotsc,n}^{a_{n}},\overbrace{\scriptstyle n-1,\dotsc,n-1}^{a_{n-1}},\cdots,\overbrace{\scriptstyle 2,\cdots,2}^{a_2},\overbrace{\scriptstyle 1,\cdots,1}^{a_1}}}\vphantom{\Big|}$\,, where $a_n\equiv \frac{3\ell-\sum_{k=1}^{n-1} k\,a_k}{n}$\,. 
When we give a name for this representation, we reverse the Dynkin label as $[a_1,a_2,\dotsc,a_{n-1}]$ and convert it into a name of the $\SL(n)$ representation by using LieART \cite{1206.6379}. 
For example, in $n=10$\,, $[0,0,0,0,0,0,1,0,0]$ corresponds to $A_{3}$ and the representation is named ${\bf 120}$, while $[0,0,1,0,0,0,0,0,0]$ corresponds to $A_{7}$ and it is named ${\bf \overline{120}}$. 

For $\OO(D,D)$\,, we convert the obtained Dynkin label directly using LieART. 
For example, in $D=10$\,, $[0,\dotsc,0,1]$ corresponds to a spinor representation ${\bf \overline{512}}$ while $[0,\dotsc,1,0]$ is a spinor representation with opposite chirality ${\bf 512}$\,. 
We denote the former/latter spinor index as $a$/$\dot{a}$\,. 
On the other hand, $[1,0,\dotsc,0]$ is the vector representation ${\bf 20}$ and the index is called $A$\,. 
For $[*,\dotsc,*,0,0]$, we named the index in a similar manner to the $\SL(n)$ mixed-symmetry potentials (e.g.~$[1,1,0,\cdots,0]$ is $A_{A_1A_2,B}$ which satisfies $A_{A_1A_2,B}\,\eta^{A_2B}=0$)\,. 
More generally, we append the spinor indices as well (e.g.~$[1,0,\cdots,0,1]$ is a vector spinor $A_{Aa}$ which follows the gamma-traceless condition). 
A bi-spinor $[0,\cdots,0,1,1]$ is understood as a $(D-1)$-form and $[0,\cdots,0,2]$ or $[0,\cdots,0,2]$ are understood as an (anti-)self-dual $D$-form. 
Some examples are summarized in Table \ref{tab:Odd-irrep}. 
{\tiny
\begin{table}[b]
\centerline{\scalebox{0.8}{
\begin{tabular}{|c||c||c|c|c|}\hline
 Dynkin labels & $\OO(D,D)$ indices & dimensions & conditions & sector
\\\hline\hline
 $[0,\dotsc,0,\overset{p\text{th}}{1},0,\dotsc,0,0,0]$& $T_{A_{1\cdots p}}$ & ${2d\choose p}$ & $1\leq p\leq D-2$ & NS--NS
\\\hline
 $[2,0,\dotsc,0,0,0]$& $T_{A,B}$ & ${2D+1\choose 2} -1$ & $D\geq 3$ & NS--NS
\\\hline
 $[1,\dotsc,0,\overset{p\text{th}}{1},0,\dotsc,0,0,0]$& $T_{A_{1\cdots p},B}$ & $p\,{2D+1\choose p+1}-{2D\choose p-1}$ & $2\leq p\leq D-2$ & NS--NS
\\\hline
 $[0,2,0,\dotsc,0,0,0]$& $T_{A_{12},B_{12}}$ & $\frac{1}{3}{2D\choose 2}{2D+1\choose 2}-{2D+1\choose 2}$ & $D\geq 4$ & NS--NS
\\\hline
 $[0,1,0,\dotsc,0,\overset{p\text{th}}{1},0,\dotsc,0,0,0]$& $T_{A_{1\cdots p},B_{12}}$ & $\frac{p-1}{p+1}{2D\choose p}{2D+1\choose 2}-(p-1){2D+1\choose p}$ & $3\leq p\leq D-2$ & NS--NS
\\\hline
 $[0,\dotsc,0,1,1]$& $T_{A_{1\cdots (D-1)}}$ & ${2D \choose D-1}$ & $D\geq 2$ & NS--NS
\\\hline
 $[0,\dotsc,0,0,1]$& $T_{a}$ & $2^{D-1}$ & $D\geq 2$ & R--R
\\\hline
 $[0,\dotsc,0,1,0]$& $T_{\dot{a}}$ & $2^{D-1}$ & $D\geq 2$ & R--R
\\\hline
 $[0,\dotsc,0,0,2]$& $T_{A_{1\cdots D}}^-$ & $\frac{1}{2}{2D\choose D}$ & $D\geq 2$ & R--R
\\\hline
 $[0,\dotsc,0,2,0]$& $T_{A_{1\cdots D}}^+$ & $\frac{1}{2}{2D\choose D}$ & $D\geq 2$ & R--R
\\\hline
 $[0,\dotsc,0,\overset{p\text{th}}{1},0,\dotsc,0,0,1]$& $T_{A_{1\cdots p}a}$ & $2^{D-1} \bigl[{2D\choose p} -{2D\choose p-1}\bigr]$ & $1\leq p\leq D-2$ & R--R
\\\hline
 $[0,\dotsc,0,\overset{p\text{th}}{1},0,\dotsc,0,1,0]$&\quad $T_{A_{1\cdots p}\dot{a}}$ & $2^{D-1} \bigl[{2D\choose p} -{2D\choose p-1}\bigr]$&$1\leq p\leq D-2$ & R--R
\\\hline
 $[2,0,\dotsc,0,0,1]$& $T_{A,Ba}$ & $2^{D-1}{2D\choose 2}$ & $D\geq 3$ & R--R
\\\hline
 $[2,0,\dotsc,0,1,0]$& $T_{A,B\dot{a}}$ & $2^{D-1}{2D\choose 2}$ & $D\geq 3$ & R--R
\\\hline
 $[1,0,\dotsc,0,\overset{p\text{th}}{1},0,\dotsc,0,0,1]$& $T_{A_{1\cdots p},Ba}$ & $2^{D-1}\,{2D+1\choose p+1}\Bigl(p-\frac{p(p+1)}{2D+2-p}\Bigr)$ & $1\leq p\leq D-2$ & R--R
\\\hline
 $[1,0,\dotsc,0,\overset{p\text{th}}{1},0,\dotsc,0,1,0]$& $T_{A_{1\cdots p},B\dot{a}}$ & $2^{D-1}\,{2D+1\choose p+1}\Bigl(p-\frac{p(p+1)}{2D+2-p}\Bigr)$ & $1\leq p\leq D-2$ & R--R
\\\hline
 $[0,2,0\dotsc,0,0,1]$& $T_{A_{12},B_{12}a}$ & $2^{D-1}\,(D+1){2D-1\choose 3}$ & $ D\geq 4$ & R--R
\\\hline
\end{tabular}
}}
\caption{Dimensions and index conventions for irreducible representations of of $\OO(D,D)$\,.}
\label{tab:Odd-irrep}
\end{table}
}

\newpage

\section{M-theory branes: $E_n\to \SL(n)$}
\label{app:M-branes}

In this appendix, we find which brane charges in M-theory are contained in each $E_n$ $U$-duality multiplet of Table \ref{tab:p-brane}. 
The first integer of the central charge $Z^{p;q,r,\cdots}$ represents that the brane is a $p$-brane. 
The remaining integers ${q,r,\cdots}$ denote the type of the $\SL(n)$ tensor. 

\subsection*{M-theory branes in $d=8$: $E_3\to \SL(3)$}

Decomposition of the particle, string, and membrane multiplets:
\begin{align*}
{\scriptsize
\begin{tabular}{|c|c|l|} \hline
 &$\ell$ & \multicolumn{1}{c|}{central charges} \\ \hline\hline
%%%%%%%%%%%%%%%%%%%%%%%%%%
 $\textcolor{red}{({\bf 3},{\bf 2})}$ 
 & $0$&
 $\SUSY{P_{1}}\,\,({\bf \overline{3}})$
\\\cline{2-3}
 & $1$&
 $\SUSY{Z^{0;2}}\,\,({\bf \overline{3}})$ 
\\\hline
\end{tabular}
}
\qquad
{\scriptsize
\begin{tabular}{|c|c|l|} \hline
 &$\ell$ & \multicolumn{1}{c|}{central charges} \\ \hline\hline
%%%%%%%%%%%%%%%%%%%%%%%%%%
 $\textcolor{red}{({\bf 3},{\bf 1})}$ 
 & $1$&
 $\SUSY{Z^{1;1}}\,\,({\bf 3})$ 
\\\hline
\end{tabular}
}
\qquad
{\scriptsize
\begin{tabular}{|c|c|l|} \hline
 &$\ell$ & \multicolumn{1}{c|}{central charges} \\ \hline\hline
%%%%%%%%%%%%%%%%%%%%%%%%%%
 $\textcolor{red}{({\bf 1},{\bf 2})}$ 
 & $1$&
 $\SUSY{Z^{2}}\,\,({\bf 1})$ 
\\\cline{2-3}
 & $2$& 
 $\SUSY{Z^{2;3}}\,\,({\bf 1})$ 
\\\hline
\end{tabular}
}
\end{align*}
Decomposition of the 3- and 4-brane multiplets:
\begin{align*}
{\scriptsize
\begin{tabular}{|c|c|l|} \hline
 &$\ell$ & \multicolumn{1}{c|}{central charges} \\ \hline\hline
%%%%%%%%%%%%%%%%%%%%%%%%%%
 $\textcolor{red}{({\bf \overline{3}},{\bf 1})}$ 
 & $2$& 
 $\SUSY{Z^{3;2}}\,\,({\bf \overline{3}})$ 
\\\hline
\end{tabular}
}
\qquad
{\scriptsize
\begin{tabular}{|c|c|l|} \hline
 &$\ell$ & \multicolumn{1}{c|}{central charges} \\ \hline\hline
%%%%%%%%%%%%%%%%%%%%%%%%%%
 $\textcolor{red}{({\bf 3},{\bf 2})}$ 
 & $2$& 
 $\SUSY{Z^{4;1}}\,\,({\bf 3})$ 
\\\cline{2-3}
 & $3$& 
 $\SUSY{Z^{4;3,1}}\,\,({\bf 3})$ 
\\\hline
\end{tabular}
}
\end{align*}
Decomposition of the 5-, 6-, and 7-brane multiplets:
\begin{align*}
{\scriptsize
\begin{tabular}{|c|c|l|} \hline
 &$\ell$ & \multicolumn{1}{c|}{central charges} \\ \hline\hline
%%%%%%%%%%%%%%%%%%%%%%%%%%
 $\textcolor{red}{({\bf 8},{\bf 1})}$ 
 & $3$& 
 $\SUSY{Z^{5;2,1}}\,\,({\bf 8})$ 
\\\hline
%%%%%%%%%
 $\textcolor{red}{({\bf 1},{\bf 3})}$ 
 & $2$& 
 $\SUSY{Z^{5}}\,\,({\bf 1})$ 
\\\cline{2-3}
 & $3$& 
 $Z^{5;3}\,\,({\bf 1})$ 
\\\cline{2-3}
 & $4$& 
 $\SUSY{Z^{5;3,3}}\,\,({\bf 1})$ 
\\\hline
%%%%%%%%%
 $({\bf 1},{\bf 1})$ 
 & $3$& 
 $Z^{5;3}\,\,({\bf 1})$ 
\\\hline
\end{tabular}
}
\qquad
{\scriptsize
\begin{tabular}{|c|c|l|} \hline
 &$\ell$ & \multicolumn{1}{c|}{central charges} \\ \hline\hline
%%%%%%%%%%%%%%%%%%%%%%%%%%
 $\textcolor{red}{({\bf 6},{\bf 2})}$ 
 & $3$& 
 $\SUSY{Z^{6;1,1}}\,\,({\bf 6})$
\\\cline{2-3}
 & $4$& 
 $\SUSY{Z^{6;3,1,1}}\,\,({\bf 6})$ 
\\\hline
%%%%%%%%%
 $({\bf \overline{3}},{\bf 2})$ 
 & $3$& 
 $Z^{6;2}\,\,({\bf \overline{3}})$ 
\\\cline{2-3}
 & $4$& 
 $Z^{6;3,2}\,\,({\bf \overline{3}})$
\\\hline
\end{tabular}
}
\qquad
{\scriptsize
\begin{tabular}{|c|c|l|} \hline
 &$\ell$ & \multicolumn{1}{c|}{central charges} \\ \hline\hline
%%%%%%%%%%%%%%%%%%%%%%%%%%
 $\textcolor{red}{({\bf 15},{\bf 1})}$ 
 & $4$& 
 $\SUSY{Z^{7;2,1,1}}\,\,({\bf 15})$
\\\hline
%%%%%%%%%
 $({\bf \overline{6}},{\bf 1})$ 
 & $4$& 
 $Z^{7;2,2}\,\,({\bf \overline{6}})$ 
\\\hline
%%%%%%%%%
 $({\bf 3},{\bf 3})$ 
 & $3$& 
 $Z^{7;1}\,\,({\bf 3})$ 
\\\cline{2-3}
 & $4$& 
 $Z^{7;3,1}\,\,({\bf 3})$ 
\\\cline{2-3}
 & $5$& 
 $Z^{7;3,3,1}\,\,({\bf 3})$ 
\\\hline
%%%%%%%%%
 $({\bf 3},{\bf 1})$ 
 & $4$& 
 $Z^{7;3,1}\,\,({\bf 3})$ 
\\\hline
\end{tabular}
}
\end{align*}

\subsection*{M-theory branes in $d=7$: $E_4\to \SL(4)$}

Decomposition of the particle, string, and membrane multiplets:
\begin{align*}
{\scriptsize
\begin{tabular}{|c|c|l|} \hline
 &$\ell$ & \multicolumn{1}{c|}{central charges} \\ \hline\hline
%%%%%%%%%%%%%%%%%%%%%%%%%%
 $\textcolor{red}{{\bf 10}}$ 
 & $0$&
 $\SUSY{P_{1}}\,\,({\bf \overline{4}})$
\\\cline{2-3}
 & $1$&
 $\SUSY{Z^{0;2}}\,\,({\bf 6})$
\\\hline
\end{tabular}
}
\qquad
{\scriptsize
\begin{tabular}{|c|c|l|} \hline
 &$\ell$ & \multicolumn{1}{c|}{central charges} \\ \hline\hline
%%%%%%%%%%%%%%%%%%%%%%%%%%
 $\textcolor{red}{{\bf 5}}$ 
 & $1$&
 $\SUSY{Z^{1;1}}\,\,({\bf 4})$
\\\cline{2-3}
 & $2$&
 $\SUSY{Z^{1}}\,\,({\bf 1})$
\\\hline
\end{tabular}
}
\qquad
{\scriptsize
\begin{tabular}{|c|c|l|} \hline
 &$\ell$ & \multicolumn{1}{c|}{central charges} \\ \hline\hline
%%%%%%%%%%%%%%%%%%%%%%%%%%
 $\textcolor{red}{{\bf \overline{5}}}$ 
 & $1$&
 $\SUSY{Z^{2}}\,\,({\bf 1})$
\\\cline{2-3}
 & $2$&
 $\SUSY{Z^{2;3}}\,\,({\bf \overline{4}})$
\\\hline
\end{tabular}
}
\end{align*}
Decomposition of the 3-, and 4-brane multiplets:
\begin{align*}
{\scriptsize
\begin{tabular}{|c|c|l|} \hline
 &$\ell$ & \multicolumn{1}{c|}{central charges} \\ \hline\hline
%%%%%%%%%%%%%%%%%%%%%%%%%%
 $\textcolor{red}{{\bf 10}}$ 
 & $2$&
 $\SUSY{Z^{3;2}}\,\,({\bf 6})$
\\\cline{2-3}
 & $3$&
 $\SUSY{Z^{3;1}}\,\,({\bf 4})$
\\\hline
\end{tabular}
}
\qquad
{\scriptsize
\begin{tabular}{|c|c|l|} \hline
 &$\ell$ & \multicolumn{1}{c|}{central charges} \\ \hline\hline
%%%%%%%%%%%%%%%%%%%%%%%%%%
 $\textcolor{red}{{\bf 24}}$ 
 & $2$&
 $\SUSY{Z^{4;1}}\,\,({\bf 4})$
\\\cline{2-3}
 & $3$&
 $\SUSY{Z^{4;3,1}}\,\,({\bf 15})$, 
 $Z^{4;4}\,\,({\bf 1})$
\\\cline{2-3}
 & $4$&
 $\SUSY{Z^{4;4,3}}\,\,({\bf \overline{4}})$
\\\hline
%%%%%%%%%
 ${\bf 1}$ 
 & $3$&
 $Z^{4;4}\,\,({\bf 1})$
\\\hline
\end{tabular}
}
\end{align*}
Decomposition of the 5- and 6-brane multiplets:
\begin{align*}
{\scriptsize
\begin{tabular}{|c|c|l|} \hline
 &$\ell$ & \multicolumn{1}{c|}{central charges} \\ \hline\hline
%%%%%%%%%%%%%%%%%%%%%%%%%%
 $\textcolor{red}{{\bf \overline{40}}}$ 
 & $3$&
 $\SUSY{Z^{5;2,1}}\,\,({\bf \overline{20}})$ 
\\\cline{2-3}
 & $4$&
 $\SUSY{Z^{5;4,1,1}}\,\,({\bf 10})$, 
 $Z^{5;4,2}\,\,({\bf 6})$
\\\cline{2-3}
 & $5$&
 $\SUSY{Z^{5,4,4,1}}\,\,({\bf 4})$
\\\hline
%%%%%%%%%
 $\textcolor{red}{{\bf \overline{15}}}$ 
 & $2$&
 $\SUSY{Z^{5}}\,\,({\bf 1})$
\\\cline{2-3}
 & $3$&
 $Z^{5;3}\,\,({\bf \overline{4}})$
\\\cline{2-3}
 & $4$&
 $\SUSY{Z^{5;3,3}}\,\,({\bf \overline{10}})$ 
\\\hline
%%%%%%%%%
 ${\bf \overline{10}}$ 
 & $3$&
 $Z^{5;3}\,\,({\bf \overline{4}})$
\\\cline{2-3}
 & $4$&
 $Z^{5;4,2}\,\,({\bf 6})$ 
\\\hline
\end{tabular}
}
\qquad
{\scriptsize
\begin{tabular}{|c|c|l|} \hline
 &$\ell$ & \multicolumn{1}{c|}{central charges} \\ \hline\hline
%%%%%%%%%%%%%%%%%%%%%%%%%%
 $\textcolor{red}{{\bf 70}}$ 
 & $3$&
 $\SUSY{Z^{6;1,1}}\,\,({\bf 10})$
\\\cline{2-3}
 & $4$&
 $\SUSY{Z^{6;3,1,1}}\,\,({\bf 36})$, 
 $Z^{6;4,1}\,\,({\bf 4})$ 
\\\cline{2-3}
 & $5$&
 $Z^{6;4,3,1}\,\,({\bf 15})$, 
 $Z^{6;4,4}\,\,({\bf 1})$
\\\cline{2-3}
 & $6$&
 $\SUSY{Z^{6;4,4,3}}\,\,({\bf \overline{4}})$
\\\hline
%%%%%%%%%
 ${\bf 45}$ 
 & $3$&
 $Z^{6;2}\,\,({\bf 6})$
\\\cline{2-3}
 & $4$&
 $Z^{6;3,2}\,\,({\bf 20})$, 
 $Z^{6;4,1}\,\,({\bf 4})$ 
\\\cline{2-3}
 & $5$&
 $Z^{6;4,3,1}\,\,({\bf 15})$
\\\hline
%%%%%%%%%
 ${\bf 5}$ 
 & $4$&
 $Z^{6;4,1}\,\,({\bf 4})$ 
\\\cline{2-3}
 & $5$&
 $Z^{6;4,4}\,\,({\bf 1})$
\\\hline
\end{tabular}
}
\end{align*}

\subsection*{M-theory branes in $d=6$: $E_5\to \SL(5)$}

Decomposition of the particle, string, and membrane multiplets:
\begin{align*}
{\scriptsize
\begin{tabular}{|c|c|l|} \hline
 &$\ell$ & \multicolumn{1}{c|}{central charges} \\ \hline\hline
%%%%%%%%%%%%%%%%%%%%%%%%%%
 $\textcolor{red}{{\bf 16}}$ 
 & $0$&
 $\SUSY{P_{1}}\,\,({\bf \overline{5}})$
\\\cline{2-3}
 & $1$&
 $\SUSY{Z^{0;2}}\,\,({\bf 10})$
\\\cline{2-3}
 & $2$&
 $\SUSY{Z^{0;5}}\,\,({\bf 1})$
\\\hline
\end{tabular}
}
\qquad
{\scriptsize
\begin{tabular}{|c|c|l|} \hline
 &$\ell$ & \multicolumn{1}{c|}{central charges} \\ \hline\hline
%%%%%%%%%%%%%%%%%%%%%%%%%%
 $\textcolor{red}{{\bf 10}}$ & $1$&
 $\SUSY{Z^{1;1}}\,\,({\bf 5})$
\\\cline{2-3}
 & $2$&
 $\SUSY{Z^{1;4}}\,\,({\bf \overline{5}})$
\\\hline
\end{tabular}
}
\qquad
{\scriptsize
\begin{tabular}{|c|c|l|} \hline
 &$\ell$ & \multicolumn{1}{c|}{central charges} \\ \hline\hline
%%%%%%%%%%%%%%%%%%%%%%%%%%
 $\textcolor{red}{{\bf \overline{16}}}$
 & $1$&
 $\SUSY{Z^{2}}\,\,({\bf 1})$
\\\cline{2-3}
 & $2$&
 $\SUSY{Z^{2;3}}\,\,({\bf \overline{10}})$
\\\cline{2-3}
 & $3$&
 $\SUSY{Z^{2;5,1}}\,\,({\bf 5})$
\\\hline
\end{tabular}
}
\end{align*}
Decomposition of the 3- and 4-brane multiplets:
\begin{align*}
{\scriptsize
\begin{tabular}{|c|c|l|} \hline
 &$\ell$ & \multicolumn{1}{c|}{central charges} \\ \hline\hline
%%%%%%%%%%%%%%%%%%%%%%%%%%
 $\textcolor{red}{{\bf 45}}$
 & $2$&
 $\SUSY{Z^{3;2}}\,\,({\bf 10})$
\\\cline{2-3}
 & $3$&
 $\SUSY{Z^{3;4,1}}\,\,({\bf 24})$, 
 $Z^{3;5}\,\,({\bf 1})$ 
\\\cline{2-3}
 & $4$&
 $\SUSY{Z^{3;5,3}}\,\,({\bf \overline{10}})$
\\\hline
%%%%%%%%%
 ${\bf 1}$
 & $3$&
 $Z^{3;5}\,\,({\bf 1})$ 
\\\hline
\end{tabular}
}
\qquad
{\scriptsize
\begin{tabular}{|c|c|l|} \hline
 &$\ell$ & \multicolumn{1}{c|}{central charges} \\ \hline\hline
%%%%%%%%%%%%%%%%%%%%%%%%%%
 $\textcolor{red}{{\bf 144}}$
 & $2$&
 $\SUSY{Z^{4;1}}\,\,({\bf 5})$
\\\cline{2-3}
 & $3$&
 $\SUSY{Z^{4;3,1}}\,\,({\bf \overline{45}})$, 
 $Z^{4;4}\,\,({\bf \overline{5}})$
\\\cline{2-3}
 & $4$&
 $\SUSY{Z^{4;4,3}}\,\,({\bf 40})$, 
 $\SUSY{Z^{4;5,1,1}}\,\,({\bf 15})$, 
 $Z^{4;5,2}\,\,({\bf 10})$ 
\\\cline{2-3}
 & $5$&
 $\SUSY{Z^{4;5,4,1}}\,\,({\bf 24})$
\\\hline
%%%%%%%%%
 ${\bf 16}$ 
 & $3$&
 $Z^{4;4}\,\,({\bf \overline{5}})$
\\\cline{2-3}
 & $4$&
 $Z^{4;5,2}\,\,({\bf 10})$
\\\cline{2-3}
 & $5$&
 $Z^{4;5,5}\,\,({\bf 1})$
\\\hline
\end{tabular}
}
\end{align*}
Decomposition of the 5-brane multiplet:
\begin{align*}
{\scriptsize
\begin{tabular}{|c|c|l|} \hline
 &$\ell$ & \multicolumn{1}{c|}{central charges} \\\hline\hline
%%%%%%%%%%%%%%%%%%%%%%%%%%
 $\textcolor{red}{{\bf 320}}$
 & $3$&
 $\SUSY{Z^{5;2,1}}\,\,({\bf \overline{40}})$
\\\cline{2-3}
 & $4$&
 $\SUSY{Z^{5;4,1,1}}\,\,({\bf 70})$, 
 $Z^{5;4,2}\,\,({\bf 45})$, 
 $Z^{5;5,1}\,\,({\bf 5})$
\\\cline{2-3}
 & $5$&
 $\SUSY{Z^{5;4,4,1}}\,\,({\bf \overline{70}})$, 
 $Z^{5;5,3,1}\,\,({\bf \overline{45}})$, 
 $Z^{5;5,4}\,\,({\bf \overline{5}})$
\\\cline{2-3}
 & $6$&
 $\SUSY{Z^{5;5,4,3}}\,\,({\bf 40})$
\\\hline
%%%%%%%%%
 $\textcolor{red}{{\bf \overline{126}}}$
 & $2$&
 $\SUSY{Z^{5}}\,\,({\bf 1})$
\\\cline{2-3}
 & $3$&
 $Z^{5;3}\,\,({\bf \overline{10}})$
\\\cline{2-3}
 & $4$&
 $\SUSY{Z^{5;3,3}}\,\,({\bf 50})$, 
 $Z^{5;5,1}\,\,({\bf 5})$
\\\cline{2-3}
 & $5$&
 $Z^{5;5,3,1}\,\,({\bf \overline{45}})$, 
\\\cline{2-3}
 & $6$&
 $\SUSY{Z^{5;5,5,1,1}}\,\,({\bf 15})$
\\\hline
%%%%%%%%%
 ${\bf 120}$
 & $3$&
 $Z^{5;3}\,\,({\bf \overline{10}})$
\\\cline{2-3}
 & $4$&
 $Z^{5;4,2}\,\,({\bf 45})$, 
 $Z^{5;5,1}\,\,({\bf 5})$
\\\cline{2-3}
 & $5$&
 $Z^{5;5,3,1}\,\,({\bf \overline{45}})$, 
 $Z^{5;5,4}\,\,({\bf \overline{5}})$
\\\cline{2-3}
 & $6$&
 $Z^{5;5,5,2}\,\,({\bf 10})$
\\\hline
%%%%%%%%%
 ${\bf 10}$
 & $4$&
 $Z^{5;5,1}\,\,({\bf 5})$
\\\cline{2-3}
 & $5$&
 $Z^{5;5,4}\,\,({\bf \overline{5}})$
\\\hline
\end{tabular}
}
\end{align*}

\subsection*{M-theory branes in $d=5$: $E_6\to \SL(6)$}

Decomposition of the particle, string, and membrane multiplets:
\begin{align*}
{\scriptsize
\begin{tabular}{|c|c|l|} \hline
 &$\ell$ & \multicolumn{1}{c|}{central charges} \\ \hline\hline
%%%%%%%%%%%%%%%%%%%%%%%%%%
 $\textcolor{red}{{\bf 27}}$ 
 & $0$&
 $\SUSY{P_{1}}\,\,({\bf \overline{6}})$
\\\cline{2-3}
 & $1$&$\SUSY{Z^{0;2}}\,\,({\bf 15})$
\\\cline{2-3}
 & $2$&
 $\SUSY{Z^{0;5}}\,\,({\bf \overline{6}})$ 
\\\hline
\end{tabular}
}
\qquad
{\scriptsize
\begin{tabular}{|c|c|l|} \hline
 &$\ell$ & \multicolumn{1}{c|}{central charges} \\ \hline\hline
%%%%%%%%%%%%%%%%%%%%%%%%%%
 $\textcolor{red}{{\bf \overline{27}}}$ 
 & $1$&
 $\SUSY{Z^{1;1}}\,\,({\bf 6})$
\\\cline{2-3}
 & $2$&
 $\SUSY{Z^{1;4}}\,\,({\bf \overline{15}})$
\\\cline{2-3}
 & $3$&
 $\SUSY{Z^{1;6,1}}\,\,({\bf 6})$
\\\hline
\end{tabular}
}
\qquad
{\scriptsize
\begin{tabular}{|c|c|l|} \hline
 &$\ell$ & \multicolumn{1}{c|}{central charges} \\ \hline\hline
%%%%%%%%%%%%%%%%%%%%%%%%%%
 $\textcolor{red}{{\bf 78}}$
 & $1$&
 $\SUSY{Z^{2}}\,\,({\bf 1})$
\\\cline{2-3}
 & $2$&
 $\SUSY{Z^{2;3}}\,\,({\bf 20})$
\\\cline{2-3}
 & $3$&
 $\SUSY{Z^{2;5,1}}\,\,({\bf 35})$, 
 $Z^{2;6}\,\,({\bf 1})$
\\\cline{2-3}
 & $4$&
 $\SUSY{Z^{2;6,3}}\,\,({\bf 20})$
\\\cline{2-3}
 & $5$&
 $\SUSY{Z^{2;6,6}}\,\,({\bf 1})$
\\\hline
%%%%%%%%%
 ${\bf 1}$
 & $3$&
 $Z^{2;6}\,\,({\bf 1})$
\\\hline
\end{tabular}
}
\end{align*}
Decomposition of the 3-brane multiplet:
\begin{align*}
{\scriptsize
\begin{tabular}{|c|c|l|} \hline
 &$\ell$ & \multicolumn{1}{c|}{central charges} \\ \hline\hline
%%%%%%%%%%%%%%%%%%%%%%%%%%
 $\textcolor{red}{{\bf 351}}$
 & $2$&
 $\SUSY{Z^{3;2}}\,\,({\bf 15})$
\\\cline{2-3}
 & $3$&
 $\SUSY{Z^{3;4,1}}\,\,({\bf \overline{84}})$, 
 $Z^{3;5}\,\,({\bf \overline{6}})$
\\\cline{2-3}
 & $4$&
 $\SUSY{Z^{3;5,3}}\,\,({\bf 105})$, 
 $\SUSY{Z^{3;6,1,1}}\,\,({\bf 21})$, 
 $Z^{3;6,2}\,\,({\bf 15})$
\\\cline{2-3}
 & $5$&
 $\SUSY{Z^{3;6,4,1}}\,\,({\bf \overline{84}})$, 
 $Z^{3;6,5}\,\,({\bf \overline{6}})$
\\\cline{2-3}
 & $6$&
 $\SUSY{Z^{3;6,6,2}}\,\,({\bf 15})$
\\\hline
%%%%%%%%%
 ${\bf 27}$ 
 & $3$&
 $Z^{3;5}\,\,({\bf \overline{6}})$
\\\cline{2-3}
 & $4$&$Z^{3;6,2}\,\,({\bf 15})$
\\\cline{2-3}
 & $5$&
 $Z^{3;6,5}\,\,({\bf \overline{6}})$ 
\\\hline
\end{tabular}
}
\end{align*}
Decomposition of the 4-brane multiplet:
\begin{align*}
{\scriptsize
\begin{tabular}{|c|c|l|} \hline
 &$\ell$ & \multicolumn{1}{c|}{central charges} \\ \hline\hline
%%%%%%%%%%%%%%%%%%%%%%%%%%
 $\textcolor{red}{{\bf \overline{1728}}}$
 & $2$& 
 $\SUSY{Z^{4;1}}\,\,({\bf 6})$
\\\cline{2-3}
 & $3$& 
 $\SUSY{Z^{4;3,1}}\,\,({\bf \overline{105}})$, 
 $Z^{4;4}\,\,({\bf \overline{15}})$ 
\\\cline{2-3}
 & $4$& 
 $\SUSY{Z^{4;4,3}}\,\,({\bf 210})$, 
 $\SUSY{Z^{4;5,1,1}}\,\,({\bf 120})$, 
 $Z^{4;5,2}\,\,({\bf 84})$, 
 $2\,Z^{4;6,1}\,\,(2\times {\bf 6})$
\\\cline{2-3}
 & $5$& 
 $\SUSY{Z^{4;5,4,1}}\,\,({\bf \overline{384}})$, 
 $2\,Z^{4;6,3,1}\,\,(2\times{\bf \overline{105}})$, 
 $2\,Z^{4;6,4}\,\,(2\times{\bf \overline{15}})$
\\\cline{2-3}
 & $6$& 
 $\SUSY{Z^{4;6,4,3}}\,\,({\bf 210})$, 
 $\SUSY{Z^{4;6,5,1,1}}\,\,({\bf 120})$, 
 $Z^{4;6,5,2}\,\,({\bf 84})$, 
 $2\,Z^{4;6,6,1}\,\,(2 \times {\bf 6})$
\\\cline{2-3}
 & $7$&
 $\SUSY{Z^{4;6,6,3,1}}\,\,({\bf \overline{105}})$, 
 $Z^{4;6,6,4}\,\,({\bf \overline{15}})$
\\\cline{2-3}
 & $8$&
 $\SUSY{Z^{4;6,6,6,1}}\,\,({\bf 6})$
\\\hline
%%%%%%%%%
 ${\bf \overline{351}}$
 & $3$& 
 $Z^{4;4}\,\,({\bf \overline{15}})$ 
\\\cline{2-3}
 & $4$& 
 $Z^{4;5,2}\,\,({\bf 84})$, 
 $Z^{4;6,1}\,\,({\bf 6})$
\\\cline{2-3}
 & $5$& 
 $Z^{4;5,1}\,\,({\bf \overline{21}})$, 
 $Z^{4;6,3,1}\,\,({\bf \overline{105}})$, 
 $Z^{4;6,4}\,\,({\bf \overline{15}})$
\\\cline{2-3}
 & $6$& 
 $Z^{4;6,5,2}\,\,({\bf 84})$, 
 $Z^{4;6,6,1}\,\,({\bf 6})$
\\\cline{2-3}
 & $7$&
 $Z^{4;6,6,4}\,\,({\bf \overline{15}})$
\\\hline
%%%%%%%%%
 ${\bf \overline{27}}$
 & $4$& 
 $Z^{4;6,1}\,\,({\bf 6})$
\\\cline{2-3}
 & $5$& 
 $Z^{4;6,4}\,\,({\bf \overline{15}})$
\\\cline{2-3}
 & $6$& 
 $Z^{4;6,6,1}\,\,({\bf 6})$ 
\\\hline
\end{tabular}
}
\end{align*}

\subsection*{M-theory branes in $d=4$: $E_7\to \SL(7)$}

Decomposition of the particle, string, and membrane multiplets:
\begin{align*}
{\scriptsize
\begin{tabular}{|c|c|l|} \hline
 &$\ell$ & \multicolumn{1}{c|}{central charges} \\ \hline\hline
%%%%%%%%%%%%%%%%%%%%%%%%%%
 $\textcolor{red}{{\bf 56}}$ 
 & $0$&
 $\SUSY{P_{1}}\,\,({\bf \overline{7}})$
\\\cline{2-3}
 & $1$&
 $\SUSY{Z^{0;2}}\,\,({\bf 21})$
\\\cline{2-3}
 & $2$&
 $\SUSY{Z^{0;5}}\,\,({\bf \overline{21}})$
\\\cline{2-3}
 & $3$&
 $\SUSY{Z^{0;7,1}}\,\,({\bf 7})$
\\\hline
\end{tabular}
}
\quad
{\scriptsize
\begin{tabular}{|c|c|l|} \hline
 &$\ell$ & \multicolumn{1}{c|}{central charges} \\ \hline\hline
%%%%%%%%%%%%%%%%%%%%%%%%%%
 $\textcolor{red}{{\bf 133}}$
 & $1$&
 $\SUSY{Z^{1;1}}\,\,({\bf 7})$
\\\cline{2-3}
 & $2$&
 $\SUSY{Z^{1;4}}\,\,({\bf \overline{35}})$
\\\cline{2-3}
 & $3$&
 $\SUSY{Z^{1;6,1}}\,\,({\bf 48})$, 
 $Z^{1;7}\,\,({\bf 1})$ 
\\\cline{2-3}
 & $4$&
 $\SUSY{Z^{1;7,3}}\,\,({\bf 35})$
\\\cline{2-3}
 & $5$&
 $\SUSY{Z^{1;7,6}}\,\,({\bf \overline{7}})$
\\\hline
%%%%%%%%%
 ${\bf 1}$
 & $3$&
 $Z^{1;7}\,\,({\bf 1})$
\\\hline
\end{tabular}
}
\quad
{\scriptsize
\begin{tabular}{|c|c|l|} \hline
 &$\ell$ & \multicolumn{1}{c|}{central charges} \\ \hline\hline
%%%%%%%%%%%%%%%%%%%%%%%%%%
 $\textcolor{red}{{\bf 912}}$
 & $1$ &
 $\SUSY{Z^{2}}\,\,({\bf 1})$
\\\cline{2-3}
 & $2$ &
 $\SUSY{Z^{2;3}}\,\,({\bf 35})$
\\\cline{2-3}
 & $3$ &
 $\SUSY{Z^{2;5,1}}\,\,({\bf \overline{140}})$, 
 $Z^{2;6}\,\,({\bf \overline{7}})$ 
\\\cline{2-3}
 & $4$ &
 $\SUSY{Z^{2;6,3}}\,\,({\bf 224})$, 
 $\SUSY{Z^{2;7,1,1}}\,\,({\bf 28})$, 
 $Z^{2;7,2}\,\,({\bf 21})$ 
\\\cline{2-3}
 & $5$ &
 $\SUSY{Z^{2;6,6}}\,\,({\bf \overline{28}})$, 
 $\SUSY{Z^{2;7,4,1}}\,\,({\bf \overline{224}})$, 
 $Z^{2;7,5}\,\,({\bf \overline{21}})$ 
\\\cline{2-3}
 & $6$ &
 $\SUSY{Z^{2;7,6,2}}\,\,({\bf 140})$, 
 $Z^{2;7,7,1}\,\,({\bf 7})$
\\\cline{2-3}
 & $7$ &
 $\SUSY{Z^{2;7,7,4}}\,\,({\bf \overline{35}})$
\\\cline{2-3}
 & $8$ &
 $\SUSY{Z^{2;7,7,7}}\,\,({\bf 1})$
\\\hline
%%%%%%%%%%%%%%%%%%%%%%%%%%
 ${\bf 56}$ 
 & $3$&
 $Z^{2;6}\,\,({\bf \overline{7}})$
\\\cline{2-3}
 & $4$&
 $Z^{2;7,2}\,\,({\bf 21})$
\\\cline{2-3}
 & $5$&
 $Z^{2;7,5}\,\,({\bf \overline{21}})$
\\\cline{2-3}
 & $6$&
 $Z^{2;7,7,1}\,\,({\bf 7})$
\\\hline
\end{tabular}
}
\end{align*}
Decomposition of the 3-brane multiplet:
\begin{align*}
{\scriptsize
\begin{tabular}{|c|c|l|} \hline
 &$\ell$ & \multicolumn{1}{c|}{central charges} \\ \hline\hline
%%%%%%%%%%%%%%%%%%%%%%%%%%
 $\textcolor{red}{{\bf 8645}}$& $2$ &
 $\SUSY{Z^{3;2}}\,\,({\bf 21})$
\\\cline{2-3}
 & $3$ &
 $\SUSY{Z^{3;4,1}}\,\,({\bf \overline{224}})$, 
 $Z^{3;5}\,\,({\bf \overline{21}})$ 
\\\cline{2-3}
 & $4$ &
 $\SUSY{Z^{3;5,3}}\,\,({\bf 588})$, 
 $\SUSY{Z^{3;6,1,1}}\,\,({\bf 189})$, 
 $Z^{3;6,2}\,\,({\bf 140})$, 
 $2\,Z^{3;7,1}\,\,(2\times{\bf 7})$ 
\\\cline{2-3}
 & $5$ & 
 $\SUSY{Z^{3;6,4,1}}\,\,({\bf \overline{1323}})$, 
 $Z^{3;6,5}\,\,({\bf \overline{112}})$, 
 $2\,Z^{3;7,3,1}\,\,(2\times{\bf \overline{210}})$, 
 $2\,Z^{3;7,4}\,\,(2\times{\bf \overline{35}})$ 
\\\cline{2-3}
 & $6$ & 
 $\SUSY{Z^{3;6,6,2}}\,\,({\bf \overline{540}})$, 
 $\SUSY{Z^{3;7,4,3}}\,\,({\bf 784})$, 
 $\SUSY{Z^{3;7,5,1,1}}\,\,({\bf 540})$, 
 $Z^{3;7,5,2}\,\,({\bf 392})$, 
 $3\,Z^{3;7,6,1}\,\,(3\times {\bf 48})$, 
 $Z^{3;7,7}\,\,({\bf 1})$
\\\cline{2-3}
 & $7$ & 
 $\SUSY{Z^{3;7,6,3,1}}\,\,({\bf 1323})$, 
 $2\,Z^{3;7,6,4}\,\,(2\times{\bf 210})$, 
 $Z^{3;7,7,2,1}\,\,({\bf 112})$, 
 $2\,Z^{3;7,7,3}\,\,(2\times{\bf 35})$
\\\cline{2-3}
 & $8$ & 
 $\SUSY{Z^{3;7,6,6,1}}\,\,({\bf \overline{189}})$, 
 $\SUSY{Z^{3;7,7,4,2}}\,\,({\bf \overline{588}})$, 
 $Z^{3;7,7,5,1}\,\,({\bf \overline{140}})$, 
 $2\,Z^{3;7,7,6}\,\,(2\times {\bf \overline{7}})$
\\
\hline
\end{tabular}
}
\end{align*}

\subsection*{M-theory branes in $d=3$: $E_8\to \SL(8)$}

Decomposition of the particle and string multiplets:
\begin{align*}
{\scriptsize
\begin{tabular}{|c|c|l|} \hline
 &$\ell$ & \multicolumn{1}{c|}{central charges} \\ \hline\hline
%%%%%%%%%%%%%%%%%%%%%%%%%%
 $\textcolor{red}{{\bf 248}}$ 
 & $0$&
 $\SUSY{P_1}\,\,({\bf \overline{8}})$
\\\cline{2-3}
 & $1$&
 $\SUSY{Z^{0;2}}\,\,({\bf 28})$
\\\cline{2-3}
 & $2$&
 $\SUSY{Z^{0;5}}\,\,({\bf \overline{56}})$
\\\cline{2-3}
 & $3$&
 $\SUSY{Z^{0;7,1}}\,\,({\bf 63})$, 
 $Z^{0;8}\,\,({\bf 1})$
\\\cline{2-3}
 & $4$&
 $\SUSY{Z^{0;8,3}}\,\,({\bf 56})$
\\\cline{2-3}
 & $5$&
 $\SUSY{Z^{0;8,6}}\,\,({\bf \overline{28}})$
\\\cline{2-3}
 & $6$&
 $\SUSY{Z^{0;8,8,1}}\,\,({\bf 8})$
\\\hline
\end{tabular}
}
\qquad
{\scriptsize
\begin{tabular}{|c|c|l|} \hline
 &$\ell$ & \multicolumn{1}{c|}{central charges} \\ \hline\hline
%%%%%%%%%%%%%%%%%%%%%%%%%%
 $\textcolor{red}{{\bf 3875}}$ 
 & $1$&
 $\SUSY{Z^{1;1}}\,\,({\bf 8})$
\\\cline{2-3}
 & $2$&
 $\SUSY{Z^{1;4}}\,\,({\bf 70})$
\\\cline{2-3}
 & $3$&
 $\SUSY{Z^{1;6,1}}\,\,({\bf \overline{216}})$, 
 $Z^{1;7}\,\,({\bf \overline{8}})$
\\\cline{2-3}
 & $4$&
 $\SUSY{Z^{1;7,3}}\,\,({\bf 420})$, 
 $\SUSY{Z^{1;8,1,1}}\,\,({\bf 36})$, 
 $Z^{1;8,2}\,\,({\bf 28})$
\\\cline{2-3}
 & $5$&
 $\SUSY{Z^{1;7,6}}\,\,({\bf \overline{168}})$, 
 $\SUSY{Z^{1;8,4,1}}\,\,({\bf \overline{504}})$, 
 $Z^{1;8,5}\,\,({\bf \overline{56}})$
\\\cline{2-3}
 & $6$&
 $\SUSY{Z^{1;8,6,2}}\,\,({\bf 720})$, 
 $2\,Z^{1;8,7,1}\,\,(2\times {\bf 63})$, 
 $Z^{1;8,8}\,\,({\bf 1})$
\\\cline{2-3}
 & $7$&
 $\SUSY{Z^{1;8,7,4}}\,\,({\bf 504})$, 
 $\SUSY{Z^{1;8,8,2,1}}\,\,({\bf 168})$, 
 $Z^{1;8,8,3}\,\,({\bf 56})$
\\\cline{2-3}
 & $8$&
 $\SUSY{Z^{1;8,7,7}}\,\,({\bf \overline{36}})$, 
 $\SUSY{Z^{1;8,8,5,1}}\,\,({\bf \overline{420}})$, 
 $Z^{1;8,8,6}\,\,({\bf \overline{28}})$
\\\cline{2-3}
 &$ 9$&
 $\SUSY{Z^{1;8,8,7,2}}\,\,({\bf 216})$, 
 $Z^{1;8,8,8,1}\,\,({\bf 8})$
\\\cline{2-3}
 &$10$&
 $\SUSY{Z^{1;8,8,8,4}}\,\,({\bf 70})$
\\\cline{2-3}
 &$11$&
 $\SUSY{Z^{1;8,8,8,7}}\,\,({\bf \overline{8}})$
\\\hline
%%%%%%%%%
 ${\bf 248}$ 
 & $3$&
 $Z^{1;7}\,\,({\bf \overline{8}})$
\\\cline{2-3}
 & $4$&
 $Z^{1;8,2}\,\,({\bf 28})$
\\\cline{2-3}
 & $5$&
 $Z^{1;8,5}\,\,({\bf \overline{56}})$
\\\cline{2-3}
 & $6$&
 $Z^{1;8,7,1}\,\,({\bf 63})$, 
 $Z^{1;8,8}\,\,({\bf 1})$
\\\cline{2-3}
 & $7$&
 $Z^{1;8,8,3}\,\,({\bf 56})$
\\\cline{2-3}
 & $8$&
 $Z^{1;8,8,6}\,\,({\bf \overline{28}})$
\\\cline{2-3}
 & $9$&
 $Z^{1;8,8,8,1}\,\,({\bf 8})$
\\\hline
%%%%%%%%%
 ${\bf 1}$
 & $6$&
 $Z^{1;8,8}\,\,({\bf 1})$
\\\hline
\end{tabular}
}
\end{align*}
Decomposition of the membrane multiplet:
{\scriptsize
\begin{longtable}[c]{|c|c|l|} \hline
 &$\ell$ & \multicolumn{1}{c|}{central charges} \\ \hline\hline
%%%%%%%%%%%%%%%%%%%%%%%%%%
 $\textcolor{red}{{\bf 147250}}$
 &$1$ & 
 $\SUSY{Z^{2}}\,\,({\bf 1})$
\\\cline{2-3}
 &$2$ & 
 $\SUSY{Z^{2;3}}\,\,({\bf 56})$
\\\cline{2-3}
 &$3$ & 
 $\SUSY{Z^{2;5,1}}\,\,({\bf \overline{420}})$, 
 $Z^{2;6}\,\,({\bf \overline{28}})$
\\\cline{2-3}
 &$4$ & 
 $\SUSY{Z^{2;6,3}}\,\,({\bf 1344})$, 
 $\SUSY{Z^{2;7,1,1}}\,\,({\bf 280})$, 
 $Z^{2;7,2}\,\,({\bf 216})$, 
 $2\,Z^{2;8,1}\,\,(2\times {\bf 8})$
\\\cline{2-3}
 &$5$ & 
 \Pbox{0.9\textwidth}{$\SUSY{Z^{2;6,6}}\,\,({\bf \overline{336}})$, 
 $\SUSY{Z^{2;7,4,1}}\,\,({\bf 3584})$, 
 $Z^{2;7,5}\,\,({\bf \overline{378}})$, 
 $2\,Z^{2;8,3,1}\,\,(2\times {\bf 378})$, 
 $2\,Z^{2;8,4}\,\,(2\times {\bf 70})$}
\\\cline{2-3}
 &$6$ & 
 \Pbox{0.9\textwidth}{$\SUSY{Z^{2;7,6,2}}\,\,({\bf \overline{4200}})$, 
 $Z^{2;7,7,1}\,\,({\bf \overline{280}})$, 
 $\SUSY{Z^{2;8,4,3}}\,\,({\bf \overline{2352}'})$, 
 $\SUSY{Z^{2;8,5,1,1}}\,\,({\bf \overline{1800}})$, 
 $Z^{2;8,5,2}\,\,({\bf \overline{1344}})$, 
 $4\,Z^{2;8,6,1}\,\,(4\times {\bf \overline{216}})$, 
 $2\,Z^{2;8,7}\,\,(2\times {\bf \overline{8}})$}
\\\cline{2-3}
 &$7$ & 
 \Pbox{0.9\textwidth}{$\SUSY{Z^{2;7,7,4}}\,\,({\bf 2100})$, 
 $\SUSY{Z^{2;8,6,3,1}}\,\,({\bf 8820})$, 
 $2\,Z^{2;8,6,4}\,\,(2\times {\bf 1512})$, 
 $2\,Z^{2;8,7,2,1}\,\,(2\times {\bf 1280})$, 
 $3\,Z^{2;8,7,3}\,\,(3\times {\bf 420})$, 
 $3\,Z^{2;8,8,1,1}\,\,(3\times {\bf 36})$, 
 $3\,Z^{2;8,8,2}\,\,(3\times {\bf 28})$}
\\
\hline
\end{longtable}
}

\newpage

\section{Type IIB branes: $E_n\to \SL(D)\times \SL(2)$}
\label{app:B-branes}

In this appendix, we decompose the $E_n$ $U$-duality multiplets of the $p$-branes given in Table \ref{tab:p-brane} in terms of brane charges in type IIB theory. 
The 10D origin of an underlined (supersymmetric) brane charge $\SUSY{Z^{p;q_1,q_2,\cdots,q_s}_{\alpha_1\cdots\alpha_r}}$ corresponds to supersymmetric branes called
\begin{align}
 (p+q_1-q_2)_{\frac{\ell_9\pm s}{2}}^{(q_s,q_{s-1}-q_s,q_{s-2}-q_{s-1},\dotsc,q_2-q_3)}\text{-brane,} 
\end{align}
where $\ell_9\equiv \frac{1+p+q_1+\cdots+q_s}{2}$\,. 

\subsection*{Type IIB branes in $d=8$: $E_3\to \SL(2)\times \SL(2)$}

Decomposition of the particle, string, and membrane multiplets:
\begin{align*}
{\scriptsize
\begin{tabular}{|c|c|l|} \hline
 &$\ell_9$ & \multicolumn{1}{c|}{central charges} \\ \hline\hline
%%%%%%%%%%%%%%%%%%%%%%%%%%
 $\textcolor{red}{({\bf \overline{3}},{\bf 2})}$
 & $0$&
 $\SUSY{P_{1}}\,\,({\bf 2},{\bf 1})$ 
\\\cline{2-3}
 & $1$&
 $\SUSY{Z^{0;1}_{\alpha}}\,\,({\bf 2},{\bf 2})$
\\\hline
\end{tabular}
}
\qquad
{\scriptsize
\begin{tabular}{|c|c|l|} \hline
 &$\ell_9$ & \multicolumn{1}{c|}{central charges} \\ \hline\hline
%%%%%%%%%%%%%%%%%%%%%%%%%%
 $\textcolor{red}{({\bf 3},{\bf 1})}$
 & $1$&
 $\SUSY{Z^{1}_{\alpha}}\,\,({\bf 1},{\bf 2})$ %[2]
\\\cline{2-3}
 & $2$&
 $\SUSY{Z^{1;2}}\,\,({\bf 1},{\bf 1})$ %[2]
\\\hline
\end{tabular}
}
\qquad
{\scriptsize
\begin{tabular}{|c|c|l|} \hline
 &$\ell_9$ & \multicolumn{1}{c|}{central charges} \\ \hline\hline
%%%%%%%%%%%%%%%%%%%%%%%%%%
 $\textcolor{red}{({\bf 1},{\bf 2})}$
 & $2$&
 $\SUSY{Z^{2;1}}\,\,({\bf 2},{\bf 1})$ %[2]
\\\hline
\end{tabular}
}
\end{align*}
Decomposition of the 3- and 4-brane multiplets:
\begin{align*}
{\scriptsize
\begin{tabular}{|c|c|l|} \hline
 &$\ell_9$ & \multicolumn{1}{c|}{central charges} \\ \hline\hline
%%%%%%%%%%%%%%%%%%%%%%%%%%
 $\textcolor{red}{({\bf \overline{3}},{\bf 1})}$
 & $2$&
 $\SUSY{Z^{3}}\,\,({\bf 1},{\bf 1})$ 
\\\cline{2-3}
 & $3$&
 $\SUSY{Z^{3;2}_{\alpha}}\,\,({\bf 1},{\bf 2})$ 
\\\hline
\end{tabular}
}
\qquad
{\scriptsize
\begin{tabular}{|c|c|l|} \hline
 &$\ell_9$ & \multicolumn{1}{c|}{central charges} \\ \hline\hline
%%%%%%%%%%%%%%%%%%%%%%%%%%
 $\textcolor{red}{({\bf 3},{\bf 2})}$
 & $3$&
 $\SUSY{Z^{4;1}_{\alpha}}\,\,({\bf 2},{\bf 2})$ 
\\\cline{2-3}
 & $4$&
 $\SUSY{Z^{4;2,1}}\,\,({\bf 2},{\bf 1})$ 
\\\hline
\end{tabular}
}
\end{align*}
Decomposition of the 5-, 6-, and 7-brane multiplets:
\begin{align*}
{\scriptsize
\begin{tabular}{|c|c|l|} \hline
 &$\ell_9$ & \multicolumn{1}{c|}{central charges} \\ \hline\hline
%%%%%%%%%%%%%%%%%%%%%%%%%%
 $\textcolor{red}{({\bf 8},{\bf 1})}$
 & $3$&
 $\SUSY{Z^{5}_{\alpha}}\,\,({\bf 1},{\bf 2})$ 
\\\cline{2-3}
 & $4$&
 $Z^{5;2}\,\,({\bf 1},{\bf 1})$, 
 $\SUSY{Z^{5;2}_{\alpha\beta}}\,\,({\bf 1},{\bf 3})$ 
\\\cline{2-3}
 & $5$&
 $\SUSY{Z^{5;2,2}_{\alpha}}\,\,({\bf 1},{\bf 2})$ %[-2]
\\\hline
%%%%%%%%%%%%%%%%%%%%%%%%%%
 $\textcolor{red}{({\bf 1},{\bf 3})}$ 
 & $4$&
 $\SUSY{Z^{5;1,1}}\,\,({\bf 3},{\bf 1})$
\\\hline
%%%%%%%%%%%%%%%%%%%%%%%%%%
 $({\bf 1},{\bf 1})$ 
 & $4$&
 $Z^{5;2}\,\,({\bf 1},{\bf 1})$
\\\hline
\end{tabular}
}
\quad
{\scriptsize
\begin{tabular}{|c|c|l|} \hline
 &$\ell_9$ & \multicolumn{1}{c|}{central charges} \\ \hline\hline
%%%%%%%%%%%%%%%%%%%%%%%%%%
 $\textcolor{red}{({\bf 6},{\bf 2})}$
 & $4$&
 $\SUSY{Z^{6;1}_{\alpha\beta}}\,\,({\bf 2},{\bf 3})$ 
\\\cline{2-3}
 & $5$&
 $Z^{6;2,1}_{\alpha}\,\,({\bf 2},{\bf 2})$ %[-2]
\\\cline{2-3}
 & $6$&
 $\SUSY{Z^{6;2,2,1}}\,\,({\bf 2},{\bf 1})$ %[0]
\\\hline
%%%%%%%%%%%%%%%%%%%%%%%%%%
 $({\bf \overline{3}},{\bf 2})$ 
 & $4$&
 $Z^{6;1}\,\,({\bf 2},{\bf 1})$
\\\cline{2-3}
 & $5$&
 $Z^{6;2,1}_{\alpha}\,\,({\bf 2},{\bf 2})$ %[-2]
\\\hline
\end{tabular}
}
\quad
{\scriptsize
\begin{tabular}{|c|c|l|} \hline
 &$\ell_9$ & \multicolumn{1}{c|}{central charges} \\ \hline\hline
%%%%%%%%%%%%%%%%%%%%%%%%%%
 $\textcolor{red}{({\bf 15},{\bf 1})}$
 & $4$&
 $\SUSY{Z^{7}_{\alpha\beta}}\,\,({\bf 1},{\bf 3})$ 
\\\cline{2-3}
 & $5$&
 $Z^{7;2}_{\alpha}\,\,({\bf 1},{\bf 2})$, % [-2]
 $\SUSY{Z^{7;2}_{\alpha\beta\gamma}}\,\,({\bf 1},{\bf 4})$ %[2]
\\\cline{2-3}
 & $6$&
 $Z^{7;2,2}\,\,({\bf 1},{\bf 1})$, % [-2]
 $Z^{7;2,2}_{\alpha\beta}\,\,({\bf 1},{\bf 3})$
\\\cline{2-3}
 & $7$&
 $\SUSY{Z^{7;2,2,2}_{\alpha}}\,\,({\bf 1},{\bf 2})$ %[2]
\\\hline
%%%%%%%%%%%%%%%%%%%%%%%%%%
 $({\bf \overline{6}},{\bf 1})$
 & $4$&
 $Z^{7}\,\,({\bf 1},{\bf 1})$ 
\\\cline{2-3}
 & $5$&
 $Z^{7;2}_{\alpha}\,\,({\bf 1},{\bf 2})$
\\\cline{2-3}
 & $6$&
 $Z^{7;2,2}_{\alpha\beta}\,\,({\bf 1},{\bf 3})$
\\\hline
%%%%%%%%%%%%%%%%%%%%%%%%%%
 $({\bf 3},{\bf 3})$
 & $5$&
 $Z^{7;1,1}_{\alpha}\,\,({\bf 3},{\bf 2})$ %[-2]
\\\cline{2-3}
 & $6$&
 $Z^{7;2,1,1}\,\,({\bf 3},{\bf 1})$ %[0]
\\\hline
%%%%%%%%%%%%%%%%%%%%%%%%%%
 $({\bf 3},{\bf 1})$ 
 & $5$&
 $Z^{7;2}_{\alpha}\,\,({\bf 1},{\bf 2})$ %[-2]
\\\cline{2-3}
 & $6$&
 $Z^{7;2,2}\,\,({\bf 1},{\bf 1})$ %[-2]
\\\hline
\end{tabular}
}
\end{align*}

\newpage

\subsection*{Type IIB branes in $d=7$: $E_4\to \SL(3)\times \SL(2)$}

Decomposition of the particle, string, and membrane multiplets:
\begin{align*}
{\scriptsize
\begin{tabular}{|c|c|l|} \hline
 &$\ell_9$ & \multicolumn{1}{c|}{central charges} \\ \hline\hline
%%%%%%%%%%%%%%%%%%%%%%%%%%
 $\textcolor{red}{{\bf \overline{10}}}$ 
 & $0$&
 $\SUSY{P_{1}}\,\,({\bf \overline{3}},{\bf 1})$ 
\\\cline{2-3}
 & $1$&
 $\SUSY{Z^{0;1}_{\alpha}}\,\,({\bf 3},{\bf 2})$
\\\cline{2-3}
 & $2$&
 $\SUSY{Z^{0;3}}\,\,({\bf 1},{\bf 1})$
\\\hline
\end{tabular}
}
\qquad
{\scriptsize
\begin{tabular}{|c|c|l|} \hline
 &$\ell_9$ & \multicolumn{1}{c|}{central charges} \\ \hline\hline
%%%%%%%%%%%%%%%%%%%%%%%%%%
 $\textcolor{red}{{\bf 5}}$ 
 & $1$&
 $\SUSY{Z^{1}_{\alpha}}\,\,({\bf 1},{\bf 2})$ %[2]
\\\cline{2-3}
 & $2$&
 $\SUSY{Z^{1;2}}\,\,({\bf \overline{3}},{\bf 1})$ %[2]
\\\hline
\end{tabular}
}
\qquad
{\scriptsize
\begin{tabular}{|c|c|l|} \hline
 &$\ell_9$ & \multicolumn{1}{c|}{central charges} \\ \hline\hline
%%%%%%%%%%%%%%%%%%%%%%%%%%
 $\textcolor{red}{{\bf \overline{5}}}$ 
 & $2$&
 $\SUSY{Z^{2;1}}\,\,({\bf 3},{\bf 1})$ %[2]
\\\cline{2-3}
 & $3$&
 $\SUSY{Z^{2;3}_{\alpha}}\,\,({\bf 1},{\bf 2})$ %[2]
\\\hline
\end{tabular}
}
\end{align*}
Decomposition of the 3- and 4-brane multiplets:
\begin{align*}
{\scriptsize
\begin{tabular}{|c|c|l|} \hline
 &$\ell_9$ & \multicolumn{1}{c|}{central charges} \\ \hline\hline
%%%%%%%%%%%%%%%%%%%%%%%%%%
 $\textcolor{red}{{\bf 10}}$ 
 & $2$&
 $\SUSY{Z^{3}}\,\,({\bf 1},{\bf 1})$ 
\\\cline{2-3}
 & $3$&
 $\SUSY{Z^{3;2}_{\alpha}}\,\,({\bf \overline{3}},{\bf 2})$ 
\\\cline{2-3}
 & $4$&
 $\SUSY{Z^{3;3,1}}\,\,({\bf 3},{\bf 1})$ 
\\\hline
\end{tabular}
}
\qquad
{\scriptsize
\begin{tabular}{|c|c|l|} \hline
 &$\ell_9$ & \multicolumn{1}{c|}{central charges} \\ \hline\hline
%%%%%%%%%%%%%%%%%%%%%%%%%%
 $\textcolor{red}{{\bf 24}}$ 
 & $3$&
 $\SUSY{Z^{4;1}_{\alpha}}\,\,({\bf 3},{\bf 2})$ 
\\\cline{2-3}
 & $4$&
 $\SUSY{Z^{4;2,1}}\,\,({\bf 8},{\bf 1})$, 
 $Z^{4;3}\,\,({\bf 1},{\bf 1})$, 
 $\SUSY{Z^{4;3}_{\alpha\beta}}\,\,({\bf 1},{\bf 3})$ 
\\\cline{2-3}
 & $5$&
 $\SUSY{Z^{4;3,2}_{\alpha}}\,\,({\bf \overline{3}},{\bf 2})$ %[-2]
\\\hline
%%%%%%%%%%%%%%%%%%%%%%%%%%
 ${\bf 1}$ 
 & $4$&
 $Z^{4;3}\,\,({\bf 1},{\bf 1})$ 
\\\hline
\end{tabular}
}
\end{align*}
Decomposition of the 5- and 6-brane multiplets:
\begin{align*}
{\scriptsize
\begin{tabular}{|c|c|l|} \hline
 &$\ell_9$ & \multicolumn{1}{c|}{central charges} \\ \hline\hline
%%%%%%%%%%%%%%%%%%%%%%%%%%
 $\textcolor{red}{{\bf \overline{40}}}$ 
 & $3$&
 $\SUSY{Z^{5}_{\alpha}}\,\,({\bf 1},{\bf 2})$ 
\\\cline{2-3}
 & $4$&
 $Z^{5;2}\,\,({\bf \overline{3}},{\bf 1})$, 
 $\SUSY{Z^{5;2}_{\alpha\beta}}\,\,({\bf \overline{3}},{\bf 3})$ 
\\\cline{2-3}
 & $5$&
 $\SUSY{Z^{5;2,2}_{\alpha}}\,\,({\bf \overline{6}},{\bf 2})$, %[-2]
 $Z^{5;3,1}_{\alpha}\,\,({\bf 3},{\bf 2})$ %[-2]
\\\cline{2-3}
 & $6$&
 $\SUSY{Z^{5;3,2,1}}\,\,({\bf 8},{\bf 1})$ %[0]
\\\hline
%%%%%%%%%%%%%%%%%%%%%%%%%%
 $\textcolor{red}{{\bf \overline{15}}}$ 
 & $4$&
 $\SUSY{Z^{5;1,1}}\,\,({\bf 6},{\bf 1})$ 
\\\cline{2-3}
 & $5$&
 $Z^{5;3,1}_{\alpha}\,\,({\bf 3},{\bf 2})$ %[-2]
\\\cline{2-3}
 & $6$&
 $\SUSY{Z^{5;3,3}_{\alpha\beta}}\,\,({\bf 1},{\bf 3})$ %[0] 
\\\hline
%%%%%%%%%%%%%%%%%%%%%%%%%%
 ${\bf \overline{10}}$ 
 & $4$&
 $Z^{5;2}\,\,({\bf \overline{3}},{\bf 1})$
\\\cline{2-3}
 & $5$&
 $Z^{5;3,1}_{\alpha}\,\,({\bf 3},{\bf 2})$ %[-2]
\\\cline{2-3}
 & $6$&
 $Z^{5;3,3}\,\,({\bf 1},{\bf 1})$ %[0] 
\\\hline
\end{tabular}
}
\qquad
{\scriptsize
\begin{tabular}{|c|c|l|} \hline
 &$\ell_9$ & \multicolumn{1}{c|}{central charges} \\ \hline\hline
%%%%%%%%%%%%%%%%%%%%%%%%%%
 $\textcolor{red}{{\bf 70}}$ 
 & $4$&
 $\SUSY{Z^{6;1}_{\alpha\beta}}\,\,({\bf 3},{\bf 3})$ 
\\\cline{2-3}
 & $5$&
 $Z^{6;2,1}_{\alpha}\,\,({\bf 8},{\bf 2})$, %[-2]
 $Z^{6;3}_{\alpha}\,\,({\bf 1},{\bf 2})$, % [-2]
 $\SUSY{Z^{6;3}_{\alpha\beta\gamma}}\,\,({\bf 1},{\bf 4})$ %[2]
\\\cline{2-3}
 & $6$&
 $\SUSY{Z^{6;2,2,1}}\,\,({\bf \overline{15}},{\bf 1})$, %[0]
 $Z^{6;3,2}_{\alpha\beta}\,\,({\bf \overline{3}},{\bf 3})$, %[0] 
 $Z^{6;3,2}\,\,({\bf \overline{3}},{\bf 1})$ % [-2]
\\\cline{2-3}
 & $7$&
 $\SUSY{Z^{6;3,2,2}_{\alpha}}\,\,({\bf \overline{6}},{\bf 2})$ %[2]
\\\hline
%%%%%%%%%%%%%%%%%%%%%%%%%%
 ${\bf 45}$ 
 & $4$&
 $Z^{6;1}\,\,({\bf 3},{\bf 1})$ 
\\\cline{2-3}
 & $5$&
 $Z^{6;2,1}_{\alpha}\,\,({\bf 8},{\bf 2})$, %[-2]
 $Z^{6;3}_{\alpha}\,\,({\bf 1},{\bf 2})$ %[-2]
\\\cline{2-3}
 & $6$&
 $Z^{6;3,1,1}\,\,({\bf 6},{\bf 1})$, %[0]
 $Z^{6;3,2}\,\,({\bf \overline{3}},{\bf 1})$, % [-2]
 $Z^{6;3,2}_{\alpha\beta}\,\,({\bf \overline{3}},{\bf 3})$ 
\\\cline{2-3}
 & $7$&
 $Z^{6;3,3,1}_{\alpha}\,\,({\bf 3},{\bf 2})$ %[0]
\\\hline
%%%%%%%%%%%%%%%%%%%%%%%%%%
 ${\bf 5}$ 
 & $5$&
 $Z^{6;3}_{\alpha}\,\,({\bf 1},{\bf 2})$ %[-2]
\\\cline{2-3}
 & $6$&
 $Z^{6;3,2}\,\,({\bf \overline{3}},{\bf 1})$ %[-2]
\\\hline
\end{tabular}
}
\end{align*}

\newpage

\subsection*{Type IIB branes in $d=6$: $E_5\to \SL(4)\times \SL(2)$}

Decomposition of the particle, string, and membrane multiplets:
\begin{align*}
{\scriptsize
\begin{tabular}{|c|c|l|} \hline
 &$\ell_9$ & \multicolumn{1}{c|}{central charges} \\ \hline\hline
%%%%%%%%%%%%%%%%%%%%%%%%%%
 $\textcolor{red}{{\bf 16}}$ 
 & $0$&
 $\SUSY{P_{1}}\,\,({\bf 4},{\bf 1})$ 
\\\cline{2-3}
 & $1$&
 $\SUSY{Z^{0;1}_{\alpha}}\,\,({\bf 4},{\bf 2})$
\\\cline{2-3}
 & $2$&
 $\SUSY{Z^{0;3}}\,\,({\bf \overline{4}},{\bf 1})$
\\\hline
\end{tabular}
}
\qquad
{\scriptsize
\begin{tabular}{|c|c|l|} \hline
 &$\ell_9$ & \multicolumn{1}{c|}{central charges} \\ \hline\hline
%%%%%%%%%%%%%%%%%%%%%%%%%%
 $\textcolor{red}{{\bf 10}}$ 
 & $1$&
 $\SUSY{Z^{2}_{\alpha}}\,\,({\bf 1},{\bf 2})$ %[2]
\\\cline{2-3}
 & $2$&
 $\SUSY{Z^{1;2}}\,\,({\bf 6},{\bf 1})$ %[2]
\\\cline{2-3}
 & $3$&
 $\SUSY{Z^{1;4}_{\alpha}}\,\,({\bf 1},{\bf 2})$ %[2]
\\\hline
\end{tabular}
}
\qquad
{\scriptsize
\begin{tabular}{|c|c|l|} \hline
 &$\ell_9$ & \multicolumn{1}{c|}{central charges} \\ \hline\hline
%%%%%%%%%%%%%%%%%%%%%%%%%%
 $\textcolor{red}{{\bf \overline{16}}}$ 
 & $2$&
 $\SUSY{Z^{2;1}}\,\,({\bf 4},{\bf 1})$ %[2]
\\\cline{2-3}
 & $3$&
 $\SUSY{Z^{2;3}_{\alpha}}\,\,({\bf \overline{4}},{\bf 2})$ %[2]
\\\cline{2-3}
 & $4$&
 $\SUSY{Z^{2;4,1}}\,\,({\bf 4},{\bf 1})$ %[2]
\\\hline
\end{tabular}
}
\end{align*}
Decomposition of the 3- and 4-brane multiplets:
\begin{align*}
{\scriptsize
\begin{tabular}{|c|c|l|} \hline
 &$\ell_9$ & \multicolumn{1}{c|}{central charges} \\ \hline\hline
%%%%%%%%%%%%%%%%%%%%%%%%%%
 $\textcolor{red}{{\bf 45}}$ 
 & $2$&
 $\SUSY{Z^{3}}\,\,({\bf 1},{\bf 1})$ 
\\\cline{2-3}
 & $3$&
 $\SUSY{Z^{3;2}_{\alpha}}\,\,({\bf 6},{\bf 2})$ 
\\\cline{2-3}
 & $4$&
 $\SUSY{Z^{3;3,1}}\,\,({\bf 15},{\bf 1})$, 
 $Z^{3;4}\,\,({\bf 1},{\bf 1})$, 
 $\SUSY{Z^{3;4}_{\alpha\beta}}\,\,({\bf 1},{\bf 3})$ 
\\\cline{2-3}
 & $5$&
 $\SUSY{Z^{3;4,2}_{\alpha}}\,\,({\bf 6},{\bf 2})$
\\\cline{2-3}
 & $6$&
 $\SUSY{Z^{3;4,4}}\,\,({\bf 1},{\bf 1})$ %[-2]
\\\hline
%%%%%%%%%%%%%%%%%%%%%%%%%%
 ${\bf 1}$
 & $4$&
 $Z^{3;4}\,\,({\bf 1},{\bf 1})$
\\\hline
\end{tabular}
}
\qquad
{\scriptsize
\begin{tabular}{|c|c|l|} \hline
 &$\ell_9$ & \multicolumn{1}{c|}{central charges} \\ \hline\hline
%%%%%%%%%%%%%%%%%%%%%%%%%%
 $\textcolor{red}{{\bf 144}}$ 
 & $3$&
 $\SUSY{Z^{4;1}_{\alpha}}\,\,({\bf 4},{\bf 2})$ 
\\\cline{2-3}
 & $4$&
 $\SUSY{Z^{4;2,1}}\,\,({\bf \overline{20}},{\bf 1})$, 
 $Z^{4;3}\,\,({\bf \overline{4}},{\bf 1})$, 
 $\SUSY{Z^{4;3}_{\alpha\beta}}\,\,({\bf \overline{4}},{\bf 3})$ 
\\\cline{2-3}
 & $5$&
 $\SUSY{Z^{4;3,2}_{\alpha}}\,\,({\bf 20},{\bf 2})$, %[-2]
 $Z^{4;4,1}_{\alpha}\,\,({\bf 4},{\bf 2})$, %[-2]
 $Z^{4;4,1}_{\alpha}\,\,({\bf 4},{\bf 2})$ %[-2]
\\\cline{2-3}
 & $6$&
 $Z^{4;4,3}\,\,({\bf \overline{4}},{\bf 1})$, %[-2]
 $\SUSY{Z^{4;4,2,1}}\,\,({\bf \overline{20}},{\bf 1})$, %[0]
 $\SUSY{Z^{4;4,3}_{\alpha\beta}}\,\,({\bf \overline{4}},{\bf 3})$ %[0] 
\\\cline{2-3}
 & $7$&
 $\SUSY{Z^{4;4,4,1}_{\alpha}}\,\,({\bf 4},{\bf 2})$ %[0]
\\\hline
 ${\bf 16}$ 
 & $4$&
 $Z^{4;3}\,\,({\bf \overline{4}},{\bf 1})$ 
\\\cline{2-3}
 & $5$&
 $Z^{4;4,1}_{\alpha}\,\,({\bf 4},{\bf 2})$
\\\cline{2-3}
 & $6$&
 $Z^{4;4,3}\,\,({\bf \overline{4}},{\bf 1})$
\\\hline
\end{tabular}
}
\end{align*}
Decomposition of the 5-brane multiplet:
\begin{align*}
{\scriptsize
\begin{tabular}{|c|c|l|} \hline
 &$\ell_9$ & \multicolumn{1}{c|}{central charges} \\ \hline\hline
%%%%%%%%%%%%%%%%%%%%%%%%%%
 $\textcolor{red}{{\bf 320}}$ 
 & $3$&
 $\SUSY{Z^{5}_{\alpha}}\,\,({\bf 1},{\bf 2})$ 
\\\cline{2-3}
 & $4$&
 $Z^{5;2}\,\,({\bf 6},{\bf 1})$, 
 $\SUSY{Z^{5;2}_{\alpha\beta}}\,\,({\bf 6},{\bf 3})$ 
\\\cline{2-3}
 & $5$&
 $\SUSY{Z^{5;2,2}_{\alpha}}\,\,({\bf 20},{\bf 2})$, %[-2]
 $Z^{5;3,1}_{\alpha}\,\,({\bf 15},{\bf 2})$, %[-2]
 $2\,Z^{5;4}_{\alpha}\,\,(2\times ({\bf 1},{\bf 2}))$, %[-2]
 $\SUSY{Z^{5;4}_{\alpha\beta\gamma}}\,\,({\bf 1},{\bf 4})$ %[2]
\\\cline{2-3}
 & $6$&
 $\SUSY{Z^{5;3,2,1}}\,\,({\bf 64},{\bf 1})$, %[0]
 $2\,Z^{5;4,2}\,\,(2\times({\bf 6},{\bf 1}))$, % [-2]
 $2\,Z^{5;4,2}_{\alpha\beta}\,\,(2\times({\bf 6},{\bf 3}))$ %[0] 
\\\cline{2-3}
 & $7$&
 $\SUSY{Z^{5;4,2,2}_{\alpha}}\,\,({\bf {20'}},{\bf 2})$, %[2]
 $Z^{5;4,3,1}_{\alpha}\,\,({\bf 15},{\bf 2})$, %[0]
 $2\,Z^{5;4,4}_{\alpha}\,\,(2\times ({\bf 1},{\bf 2}))$, % [-2]
 $\SUSY{Z^{5;4,4}_{\alpha\beta\gamma}}\,\,({\bf 1},{\bf 4})$ %[2]
\\\cline{2-3}
 & $8$&
 $Z^{5;4,4,2}\,\,({\bf 6},{\bf 1})$, %[0]
 $\SUSY{Z^{5;4,4,2}_{\alpha\beta}}\,\,({\bf 6},{\bf 3})$ %[2]
\\\cline{2-3}
 & $9$&
 $\SUSY{Z^{5;4,4,4}_{\alpha}}\,\,({\bf 1},{\bf 2})$ %[2]
\\
\hline
\end{tabular}
}
\end{align*}

\newpage

\subsection*{Type IIB branes in $d=5$: $E_6\to \SL(5)\times \SL(2)$}

Decomposition of the particle, string, and membrane multiplets:
\begin{align*}
{\scriptsize
\begin{tabular}{|c|c|l|} \hline
 &$\ell_9$ & \multicolumn{1}{c|}{central charges} \\ \hline\hline
%%%%%%%%%%%%%%%%%%%%%%%%%%
 $\textcolor{red}{{\bf 27}}$ 
 & $0$&
 $\SUSY{P_{1}}\,\,({\bf \overline{5}},{\bf 1})$ 
\\\cline{2-3}
 & $1$&
 $\SUSY{Z^{0;1}_{\alpha}}\,\,({\bf 5},{\bf 2})$
\\\cline{2-3}
 & $2$&
 $\SUSY{Z^{0;3}}\,\,({\bf \overline{10}},{\bf 1})$
\\\cline{2-3}
 & $3$&
 $\SUSY{Z^{0;5}_{\alpha}}\,\,({\bf 1},{\bf 2})$
\\\hline
\end{tabular}
}
\quad
{\scriptsize
\begin{tabular}{|c|c|l|} \hline
 &$\ell_9$ & \multicolumn{1}{c|}{central charges} \\ \hline\hline
%%%%%%%%%%%%%%%%%%%%%%%%%%
 $\textcolor{red}{{\bf \overline{27}}}$ 
 & $1$&
 $\SUSY{Z^{1}_{\alpha}}\,\,({\bf 1},{\bf 2})$ %[2]
\\\cline{2-3}
 & $2$&
 $\SUSY{Z^{1;2}}\,\,({\bf 10},{\bf 1})$ %[2]
\\\cline{2-3}
 & $3$&
 $\SUSY{Z^{1;4}_{\alpha}}\,\,({\bf \overline{5}},{\bf 2})$ %[2]
\\\cline{2-3}
 & $4$&
 $\SUSY{Z^{1;5,1}}\,\,({\bf 5},{\bf 1})$ %[2]
\\\hline
\end{tabular}
}
\quad
{\scriptsize
\begin{tabular}{|c|c|l|} \hline
 &$\ell_9$ & \multicolumn{1}{c|}{central charges} \\ \hline\hline
%%%%%%%%%%%%%%%%%%%%%%%%%%
 $\textcolor{red}{{\bf 78}}$ 
 & $2$&
 $\SUSY{Z^{2;1}}\,\,({\bf 5},{\bf 1})$ %[2]
\\\cline{2-3}
 & $3$&
 $\SUSY{Z^{2;3}_{\alpha}}\,\,({\bf \overline{10}},{\bf 2})$ %[2]
\\\cline{2-3}
 & $4$&
 $\SUSY{Z^{2;4,1}}\,\,({\bf 24},{\bf 1})$, %[2]
 $Z^{2;5}\,\,({\bf 1},{\bf 1})$, %[0]
 $\SUSY{Z^{2;5}_{\alpha\beta}}\,\,({\bf 1},{\bf 3})$ %[2]
\\\cline{2-3}
 & $5$&
 $\SUSY{Z^{2;5,2}_{\alpha}}\,\,({\bf 10},{\bf 2})$ %[2]
\\\cline{2-3}
 & $6$&
 $\SUSY{Z^{2;5,4}}\,\,({\bf \overline{5}},{\bf 1})$ %[2]
\\\hline
 ${\bf 1}$ 
 & $4$&
 $Z^{2;5}\,\,({\bf 1},{\bf 1})$ %[0]
\\\hline
\end{tabular}
}
\end{align*}
Decomposition of the 3-brane multiplet:
\begin{align*}
{\scriptsize
\begin{tabular}{|c|c|l|} \hline
 &$\ell_9$ & \multicolumn{1}{c|}{central charges} \\ \hline\hline
%%%%%%%%%%%%%%%%%%%%%%%%%%
 $\textcolor{red}{{\bf 351}}$ 
 & $2$&
 $\SUSY{Z^{3}}\,\,({\bf 1},{\bf 1})$ 
\\\cline{2-3}
 & $3$&
 $\SUSY{Z^{3;2}_{\alpha}}\,\,({\bf 10},{\bf 2})$ 
\\\cline{2-3}
 & $4$&
 $\SUSY{Z^{3;3,1}}\,\,({\bf \overline{45}},{\bf 1})$, 
 $Z^{3;4}\,\,({\bf \overline{5}},{\bf 1})$, 
 $\SUSY{Z^{3;4}_{\alpha\beta}}\,\,({\bf \overline{5}},{\bf 3})$ 
\\\cline{2-3}
 & $5$&
 $\SUSY{Z^{3;4,2}_{\alpha}}\,\,({\bf 45},{\bf 2})$, 
 $2\,Z^{3;5,1}_{\alpha}\,\,(2\times({\bf 5},{\bf 2}))$
\\\cline{2-3}
 & $6$&
 $\SUSY{Z^{3;4,4}}\,\,({\bf \overline{15}},{\bf 1})$, %[-2]
 $\SUSY{Z^{3;5,2,1}}\,\,({\bf \overline{40}},{\bf 1})$, %[-2]
 $Z^{3;5,3}\,\,({\bf \overline{10}},{\bf 1})$, %[-2]
 $\SUSY{Z^{3;5,3}_{\alpha\beta}}\,\,({\bf \overline{10}},{\bf 3})$
\\\cline{2-3}
 & $7$&
 $\SUSY{Z^{3;5,4,1}_{\alpha}}\,\,({\bf 24},{\bf 2})$, %[-2]
 $Z^{3;5,5}_{\alpha}\,\,({\bf 1},{\bf 2})$ 
\\\cline{2-3}
 & $8$&
 $\SUSY{Z^{3;5,5,2}}\,\,({\bf 10},{\bf 1})$ %[0]
\\\hline
 ${\bf 27}$ 
 & $3$&
 $Z^{3;4}\,\,({\bf \overline{5}},{\bf 1})$ 
\\\cline{2-3}
 & $4$&
 $Z^{3;5,1}_{\alpha}\,\,({\bf 5},{\bf 2})$
\\\cline{2-3}
 & $5$&
 $Z^{3;5,3}\,\,({\bf \overline{10}},{\bf 1})$
\\\cline{2-3}
 & $6$&
 $Z^{3;5,5}_{\alpha}\,\,({\bf 1},{\bf 2})$
\\\hline
\end{tabular}
}
\end{align*}
Decomposition of the 4-brane multiplet:
\begin{align*}
{\scriptsize
\begin{tabular}{|c|c|l|} \hline
 &$\ell_9$ & \multicolumn{1}{c|}{central charges} \\ \hline\hline
%%%%%%%%%%%%%%%%%%%%%%%%%%
 $\textcolor{red}{{\bf \overline{1728}}}$ 
 & $3$&
 $\SUSY{Z^{4;1}_{\alpha}}\,\,({\bf 5},{\bf 2})$ 
\\\cline{2-3}
 & $4$&
 $\SUSY{Z^{4;2,1}}\,\,({\bf \overline{40}},{\bf 1})$, 
 $Z^{4;3}\,\,({\bf \overline{10}},{\bf 1})$, 
 $\SUSY{Z^{4;3}_{\alpha\beta}}\,\,({\bf \overline{10}},{\bf 3})$ 
\\\cline{2-3}
 & $5$&
 $\SUSY{Z^{4;3,2}_{\alpha}}\,\,({\bf 75},{\bf 2})$, %[-2]
 $2\,Z^{4;4,1}_{\alpha}\,\,(2\times({\bf 24},{\bf 2}))$, %[-2]
 $2\,Z^{4;5}_{\alpha}\,\,(2\times({\bf 1},{\bf 2}))$, %[-2]
 $\SUSY{Z^{4;5}_{\alpha\beta\gamma}}\,\,({\bf 1},{\bf 4})$ %[2]
\\\cline{2-3}
 & $6$&
 \Pbox{0.9\textwidth}{ $\SUSY{Z^{4;4,2,1}}\,\,({\bf 175},{\bf 1})$, %[0]
 $Z^{4;4,3}\,\,({\bf 40},{\bf 1})$, %[-2]
 $\SUSY{Z^{4;4,3}_{\alpha\beta}}\,\,({\bf 40},{\bf 3})$, %[0]
 $Z^{4;5,1,1}\,\,({\bf 15},{\bf 1})$, %[0]
 $3\,Z^{4;5,2}\,\,(3\times({\bf 10},{\bf 1}))$, % [-2]
 $2\,Z^{4;5,2}_{\alpha\beta}\,\,(2\times({\bf 10},{\bf 3}))$} %[0] 
\\\cline{2-3}
 & $7$&
 \Pbox{0.9\textwidth}{$\SUSY{Z^{4;4,4,1}_{\alpha}}\,\,({\bf \overline{70}},{\bf 2})$, %[0]
 $\SUSY{Z^{4;5,2,2}_{\alpha}}\,\,({\bf \overline{50}},{\bf 2})$, %[2]
 $2\,Z^{4;5,3,1}_{\alpha}\,\,(2\times({\bf \overline{45}},{\bf 2}))$, %[0]
 $3\,Z^{4;5,4}_{\alpha}\,\,(3\times ({\bf \overline{5}},{\bf 2}))$, % [-2]
 $\SUSY{Z^{4;5,4}_{\alpha\beta\gamma}}\,\,({\bf \overline{5}},{\bf 4})$} %[2]
\\\cline{2-3}
 & $8$&
 \Pbox{0.9\textwidth}{$\SUSY{Z^{4;5,4,1,1}}\,\,({\bf 70},{\bf 1})$, %[2]
 $2\,Z^{4;5,4,2}\,\,(2\times({\bf 45},{\bf 1}))$, %[0]
 $\SUSY{Z^{4;5,4,2}_{\alpha\beta}}\,\,({\bf 45},{\bf 3})$, %[2]
 $2\,Z^{4;5,5,1}\,\,(2\times ({\bf 5},{\bf 1}))$, %[-2]
 $Z^{4;5,5,1}_{\alpha\beta}\,\,({\bf 5},{\bf 3})$} %[0]
\\\cline{2-3}
 & $9$&
 $\SUSY{Z^{4;5,4,4}_{\alpha}}\,\,({\bf \overline{15}},{\bf 2})$, %[2]
 $\SUSY{Z^{4;5,5,2,1}_{\alpha}}\,\,({\bf \overline{40}},{\bf 2})$, %[2]
 $Z^{4;5,5,3}_{\alpha}\,\,({\bf \overline{10}},{\bf 2})$ %[0]
\\
\hline
\end{tabular}
}
\end{align*}

\newpage

\subsection*{Type IIB branes in $d=4$: $E_7\to \SL(6)\times \SL(2)$}

Decomposition of the particle and string multiplets:
\begin{align*}
{\scriptsize
\begin{tabular}{|c|c|l|} \hline
 &$\ell_9$ & \multicolumn{1}{c|}{central charges} \\ \hline\hline
%%%%%%%%%%%%%%%%%%%%%%%%%%
 $\textcolor{red}{{\bf 56}}$ 
 & $0$&
 $\SUSY{P_{1}}\,\,({\bf \overline{6}},{\bf 1})$ 
\\\cline{2-3}
 & $1$&
 $\SUSY{Z^{0;1}_{\alpha}}\,\,({\bf 6},{\bf 2})$
\\\cline{2-3}
 & $2$&
 $\SUSY{Z^{0;3}}\,\,({\bf 20},{\bf 1})$
\\\cline{2-3}
 & $3$&
 $\SUSY{Z^{0;5}_{\alpha}}\,\,({\bf \overline{6}},{\bf 2})$
\\\cline{2-3}
 & $4$&
 $\SUSY{Z^{0;6,1}}\,\,({\bf 6},{\bf 1})$ 
\\\hline
\end{tabular}
}
\qquad
{\scriptsize
\begin{tabular}{|c|c|l|} \hline
 &$\ell_9$ & \multicolumn{1}{c|}{central charges} \\ \hline\hline
%%%%%%%%%%%%%%%%%%%%%%%%%%
 $\textcolor{red}{{\bf 133}}$ 
 & $1$&
 $\SUSY{Z^{1}_{\alpha}}\,\,({\bf 1},{\bf 2})$ %[2]
\\\cline{2-3}
 & $2$&
 $\SUSY{Z^{1;2}}\,\,({\bf 15},{\bf 1})$ %[2]
\\\cline{2-3}
 & $3$&
 $\SUSY{Z^{1;4}_{\alpha}}\,\,({\bf \overline{15}},{\bf 2})$ %[2]
\\\cline{2-3}
 & $4$&
 $\SUSY{Z^{1;5,1}}\,\,({\bf 35},{\bf 1})$, %[2]
 $Z^{1;6}\,\,({\bf 1},{\bf 1})$, %[0]
 $\SUSY{Z^{1;6}_{\alpha\beta}}\,\,({\bf 1},{\bf 3})$ %[2]
\\\cline{2-3}
 & $5$&
 $\SUSY{Z^{1;6,2}_{\alpha}}\,\,({\bf 15},{\bf 2})$ %[2]
\\\cline{2-3}
 & $6$&
 $\SUSY{Z^{1;6,4}}\,\,({\bf \overline{15}},{\bf 1})$ %[2]
\\\cline{2-3}
 & $7$&
 $\SUSY{Z^{1;6,6}_{\alpha}}\,\,({\bf 1},{\bf 2})$ %[2]
\\\hline
%%%%%%%%%%%%%%%%%%%%%%%%%%
 ${\bf 1}$ 
 & $4$&
 $Z^{1;6}\,\,({\bf 1},{\bf 1})$ %[0]
\\\hline
\end{tabular}
}
\end{align*}
Decomposition of the membrane multiplet:
\begin{align*}
{\scriptsize
\begin{tabular}{|c|c|l|} \hline
 &$\ell_9$ & \multicolumn{1}{c|}{central charges} \\ \hline\hline
%%%%%%%%%%%%%%%%%%%%%%%%%%
 &$\ell_9$ & \multicolumn{1}{c|}{Decomposition of the membrane multiplet in $d=4$} \\ \hline\hline
 $\textcolor{red}{{\bf 912}}$ 
 & $2$&
 $\SUSY{Z^{2;1}}\,\,({\bf 6},{\bf 1})$ %[2]
\\\cline{2-3}
 & $3$&
 $\SUSY{Z^{2;3}_{\alpha}}\,\,({\bf 20},{\bf 2})$ %[2]
\\\cline{2-3}
 & $4$&
 $\SUSY{Z^{2;4,1}}\,\,({\bf \overline{84}},{\bf 1})$, %[2]
 $Z^{2;5}\,\,({\bf \overline{6}},{\bf 1})$, %[0]
 $\SUSY{Z^{2;5}_{\alpha\beta}}\,\,({\bf \overline{6}},{\bf 3})$ %[2]
\\\cline{2-3}
 & $5$&
 $\SUSY{Z^{2;5,2}_{\alpha}}\,\,({\bf 84},{\bf 2})$, %[2]
 $Z^{2;6,1}_{\alpha}\,\,({\bf 6},{\bf 2})$, %[0]
 $Z^{2;6,1}_{\alpha}\,\,({\bf 6},{\bf 2})$ %[0]
\\\cline{2-3}
 & $6$&
 $\SUSY{Z^{2;5,4}}\,\,({\bf \overline{70}},{\bf 1})$, %[2]
 $Z^{2;6,3}\,\,({\bf 20},{\bf 1})$ %[0]
 $\SUSY{Z^{2;6,2,1}}\,\,({\bf 70},{\bf 1})$ %[2]
 $\SUSY{Z^{2;6,3}_{\alpha\beta}}\,\,({\bf 20},{\bf 3})$ %[2]
\\\cline{2-3}
 & $7$&
 $\SUSY{Z^{2;6,4,1}_{\alpha}}\,\,({\bf \overline{84}},{\bf 2})$, %[2]
 $Z^{2;6,5}_{\alpha}\,\,({\bf \overline{6}},{\bf 2})$ %[0]
 $Z^{2;6,5}_{\alpha}\,\,({\bf \overline{6}},{\bf 2})$ %[0]
\\\cline{2-3}
 & $8$&
 $\SUSY{Z^{2;6,5,2}}\,\,({\bf 84},{\bf 1})$ %[2]
 $Z^{2;6,6,1}\,\,({\bf 6},{\bf 1})$ %[0]
 $\SUSY{Z^{2;6,6,1}_{\alpha\beta}}\,\,({\bf 6},{\bf 3})$ %[2]
\\\cline{2-3}
 & $9$&
 $\SUSY{Z^{2;6,6,3}_{\alpha}}\,\,({\bf 20},{\bf 2})$ %[2]
\\\cline{2-3}
 &$10$&
 $\SUSY{Z^{2;6,6,5}}\,\,({\bf \overline{6}},{\bf 1})$ %[2]
\\\hline
%%%%%%%%%%%%%%%%%%%%%%%%%%
 ${\bf 56}$ 
 & $4$&
 $Z^{2;5}\,\,({\bf \overline{6}},{\bf 1})$ 
\\\cline{2-3}
 & $5$&
 $Z^{2;6,1}_{\alpha}\,\,({\bf 6},{\bf 2})$
\\\cline{2-3}
 & $6$&
 $Z^{2;6,3}\,\,({\bf 20},{\bf 1})$
\\\cline{2-3}
 & $7$&
 $Z^{2;6,5}_{\alpha}\,\,({\bf \overline{6}},{\bf 2})$
\\\cline{2-3}
 & $8$&
 $\SUSY{Z^{2;7,6,1}}\,\,({\bf 6},{\bf 1})$ 
\\\hline
\end{tabular}
}
\end{align*}
Decomposition of the 3-brane multiplet:
\begin{align*}
{\scriptsize
\begin{tabular}{|c|c|l|} \hline
 &$\ell_9$ & \multicolumn{1}{c|}{central charges} \\ \hline\hline
%%%%%%%%%%%%%%%%%%%%%%%%%%
 $\textcolor{red}{{\bf 8645}}$ 
 & $2$&
 $\SUSY{Z^{3}}\,\,({\bf 1},{\bf 1})$ 
\\\cline{2-3}
 & $3$&
 $\SUSY{Z^{3;2}_{\alpha}}\,\,({\bf 15},{\bf 2})$ 
\\\cline{2-3}
 & $4$&
 $\SUSY{Z^{3;3,1}}\,\,({\bf \overline{105}},{\bf 1})$, 
 $Z^{3;4}\,\,({\bf \overline{15}},{\bf 1})$, 
 $\SUSY{Z^{3;4}_{\alpha\beta}}\,\,({\bf \overline{15}},{\bf 3})$ 
\\\cline{2-3}
 & $5$&
 $\SUSY{Z^{3;4,2}_{\alpha}}\,\,({\bf 189},{\bf 2})$, 
 $2\,Z^{3;5,1}_{\alpha}\,\,(2\times ({\bf 35},{\bf 2}))$, 
 $2\,Z^{3;6}_{\alpha}\,\,(2\times ({\bf 1},{\bf 2}))$, %[-2]
 $\SUSY{Z^{3;6}_{\alpha\beta\gamma}}\,\,({\bf 1},{\bf 4})$ %[2]
\\\cline{2-3}
 & $6$&
 \Pbox{0.9\textwidth}{$\SUSY{Z^{3;4,4}}\,\,({\bf 105'},{\bf 1})$, %[-2]
 $\SUSY{Z^{3;5,2,1}}\,\,({\bf 384},{\bf 1})$, %[-2]
 $Z^{3;5,3}\,\,({\bf 105},{\bf 1})$, %[-2]
 $\SUSY{Z^{3;5,3}_{\alpha\beta}}\,\,({\bf 105},{\bf 3})$, 
 $Z^{3;6,1,1}\,\,({\bf 21},{\bf 1})$, %[0]
 $3\,Z^{3;6,2}\,\,(3\times({\bf 15},{\bf 1}))$, % [-2]
 $2\,Z^{3;6,2}_{\alpha\beta}\,\,(2\times({\bf 15},{\bf 3}))$} %[0]
\\\cline{2-3}
 & $7$&
 \Pbox{0.9\textwidth}{$\SUSY{Z^{3;5,4,1}_{\alpha}}\,\,({\bf \overline{384}},{\bf 2})$, %[-2]
 $Z^{3;5,5}_{\alpha}\,\,({\bf \overline{21}},{\bf 2})$, 
 $\SUSY{Z^{3;6,2,2}_{\alpha}}\,\,({\bf \overline{105'}},{\bf 2})$, %[2]
 $2\,Z^{3;6,3,1}_{\alpha}\,\,(2\times ({\bf \overline{105}},{\bf 2}))$, %[0]
 $4\,Z^{3;6,4}_{\alpha}\,\,(4\times ({\bf \overline{15}},{\bf 2}))$, % [-2]
 $\SUSY{Z^{3;6,4}_{\alpha\beta\gamma}}\,\,({\bf \overline{15}},{\bf 4})$} %[2]
\\
\hline
\end{tabular}
}
\end{align*}

\subsection*{Type IIB branes in $d=3$: $E_8\to \SL(7)\times \SL(2)$}

Decomposition of the particle multiplet:
\begin{align*}
{\scriptsize
\begin{tabular}{|c|c|l|} \hline
 &$\ell_9$ & \multicolumn{1}{c|}{central charges} \\ \hline\hline
%%%%%%%%%%%%%%%%%%%%%%%%%%
 $\textcolor{red}{{\bf 248}}$ 
 & $0$&
 $\SUSY{P_1}\,\,({\bf \overline{7}},{\bf 1})$ % [-4]
\\\cline{2-3}
 & $1$&
 $\SUSY{Z^{0;1}_{\alpha}}\,\,({\bf 7},{\bf 2})$ % [-4]
\\\cline{2-3}
 & $2$&
 $\SUSY{Z^{0;3}}\,\,({\bf 35},{\bf 1})$ % [-4]
\\\cline{2-3}
 & $3$&
 $\SUSY{Z^{0;5}_{\alpha}}\,\,({\bf \overline{21}},{\bf 2})$ % [-4]
\\\cline{2-3}
 & $4$&
 $\SUSY{Z^{0;6,1}}\,\,({\bf 48},{\bf 1})$, % [-4]
 $Z^{0;7}\,\,({\bf 1},{\bf 1})$, % [-6]
 $\SUSY{Z^{0;7}_{\alpha\beta}}\,\,({\bf 1},{\bf 3})$ % [-4]
\\\cline{2-3}
 & $5$&
 $\SUSY{Z^{0;7,2}_{\alpha}}\,\,({\bf 21},{\bf 2})$ % [-4]
\\\cline{2-3}
 & $6$&
 $\SUSY{Z^{0;7,4}}\,\,({\bf \overline{35}},{\bf 1})$ % [-4]
\\\cline{2-3}
 & $7$&
 $\SUSY{Z^{0;7,6}_{\alpha}}\,\,({\bf \overline{7}},{\bf 2})$ % [-4]
\\\cline{2-3}
 & $8$&
 $\SUSY{Z^{0;7,7,1}}\,\,({\bf 7},{\bf 1})$% [-4]
\\\hline
\end{tabular}
}
\end{align*}
Decomposition of the string multiplet:
\begin{align*}
{\scriptsize
\begin{tabular}{|c|c|l|} \hline
 &$\ell_9$ & \multicolumn{1}{c|}{central charges} \\ \hline\hline
%%%%%%%%%%%%%%%%%%%%%%%%%%
 $\textcolor{red}{{\bf 3875}}$ 
 & $1$&
 $\SUSY{Z^{1}_{\alpha}}\,\,({\bf 1},{\bf 2})$ % [-2]
\\\cline{2-3}
 & $2$&
 $\SUSY{Z^{1;2}}\,\,({\bf 21},{\bf 1})$ % [-2]
\\\cline{2-3}
 & $3$&
 $\SUSY{Z^{1;4}_{\alpha}}\,\,({\bf \overline{35}},{\bf 2})$ % [-2]
\\\cline{2-3}
 & $4$&
 $\SUSY{Z^{1;5,1}}\,\,({\bf \overline{140}},{\bf 1})$, % [-2]
 $Z^{1;6}\,\,({\bf \overline{7}},{\bf 1})$, % [-4]
 $\SUSY{Z^{1;6}}_{\alpha\beta}\,\,({\bf \overline{7}},{\bf 3})$ % [-2]
\\\cline{2-3}
 & $5$&
 $\SUSY{Z^{1;6,2}_{\alpha}}\,\,({\bf 140},{\bf 2})$, % [-2]
 $2\,Z^{1;7,1}_{\alpha}\,\,(2\times ({\bf 7},{\bf 2}))$ % [-4]
\\\cline{2-3}
 & $6$&
 $\SUSY{Z^{1;6,4}}\,\,({\bf 210},{\bf 1})$, % [-2]
 $\SUSY{Z^{1;7,2,1}}\,\,({\bf 112},{\bf 1})$, % [-2]
 $Z^{1;7,3}\,\,({\bf 35},{\bf 1})$, % [-4]
 $\SUSY{Z^{1;7,3}_{\alpha\beta}}\,\,({\bf 35},{\bf 3})$ % [-2]
\\\cline{2-3}
 & $7$&
 $\SUSY{Z^{1;6,6}_{\alpha}}\,\,({\bf \overline{28}},{\bf 2})$, % [-2]
 $\SUSY{Z^{1;7,4,1}_{\alpha}}\,\,({\bf \overline{224}},{\bf 2})$, % [-2]
 $2\,Z^{1;7,5}_{\alpha}\,\,(2\times ({\bf \overline{21}},{\bf 2}))$ % [-4]
\\
\hline
\end{tabular}
}
\end{align*}
\newpage
Decomposition of the membrane multiplet:
{\tiny
\begin{longtable}{|c|c|l|} \hline
 &$\ell_9$ & \multicolumn{1}{c|}{central charges} \\ \hline\hline
%%%%%%%%%%%%%%%%%%%%%%%%%%
 $\textcolor{red}{{\bf 147250}}$ 
 & $2$&
 $\SUSY{Z^{2;1}}\,\,({\bf 7},{\bf 1})$ %[2]
\\\cline{2-3}
 & $3$&
 $\SUSY{Z^{2;3}_{\alpha}}\,\,({\bf 35},{\bf 2})$ %[2]
\\\cline{2-3}
 & $4$&
 \Pbox{0.85\textwidth}{$\SUSY{Z^{2;4,1}}\,\,({\bf \overline{224}},{\bf 1})$, %[2]
 $Z^{2;5}\,\,({\bf \overline{21}},{\bf 1})$, %[0]
 $\SUSY{Z^{2;5}_{\alpha\beta}}\,\,({\bf \overline{21}},{\bf 3})$} %[2]
\\\cline{2-3}
 & $5$&
 \Pbox{0.85\textwidth}{$\SUSY{Z^{2;5,2}_{\alpha}}\,\,({\bf 392},{\bf 2})$, %[2]
 $2\,Z^{2;6,1}_{\alpha}\,\,(2\times ({\bf 48},{\bf 2}))$, %[0]
 $2\,Z^{2;7}_{\alpha}\,\,(2\times ({\bf 1},{\bf 2}))$, %[-2]
 $\SUSY{Z^{2;7}_{\alpha\beta\gamma}}\,\,({\bf 1},{\bf 4})$} %[2]
\\\cline{2-3}
 & $6$&
 \Pbox{0.85\textwidth}{$\SUSY{Z^{2;5,4}}\,\,({\bf 490},{\bf 1})$, %[2]
 $\SUSY{Z^{2;6,2,1}}\,\,({\bf 735},{\bf 1})$, %[2]
 $Z^{2;6,3}\,\,({\bf 224},{\bf 1})$, %[0]
 $\SUSY{Z^{2;6,3}_{\alpha\beta}}\,\,({\bf 224},{\bf 3})$, %[2]
 $Z^{2;7,1,1}\,\,({\bf 28},{\bf 1})$, %[0]
 $3\,Z^{2;7,2}\,\,(3\times ({\bf 21},{\bf 1}))$, %[-2]
 $2\,Z^{2;7,2}_{\alpha\beta}\,\,(2\times ({\bf 21},{\bf 3}))$} %[0]
\\\cline{2-3}
 & $7$&
 \Pbox{0.85\textwidth}{$\SUSY{Z^{2;6,4,1}_{\alpha}}\,\,({\bf \overline{1323}},{\bf 2})$, %[2]
 $2\,Z^{2;6,5}_{\alpha}\,\,(2\times ({\bf \overline{112}},{\bf 2}))$, %[0]
 $\SUSY{Z^{2;7,2,2}_{\alpha}}\,\,({\bf \overline{196}},{\bf 2})$, %[2]
 $2\,Z^{2;7,3,1}_{\alpha}\,\,(2\times ({\bf \overline{210}},{\bf 2}))$, %[0]
 $4\,Z^{2;7,4}_{\alpha}\,\,(4\times ({\bf \overline{35}},{\bf 2}))$, %[-2]
 $\SUSY{Z^{2;7,4}_{\alpha\beta\gamma}}\,\,({\bf \overline{35}},{\bf 4})$} %[2]
\\
\hline
\end{longtable}
}

\newpage

\section{M-theory potentials: $E_n\to \SL(n)$ and 11D uplifts}
\label{app:M-potential}

In this appendix, we summarize the mixed-symmetry potentials in $d$ dimensions. 
For each of the $d$-dimensional potentials, we also identify the 11D origin. 
The same analysis in type IIB theory is worked out in Appendix \ref{app:B-potential}. 

\subsubsection*{M-theory: $d=9$}
\begin{align*}
\centerline{\tiny
\begin{tabular}{|c|c||p{1.cm}|p{1.cm}|p{0.8cm}|p{1.cm}|p{1.cm}|p{1.2cm}|p{1.4cm}|p{1.4cm}|p{1.5cm}|}\hline
 $\ell$ & 11D origin & \multicolumn{1}{c|}{$p=1$} & \multicolumn{1}{c|}{$p=2$} & \multicolumn{1}{c|}{$p=3$} & \multicolumn{1}{c|}{$p=4$} & \multicolumn{1}{c|}{$p=5$} & \multicolumn{1}{c|}{$p=6$} & \multicolumn{1}{c|}{$p=7$} & \multicolumn{1}{c|}{$p=8$} & \multicolumn{1}{c|}{$p=9$} \\\hline\hline
 $0$ & $\SUSY{A_1^1}$ 
 &%p=1
 $\SUSY{A_{1}^1}\,\,({\bf 2})$
 &%p=2
 &%p=3
 &%p=4
 &%p=5
 &%p=6
 &%p=7
 &%p=8
 &%p=9
\\\hline
\rowcolor{Gray}
 $1$ & $\SUSY{A_{3}}$ 
 &%p=1
 $\SUSY{A_{1;2}}\,\,({\bf 1})$
 &%p=2
 $\SUSY{A_{2;1}}\,\,({\bf 2})$
 &%p=3
 $\SUSY{A_{3}}\,\,({\bf 1})$
 &%p=4
 &%p=5
 &%p=6
 &%p=7
 &%p=8
 &%p=9
\\\hline
 $2$ & $\SUSY{A_{6}}$ 
 &%p=1
 &%p=2
 &%p=3
 &%p=4
 $\SUSY{A_{4;2}}\,\,({\bf 1})$
 &%p=5
 $\SUSY{A_{5;1}}\,\,({\bf 2})$
 &%p=6
 $\SUSY{A_{6}}\,\,({\bf 1})$
 &%p=7
 &%p=8
 &%p=9
\\\hline
\rowcolor{Gray}
 $3$ & $\SUSY{A_{8;1}}$ 
 &%p=1
 &%p=2
 &%p=3
 &%p=4
 &%p=5
 &%p=6
 $\SUSY{A_{6;2,1}}\,\,({\bf 2})$
 &%p=7
 $\SUSY{A_{7;1,1}}\,\,({\bf 3})$, 
 $A_{7;2}\,\,({\bf 1})$
 &%p=8
 $\SUSY{A_{8;1}}\,\,({\bf 2})$
 &%p=9
\\\hline
 $4$ & $\SUSY{A_{10,1,1}}$ 
 &%p=1
 &%p=2
 &%p=3
 &%p=4
 &%p=5
 &%p=6
 &%p=7
 &%p=8
 $\SUSY{A_{8;2,1,1}}\,\,({\bf 3})$
 &%p=9
 $\SUSY{A_{9;1,1,1}}\,\,({\bf 4})$, 
 $A_{9;2,1}\,\,({\bf 2})$
\\
\rowcolor{Gray}
 & $A_{11,1}$ 
 &%p=1
 &%p=2
 &%p=3
 &%p=4
 &%p=5
 &%p=6
 &%p=7
 &%p=8
 &%p=9
 $A_{9;2,1}\,\,({\bf 2})$
\\
\hline
\end{tabular}
}
\end{align*}

\subsubsection*{M-theory: $d=8$}
\begin{align*}
\centerline{\tiny
%
}
\end{align*}

\subsubsection*{M-theory: $d=7$}
\begin{align*}
\centerline{\tiny
%
}
\end{align*}

\subsubsection*{M-theory: $d=6$}

\begin{align*}
\centerline{\tiny
%
}
\end{align*}

\subsubsection*{M-theory: $d=5$}

\begin{align*}
\centerline{\tiny
%
}
\end{align*}

\subsubsection*{M-theory: $d=4$}

\begin{align*}
\centerline{\tiny
%
}
\end{align*}

\subsubsection*{M-theory: $d=3$}

In $d=3$\,, the task is rather complicated. 
The 1-form and 2-form gauge potentials are determined in \cite{0705.0752} but the details of the 3-form have not been given in the literature. 
As mentioned in the introduction, in order to reproduce all of the potentials in the 3-form multiplets, we need to determine the $E_{11}$ generators up to level $\ell=17$ but it is not easy even if using SimpLie. 
On the other hand, if we consider the level decomposition of \cite{0705.1304} by deleting the node $\alpha_3$\,, we can find that the 3-forms are in the $\textcolor{red}{\bf 147250}+{\bf 3875}+{\bf 248}$ representations of $E_8$\,. 

\newpage

We decompose the $E_8$ representations into the irreducible representations of the subgroup $\SL(8)$\,, and then the following set of 3-form potentials are obtained.
\begin{align*}
{\scriptsize

}
\end{align*}
By uplifting these potentials into $d=11$, we can identify all of the mixed-symmetry potentials in 11D up to level $\ell=17$ that can appear in $d=3$. 
The result is as follows.

\newpage

{\tiny
%
}

\section{Type IIB potentials: $E_n\to \SL(D)\times \SL(2)$ and 10D uplifts}
\label{app:B-potential}

\subsubsection*{Type IIB theory: $d=9$}

\begin{align*}
\centerline{\tiny
%
}
\end{align*}

\subsubsection*{Type IIB theory: $d=8$}

\begin{align*}
\centerline{\tiny
%
}
\end{align*}

\subsubsection*{Type IIB theory: $d=7$}

\begin{align*}
\centerline{\tiny
%
}
\end{align*}

\subsubsection*{Type IIB theory: $d=6$}

\begin{align*}
\centerline{\tiny
%
}
\end{align*}

\subsubsection*{Type IIB theory: $d=5$}

\begin{align*}
\centerline{\tiny
%
}
\end{align*}

\subsubsection*{Type IIB theory: $d=4$}

{\tiny
%
}

\newpage

\subsubsection*{Type IIB theory: $d=3$}

{\tiny
%
}

\newpage

\section{$\OO(10,10)$ potentials with level $\NN=5$ and $\NN=6$}
\label{app:O(1010)-G-H}

In this appendix, we determine all of the $\OO(10,10)$ potentials with level $\NN=5$ and $\NN=6$ that are predicted by the $E_{11}$ conjecture. 
Only the multiplicity of the $\OO(10,10)$ singlet $H$ ($\NN=6$) is not determined. 

\subsection{Potential $G$ ($\NN=5$): Type IIA theory}

All of the $\OO(10,10)$ potentials $G$ with level $\NN=5$ have a spinor index $a$ or $\dot{a}$\,, and they give different mixed-symmetry potentials in type IIA theory and type IIB theory. 
We here show the decomposition in type IIA theory. 

{\footnotesize
\begin{longtable}{|C{1.5cm}||L{13.5cm}|}\hline
 $\OO(10,10)$ tensors & \multicolumn{1}{c|}{Type IIA potentials} 
\\
\hline
\end{longtable}
}

\subsection{Potential $G$ ($\NN=5$): Type IIB theory}

Here, we here show the decomposition of the $\OO(10,10)$ tensors $G$ in type IIB theory. 

{\footnotesize
\begin{longtable}{|C{1.5cm}||L{13.5cm}|}\hline
 $\OO(10,10)$ tensors & \multicolumn{1}{c|}{Type IIB potentials} 
\\
\hline
\end{longtable}
}

\newpage

\subsection{Potential $H$ ($\NN=6$): Type II theories}

Here, we show all potentials with level $\NN=6$\,. 
Unlike potentials with level $\NN \leq 5$\,, the current personal computers do not provide the level-$\NN=6$ potentials from a naive level decomposition by SimpLie. 
By taking an indirect approach, we determine all of the potentials with level $\NN=6$ (except the multiplicity of the $\OO(10,10)$ singlet $H$). 

By using SimpLie, we can first determine a set of the allowed $\OO(10,10)$ potentials, whose multiplicities are not yet determined. 
We then decompose all of the $\OO(10,10)$ potentials into type IIA/type IIB potentials. 
Then, since the type IIB potentials with level $\ell_9\leq 14$ are already obtained in Table \ref{tab:IIB-potentials-14}, we can determine the multiplicities of the $\OO(10,10)$ tensors such that Table \ref{tab:IIB-potentials-14} is reproduced after decomposing the $\OO(10,10)$ potentials into the type IIB potentials. 
Unfortunately, the $\OO(10,10)$ singlet $H$ contains only a generator $H_{10,10,10}$ with level $\ell_9=15$\,, whose multiplicity has not been determined. 
Therefore, the multiplicity for the singlet is not determined in this paper. 

There is an additional subtlety other than the $\OO(10,10)$ singlet.
The dimensions of three (underlined) representations, ${\bf 354411750}$, ${\bf 382764690}$, and ${\bf 330752000}$, are extremely large, and we could not identify all weights in those representations; doing so is important in decomposing each $\OO(10,10)$ tensor into $\SL(10)$ tensors. 
At least, we identified all of the weights that correspond to the potentials with level $\ell_9\leq 14$\,. 
By using the symmetry \eqref{eq:index-dual2}, we also identified all generators with level $16\leq \ell_9$\,. 
On the other hand, regarding the potentials with level $\ell_9=15$\,, we have determined most of them, but there are still weights missing that are needed in order to reproduce the dimension of the $\OO(10,10)$ representations. 
Concretely, in representations, ${\bf 354411750}$, ${\bf 382764690}$, and ${\bf 330752000}$, 199, 198, and 1430 weights were missing. 
The only allowed potentials with level $\ell_9=15$ and dimension less than $1925$ are $H_{10,10,9,1}\,\,({\bf 99})$ and $H_{10,10,10}\,\,({\bf 1})$\,, and the missing weights should be coming from them. 
Then, we assumed the missing weights in ${\bf 354411750}$, ${\bf 382764690}$, and ${\bf 330752000}$, respectively, as \textcolor{blue}{$2\,H_{10,10,9,1}\,\,(2\times {\bf 99})$} and \textcolor{blue}{$H_{10,10,10}\,\,({\bf 1})$}\,, \textcolor{blue}{$2\,H_{10,10,9,1}\,\,(2\times {\bf 99})$}\,, and \textcolor{blue}{$14\,H_{10,10,9,1}\,\,(14\times {\bf 99})$} and \textcolor{blue}{$44\,H_{10,10,10}\,\,(44\times{\bf 1})$}\,. 
In order to note that they are assumed, we colored the results in blue. 
We could check whether the assumption is correct if Table \ref{tab:IIB-potentials-14} was extended to level $\ell_9=15$\,. 
In that case, we could also determine the multiplicity of the $\OO(10,10)$ singlet $H$\,. 
In the following, we show the results, and there the multiplicities of the mixed-symmetry potentials, other than those colored in blue, will be reliable. 

\newpage

We first show the potentials $H$ that contribute to Table \ref{tab:p-form}. 
All of them are in the generalized NS--NS sector.
{\footnotesize

%\caption{List of the $\OO(10,10)$ tensors with level $0\leq n\leq 5$\,. Their decompositions into type IIA/IIB tensors are also shown.}
}\newpage\noindent
Potentials $H$ (in the generalized NS--NS sector) which do not contribute to Table \ref{tab:p-form} are as follows:
{\footnotesize
%
}\newpage\noindent
Potentials $H$ (in the generalized R--R sector) which do not contribute to Table \ref{tab:p-form} are as follows:
{\footnotesize
\begin{longtable}{|C{1.7cm} || L{6.5cm} | L{6.5cm} |}\hline
%%%%%%%%%%%%%%%%%%%%%%%%%%%%%%%%%%%%%%%%%%%%%%%%%%%
 $\OO(10,10)$ tensors & \multicolumn{1}{c|}{Type IIA potentials} & \multicolumn{1}{c|}{Type IIB potentials}
\\
\hline
\end{longtable}
}

\newpage

\section{Decomposition of $p$-forms: $E_n \to \OO(D,D) \to \SL(D)$}
\label{app:En2ODD}

In this appendix, we decompose the $E_n$ $U$-duality multiplets of $p$-form given in Table \eqref{tab:p-form} into $\OO(D,D)$ potentials. 
We further decompose the $\OO(D,D)$ potentials into type II mixed-symmetry potentials in $d$ dimensions. 
By uplifting the obtained type II potentials to 10D, we obtain Table \ref{tab:IIA-IIB-list}. 

\subsection*{$d=8$}

The 1-form multiplet, $\textcolor{red}{(\overline{\bf 3},{\bf 2})}$ of $E_3$, is decomposed into the following $\OO(2,2)$ tensors:
\begin{align*}
\centerline{\tiny

}
\end{align*}
The 2-form multiplet, $\textcolor{red}{({\bf 3},{\bf 1})}$ of $E_3$, is decomposed into the following $\OO(2,2)$ tensors:
\begin{align*}
\centerline{\tiny
%
}
\end{align*}
The 3-form multiplet, $\textcolor{red}{({\bf 1},{\bf 2})}$ of $E_3$, is decomposed into the following $\OO(2,2)$ tensors:
\begin{align*}
\centerline{\tiny
%
}
\end{align*}
The 4-form multiplet, $\textcolor{red}{(\overline{\bf 3},{\bf 1})}$ of $E_3$, is decomposed into the following $\OO(2,2)$ tensors:
\begin{align*}
\centerline{\tiny
%
}
\end{align*}
The 5-form multiplet, $\textcolor{red}{({\bf 3},{\bf 2})}$ of $E_3$, is decomposed into the following $\OO(2,2)$ tensors:
\begin{align*}
\centerline{\tiny
%
}
\end{align*}
The 7-form multiplet, $(\overline{\bf 3},{\bf 2})$ of $E_3$, is decomposed into the following $\OO(2,2)$ tensors:
\begin{align*}
\centerline{\tiny
%
}
\end{align*}
The 8-form multiplet, $({\bf 3},{\bf 3})$ of $E_3$, is decomposed into the following $\OO(2,2)$ tensors:
\begin{align*}
\centerline{\tiny
%
}
\end{align*}
The 8-form multiplet, $2\times({\bf 3},{\bf 1})$ of $E_3$, is decomposed into the following $\OO(2,2)$ tensors:
\begin{align*}
\centerline{\tiny
%
}
\end{align*}

\subsection*{$d=7$}

The 1-form multiplet, $\textcolor{red}{\overline{\bf 10}}$ of $E_4$, is decomposed into the following $\OO(3,3)$ tensors:
\begin{align*}
\centerline{\tiny
%
}
\end{align*}
The 2-form multiplet, $\textcolor{red}{\bf 5}$ of $E_4$, is decomposed into the following $\OO(3,3)$ tensors:
\begin{align*}
\centerline{\tiny
%
}
\end{align*}
The 3-form multiplet, $\textcolor{red}{\overline{\bf 5}}$ of $E_4$, is decomposed into the following $\OO(3,3)$ tensors:
\begin{align*}
\centerline{\tiny
%
}
\end{align*}
The 4-form multiplet, $\textcolor{red}{\bf 10}$ of $E_4$, is decomposed into the following $\OO(3,3)$ tensors:
\begin{align*}
\centerline{\tiny
%
}
\end{align*}
The 5-form multiplet, $\textcolor{red}{\bf 24}$ of $E_4$, is decomposed into the following $\OO(3,3)$ tensors:
\begin{align*}
\centerline{\tiny
%
}
\end{align*}
The 6-form multiplet, $\textcolor{red}{\overline{\bf 40}}$ of $E_4$, is decomposed into the following $\OO(3,3)$ tensors:
\begin{align*}
\centerline{\tiny
%
}
\end{align*}
The 6-form multiplet, $\textcolor{red}{\overline{\bf 15}}$ of $E_4$, is decomposed into the following $\OO(3,3)$ tensors:
\begin{align*}
\centerline{\tiny
%
}
\end{align*}
The 7-form multiplet, ${\bf 45}$ of $E_4$, is decomposed into the following $\OO(3,3)$ tensors:
\begin{align*}
\centerline{\tiny
%
}
\end{align*}
The 7-form multiplet, ${\bf 5}$ of $E_4$, is decomposed into the following $\OO(3,3)$ tensors:
\begin{align*}
\centerline{\tiny
%
}
\end{align*}

\subsection*{$d=6$}

The 1-form multiplet, $\textcolor{red}{\bf 16}$ of $E_5$, is decomposed into the following $\OO(4,4)$ tensors:
\begin{align*}
\centerline{\tiny
%
}
\end{align*}
The 6-form multiplet, ${\bf 10}$ of $E_5$, is decomposed into the following $\OO(4,4)$ tensors:
\begin{align*}
\centerline{\tiny
%
}
\end{align*}

\subsection*{$d=5$}

The 1-form multiplet, $\textcolor{red}{\bf 27}$ of $E_6$, is decomposed into the following $\OO(5,5)$ tensors:
\begin{align*}
\centerline{\tiny
%
}
\end{align*}
The 5-form multiplet, $\overline{\bf 27}$ of $E_6$, is decomposed into the following $\OO(5,5)$ tensors:
\begin{align*}
\centerline{\tiny
%
}
\end{align*}

\subsection*{$d=4$}

The 1-form multiplet, $\textcolor{red}{\bf 56}$ of $E_7$, is decomposed into the following $\OO(6,6)$ tensors:
\begin{align*}
\centerline{\tiny
%
}
\end{align*}

\subsection*{$d=3$}

The 1-form multiplet, $\textcolor{red}{\bf 248}$ of $E_8$, is decomposed into the following $\OO(7,7)$ tensors:
\begin{align*}
\centerline{\tiny
%
}
\end{align*}

\newpage

The 3-form multiplet, $\textcolor{red}{\bf 147250}$ of $E_8$, is decomposed into the following $\OO(7,7)$ tensors:
\setlength{\LTleft}{-20cm plus -1fill}
\setlength{\LTright}{\LTleft}
{\tiny
%
}%
\noindent The 3-form multiplet, ${\bf 3875}$ of $E_8$, is decomposed into the following $\OO(7,7)$ tensors:
\begin{align*}
\centerline{\tiny
%
}
\end{align*}

\newpage

\section{SUSY potentials from $B$ to $Z$}
\label{app:BtoZ}

As discussed in Appendix \ref{app:O(1010)-G-H}, $\OO(10,10)$ potentials with level $\NN=6$ are the maximal we can determine through the level decomposition of $E_{11}$\,. 
If we do not care about the multiplicity, we can determine all of the allowed $\OO(10,10)$ potentials up to level $\NN=34$\,. 
As we observed at the low levels, the underlined potentials are always coming with multiplicity 1, but some of the allowed potentials may have multiplicity 0. 
Thus, in order to check whether such $\OO(10,10)$ potentials really appear in the adjoint representation of $E_{11}$\,, we take a different approach. 

In \cite{1805.12117}, the $E_8$ $U$-duality multiplets of exotic branes are searched by using the $T$- and $S$-duality, and we take a similar method here.\footnote{Here, we perform the $T$-duality even in the timelike direction, and under that $T$-duality, type II theory may be mapped to type II$^*$ theory, but we do not distinguish type II and type II$^*$ carefully. At least, the results obtained here are totally consistent with the $E_{11}$ conjecture.} 
Starting with a standard potential, such as the R-R potential, we repeatedly perform $T$-duality \eqref{eq:T-rule} and $S$-duality \eqref{eq:S-rule} and obtain a chain of the underlined potentials, which we here call ``SUSY potentials''. 
Of course, this procedure generates infinitely many SUSY potentials, and we need to cut off the potentials at a certain level. 
By using the ancillary files \cite{anc}, we have checked that we can determine all of the SUSY potentials, at least up to level $\NN=36$\,.\footnote{A similar result up to level $\NN=25$ has been obtained \cite{1903.10247} but the reported number of potentials is much smaller than that obtained here, and their analysis may be restricted to a certain subclass. For example, as explained in footnote \ref{ref:10247}, $L_{10,10,7,1}$ (which is predicted from $E_{11}$) was not considered there.} 
Then, among the allowed $\OO(10,10)$ potentials generated by SimpLie, we have identified which $\OO(10,10)$ potentials indeed appear with multiplicity 1, by comparing them with a list of mixed-symmetry potentials. 

We have determined all of the SUSY potentials up to $\NN=36$\,, but here we show the results only up to level $\NN=11$\,. 
The results up to level $\NN=24$ (potential $Z$) are given in the ancillary files \cite{anc}. 
In addition, for potentials with level $\NN\geq1$\,, we display only the type II mixed-symmetry potentials associated with the highest-weights and the number of the potentials (with different tensor structures) contained in the $\OO(10,10)$ multiplet is given in the square bracket. 
For example, at level $\NN=1$, there is only one $\OO(10,10)$ multiplet $C_{\dot{a}}$ and it contains five type IIA potentials $C_{1}$, $C_{3}$, $C_{5}$, $C_{7}$, $C_{9}$, and six type IIB potentials, $C_{0}$, $C_{2}$, $C_{4}$, $C_{6}$, $C_{8}$, $C_{10}$\,. 
The detailed contents for the higher levels can be easily generated by using the ancillary files \cite{anc}. 

In this paper, we determined all of the SUSY potentials up to level $\NN=6$\,, and they are precisely the same as those obtained here. 
Namely, at least for $\NN\leq 6$\,, all of the underlined $E_{11}$ potentials are in a single orbit of Weyl reflections ($T$-/$S$-dualities). 
We expect that this is the case also for the higher-level potentials, and the SUSY potentials obtained in this appendix will be all of the SUSY potentials (with $\NN\leq 24$) predicted from the $E_{11}$ conjecture. 

\newpage

{\tiny
\begin{longtable}{|c||c|c|c|c|}\hline
 $\NN$ & $\OO(10,10)$ rep. & $E_{11}$ roots $\alpha_i$ & type IIA & type IIB \\
\hline\hline
0
 & [0,1,0,0,0,0,0,0,0,0] & [-1,-2,-2,-2,-2,-2,-2,-2,-1,0,-1] & \multicolumn{2}{c|}{$\beta^{2}$, $B_1^1$, $B_{2}$} \\
\hline
\rowcolor{Gray}
1
 & [0,0,0,0,0,0,0,0,1,0] & [0,0,0,0,0,0,0,0,0,1,0] & $C_{1}$ [5] & $C_{0}$ [6] \\
\hline
2
 & [0,0,0,1,0,0,0,0,0,0] & [0,0,0,0,1,2,3,4,3,2,2] & \multicolumn{2}{c|}{$D_{6}$ [5]} \\
\hline
\rowcolor{Gray}
3
 & [0,1,0,0,0,0,0,0,1,0] & [0,0,1,2,3,4,5,6,4,3,3] & $E_{8,1}$ [13] & $E_{8}$ [14] \\
\hline
4
 & [1,0,0,0,0,0,1,0,0,0] & [0,1,2,3,4,5,6,8,6,4,4] & \multicolumn{2}{c|}{$F_{9,3}$ [14]} \\
\rowcolor{Gray}
 & [0,1,0,1,0,0,0,0,0,0] & [0,0,1,2,4,6,8,10,7,4,5] & \multicolumn{2}{c|}{$F_{8,6}$ [9]} \\
\cline{2-5}
 & [0,0,0,0,0,0,0,0,2,0] & [1,2,3,4,5,6,7,8,5,4,4] & $F_{10,1,1}$ [5] & $F_{10}$ [6] \\
\hline
\rowcolor{Gray}
5
 & [0,0,0,0,0,1,0,0,1,0] & [1,2,3,4,5,6,8,10,7,5,5] & $G_{10,4,1}$ [17] & $G_{10,4}$ [18] \\
 & [1,0,0,1,0,0,0,0,0,1] & [0,1,2,3,5,7,9,11,8,5,5] & $G_{9,6}$ [28] & $G_{9,6,1}$ [28] \\
\rowcolor{Gray}
 & [0,2,0,0,0,0,0,0,1,0] & [0,0,2,4,6,8,10,12,8,5,6] & $G_{8,8,1}$ [13] & $G_{8,8}$ [14] \\
\hline
6
 & [0,0,0,1,0,0,0,1,0,0] & [1,2,3,4,6,8,10,12,9,6,6] & \multicolumn{2}{c|}{$H_{10,6,2}$ [25]} \\
\rowcolor{Gray}
 & [1,1,0,0,0,0,0,0,1,1] & [0,1,3,5,7,9,11,13,9,6,6] & \multicolumn{2}{c|}{$H_{9,8,1}$ [32]} \\
 & [1,0,1,0,0,1,0,0,0,0] & [0,1,2,4,6,8,11,14,10,6,7] & \multicolumn{2}{c|}{$H_{9,7,4}$ [24]} \\
\rowcolor{Gray}
 & [0,2,0,1,0,0,0,0,0,0] & [0,0,2,4,7,10,13,16,11,6,8] & \multicolumn{2}{c|}{$H_{8,8,6}$ [9]} \\
\hline
7
 & [0,1,0,0,0,0,0,1,1,0] & [1,2,4,6,8,10,12,14,10,7,7] & $I_{10,8,2,1}$ [31] & $I_{10,8,2}$ [32] \\
\rowcolor{Gray}
 & [0,0,1,0,0,1,0,0,0,1] & [1,2,3,5,7,9,12,15,11,7,7] & $I_{10,7,4}$ [40] & $I_{10,7,4,1}$ [40] \\
 & [2,0,0,0,0,0,1,0,0,1] & [0,2,4,6,8,10,12,15,11,7,7] & $I_{9,9,3}$ [28] & $I_{9,9,3,1}$ [28] \\
\rowcolor{Gray}
 & [0,0,0,2,0,0,0,0,1,0] & [1,2,3,4,7,10,13,16,11,7,8] & $I_{10,6,6,1}$ [17] & $I_{10,6,6}$ [18] \\
 & [1,1,0,0,1,0,0,0,1,0] & [0,1,3,5,7,10,13,16,11,7,8] & $I_{9,8,5,1}$ [48] & $I_{9,8,5}$ [48] \\
\rowcolor{Gray}
 & [1,0,2,0,0,0,0,0,0,1] & [0,1,2,5,8,11,14,17,12,7,8] & $I_{9,7,7}$ [24] & $I_{9,7,7,1}$ [24] \\
 & [0,3,0,0,0,0,0,0,1,0] & [0,0,3,6,9,12,15,18,12,7,9] & $I_{8,8,8,1}$ [13] & $I_{8,8,8}$ [14] \\
\hline
\rowcolor{Gray}
8
 & [0,1,0,0,1,0,0,0,1,1] & [1,2,4,6,8,11,14,17,12,8,8] & \multicolumn{2}{c|}{$J_{10,8,5,1}$ [60]} \\
 & [1,0,0,0,0,0,1,1,0,0] & [1,3,5,7,9,11,13,16,12,8,8] & \multicolumn{2}{c|}{$J_{10,9,3,2}$ [28]} \\
\rowcolor{Gray}
 & [0,0,1,1,0,0,1,0,0,0] & [1,2,3,5,8,11,14,18,13,8,9] & \multicolumn{2}{c|}{$J_{10,7,6,3}$ [32]} \\
 & [0,1,0,0,0,2,0,0,0,0] & [1,2,4,6,8,10,14,18,13,8,9] & \multicolumn{2}{c|}{$J_{10,8,4,4}$ [15]} \\
\rowcolor{Gray}
 & [1,1,1,0,0,0,0,1,0,0] & [0,1,3,6,9,12,15,18,13,8,9] & \multicolumn{2}{c|}{$J_{9,8,7,2}$ [48]} \\
 & [2,0,0,0,1,0,1,0,0,0] & [0,2,4,6,8,11,14,18,13,8,9] & \multicolumn{2}{c|}{$J_{9,9,5,3}$ [30]} \\
\rowcolor{Gray}
 & [1,1,0,1,1,0,0,0,0,0] & [0,1,3,5,8,12,16,20,14,8,10] & \multicolumn{2}{c|}{$J_{9,8,6,5}$ [24]} \\
 & [0,3,0,1,0,0,0,0,0,0] & [0,0,3,6,10,14,18,22,15,8,11] & \multicolumn{2}{c|}{$J_{8,8,8,6}$ [9]} \\
\rowcolor{Gray}
 & [1,0,3,0,0,0,0,0,0,0] & [0,1,2,6,10,14,18,22,15,8,11] & \multicolumn{2}{c|}{$J_{9,7,7,7}$ [6]} \\
\cline{2-5}
 & [0,0,0,0,0,0,0,1,2,0] & [2,4,6,8,10,12,14,16,11,8,8] & $J_{10,10,2,1,1}$ [13] & $J_{10,10,2}$ [14] \\
\rowcolor{Gray}
 & [0,0,2,0,0,0,0,0,0,2] & [1,2,3,6,9,12,15,18,13,8,8] & $J_{10,7,7}$ [16] & $J_{10,7,7,1,1}$ [16] \\
 & [2,0,0,1,0,0,0,0,0,2] & [0,2,4,6,9,12,15,18,13,8,8] & $J_{9,9,6}$ [28] & $J_{9,9,6,1,1}$ [28] \\
\rowcolor{Gray}
 & [2,0,0,1,0,0,0,0,2,0] & [0,2,4,6,9,12,15,18,12,8,9] & $J_{9,9,6,1,1}$ [28] & $J_{9,9,6}$ [28] \\
\hline
9
 & [0,0,0,0,0,1,0,1,1,0] & [2,4,6,8,10,12,15,18,13,9,9] & $K_{10,10,4,2,1}$ [31] & $K_{10,10,4,2}$ [32] \\
\rowcolor{Gray}
 & [1,0,0,1,0,0,0,0,2,1] & [1,3,5,7,10,13,16,19,13,9,9] & $K_{10,9,6,1,1}$ [48] & $K_{10,9,6,1}$ [48] \\
 & [0,1,1,0,0,0,0,1,0,1] & [1,2,4,7,10,13,16,19,14,9,9] & $K_{10,8,7,2}$ [54] & $K_{10,8,7,2,1}$ [54] \\
\rowcolor{Gray}
 & [1,0,0,0,1,0,1,0,0,1] & [1,3,5,7,9,12,15,19,14,9,9] & $K_{10,9,5,3}$ [60] & $K_{10,9,5,3,1}$ [60] \\
 & [2,1,0,0,0,0,0,0,1,2] & [0,2,5,8,11,14,17,20,14,9,9] & $K_{9,9,8,1}$ [32] & $K_{9,9,8,1,1}$ [32] \\
\rowcolor{Gray}
 & [0,2,0,0,0,0,0,0,3,0] & [1,2,5,8,11,14,17,20,13,9,10] & $K_{10,8,8,1,1,1}$ [13] & $K_{10,8,8}$ [14] \\
 & [0,1,0,1,0,1,0,0,1,0] & [1,2,4,6,9,12,16,20,14,9,10] & $K_{10,8,6,4,1}$ [67] & $K_{10,8,6,4}$ [68] \\
\rowcolor{Gray}
 & [2,0,1,0,0,0,1,0,1,0] & [0,2,4,7,10,13,16,20,14,9,10] & $K_{9,9,7,3,1}$ [60] & $K_{9,9,7,3}$ [60] \\
 & [0,0,2,0,1,0,0,0,0,1] & [1,2,3,6,9,13,17,21,15,9,10] & $K_{10,7,7,5}$ [36] & $K_{10,7,7,5,1}$ [36] \\
\rowcolor{Gray}
 & [1,2,0,0,0,1,0,0,0,1] & [0,1,4,7,10,13,17,21,15,9,10] & $K_{9,8,8,4}$ [50] & $K_{9,8,8,4,1}$ [50] \\
 & [2,0,0,1,1,0,0,0,0,1] & [0,2,4,6,9,13,17,21,15,9,10] & $K_{9,9,6,5}$ [48] & $K_{9,9,6,5,1}$ [48] \\
\rowcolor{Gray}
 & [1,1,1,1,0,0,0,0,1,0] & [0,1,3,6,10,14,18,22,15,9,11] & $K_{9,8,7,6,1}$ [56] & $K_{9,8,7,6}$ [56] \\
 & [0,4,0,0,0,0,0,0,1,0] & [0,0,4,8,12,16,20,24,16,9,12] & $K_{8,8,8,8,1}$ [13] & $K_{8,8,8,8}$ [14] \\
\hline
\rowcolor{Gray}
10
 & [0,0,0,0,1,1,0,0,1,1] & [2,4,6,8,10,13,17,21,15,10,10] & \multicolumn{2}{c|}{$L_{10,10,5,4,1}$ [48]} \\
 & [0,0,0,1,0,0,0,2,0,0] & [2,4,6,8,11,14,17,20,15,10,10] & \multicolumn{2}{c|}{$L_{10,10,6,2,2}$ [25]} \\
\rowcolor{Gray}
 & [1,0,1,0,0,0,1,0,1,1] & [1,3,5,8,11,14,17,21,15,10,10] & \multicolumn{2}{c|}{$L_{10,9,7,3,1}$ [90]} \\
 & [0,1,1,0,1,0,0,1,0,0] & [1,2,4,7,10,14,18,22,16,10,11] & \multicolumn{2}{c|}{$L_{10,8,7,5,2}$ [72]} \\
\rowcolor{Gray}
 & [1,0,0,1,0,1,1,0,0,0] & [1,3,5,7,10,13,17,22,16,10,11] & \multicolumn{2}{c|}{$L_{10,9,6,4,3}$ [48]} \\
 & [2,0,1,1,0,0,0,0,1,1] & [0,2,4,7,11,15,19,23,16,10,11] & \multicolumn{2}{c|}{$L_{9,9,7,6,1}$ [72]} \\
\rowcolor{Gray}
 & [2,1,0,0,0,1,0,1,0,0] & [0,2,5,8,11,14,18,22,16,10,11] & \multicolumn{2}{c|}{$L_{9,9,8,4,2}$ [60]} \\
 & [0,0,3,0,0,0,1,0,0,0] & [1,2,3,7,11,15,19,24,17,10,12] & \multicolumn{2}{c|}{$L_{10,7,7,7,3}$ [20]} \\
\rowcolor{Gray}
 & [0,1,0,2,0,1,0,0,0,0] & [1,2,4,6,10,14,19,24,17,10,12] & \multicolumn{2}{c|}{$L_{10,8,6,6,4}$ [27]} \\
 & [1,2,0,1,0,0,1,0,0,0] & [0,1,4,7,11,15,19,24,17,10,12] & \multicolumn{2}{c|}{$L_{9,8,8,6,3}$ [48]} \\
\rowcolor{Gray}
 & [2,0,1,0,1,1,0,0,0,0] & [0,2,4,7,10,14,19,24,17,10,12] & \multicolumn{2}{c|}{$L_{9,9,7,5,4}$ [36]} \\
 & [1,1,2,0,1,0,0,0,0,0] & [0,1,3,7,11,16,21,26,18,10,13] & \multicolumn{2}{c|}{$L_{9,8,7,7,5}$ [24]} \\
\rowcolor{Gray}
 & [2,0,0,3,0,0,0,0,0,0] & [0,2,4,6,11,16,21,26,18,10,13] & \multicolumn{2}{c|}{$L_{9,9,6,6,6}$ [8]} \\
 & [0,4,0,1,0,0,0,0,0,0] & [0,0,4,8,13,18,23,28,19,10,14] & \multicolumn{2}{c|}{$L_{8,8,8,8,6}$ [9]} \\
\cline{2-5}
\rowcolor{Gray}
 & [0,0,1,0,0,0,0,0,3,1] & [2,4,6,9,12,15,18,21,14,10,10] & $L_{10,10,7,1,1,1}$ [28] & $L_{10,10,7,1}$ [28] \\
 & [0,2,0,0,0,1,0,0,0,2] & [1,2,5,8,11,14,18,22,16,10,10] & $L_{10,8,8,4}$ [38] & $L_{10,8,8,4,1,1}$ [37] \\
\rowcolor{Gray}
 & [1,0,0,1,1,0,0,0,0,2] & [1,3,5,7,10,14,18,22,16,10,10] & $L_{10,9,6,5}$ [48] & $L_{10,9,6,5,1,1}$ [48] \\
 & [3,0,0,0,0,0,1,0,0,2] & [0,3,6,9,12,15,18,22,16,10,10] & $L_{9,9,9,3}$ [28] & $L_{9,9,9,3,1,1}$ [28] \\
\rowcolor{Gray}
 & [0,2,0,0,0,1,0,0,2,0] & [1,2,5,8,11,14,18,22,15,10,11] & $L_{10,8,8,4,1,1}$ [37] & $L_{10,8,8,4}$ [38] \\
 & [1,0,0,1,1,0,0,0,2,0] & [1,3,5,7,10,14,18,22,15,10,11] & $L_{10,9,6,5,1,1}$ [48] & $L_{10,9,6,5}$ [48] \\
\rowcolor{Gray}
 & [3,0,0,0,0,0,1,0,2,0] & [0,3,6,9,12,15,18,22,15,10,11] & $L_{9,9,9,3,1,1}$ [28] & $L_{9,9,9,3}$ [28] \\
 & [1,2,1,0,0,0,0,0,0,2] & [0,1,4,8,12,16,20,24,17,10,11] & $L_{9,8,8,7}$ [32] & $L_{9,8,8,7,1,1}$ [32] \\
\rowcolor{Gray}
 & [1,2,1,0,0,0,0,0,2,0] & [0,1,4,8,12,16,20,24,16,10,12] & $L_{9,8,8,7,1,1}$ [32] & $L_{9,8,8,7}$ [32] \\
\hline
11
 & [0,0,1,0,0,1,0,0,2,1] & [2,4,6,9,12,15,19,23,16,11,11] & $M_{10,10,7,4,1,1}$ [64] & $M_{10,10,7,4,1}$ [64] \\
\rowcolor{Gray}
 & [0,1,0,0,0,0,0,2,1,0] & [2,4,7,10,13,16,19,22,16,11,11] & $M_{10,10,8,2,2,1}$ [31] & $M_{10,10,8,2,2}$ [32] \\
 & [2,0,0,0,0,0,1,0,2,1] & [1,4,7,10,13,16,19,23,16,11,11] & $M_{10,9,9,3,1,1}$ [42] & $M_{10,9,9,3,1}$ [42] \\
\rowcolor{Gray}
 & [0,0,0,1,1,0,0,1,0,1] & [2,4,6,8,11,15,19,23,17,11,11] & $M_{10,10,6,5,2}$ [60] & $M_{10,10,6,5,2,1}$ [60] \\
 & [0,0,1,0,0,0,2,0,0,1] & [2,4,6,9,12,15,18,23,17,11,11] & $M_{10,10,7,3,3}$ [40] & $M_{10,10,7,3,3,1}$ [40] \\
\rowcolor{Gray}
 & [1,0,1,1,0,0,0,0,1,2] & [1,3,5,8,12,16,20,24,17,11,11] & $M_{10,9,7,6,1}$ [72] & $M_{10,9,7,6,1,1}$ [72] \\
 & [1,1,0,0,0,1,0,1,0,1] & [1,3,6,9,12,15,19,23,17,11,11] & $M_{10,9,8,4,2}$ [90] & $M_{10,9,8,4,2,1}$ [90] \\
\rowcolor{Gray}
 & [0,2,1,0,0,0,0,0,0,3] & [1,2,5,9,13,17,21,25,18,11,11] & $M_{10,8,8,7}$ [24] & $M_{10,8,8,7,1,1,1}$ [24] \\
 & [3,0,0,1,0,0,0,0,0,3] & [0,3,6,9,13,17,21,25,18,11,11] & $M_{9,9,9,6}$ [28] & $M_{9,9,9,6,1,1,1}$ [28] \\
\rowcolor{Gray}
 & [0,0,0,0,0,0,0,0,5,0] & [3,6,9,12,15,18,21,24,15,11,12] & $M_{10,10,10,1,1,1,1,1}$ [5] & $M_{10,10,10}$ [6] \\
 & [0,0,0,2,0,0,0,0,3,0] & [2,4,6,8,12,16,20,24,16,11,12] & $M_{10,10,6,6,1,1,1}$ [17] & $M_{10,10,6,6}$ [18] \\
\rowcolor{Gray}
 & [1,1,0,0,1,0,0,0,3,0] & [1,3,6,9,12,16,20,24,16,11,12] & $M_{10,9,8,5,1,1,1}$ [48] & $M_{10,9,8,5}$ [48] \\
 & [0,0,0,1,0,2,0,0,1,0] & [2,4,6,8,11,14,19,24,17,11,12] & $M_{10,10,6,4,4,1}$ [37] & $M_{10,10,6,4,4}$ [38] \\
\rowcolor{Gray}
 & [0,2,0,1,0,0,0,1,1,0] & [1,2,5,8,12,16,20,24,17,11,12] & $M_{10,8,8,6,2,1}$ [67] & $M_{10,8,8,6,2}$ [68] \\
 & [1,0,1,0,1,0,1,0,1,0] & [1,3,5,8,11,15,19,24,17,11,12] & $M_{10,9,7,5,3,1}$ [108] & $M_{10,9,7,5,3}$ [108] \\
\rowcolor{Gray}
 & [2,1,1,0,0,0,0,0,2,1] & [0,2,5,9,13,17,21,25,17,11,12] & $M_{9,9,8,7,1,1}$ [56] & $M_{9,9,8,7,1}$ [56] \\
 & [3,0,0,0,1,0,0,1,1,0] & [0,3,6,9,12,16,20,24,17,11,12] & $M_{9,9,9,5,2,1}$ [60] & $M_{9,9,9,5,2}$ [60] \\
\rowcolor{Gray}
 & [0,1,2,0,0,0,1,0,0,1] & [1,2,4,8,12,16,20,25,18,11,12] & $M_{10,8,7,7,3}$ [60] & $M_{10,8,7,7,3,1}$ [60] \\
 & [0,2,0,0,1,1,0,0,0,1] & [1,2,5,8,11,15,20,25,18,11,12] & $M_{10,8,8,5,4}$ [60] & $M_{10,8,8,5,4,1}$ [60] \\
\rowcolor{Gray}
 & [1,0,0,2,0,1,0,0,0,1] & [1,3,5,7,11,15,20,25,18,11,12] & $M_{10,9,6,6,4}$ [60] & $M_{10,9,6,6,4,1}$ [60] \\
 & [2,1,0,1,0,0,1,0,0,1] & [0,2,5,8,12,16,20,25,18,11,12] & $M_{9,9,8,6,3}$ [96] & $M_{9,9,8,6,3,1}$ [96] \\
\rowcolor{Gray}
 & [3,0,0,0,0,2,0,0,0,1] & [0,3,6,9,12,15,20,25,18,11,12] & $M_{9,9,9,4,4}$ [30] & $M_{9,9,9,4,4,1}$ [30] \\
 & [0,1,1,1,1,0,0,0,1,0] & [1,2,4,7,11,16,21,26,18,11,13] & $M_{10,8,7,6,5,1}$ [72] & $M_{10,8,7,6,5}$ [72] \\
\rowcolor{Gray}
 & [1,3,0,0,0,0,1,0,1,0] & [0,1,5,9,13,17,21,26,18,11,13] & $M_{9,8,8,8,3,1}$ [48] & $M_{9,8,8,8,3}$ [48] \\
 & [2,0,2,0,0,1,0,0,1,0] & [0,2,4,8,12,16,21,26,18,11,13] & $M_{9,9,7,7,4,1}$ [60] & $M_{9,9,7,7,4}$ [60] \\
\rowcolor{Gray}
 & [2,1,0,0,2,0,0,0,1,0] & [0,2,5,8,11,16,21,26,18,11,13] & $M_{9,9,8,5,5,1}$ [48] & $M_{9,9,8,5,5}$ [48] \\
 & [1,2,1,0,1,0,0,0,0,1] & [0,1,4,8,12,17,22,27,19,11,13] & $M_{9,8,8,7,5}$ [72] & $M_{9,8,8,7,5,1}$ [72] \\
\rowcolor{Gray}
 & [2,0,1,2,0,0,0,0,0,1] & [0,2,4,7,12,17,22,27,19,11,13] & $M_{9,9,7,6,6}$ [42] & $M_{9,9,7,6,6,1}$ [42] \\
 & [0,0,4,0,0,0,0,0,1,0] & [1,2,3,8,13,18,23,28,19,11,14] & $M_{10,7,7,7,7,1}$ [16] & $M_{10,7,7,7,7}$ [16] \\
\rowcolor{Gray}
 & [0,5,0,0,0,0,0,0,1,0] & [0,0,5,10,15,20,25,30,20,11,15] & $M_{8,8,8,8,8,1}$ [13] & $M_{8,8,8,8,8}$ [14] \\
\hline
\end{longtable}
}

\end{document}